\documentclass[11pt]{article}
\usepackage{amsmath}
\usepackage{amssymb}
\usepackage{amsfonts}
\usepackage{amscd}
\usepackage{accents}
\usepackage{pst-all}
\usepackage{epsfig}
\usepackage{graphicx,amssymb,amsmath,wasysym}
\date{\today}
\newcommand{\bmat}{\left(\begin{array}}
\newcommand{\emat}{\end{array}\right)}
\newcommand{\be}{\begin{equation}}
\newcommand{\ee}{\end{equation}}
\newcommand{\ba}{\begin{eqnarray}}
\newcommand{\ea}{\end{eqnarray}}

\def\lsim{\raise0.3ex\hbox{$\;<$\kern-0.75em\raise-1.1ex\hbox{$\sim\;$}}}
\def\gsim{\raise0.3ex\hbox{$\;>$\kern-0.75em\raise-1.1ex\hbox{$\sim\;$}}}

\def\be{\beta}

\def\nn{\nonumber}

\usepackage{graphicx}
\usepackage[tight]{subfigure}

\def\beq{\begin{equation}}
\def\eeq{\end{equation}}
\def\bal#1\eal{\begin{align}#1\end{align}}

\def\round#1{\left({#1}\right)}

\newcommand{\Mcal}{{\cal M}}

\newcommand{\Lcal}{{\cal L}}

\newcommand{\lif}{\lambda_\phi}

\newcommand{\lih}{\lambda_h}

\newcommand{\bea}{\begin{eqnarray}}
\newcommand{\eea}{\end{eqnarray}}
\newcommand{\besub}{\begin{subequations}}
\newcommand{\eesub}{\end{subequations}}
 \newcommand{\bi}{\begin{itemize}}
\newcommand{\ei}{\end{itemize}}

\usepackage{color}

\begin{document}

\begin{titlepage}

\begin{center}
{\huge \bf The Higgs Portal to Cosmology}
\end{center}
 
 \vspace{1cm}

 \begin{center}
{ \bf Oleg Lebedev}
\end{center}
 
  \begin{center}
 \vspace*{0.15cm}
\it{Department of Physics and Helsinki Institute of Physics, \\
Gustaf H\"allstr\"omin katu 2a, FI-00014 Helsinki, Finland
}
 \end{center}
 
 \vspace{3cm}

\begin{center} {\bf Abstract} \end{center}
 \noindent
The discovery of the Higgs boson has opened a new avenue for exploring  physics beyond the Standard Model.
 In this review, we discuss 
  cosmological aspects of the simplest Higgs couplings to the hidden sector, known as the Higgs portal.
 We focus on implications of such couplings 
 for inflation, vacuum stability and dark matter, with the latter including
 both the traditional weakly interacting massive particles as well as feebly interacting scalars. The cosmological impact 
 of the Higgs portal can be important even for tiny values of the couplings.

\end{titlepage}

\tableofcontents

 \vspace{1cm}
 
 \section{Introduction}
   
   The Standard Model (SM) of particle physics provides us with a successful and precise framework to describe 
   microscopic processes at particle accelerators. It was made complete with the discovery of 
   the Higgs boson in 2012   \cite{Chatrchyan:2012ufa},\cite{Aad:2012tfa}  predicted by Brout and Englert\;\cite{Englert:1964et}, Higgs\;\cite{Higgs:1964pj}, and Guralnik, Hagen and   Kibble\;\cite{Guralnik:1964eu}.
  Even though the Standard Model is exceptionally  successful, it leaves a number of important questions unanswered. Among them, 
  some are of cosmological nature. These include the origin of dark matter (DM), inflation as well as matter--antimatter asymmetry. Addressing these fundamental
  questions requires physics beyond the Standard Model in some form.
  
  The discovery of the Higgs boson  has opened an important avenue for exploring physics beyond the Standard Model. 
  Its current mass stands at  \cite{Zyla:2020zbs}
   \begin{equation}
  m_{h_0} = 125.10 \pm 0.14  \;{\rm GeV}
  \end{equation}
 and its properties appear to match the corresponding 
   SM predictions within the experimental uncertainties. Nevertheless,   this does not preclude the Higgs boson from having additional interactions
   with  fields beyond the SM. These can belong to the ``hidden sector''  in the sense that they have no SM quantum numbers and  thus   escape detection.   
    Such fields can, however, have an important cosmological role: they can drive inflation or constitute dark matter.
  
  In fact, on general grounds, one expects couplings between the Higgs field  $H$ and the hidden sector scalars in quantum field theory (QFT).
One of the  special features of the Higgs field    in the Standard Model is that 
 \begin{equation}
 {\cal O}_2 = H^\dagger H
 \end{equation}
 is the only Lorentz and gauge invariant operator of dimension 2. Therefore, given  a scalar $\phi$, the coupling $H^\dagger  H \phi^2$ or
 $H^\dagger  H \phi^\dagger \phi$, if the scalar transforms under some symmetry, is renormalizable and consistent with all the symmetries.
  As such, it must be included in the Lagrangian, although not much can be said of the coupling strength. In much of the present review,
  we will consider the simplest possibility that the scalar $\phi$ is real and has no quantum numbers. In this case, the allowed renormalizable   terms are 
  \begin{equation}
 V_{\phi h} = {1\over 2} \lambda_{\phi h} H^\dagger H \,\phi^2 + \sigma_{\phi h} H^\dagger H \,\phi \;.
 \label{higgs-portal-V}
 \end{equation}
$\phi$ can play the role of an inflaton, in which case 
 this interaction has important implications for inflation, reheating and vacuum stability. 
  If the system is endowed with a $Z_2$ symmetry, $\phi \rightarrow -\phi$,
 the trilinear term is forbidden and 
 the scalar becomes a viable dark matter candidate. Depending on the coupling, it can be a traditional  weakly interacting massive particle (WIMP)
 or constitute feebly interacting dark matter. We will also consider the possibility that $\phi$ is a multiplet transforming under some hidden sector symmetry
 and plays the role of a messenger between the observable and dark sectors.
 
 The Higgs coupling to the hidden sector scalar was first considered by Silveira and Zee in the context of dark matter  in Ref.\,\cite{Silveira:1985rk}, and as an auxiliary tool in Ref.\,\cite{Veltman:1989vw}.
Such a coupling can also induce a Higgs--singlet mixing leading to specific signatures at 
  the Large Hadron Collider   (LHC) \cite{Schabinger:2005ei}.   Renewed interest in this framework was triggered by Patt and Wilczek's paper \cite{Patt:2006fw}, where the phrase ``Higgs portal''
  was coined.
  The common definition of the Higgs portal, also adopted in this review, is the coupling of the Higgs bilinear  ${\cal O}_2$ to fields {\it neutral} under the SM symmetries,
  a prime example of which is given by Eq.\,\ref{higgs-portal-V}.

 In what follows, we review, at times complicated, physics of  simple Higgs portal  couplings, focussing on their cosmological implications.

 \section{Generalities}
 \label{generalities}
 
 In this review, we focus on  renormalizable interactions between the Higgs field and the hidden sector. On general grounds, these are expected to be present in QFT. In addition,
 we take into account  non--minimal scalar couplings to gravity \cite{Chernikov:1968zm}. Although these are effectively higher dimensional operators, the couplings are dimensionless   and their presence 
  can be motivated by scale invariance of the theory at large field values \cite{Bezrukov:2007ep}.
 Furthermore, their impact is  important in the Early Universe, when the space--time curvature is significant \cite{Spokoiny:1984bd},\cite{Fakir:1990eg},\cite{Salopek:1988qh}.
 
Consider the system of the Higgs field and a real scalar $\phi$. The scalar may, for example, be an inflaton.
 The action that includes the most general renormalizable potential and lowest order non--minimal couplings to gravity has the form \cite{Ema:2017ckf}
 \begin{equation}
{\cal L}_{J} = \sqrt{-\hat g} \left(   -{1\over 2}  \Omega  \hat R \,  
 +  {1\over 2 } \, \partial_\mu \phi \partial^\mu \phi +  \,  (D_\mu H)^\dagger D^\mu  H  - {V(\phi,H)  }\right) \;.
\label{L-J}
\end{equation}
 For many purposes, it is convenient to use the unitary gauge
 \begin{equation}
H(x)= {1\over \sqrt{2}} \left(
\begin{matrix}
0\\
h(x)
\end{matrix}
\right)\;,
\end{equation}
 in which the potential takes the form
 \begin{equation}
V(\phi, h) = {1\over 4} \lambda_h h^4 + {1\over 4} \lambda_{\phi h }h^2 \phi^2 + {1\over 4} \lambda_\phi \phi^4 + {1\over 2} \sigma_{\phi h} \phi h^2 + {1\over 3} b_3 \phi^3 +
{1\over 2} m_h^2 h^2 + {1\over 2} m_\phi^2 \phi^2  + b_1 \phi\;.
\label{potential}
\end{equation}
We assume that all the dimensionful parameters are far below the Planck scale.
The function $\Omega $ has the general form $\Lambda^2 + \Lambda^\prime \phi + \xi_\phi \phi^2 + \xi_h h^2 $. The linear term can be eliminated by field redefinition.
Then, the constant term can be identified with the Planck mass squared, as long as the vacuum average of $\xi_\phi \phi^2 + \xi_h h^2$ can be neglected.
 Therefore, we may take
\begin{equation}
 \Omega = M_{\rm Pl}^2+  \xi_h h^2 + \xi_\phi \phi^2 \;.
 \label{Omega}
 \end{equation}
In the presence 
of odd in $\phi$ terms in the potential, the $\phi \hat R$ coupling  is still generated at loop level. However, if the above form is enforced at the high scale, say the Planck scale,
the resulting coupling at the inflationary scale is suppressed by a loop factor as well as by the dimensionful couplings $\sigma_{\phi h}$, $b_i$ which we assume to be far below the Planck scale.
  On the other hand,  at late stages of the Universe evolution,  the impact of the non--minimal couplings to gravity is negligible so they can be dropped altogether. Thus, it is sufficient for our purposes to adhere to $\Omega$ of the form
  (\ref{Omega}).\footnote{Some effects of the linear in $\phi$ coupling were explored in \cite{Lee:2018esk}.}

 It is important to remember that the unitary gauge is singular at $\langle h \rangle =0$, and all 4 Higgs degrees of freedom have to be considered in this case. Equivalently, at energies far above 
 $\langle h \rangle $, effects of the longitudinal components of $W$ and $Z$ must be taken into account.
 In loop computations,
  it is   necessary to include 
  the Goldstone contributions, which are made obscure by the unitary gauge.
 
 One rather general consequence of the potential (\ref{potential}) is that a Higgs--inflaton mixing is expected  \cite{Ema:2017ckf}. This is due to the presence of the trilinear $\phi h^2$ term coupled with the fact that the Higgs develops a non--zero VEV. The  $\phi h^2$ interaction itself is also quite generic: if the inflaton decays into SM fields, as assumed in most models of reheating, this coupling is generated 
 radiatively \cite{Gross:2015bea}.
 The effects of the  Higgs--inflaton mixing may be too small to be observed depending on the size of the model--dependent mixing angle. 
 Having said that, even a small  mixing  can have a major impact on vacuum stability 
 by increasing the Higgs self--coupling.
 
 In what follows, we will discuss the most important properties of the system.
 
 \subsection{Renormalization group evolution}
 \label{RG}
 
 The couplings that appear in the above Lagrangian enter observables which are associated with vastly different energy scales. It is therefore necessary to take the scale dependence of the couplings into account.   For our purposes, it suffices to use the leading log corrections encoded in the 1--loop renormalization group  (RG) equations.
 
 The 1--loop renormalization group running of the relevant couplings in the potential 
 and  the  SM couplings
  is given by  \cite{Ema:2017ckf}\footnote{Note a different convention for $\lambda_{\phi h}$ and $\sigma_{\phi h}$ in  \cite{Ema:2017ckf}.}
 \bal
\label{eq:rges}
\nn
16\pi^2 {\dfrac{d\lambda_h}{dt}}
&=
24 \lambda_h^2 -6 y_t^4 + \dfrac38
\left( 2 g^4 + (g^2 + g^{\prime 2})^2 \right) 
+ (12 y_t^2 -9 g^2 -3 g^{\prime 2}) \lambda_h + {1\over 2} \lambda_{\phi h}^2 \;,
\\ \nn
16\pi^2 \dfrac{d\lambda_{\phi h}}{dt} 
&=
4 \lambda_{\phi h}^2 + 12 \lambda_h \lambda_{\phi h}
-\dfrac32 (3 g^2 + g^{\prime 2}) \lambda_{\phi h} 
+ 6 y_t^2 \lambda_{\phi h} + 6 \lif\lambda_{\phi h} \;,
\\ \nn
 16\pi^2 \dfrac{d \lif}{dt} 
 &= 
 2 \lambda_{\phi h}^2 + 18 \lif^2 \;,
\\ 
 16\pi^2 \dfrac{d \sigma_{\phi h}}{dt} 
 &= 
 \sigma_{\phi h}\round{12\lih + 4\lambda_{\phi h} - \frac{3g^{\prime 2}}{2} - \frac{9g^2}{2} 
 + 6 y_t^2} + 2\lambda_{\phi h} b_3
     \;,
\\ \nn
 16\pi^2 \dfrac{db_3}{dt} 
 &= 
 6\sigma_{\phi h}\lambda_{\phi h} + 18\lif b_3 \;,
\\ \nn
16\pi^2\dfrac{d y_t}{dt}&= y_t \left(\dfrac92 y_t^2 - \dfrac{17}{12} g'^2 - \dfrac94 g^2 - 8 g_3^2 \right) \;,
\\ \nn
16\pi^2\dfrac{d g_i}{dt}&= c_i \, g_i^3 \quad \textrm{with} \quad (c_1,c_2,c_3)=(41/6,-19/6,-7) \;,
\eal
where  $t= \ln \mu$ with $\mu$ being the RG energy scale,    $g_i=(g',g,g_3)$ denote the gauge couplings and $y_t$ is the top quark Yukawa coupling. 
The standard input values of the couplings at the top quark mass scale $M_t$ are 
  $g(M_t)=0.64,\; g'(M_t)=0.35,\; g_3(M_t)=1.16$ and $y_t(M_t)=0.93$, although $y_t$ is subject to tangible uncertainties.
  As emphasised  above, the RG running includes the Goldstone contributions as well.

 The non--minimal couplings to gravity also receive significant leading log corrections. Such couplings are generated even if their tree level values are zero, while in the conformal limit 
 $\xi_i=-1/6$  they remain scale--invariant.
The 1--loop RG running  is given by \cite{Lerner:2009xg}
\begin{eqnarray}
&& 16\pi^2 {d \xi_h \over dt} = \left(  \xi_h + {1\over 6 } \right) \Bigl( 12 \lambda_h + 6 y_t^2 -{3\over 2} (3 g^2 + g^{\prime 2}) 
\Bigr) + \left(  \xi_\phi + {1\over 6 } \right) \lambda_{\phi h} \;,\nonumber \\
&& 16\pi^2 {d \xi_\phi \over dt} = \left(  \xi_\phi + {1\over 6 } \right) 6 \lambda_\phi + 
 \left(  \xi_h + {1\over 6 } \right) 4 \lambda_{\phi h } \;.
\end{eqnarray}
 These corrections do not include gravitons in the loop, so their computation parallels that for the scalar couplings. For  example, one notices the same combination of the SM couplings
 that contributes to the beta functions of $\lambda_{\phi h}$ and $\xi_h$.
 
 The above results are valid at low scalar background values,  $\xi_\phi \phi^2 + \xi_h h^2 \ll M_{\rm Pl}^2$. If this is not the case, the scalar propagators get modified by the curvature term and the RG
 equations receive corrections. However, connecting the small and large field  regimes can be  challenging due to the unitarity issues to be discussed in Section\;\ref{uni}, hence we focus in this section on the former.

 The effect of the running can be crucial. A widely appreciated example is provided by the Higgs self--interaction: the RG running can turn the coupling negative indicating vacuum metastability.
 Analogous corrections to other couplings can also play an important role. For example, the Higgs portal coupling $\lambda_{\phi h}$ generates the inflaton self--coupling $\lambda_\phi$ at one loop,
 so the latter cannot be too small, at least in a substantial interval of energy scales. 
 
 On the other hand, for  many applications in this review, what is relevant is the coupling value at a particular scale. For instance, the non--minimal couplings to gravity are 
 important during inflation or preheating, but not at low energies. Unless specified otherwise, in what follows, the couplings are to be understood as the running   couplings
 evaluated at the scale of the considered process.

 \subsection{Radiative generation of the Higgs--inflaton couplings }
 
 The Higgs--inflaton interactions are expected on general grounds in QFT. Specifically, Lorentz and gauge invariant interactions of dimension 4 or lower must be included in a renormalizable theory, and
 the Higgs--portal couplings belong to this class (unless additional  symmetries are imposed on the inflaton sector). In cosmology, there is a further  argument in favor of these couplings  \cite{Gross:2015bea}.
 The inflaton energy must be transferred to the SM sector at the end of inflation due to the process known as reheating.   In most models of reheating, the transfer  is accomplished at least partly via inflaton decay into SM matter.
 This implies a linear inflaton coupling to some SM fields or other fields that couple to the SM. Loop effects  then induce a coupling between the Higgs and the inflaton. Typically, such loop corrections are divergent and require the corresponding counterterms. This means that the Higgs--inflaton couplings are {\it a priori} arbitrary  parameters and cannot be ignored.
   
   \begin{figure}[t] 
\centering{
\includegraphics[scale=0.81]{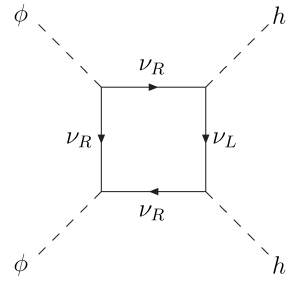}~~~~
\includegraphics[scale=0.82]{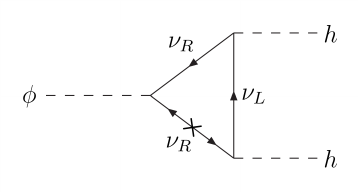}~~~~~~~~
\includegraphics[scale=0.81]{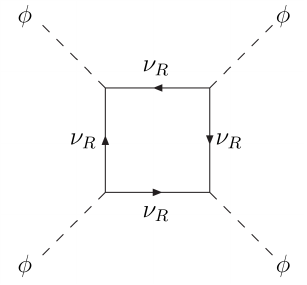}
}
\caption{ \label{RH}
Higgs--inflaton couplings  and inflaton self--interaction generated via
  the right--handed neutrinos \cite{Gross:2015bea}. (Other diagrams with the same topology are not shown). 
}
\end{figure}

 To illustrate this point, let us set the tree level Higgs--inflaton couplings to zero and 
  consider a set--up in which reheating occurs due to the inflaton coupling to heavy right--handed Majorana neutrinos $\nu_R$. The neutrinos produced in inflaton decay
 subsequently decay into leptons and Higgses, leading to thermalization of the SM fields.
  The relevant neutrino interactions read
\begin{equation}
-\Delta {\cal L}_\nu = {\lambda_\nu \over 2} ~\phi\; \nu_R \nu_R + {y_\nu } ~
\bar l_L \! \cdot \! H^* \, \nu_R + {M\over 2} ~ \nu_R \nu_R + {\rm h.c.}~, 
\end{equation}
where $l_L$ is the lepton doublet and  $M$ is taken  to be real for simplicity. At 1 loop,
this Lagrangian generates  various scalar couplings, including Higgs--inflaton interactions (Fig.\,\ref{RH}).
Such corrections are divergent and require a renormalization condition. Setting the couplings to zero at the Planck scale,
  at the inflationary scale $H$ we have 
 \cite{Gross:2015bea}:
\begin{eqnarray}
\lambda_{\phi h}&\simeq& { \vert \lambda_\nu y_\nu \vert^2 \over 2 \pi^2} \ln {M_{\rm Pl}\over H} \;, 
\nonumber\\
\sigma_{\phi h}&\simeq& -{ M \vert y_\nu \vert^2 {\rm Re} \lambda_\nu \over 2 \pi^2} \ln {M_{\rm Pl}\over 
H} \;,
\end{eqnarray}
where we keep only the leading log terms.
An analogous  correction to the inflaton self--coupling is also generated,
\begin{equation}
\Delta \lambda_{\phi} \simeq { \vert \lambda_\nu \vert^4 \over 4 \pi^2} \ln {M_{\rm Pl}\over H} \;.
\end{equation}
The latter is important since it sets an upper bound on $\lambda_\nu$, depending on the model of inflation.
 The radiative contribution to the inflaton potential should not spoil its  flatness.
  For example, for chaotic $\phi^2$--inflation, one requires $m_\phi^2 \phi^2 /2 \gg \Delta \lambda_\phi \phi^4 /4$
during the last  60 e--folds. This results in $\lambda_\nu < 10^{-3}$.
Assuming $M< 10^{13}$ GeV and $y_\nu <0.6$ as required by the bound on the neutrino masses in seesaw models,
one finds $\lambda_{\phi h } \lesssim 10^{-7}$ and $\sigma_{\phi h } \lesssim 10^{-4} M$.  
These results are model--dependent and 
in other models such  as Higgs--like inflation the couplings are allowed to be  much larger. The radiative corrections 
control the $natural$ size of the couplings so that their smaller values would require cancellations and thus finetuning.

 Similar conclusions apply to other popular reheating mechanisms.  For instance, 
 inflaton decay may proceed through non--renormalizable interactions
 \begin{equation}
O_1 = {1\over \Lambda_1} \phi\; \bar q_L \! \cdot \!  H^* \, t_R ~~,~~
O_2 = {1\over \Lambda_2} \phi\; G_{\mu\nu} G^{\mu\nu} ~, 
\end{equation} 
where $\Lambda_{1,2}$ are some scales, $G_{\mu\nu}$ is the gluon field strength and $q_L,t_R$ are the third generation quarks.  
 Closing the quarks or gluons in the loops, one finds Higgs--inflaton couplings of the size similar to that obtained above and 
 analogous conclusions apply \cite{Gross:2015bea}.

  Even small Higgs--inflaton couplings can have a significant impact on the Early Universe physics. In particular, these terms affect  stability of the Higgs potential
  when the inflaton field takes on large values. Furthermore, non--perturbative Higgs production during preheating is efficient even for tiny couplings, e.g.
  $\lambda_{\phi h} < 10^{-8}$. It is therefore necessary to account for  the Higgs--inflaton interactions in a realistic setting.

 \subsection{$Z_2$ symmetric scalar potential} 
 \label{sec-z2}
 
 The vacuum structure   in the most general Higgs--inflaton system is rather complicated \cite{Espinosa:2011ax}. A useful simpler limit to consider is the $Z_2$ symmetric scalar potential,
 that is, the potential invariant under    
  $\phi \rightarrow -\phi$. Although this eliminates some of the couplings which can be important for reheating, it provides a helpful perspective on the 
 Higgs--inflaton mixing  and applies more generally to a singlet extended Standard Model. A special case of this  potential with no dimensionful parameters is relevant to the conformal extension of the 
 Standard Model \cite{Meissner:2006zh},\cite{Hur:2011sv},\cite{Steele:2013fka}.
  
 Consider the $Z_2$ symmetric potential
 \begin{equation}
V(\phi, h) = {1\over 4} \lambda_h h^4 + {1\over 4} \lambda_{\phi h }h^2 \phi^2 + {1\over 4} \lambda_\phi \phi^4 + {1\over 2} m_h^2 h^2 + {1\over 2} m_\phi^2 \phi^2 \;.
\label{Z2-pot}
\end{equation}
 Denoting the vacuum expectation values
 \begin{equation}
\langle h \rangle \equiv v ~~,~~ \langle \phi \rangle \equiv w,
\end{equation}
one finds the following classification of the minima \cite{Lebedev:2011aq}:
\begin{eqnarray}
&& v\not=0, w\not=0~:~ \lambda_{\phi h} m_\phi^2 - 2 \lambda_\phi m_h^2 >0~,~ \lambda_{\phi h} m_h^2 - 2 \lambda_h m_\phi^2>0 ~, \nonumber \\
&& v\not=0, w=0~:~  \lambda_{\phi h} m_\phi^2 - 2 \lambda_\phi m_h^2<0 ~,~ m_h^2 <0 ~, \nonumber \\
&& v=0, w\not=0~:~ \lambda_{\phi h} m_\phi^2 - 2 \lambda_\phi m_h^2 <0~,~m_\phi^2 <0 ~, \nonumber \\
&& v=0, w=0~:~ m_h^2 >0~,~ m_\phi^2 >0 ~.
\end{eqnarray}
These conditions are mutually exclusive such that there is only one local minimum at tree level (barring the reflected minimum $\phi \rightarrow -\phi$)\footnote{The couplings are assumed to satisfy the
``positivity'' constraints (\ref{loc-min}).}. If $w=0$, there is no Higgs--inflaton mixing and 
the system is straightforward to analyze. Let us consider the case $v,w \not= 0$ in detail. The stationary point condition requires
\begin{eqnarray}
v^2=\frac{2  \lambda_{\phi h}  m_\phi^2 - 4 \lambda_\phi m_h^2}{4 \lambda_h \lambda_\phi - \lambda_{\phi h}^2}  \,,
\qquad
w^2=\frac{2  \lambda_{\phi h}  m_h^2 - 4 \lambda_h m_\phi^2}{4 \lambda_h \lambda_\phi - \lambda_{\phi h}^2 }   \,.
\end{eqnarray}
The corresponding mass matrix is
\begin{equation}
\Mcal^2= \left( \begin{array}{cc}
2 \lambda_h v^2 & \lambda_{\phi h} v w \\
\lambda_{\phi h} v w & 2 \lambda_\phi w^2
\end{array} \right) \,.
\end{equation}
 Its eigenvalues are positive if
 \begin{equation}
\lambda_{h}, \lambda_\phi >0 ~~,~~ 4 \lambda_h \lambda_\phi - \lambda_{\phi h}^2 >0\;.
\label{loc-min}
\end{equation}
 The mass matrix  can be diagonalised by the orthogonal transformation
$O^T \Mcal^2 O = \textrm{diag}(m_{1}^2,m_{2}^2) ,$
where
\begin{equation}
O= \left( \begin{array}{cc}
\cos \theta & \sin \theta \\
- \sin \theta & \cos \theta
\end{array} \right) \;
\end{equation}
and the angle $\theta$ is given by 
 \begin{equation}
\tan 2 \theta = \frac{ \lambda_{\phi h} v w}{\lambda_\phi w^2-\lambda_h v^2} \,. \label{tantwotheta}
 \end{equation}
The mass squared eigenvalues are 
\begin{equation} 
m_{{1,2}}^2 = \lambda_h v^2+\lambda_\phi w^2 \mp \frac{ \lambda_\phi w^2 - \lambda_h v^2}{\cos 2 \theta} \,.
\label{evalues}
  \end{equation}
 In our convention, $\textrm{sign}(m_{2}^2-m_{1}^2)=\textrm{sign}(\cos 2 \theta) \,\textrm{sign}(\lambda_\phi w^2 - \lambda_h v^2)$. The mass eigenstates $h_1,h_2$
are given by 
 \begin{equation}
\left( \begin{array}{c}
h_1\\
h_2
\end{array} \right)
=
 \left( \begin{array}{c}
  \left (h - v\right ) \cos \theta -  \left (\phi - w \right )  \sin \theta \\
\left (\phi - w \right ) \cos \theta  +\left (h - v \right )  \sin \theta 
\end{array} \right)\,.
\end{equation}
 
 Finally, it is important to note that the leading log corrections can be included by using the running couplings evaluated at the relevant scale. These are particularly significant
 for the conditions imposed at large field values, when the logarithms are large. Such corrections are relevant to the asymptotic behaviour of the potential, especially  in regard to 
 vacuum stability. Clearly, the quartic couplings must be positive (semidefinite) for the potential to be bounded from below, while the constraint on the Higgs portal coupling depends on the sign of $\lambda_{\phi h }$.   Neglecting the quadratic terms, one finds
 \begin{equation}
 V\simeq {1\over 4 } \Bigl[  \left(\sqrt{\lambda_h} h^2 - \sqrt{\lambda_\phi} \phi^2\right)^2 +  \left(\lambda_{\phi h} +2 \sqrt{\lambda_h \lambda_\phi} \right) h^2 \phi^2 \Bigr] \;.
\end{equation}
 Thus,  for $\lambda_{\phi h} >0$ there is no extra constraint, whereas at negative $\lambda_{\phi h} $ there exists a run--away direction unless $|\lambda_{\phi h} |<2 \sqrt{\lambda_h \lambda_\phi} $.
 As a result, one finds the following constraints on the running couplings
\begin{equation}
\lambda_{h}(\mu), \lambda_\phi (\mu)>0 ~~,~~ 4 \lambda_h(\mu) \lambda_\phi(\mu) + {\rm sign} \left[ \lambda_{\phi h }(\mu)\right]\; \lambda_{\phi h}(\mu)^2 >0\;,
\end{equation}
where $\mu$ is some large energy scale such as the Planck scale. To avoid deep minima at other scales, one imposes this condition for all $\mu \gg v,w$.
Note, however, that the above  constraint may be too restrictive: our vacuum can be metastable and one can still
have a well defined theory  at small field values. In particular, $\lambda_h({\rm M_{Pl}}) <0$ does not pose any fundamental problems although certain cosmological issues do arise.
 
 \subsubsection{Some phenomenological implications}
 \label{Z2-pheno}
 
For phenomenological applications, it is often convenient to analyze the model in terms of the set $(m_{1}^2, m_{2}^2,\sin \theta,v,\lambda_{\phi h})$ instead of the input
parameters in the scalar potential \cite{Falkowski:2015iwa}. 
The self--couplings  can then be found via
\begin{eqnarray}
\lambda_h &=&\frac{m_{1}^2}{2 v^2} + \sin^2 \theta \ \frac{m_{2}^2-m_{1}^2}{2 v^2} \label{lhLE} \;, \nonumber
\\
\lambda_\phi &= &\frac{ 2 \, \lambda_{\phi h}^2}{\sin^2 2 \theta } \frac{v^2}{m_{2}^2-m_{1}^2} \left(\frac{m_{2}^2}{m_{2}^2-m_{1}^2} -\sin^2 \theta \right) \;. \label{lsLE}
\end{eqnarray}
 These are useful for analyzing perturbativity and stability of the model.  The most important couplings of the mass eigenstates to the SM matter are 
 \begin{equation} 
\Lcal \supset { h_1 \cos \theta + h_2 \sin \theta \over v} \left [ 2 m_W^2 W_\mu^+ W^{\mu -} + m_Z^2 Z_\mu Z^\mu - \sum_f m_f \bar f f \right] \,.
 \end{equation}
 Compared to the SM Higgs boson partial decay  widths, these  for the $h_1$ and $h_2$ scalars  are universally suppressed by $\cos^2 \theta$ and $\sin^2 \theta$, respectively.
 The trilinear couplings between $h_1$ and $h_2$ are also important since they can lead to the decays 
 $h_2 \to h_1 h_1$ and $h_1 \to h_2 h_2$, if kinematically allowed. We define them as
  \begin{equation}
\Lcal \supset -{\kappa_{112} \over 2} v \sin \theta ~h_1^2 h_2 -{\kappa_{221} \over 2} v \cos \theta ~h_2^2 h_1 ~,
\end{equation}
with
\besub
\bal
\kappa_{112} &= 
 {2 m_{1}^2 + m_{2}^2 \over v^2 } \left ( \cos^2 \theta + { \lambda_{\phi h} v^2 \over m_{2}^2 - m_{1}^2} \right )~,
\\ 
\kappa_{221} &= 
 {2 m_{2}^2 + m_{1}^2 \over v^2 } \left ( \sin^2 \theta + { \lambda_{\phi h} v^2 \over m_{1}^2 - m_{2}^2} \right ) \,.
\eal
\eesub
In the kinematically allowed regime, the decay widths are given by
\besub
\bal \label{gamma211}
\Gamma(h_2 \to h_1 h_1) &= {\sin^2 \theta ~\kappa_{112}^2 v^2 \over 32 \pi \, m_{2}} \sqrt{1 - {4 m_{1}^2 \over m_{2}^2}}~,
\\ 
\Gamma(h_1 \to h_2 h_2) &= {\cos^2 \theta ~\kappa_{122}^2 v^2 \over 32 \pi \, m_{1}} \sqrt{1 - {4 m_{2}^2 \over m_{1} ^2}} \,.
\eal
\eesub

The model is subject to a range of constraints from various particle experiments. Both $h_1$ and $h_2$ behave as Higgs--like particles, and $h_1$ is identified with the scalar observed at the LHC, given that 
$\cos\theta$ is close to one. This, together with the correct electroweak symmetry breaking requires
\begin{equation}
m_1 = 125 \; {\rm GeV} ~~, ~~ v=246\; {\rm GeV}\;,
\end{equation}
which leaves only three free parameters in the model: $m_2, \sin\theta$ and $\lambda_{\phi h}$. A comprehensive analysis of the relevant constraints is presented in \cite{Falkowski:2015iwa} and
here we only outline the most important bounds.

Combined measurements of the Higgs couplings \cite{Englert:2014uua}, encapsulated in the ``signal strength'' $\mu > 0.92$ (ATLAS) \cite{ATLAS:2020qdt} and $\mu>0.90$ (CMS) \cite{CMS:2020gsy},
 impose an upper bound on the mixing angle:
\begin{equation}
| \sin\theta| \lesssim 0.3  
\end{equation}
at the 2$\sigma$ level.
This bound is independent of $m_2$, while if
  $h_2$ is light, it is superseded by other constraints. The LEP Higgs searches require $ | \sin\theta| \lesssim 0.1-0.2$ for $m_2 <90$ GeV, whereas B--physics imposes a strong bound $ | \sin\theta| \lesssim 10^{-2}-10^{-3}$ for $m_2 < 5 $ GeV.
If $h_2$ is very heavy, the constraints are also strong: the electroweak precision measurements  set an upper bound on $|\sin\theta|$ that scales   as 
 \begin{equation}
| \sin\theta| \lesssim 0.3 / \sqrt{1+\ln (m_2/ {\rm TeV}) } \;,
\label{theta-log}
\end{equation}
with $m_2 \gsim 1$ TeV.
 This behaviour may appear counterintuitive:  the constraint gets stronger for larger $m_2$.  It is nevertheless easily understandable since the  Higgs contribution to the gauge boson propagators gets reduced by $\cos^2\theta$, so $h_2$ should be light enough to make up  the deficit. The scaling then follows from the corresponding bound  on the  $S,T$ Peskin--Takeuchi variables 
 \cite{Peskin:1991sw}.\footnote{For light and moderate $m_2$, the $S,T$ variable approximation is inadequate and one should use the full corrections as in  \cite{Falkowski:2015iwa}.
 This reference performs a global electroweak fit to 15 observables including the $Z$--pole and $W$--data.}
 We conclude that, unless $h_2$ is too heavy or too light, a significant Higgs--inflaton mixing up to $| \sin\theta | \simeq 0.3$ is allowed by  experiment \cite{Falkowski:2015iwa}.

Some parameter space of the model can  further be probed at the LHC \cite{Robens:2016xkb}. 
 The general predictions  are (i) a universal reduction of the SM couplings of $h_1$ compared to those of the Higgs, (ii) the existence of the Higgs--like resonance $h_2$, (iii) possible resonant di--Higgs 
 production $h_2 \rightarrow h_1 h_1$ or  $h_1 \rightarrow h_2 h_2$, if kinematically allowed.
 In particular, for $500\,{\rm GeV}> m_2 > 2 m_1$, the di--Higgs production rate can exceed the SM prediction by an order of magnitude \cite{Chen:2014ask},\cite{Falkowski:2015iwa},\cite{Lewis:2017dme}.
 The Higgs couplings are expected to be measured within about 5\% at HL-LHC \cite{Dawson:2018dcd}, which would tighten the bound on $\sin \theta$ to about 0.2. Searches for the ``heavy Higgs'' $h_2$ are likely to cover the mass range up to about 1 TeV, unless the mixing angle is very small.
 
  A recent  analysis of the singlet scalar model without the $Z_2$ symmetry can be found in \cite{Dawson:2021jcl}. Many of the above conclusions apply to this more general set--up as well.

 \section{Higgs portal inflation}
 \label{HiggsPortalInflation}
 
 Inflation   is one of the cornerstones of modern cosmology \cite{Starobinsky:1980te},\cite{Guth:1980zm},\cite{Linde:1981mu}. It provides us with a compelling explanation why the Universe is so big and flat, why causally disconnected regions happen to have the same temperature and also seeds fluctuations for structure formation  \cite{Mukhanov:1981xt} and the Cosmic Microwave Background (CMB) (see 
 \cite{Linde:2005ht},\cite{Lyth:1998xn} for  reviews). 
 To realize inflation one normally needs a scalar field with a flat enough potential,
 \begin{equation}
{V^\prime \over V} \, M_{\rm Pl } \ll 1~~,~~ {V^{\prime\prime} \over V} \, M_{\rm Pl }^2 \ll 1 ~~,
 \end{equation}
 where the prime denotes a derivative with respect to the field. 
 The Universe dominated by the potential energy of this field expands exponentially with an almost constant Hubble rate $H=\dot a /a$, where $a$ is the scale factor, while the associated quantum fluctuations 
 eventually lead to the observed CMB spectrum. The current CMB data disfavor simple polynomial potentials and tend to prefer   concave ones, 
   in the relevant field range \cite{Akrami:2018odb}.

 The Higgs portal framework provides an excellent setting for inflationary model building. Even though the polynomial terms in the scalar potential cannot fit the data themselves, the presence of the  non--minimal couplings to gravity 
 changes the situation \cite{Fakir:1990eg}. 
 As in the case of well known Higgs inflation \cite{Bezrukov:2007ep}, 
   the potential becomes exponentially close to a flat one at large field values, which is favored by the inflationary data. 
  In what follows, we apply this idea to the   Higgs--singlet system where the role of the inflaton can be played by a combination of the Higgs and the singlet \cite{Lebedev:2011aq},
 in addition to the Higgs  or the singlet itself \cite{Lerner:2009xg},\cite{Clark:2009dc}.  
  For simplicity, we consider the 
    $Z_2$ symmetric version  of the scalar potential, although the main results apply  more generally (see Section\;\ref{general-mixing}). A special case of this system has been 
    studied in the context of the $\nu$MSM \cite{Shaposhnikov:2006xi},\cite{Anisimov:2008qs}.

 In the unitary gauge 
 \begin{equation}
H(x)= {1\over \sqrt{2}} \left(
\begin{matrix}
0\\
h(x)
\end{matrix}
\right)\;,
\end{equation}
the Lagrangian in the Jordan frame reads
\begin{equation}
{\cal L}_{J} = \sqrt{-\hat g} \left(   -{1\over 2}    \hat R \, ( M_{\rm Pl}^2   + \xi_h h^2 + \xi_\phi \phi^2) 
 +  {1\over 2 } \, \partial_\mu \phi \partial^\mu \phi +  {1\over 2}\,  \partial_\mu h \partial^\mu  h  - {V(\phi,h)  }\right) \;,
\end{equation}
where we assume for simplicity a $Z_2$--symmetric scalar potential
\begin{equation}
V(\phi, h) = {1\over 4} \lambda_h h^4 + {1\over 4} \lambda_{\phi h }h^2 \phi^2 + {1\over 4} \lambda_\phi \phi^4 + {1\over 2} m_h^2 h^2 + {1\over 2} m_\phi^2 \phi^2 \;.
\end{equation}
Here $\hat R$ is the scalar curvature based on the Jordan frame metric $\hat g_{\mu\nu}$. To avoid a singularity at large field values, we take $\xi_h , \xi_\phi >0$.
Although there are indications that $\lambda_h$ in the pure SM is negative at high energies, here we assume that both $\lambda_h$ and $\lambda_\phi$ are positive at the
inflationary scale. This may be due to the RG effects associated with the singlet, Higgs--singlet mixing (Section\;\ref{section-mixing})  or simply a somewhat  lower top quark mass.
The precise mechanism of vacuum stabilization is unimportant for our purposes.

 It is often more convenient to work in the Einstein frame, where the ``Planck mass'' is constant, that is, 
 the only coupling to the scalar curvature is $-1/2 M_{\rm Pl}^2 R$. 
 To simplify formulas, let us choose the Planck units in this section, 
 \begin{equation}
 M_{\rm Pl}=1 \;.
 \end{equation}
 Then, 
 the transition to the Einstein frame  is accomplished by rescaling the metric
 \begin{equation}
 g_{\mu\nu} = \Omega \, \hat g_{\mu\nu} ~~\;,~~ \Omega = 1+  \xi_h h^2 + \xi_\phi \phi^2 \;.
\end{equation} 
This transformation affects the kinetic terms making them non--canonical and also rescales the scalar potential.
For a general $\Omega$, the  kinetic function multiplying $1/2 \, g^{\mu \nu}  \partial_\mu \phi_i \, \partial _\nu \phi_j$  and the potential become  \cite{Salopek:1988qh}:
\begin{eqnarray}
 &&  K^{ij}= {3\over 2}\; {\partial \log \Omega \over \partial \phi_i} \, {\partial \log \Omega \over \partial \phi_j} + {\delta^{ij} \over \Omega}  \;, \nonumber\\
 && V_E = {V \over \Omega^2} \;,
 \end{eqnarray}
 where $i,j$ label scalar fields.
 Consider now large field values such that 
  \begin{equation}
   \xi_h h^2 + \xi_\phi \phi^2 \gg 1 \;.
\end{equation} 
 In this case,   $\Omega \simeq  \xi_h h^2 + \xi_\phi \phi^2$ and  the Lagrangian in the Einstein frame is given by  
   \begin{equation}
 {\cal L} = {3\over 4} \Bigl(\partial_\mu \ln ( \xi_h h^2 + \xi_\phi \phi^2 )\Bigr)^2 + {1\over 2} {1\over  \xi_h h^2 + \xi_\phi \phi^2} \Bigl(  (\partial_\mu h)^2 + (\partial_\mu \phi)^2 \Bigr)
 - {V \over ( \xi_h h^2 + \xi_\phi \phi^2)^2} \;.
 \end{equation} 
 Introduce new variables \cite{Lebedev:2011aq}
 \begin{eqnarray}
 && \chi= \sqrt{3\over 2} \ln ( \xi_h h^2 + \xi_\phi \phi^2) \;, \nonumber\\
 && \tau = {h \over \phi} \;.
 \end{eqnarray}
 In terms of $\chi$ and $\tau$, the kinetic terms read
 \begin{eqnarray}
   {\cal L}_{\rm kin} &=& {1\over 2} \left(    1+{1\over 6} { \tau^2 +1 \over \xi_h \tau^2 + \xi_\phi}   \right) \, (\partial_\mu \chi)^2 +
 {1\over \sqrt{6}}  {(\xi_\phi -\xi_h)\tau \over  ( \xi_h \tau^2 + \xi_\phi)^2 }\,  (\partial_\mu \chi)(\partial^\mu \tau)  \nonumber\\
 &+&    {1\over 2} {  \xi_h^2 \tau^2 + \xi_\phi^2 \over (   \xi_h \tau^2 + \xi_\phi )^3   } \, (\partial_\mu \tau)^2  \;.
 \label{full-kinetic}
 \end{eqnarray}
 Since the kinetic functions depend on $\tau$ only, the mixing term can be eliminated by the shift
 \begin{equation}
 \chi \rightarrow \chi + f(\tau) \;.
 \end{equation}
 The explicit form of $f(\tau) $ will not be necessary for our applications. Let us focus on a few special cases in which the mixing 
 vanishes and the kinetic functions simplify:
 $\xi_\phi +\xi_h \gg 1$, $\tau =0$ and  $\tau= \infty$. 
 This also applies to the case $\xi_h = \xi_\phi$, however, this relation is not radiatively stable.  
 
 \subsection{Large non--minimal couplings to gravity}
 \label{large-non-min}
 
 Consider the limit  $\xi \equiv \xi_\phi +\xi_h \gg 1$, i.e. at least one of the non--minimal couplings to gravity is large, and expand the kinetic terms in ``$1/\xi$''.
 The term $(\partial_\mu \tau)^2 $ scales as $1/ \xi$, so does the mixing term  $(\partial_\mu \chi)(\partial^\mu \tau)$.
 Therefore, in terms of canonically normalized variables, the mixing is suppressed and can be neglected. To leading order in $1/ \xi$, we have
   \begin{equation}
  {\cal L}_{\rm kin} = {1\over 2} (\partial_\mu \chi)^2 + {1\over 2} {  \xi_h^2 \tau^2 + \xi_\phi^2 \over (   \xi_h \tau^2 + \xi_\phi )^3   } \, (\partial_\mu \tau)^2 \;.
   \end{equation} 
 We see that variables $\chi$ and $\tau$ fully separate, while $\chi$ is already canonically normalized. One may introduce a canonically normalized $\tau^\prime$ by integrating the kinetic function, however 
 it cannot be written in closed form nor is it necessary. In a few interesting limits, the expression for $\tau^\prime$ simplifies:
  \begin{eqnarray}
&&    \xi_\phi \gg \xi_h ~~{\rm or}~~\tau \rightarrow 0~:~~~\tau^\prime = {\tau \over \sqrt{\xi_\phi}}~, \nonumber \\
&&    \xi_h \gg \xi_\phi ~~{\rm or}~~\tau \rightarrow \infty~:~~\tau^\prime = {1 \over \sqrt{\xi_h} \tau}~, \nonumber \\
&& \xi_\phi =\xi_h ~:~~~~~~~~~~~~~~~~~~~ \tau^\prime = {1 \over \sqrt{\xi_h}  } \arctan \,\tau~ .
     \end{eqnarray}
 In order to study  certain properties of the scalar potential, e.g. stability, using the canonically normalized $\tau^\prime$ is not necessary and we may work with $\tau$.
 Let us consider large  $h,\phi$ field values and neglect the mass terms. Then, the potential in the Einstein frame $V_E$ becomes
\begin{equation}
V_E = { \lambda_h \tau^4 + \lambda_{\phi h } \tau^2 +\lambda_\phi  \over 4 ( \xi_h \tau^2 + \xi_\phi)^2 } \;.
\end{equation}
 Its minima are classified according to \cite{Lebedev:2011aq}\footnote{We are assuming no special relations among the couplings, e.g. $2 \lambda_h \xi_\phi = \lambda_{\phi h } \xi_h$. Such relations are not radiatively stable.}
   \begin{eqnarray}
 && (1) ~ 2 \lambda_h \xi_\phi - \lambda_{\phi h } \xi_h >0, ~2\lambda_\phi \xi_h - \lambda_{\phi h } \xi_\phi >0 ~, ~~~ \tau_{\rm min}= \sqrt{  2 \lambda_\phi \xi_h -       \lambda_{\phi h } \xi_\phi  
 \over   2 \lambda_h \xi_\phi -       \lambda_{\phi h } \xi_h  } \;, \nonumber\\
 &&  (2) ~ 2 \lambda_h \xi_\phi - \lambda_{\phi h } \xi_h >0, ~2\lambda_\phi \xi_h - \lambda_{\phi h } \xi_\phi <0 ~, ~~~ \tau_{\rm min}= 0 \;, \nonumber\\
  &&  (3) ~ 2 \lambda_h \xi_\phi - \lambda_{\phi h } \xi_h <0, ~2\lambda_\phi \xi_h - \lambda_{\phi h } \xi_\phi >0 ~, ~~~ \tau_{\rm min}= \infty \;, \nonumber\\
   &&  (4) ~ 2 \lambda_h \xi_\phi - \lambda_{\phi h } \xi_h <0, ~2\lambda_\phi \xi_h - \lambda_{\phi h } \xi_\phi <0 ~, ~~~ \tau_{\rm min}= 0,\infty \;.
      \end{eqnarray}
Cases (2) and (3) correspond to the singlet and Higgs inflation, respectively while both are possible in case (4) due to the existence of two local minima. 
 
 Let us focus on case (1), where inflation is driven by  a combination of $\phi$ and $h$ with their ratio being fixed. The potential takes on the value
 \begin{equation}
V_E\Bigl\vert_{\rm min\; (1)} = {1\over 16} { 4 \lambda_\phi \lambda_h - \lambda_{\phi h}^2 \over
\lambda_\phi \xi_h^2 + \lambda_h \xi_\phi^2 - \lambda_{\phi h} \xi_\phi \xi_h } \;.
\end{equation}
 This energy density is positive: 
 the numerator is positive as required by the absence of run--away directions in the scalar potential, while the denominator is positive according to   conditions (1).
 For large non--minimal couplings to gravity, the field $\tau$ is a spectator during inflation. Indeed, the Hubble rate scales as $\sqrt{V_E} \sim 1/\xi $, while the mass for the canonically normalized $\tau^\prime 
 \sim \tau/ \sqrt{\xi}$
 scales as $1/\sqrt{\xi}$. Hence,
 \begin{equation}
m_{\tau^\prime}^2 \gg H^2 
 \end{equation}
and   it evolves quickly to the minimum. In other words, $\tau$  can be ``integrated out''.

 In this framework, inflation is driven by the $\chi$ field. At leading order in large $\xi$ and $ \xi_h h^2 + \xi_\phi \phi^2$, the potential is flat with respect to $\chi$.
 At next-to-leading order, mild $\chi$--dependence appears. 
 Retaining the   $1/ (  \xi_h h^2 + \xi_\phi \phi^2) = \exp(- {2} \chi /\sqrt{6})$ term in $\Omega$, one finds that the potential is modified in two ways:
 first, the rescaling $V_E=V/\Omega^2$ introduces $\chi$--dependence; second, $\tau = h/s$ mixes kinetically with $\chi$ thereby bringing
 additional $\chi$--dependence in. However, at the minimum of the potential,
  \begin{equation}
{ \partial V_E \over \partial \tau} \Bigl\vert_{\rm min} =0 \;,
 \end{equation}
 and the shift in $\tau$ does not affect the value of the potential. Thus, the $\chi$--dependence at this order comes from  the first source and 
   \begin{equation}
V_E = {\lambda_{\rm eff} \over 4 \xi_h^2} \left( 1+  \exp \left(     - {2\chi \over \sqrt{6}}  \right) \right)^{-2} \;,
  \label{VE}
 \end{equation}
where 
 \begin{equation}
 \lambda_{\rm eff} = { 1\over 4} {  4 \lambda_\phi \lambda_h - \lambda_{\phi h}^2 \over \lambda_\phi + \lambda_h x^2 - \lambda_{\phi h } x}
 \end{equation}
and 
 \begin{equation}
x = {\xi_\phi \over \xi_h} \;.
 \end{equation}
 This notation is convenient to draw a parallel with Higgs inflation \cite{Bezrukov:2007ep}. 
 We observe that the inflationary potential is exponentially close to the flat  one at large $\chi$. As we show below,  potentials of this type are favored by cosmological data.
 
 The above considerations can trivially be extended to  $\tau_{\rm min}=0$ and $\infty$  by replacing the potential value at the minimum by $\lambda_\phi /(4 \xi^2_\phi)$ and $\lambda_h /(4 \xi^2_h)$, respectively.
 
 \subsubsection{Inflationary predictions}
 \label{inf-predictions}
 
 The  potential (\ref{VE}) leads to specific inflationary predictions. The slow roll parameters are given by
  \begin{eqnarray}
&&    \epsilon = {1\over 2} \left( {\partial V_E / \partial \chi \over V_E }    \right)^2 \simeq {4\over 3} \exp ( -4 \chi / \sqrt{6})  \;, \nonumber\\
&& \eta = {\partial^2 V_E / \partial \chi^2 \over V_E } \simeq -  {4\over 3}   \exp ( -2 \chi / \sqrt{6})  \;,
      \end{eqnarray}
      where we have neglected $ \exp ( -2 \chi / \sqrt{6}) $ compared to 1.
 During inflation $e^\chi \gg 1$ and $\epsilon, \eta  \ll 1$. Inflation ends when $\epsilon$ approaches 1. This corresponds to 
  \begin{equation}
\chi_{\rm end} \simeq \sqrt{3\over 8 } \ln {4\over 3} \;.
 \end{equation}
The number of e--folds is found through
    \begin{equation}
N= \int_{\rm in}^{\rm end } H \, dt = - \int^{\rm end }_{\rm in} {V_E \over \partial V_E / \partial \chi }\, d \chi \simeq {3\over 4} \exp ( 2 \chi_{\rm in} / \sqrt{6}) \;,
 \end{equation}
which implies that the initial field value is
 \begin{equation}
\chi_{\rm in} \simeq {\sqrt{6}\over 2 } \ln {4 N\over 3} \;.
 \end{equation}
 The last 60 or so e--folds of inflation are constrained by the observed inflationary perturbations. In particular, 
the COBE constraint  requires $V_E / \epsilon = 0.027^4$ at $\chi_{\rm in}$ \cite{Lyth:1998xn}, which  implies
\begin{equation}
 {\lambda_{\rm eff} \over 4 \xi_h^2} = 4\times 10^{-7} \, {1\over N^2 } \;.
 \label{COBE}
 \end{equation}
 For $N=60$, this gives ${\lambda_{\rm eff} / (4 \xi_h^2)} \simeq 10^{-10}$. Therefore, either the coupling has to be small or $\xi$ large enough.
 If inflation is Higgs--like, $\tau_{\rm min} \gg 1$,  the non--minimal coupling $\xi_h$ is required to be very  large, $\xi_h \sim   10^4$. 
   On the other hand, for mostly--singlet inflation, $\tau_{\rm min} \ll 1$, the COBE constraint can be satisfied with a small $\lambda_\phi$ and moderate
   $\xi$'s.  Models of this type can be embedded into 
    interesting particle physics frameworks which address various problems of the SM \cite{Ballesteros:2016euj},\cite{Ballesteros:2016xej}.

      \begin{figure}[h] 
\begin{center}
\includegraphics[scale=0.25]{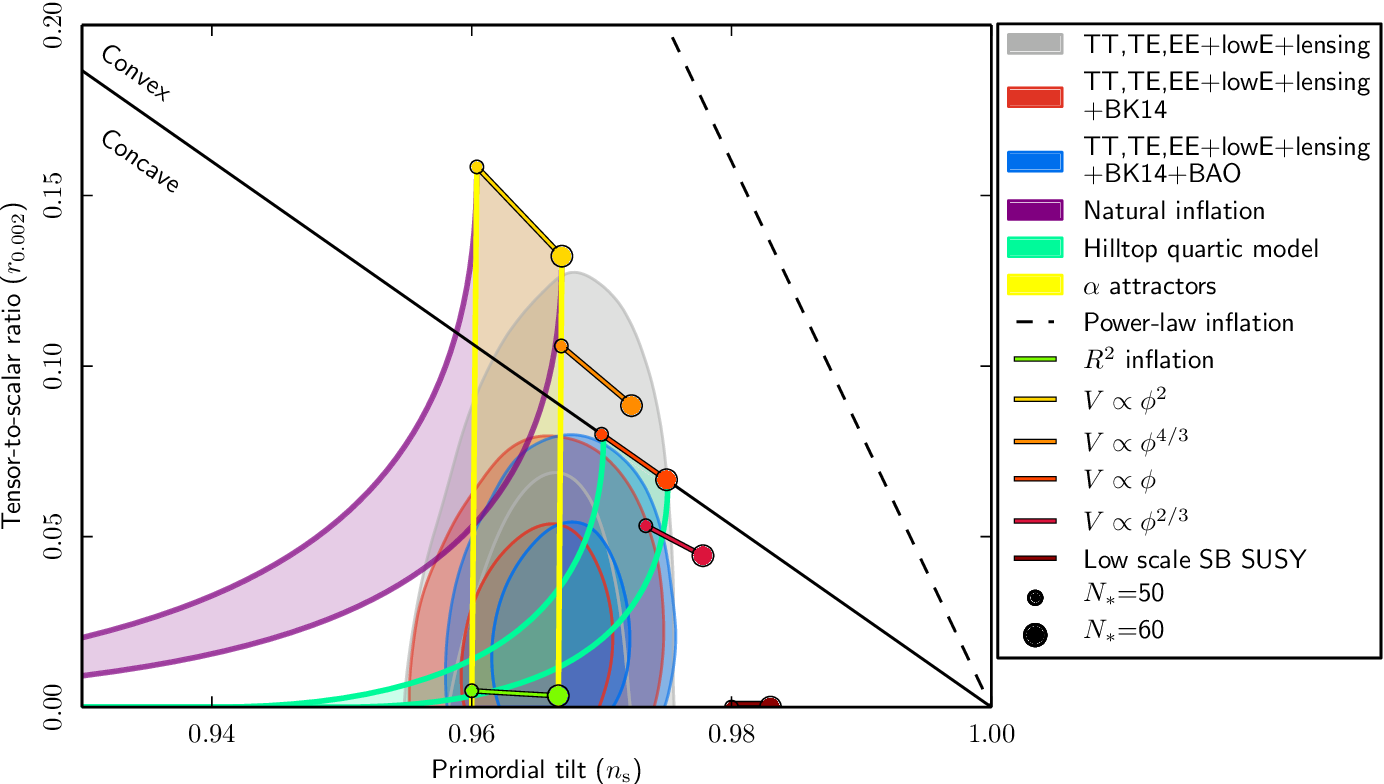}
\end{center}
\caption{ \label{PLANCK} PLANCK constraints on inflationary models \cite{Akrami:2018odb}. The $R^2$ (Starobinsky) inflation is largely equivalent to
inflation driven by a non--minimal scalar coupling to gravity. {\it Figure credit:} 
Y.~Akrami \textit{et al.} [Planck],
Astron. Astrophys. \textbf{641}, A10 (2020); reproduced with permission \copyright  \;ESO.}
\end{figure}

  The predictions of the model are conveniently formulated in terms of the number of e--folds $N$. Using
      \begin{eqnarray}
&&    \epsilon \simeq {3\over 4 N^2} \;, \nonumber\\
&&  \eta \simeq - {1\over N}
      \end{eqnarray}
at $\chi_{\rm in}$, 
  the spectral index $n$ and and the tensor to scalar perturbation ratio are computed via
    \begin{eqnarray}
&&     n = 1-6 \epsilon + 2 \eta \simeq 1-2/N \simeq 0.97  \;, \nonumber\\
&&  r =16 \epsilon \simeq 12/N^2 \simeq 3 \times 10^{-3} \;.
      \end{eqnarray}
   Note that $\epsilon \ll |\eta|$ and the gravitational wave production is suppressed. These values are  consistent with and even preferred  by  the PLANCK cosmological data \cite{Akrami:2018odb}.
   Figure\;\ref{PLANCK} shows constraints on $n,r$ for different inflationary models. The predictions of models based on the non--minimal scalar coupling to gravity are largely equivalent 
   to those of Starobinsky $R^2$ inflation \cite{Starobinsky:1980te} given by the green line and are well within the preferred region.

   \subsection{General non--minimal couplings at $\tau_{\rm }=0, \infty$}
   
   At specific points in field space such as $\tau_{\rm }=0 $ and $  \infty$, the analysis can be performed for arbitrary non--minimal couplings. 
   These correspond to singlet and Higgs inflation, respectively.
   One needs to make sure, however,   that these points are stable and, 
   in order to have  single--field inflation, the $\tau$ variable is sufficiently heavy. 
   
  Consider  the large field limit $   \xi_h h^2 + \xi_\phi \phi^2 \gg 1$. If the scalar potential has a minimum at $\tau=0$ or $\tau=\infty$, the kinetic terms simplify.
    At $\tau =0$, the mixing term disappears and the  canonically normalized variables are given by
 \begin{equation}
    \chi^\prime = {\chi \; \sqrt{1+{1\over 6\xi_\phi} }}~~,~~ \tau^\prime={\tau \over \sqrt{\xi_\phi}} \;.  
       \end{equation}
Let us now determine under what circumstances $\tau =0$ is a local minimum and $\tau$ can be integrated out.
The scalar potential at large $\chi$ reads
 \begin{equation}
   V_E={   \lambda_h \tau^4 +\lambda_{\phi h} \tau^2 + \lambda_\phi \over 4 (\xi_h \tau^2 +\xi_\phi)^2 } \;.
          \end{equation}
 Expanding it at small $\tau$, one finds that for 
  \begin{equation}
   \lambda_{\phi h} \xi_\phi - 2 \lambda_\phi \xi_h >0 \;,
          \end{equation}
 $\tau =0$ is a local minimum. The mass squared of the canonically normalized $\tau'$ is then given by
  \begin{equation}
   m_{\tau^\prime}^2=  {  \lambda_{\phi h}  \xi_\phi -2  \lambda_\phi  \xi_h \over 2 \xi_\phi^2} \;.
             \end{equation}
 On the other hand, the Hubble rate is found from 
  \begin{equation}
3H^2= V_E(\tau=0)={\lambda_\phi \over 4 \xi_\phi^2 }. 
   \end{equation}
 The $\tau^\prime$ variable can be integrated out if $m_{\tau^\prime}^2 \gg H^2$ or
 \begin{equation}
   \lambda_{h\phi} \gg \lambda_\phi \; {12 \xi_h +1 \over 6 \xi_\phi} \;.
             \end{equation}
 In this case, inflation is driven by $\chi$ and $\tau^\prime$ is a static heavy spectator.  Note that small as well as  order one 
  non--minimal couplings are consistent with this inequality.

 The inflationary potential is derived along lines of the previous subsection. The $\chi$--dependent correction to the potential comes from the 
 Weyl rescaling $V_E = V/\Omega^2$, where the 
 subleading 
 $1/(  \xi_h h^2 + \xi_\phi \phi^2)$  term at large $ \xi_h h^2 + \xi_\phi \phi^2$ is retained. The kinetic mixing does not contribute at this order and, in terms of the canonically normalized $\chi^\prime$, we get
   \begin{equation}
V_E = {\lambda_{\phi} \over 4 \xi_\phi^2} \left( 1+  \exp \left(     - {2 \gamma \chi^\prime \over \sqrt{6}}  \right) \right)^{-2} \;,
  \label{VE1}
 \end{equation}
 where 
  \begin{equation}
  \gamma = \sqrt{6 \xi_\phi \over 6\xi_\phi +1  } \;.
 \end{equation}
Since $\gamma <1$, the potential is less steep than $V_E$ at large non--minimal couplings.  

The potential of this form was considered  in Section~\ref{inf-predictions} except for the $\gamma$ factor. 
Repeating the steps described in that section, one finds
that the COBE constraint becomes 
\begin{equation}
 {\lambda_{\phi} \over 4 \xi_\phi^2} = 4\times 10^{-7} \, {1\over \gamma^2 N^2 } \;.
 \end{equation}
The slow roll parameters are now given by  
    \begin{eqnarray}
&&    \epsilon \simeq {3\over 4 \gamma^ 2 N^2} \;, \nonumber\\
&&  \eta \simeq - {1\over N} \;,
      \end{eqnarray}
such that 
   \begin{eqnarray}
&&     n  \simeq 1-{2\over N}  - {9\over 2 \gamma^2 N^2 }  \;, \nonumber\\
&&  r \simeq {12\over \gamma^2 N^2}   \;.
      \end{eqnarray}
The resulting modifications of the inflationary predictions can be significant while still well within  the PLANCK bounds. 
They tend to increase the tensor to scalar ratio and decrease the spectral index.
Note, however, that $\gamma \ll 1$ is not consistent with our approximations: in this case, inflation proceeds at least partly in the region 
$e^{-2 \gamma \chi^\prime / \sqrt{6} } \gsim 1$.

 Inspection  of the $\tau = \infty$ point proceeds analogously with the help  of the inversion $\tau \rightarrow 1/\tau$. The main difference is that the Higgs coupling  that appears in the inflationary potential cannot be too small, barring accidental cancellations.\footnote{For special values of $m_t$ and $m_h$, the SM Higgs potential develops a plateaux at large field values \cite{Masina:2011aa},\cite{Hamada:2013mya},\cite{Hamada:2014xka}, in which case a separate analysis is required.  }
 Indeed, in the SM one typically has $\lambda_h \sim {\cal O}(10^{-2})$ at relevant energy scales, although this number is sensitive to the top quark mass. The coupling to the singlet tends to increase $\lambda_h$ via renormalization group running. Therefore, $\xi_h$ has to be large to fit the COBE normalization and the corresponding $\gamma$ is close to 1.

 We see that the Higgs portal framework endowed with non--minimal couplings to gravity provides an excellent setting for inflationary model building. The resulting inflaton potential is concave and exponentially close to the flat one, which fits very well with the PLANCK constraints.

 \subsection{Unitarity issues}
 \label{uni}
 
 The non--minimal coupling to gravity generates non--renormalizable operators leading to scattering amplitudes growing with energy. This signifies the effective field theory nature of our description which
 breaks down above a certain energy scale. Since we are interested in the inflationary dynamics, the corresponding
  cut--off must lie above the inflationary energy scale. As we show below, this issue becomes critical for large $\xi_i$ characteristic of Higgs inflation \cite{Burgess:2009ea},\cite{Barbon:2009ya}. 
 
 Let us determine the cut--off of our theory in the presence of a large non--minimal Higgs coupling to gravity, 
  $\xi \gg 1$. To make the discussion more transparent, restore   the Planck scale in our formulae and consider 
   linearized gravity,
    \begin{equation}
 g_{\mu\nu}= \eta_{\mu\nu} + h_{\mu\nu}/ { M_{\rm Pl}} \;,
 \end{equation}
    where $\eta_{\mu\nu}$ is the Minkowski metric and $h_{\mu\nu}$ is a small perturbation.  
    In the Jordan frame, the coupling $\xi h^2 R$ leads to the dimension--5 operator
  \begin{equation}
 {\xi\over M_{\rm Pl} } h^2 \eta^{\mu \nu}\partial^2 h_{\mu\nu} \;.
  \end{equation}
 The scattering processes mediated by this operator, e.g. $h+ h_{\mu\nu} \rightarrow h+ h_{\mu\nu}$, exhibit rapid growth of the rate with energy. This leads to unitarity violation \cite{Burgess:2009ea},\cite{Barbon:2009ya}
 at energies above 
   \begin{equation}
   \Lambda_U \sim {M_{\rm Pl}\over \xi} \;.
   \label{uni-cutoff}
   \end{equation}

 The same conclusion is reached in the Einstein frame. The conformal transformation of the metric $ g_{\mu\nu} \rightarrow  \Omega \,g_{\mu\nu}$,  with 
 $\Omega =1 + \xi h^2 / M_{\rm Pl}^2 $,  generates the term 
 $(\partial_\mu \, \ln \Omega)^2$, i.e. the dim--6 operator 
     \begin{equation}
  {\xi^2 \over M_{\rm Pl}^2} \, {h^2 \over ( 1 + \xi h^2 / M_{\rm Pl}^2)^2} \, (\partial_\mu h)^2 \;.
 \end{equation}
 The cutoff is again $M_{\rm Pl}/\xi$. 
 
 In these  considerations, we are expanding the fields around the flat background with a zero Higgs expectation value. The presence of a non--trivial background affects these results \cite{Bezrukov:2010jz}. In this case,
 one expands the fields in terms of the average values and fluctuations,
 \begin{equation}
  g_{\mu\nu}= \bar g_{\mu\nu} + h_{\mu\nu}/ { M_{\rm Pl}} ~~,~~ h = \bar h +\delta h \;.
 \end{equation}
  At large values of the background, the unitarity cutoff changes. This is clearly seen in the Einstein frame for $\xi h^2 \gg  M_{\rm Pl}$: in this case, the results of Section~\ref{large-non-min} apply
  and, for a canonically normalized $\chi$ and $\tau =\infty $,  the higher dimensional operators are encoded in the potential $V_E \propto 1-2 \exp \Bigl(-2\chi /(\sqrt{6} M_{\rm Pl})\Bigr)$.
  Expanding it around $\bar \chi \gg M_{\rm Pl}$, one finds a Taylor series that behaves like
     \begin{equation}
   \exp \Bigl(-2\bar \chi /(\sqrt{6} M_{\rm Pl})\Bigr) \; \left( {\delta \chi \over  M_{\rm Pl}}\right)^n\;.
 \end{equation}
  At sufficiently large $n$, the overall prefactor is unimportant and the unitarity cutoff is
  \begin{equation}
   \Lambda_U^{\rm infl} \sim {M_{\rm Pl} } \;.
   \end{equation}
The corresponding bound in the Jordan frame, $\sqrt{\xi} h$, can be obtained by
  noting that the cutoffs are related via  the metric rescaling $\Omega$: $\Lambda_U\rightarrow \Omega^{1/2} \Lambda_U$. Since in the Jordan frame
the scalar curvature is multiplied by $M_{\rm Pl}^2 + \xi h^2 $, the cutoff of the theory in both cases coincides with the cutoff of the gravitational sector \cite{Bezrukov:2010jz}.  This can be compared to the scale of 
inflation,
    \begin{equation}
  V_E^{1/4} \sim \lambda_h^{1/4} \, {M_{\rm Pl} \over \sqrt{\xi} }\;.
   \end{equation}
We thus conclude that inflation proceeds in a controllable manner below the unitarity cutoff.

  Although during inflation the unitarity cutoff is high, it relaxes to (\ref{uni-cutoff}) as the inflaton background value drops at the end of inflation. In the inflaton oscillation epoch, violent particle production takes place. At large $\xi$,   the particle momentum can be as high as $\sqrt{\lambda_h} M_{\rm Pl}$ \cite{Ema:2016dny},\cite{DeCross:2016cbs}.  This exceeds the cutoff casting doubt on the validity of 
  our approach. While the theory is well--behaved at large and small field values, the intermediate field range is problematic calling for a UV completion. 
  These problems do not arise in the singlet--driven inflation at   smaller $\xi_\phi$ and $\lambda_\phi$ subject to the COBE relation (\ref{COBE}).
  
 Quantum corrections \cite{Barvinsky:2008ia},\cite{DeSimone:2008ei},\cite{Bezrukov:2008ej},\cite{Barvinsky:2009fy} are also controllable at small and large field values. For $\xi h^2 \ll M_{\rm Pl}^2$, we recover the Standard Model, while at $\xi h^2 \gg M_{\rm Pl}^2$, one can use 
 the approximate shift symmetry in the Einstein frame,
    \begin{equation}
 \chi \rightarrow \chi + {\rm const}~,
    \end{equation}
 to organize the perturbation series in the effective field theory (EFT) form \cite{Bezrukov:2010jz}. The effective Lagrangian is expanded in powers of the inverse cutoff for fluctuations, $1/M_{\rm Pl}$, and 
 the symmetry violating corrections $e^{-n \chi/M_{\rm Pl}}$. The theory  is then  renormalizable in the EFT sense. However, in the intermediate field range,
 there is no organizing principle to control  the quantum corrections. Again, one concludes that a UV completion is needed for a consistent description of the entire field range
 \cite{Bezrukov:2010jz},\cite{George:2015nza},\cite{Fumagalli:2016lls}.
  A  discussion of related  issues and their possible solutions can be found in \cite{Giudice:2010ka},\cite{Bauer:2010jg},\cite{Barbon:2015fla},\cite{Shaposhnikov:2020fdv}.
 
 In conclusion, the Higgs portal allows us to build viable inflationary models, where the role of the inflaton is played by a combination of the Higgs and singlet fields. Such models 
 fit the PLANCK data  and satisfy the tree level unitarity constraint as long as the effective quartic coupling is sufficiently small.

 \section{Vacuum stability and  inflation}

 The issue of vacuum stability has become one of the central questions in Higgs physics in recent years. The current data favor vacuum metastability, which entails a number of cosmological puzzles.
 Even if our vacuum  is very long lived, one should explain how the Universe ended up in this energetically disfavored state in the first place. In what follows, we formulate the problems and 
 discuss their possible solutions within the Higgs portal framework.

          \begin{figure}[h] 
\begin{center}
\includegraphics[scale=1.15]{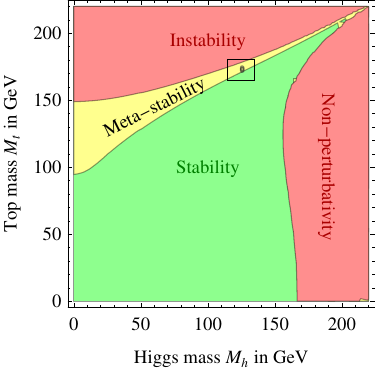}
\end{center}
\caption{ \label{SMstability} 
Vacuum stability in the Standard Model in terms of the top quark and Higgs masses. 
The measured values are marked by a dot.
The figure is from Ref.\;\cite{Degrassi:2012ry} \copyright\, CC--BY.}
\end{figure}

 \subsection{Higgs potential in the Standard Model and quantum fluctuations}
 \label{SM-stability}
 
 The value of the Higgs mass is intimately related to stability of the electroweak (EW) vacuum in the Standard Model. If the Higgs is light, the corresponding quartic coupling is small and  driven negative 
 at high energy
 by the top quark loop \cite{Cabibbo:1979ay},\cite{Hung:1979dn},\cite{Grzadkowski:1986zw}. This implies that the potential turns negative at large field values and the EW vacuum is not absolutely stable \cite{Arnold:1989cb}. The current Higgs   and top quark masses 
 $m_h \simeq 125$ GeV, $m_t \simeq 173$ GeV\;\footnote{The current LHC top quark mass measurements tend to give a slightly lower value
 than the world average, i.e. close to 172.5 GeV, as reviewed in \cite{Hoang:2020iah}.}
favor its metastability (Fig.\;\ref{SMstability}), meaning that  the vacuum decays on a long time scale by tunnelling to the energetically favored state.
Its lifetime is controlled  by the Lee-Weinberg bounce  \cite{Lee:1985uv} with the Euclidean  action $S_E = 8 \pi^2 / (3 |\lambda_h(\mu)|)$, where $\mu$ is such that $\beta_{\lambda_h}(\mu)=0$ \cite{Arnold:1989cb}. The result is that the decay takes much longer than the age of the Universe, $\tau \gg 10^{100}$ years, so it can be considered stable for most practical purposes.\footnote{Vacuum decay can be catalyzed by primordial black holes \cite{Burda:2015isa},\cite{Burda:2015yfa}.}

 The absolute vacuum stability bound on the Higgs mass according to   \cite{Buttazzo:2013uya} reads
   \begin{equation}
 m_h > 129.6~ {\rm GeV} + 2.0(m_t - 173.34 ~{\rm GeV}) - 0.5~{\rm GeV} ~ {\alpha_3 (M_Z) - 0.1184 \over 0.0007} \pm 0.3 ~{\rm GeV} \;,
   \end{equation}
   where $m_t$ is the top quark mass and $\alpha_3=g_3^2/4\pi$ is the color structure constant. The uncertainty of this lower bound  quoted in  \cite{Buttazzo:2013uya} is $\pm 1.5$ GeV meaning
   that absolute stability of the EW vacuum is excluded at around 3$\sigma$.
   Other evaluations of the stability bound confirm that 
  the currently preferred top quark and Higgs masses lead to a metastable electroweak vacuum, although the error bars and the statistical preference for metastability vary \cite{Andreassen:2014gha},\cite{Andreassen:2017rzq},\cite{Bednyakov:2015sca} with the former reaching 6 GeV in \cite{Bezrukov:2012sa},\cite{Alekhin:2012py}. A notable complication in these calculations  concerns gauge dependence 
  \cite{DiLuzio:2014bua},\cite{Espinosa:2016nld}.

 It should be noted that the  decay rate of the EW vacuum is sensitive to Planck suppressed operators  $h^n/{M_{\rm Pl}^{n-4}}$ as well as the Higgs non--minimal coupling to gravity
  \cite{Branchina:2013jra},\cite{Branchina:2019tyy}. Such  UV sensitivity is due to  the tunnelling rate being  an exponential function of the potential.
   Since non--renormalizable operators   can be generated by quantum gravity, this creates an additional source of uncertainty in our considerations. The state-of-the-art calculation of the decay rate in the pure Standard Model is presented in
  \cite{Chigusa:2017dux},\cite{Chigusa:2018uuj}.

  The Standard Model Higgs potential at large field values can be approximated by $ V(h) \simeq \lambda_h (h) h^4/4$, where $\lambda_h(h)$ is the running coupling evaluated at the scale $\mu \sim h$
  \cite{Altarelli:1994rb}. A more careful
  analysis shows that the effective potential around the value $h_{\rm max}$, where the potential is maximized, reads \cite{Espinosa:2015qea}
   \begin{equation}
 V(h) \simeq -b \ln \left( h^2 \over h^2_{\rm max}  \sqrt{e} \right) \; {h^4 \over 4} \;,
   \end{equation}
  with $b \simeq 0.16/(4\pi^2)$.   $h_{\rm max}  $ defines the position of the barrier separating the two vacua, with its typical value being  $h_{\rm max}  \sim 5\times 10^{10}$ GeV.
  This form of the potential is important for analyzing stability of the EW vacuum under fluctuations.

  Even though vacuum metastability does not pose any immediate danger in the current epoch, the situation was different in the Early Universe when the field fluctuations were large \cite{Espinosa:2007qp}. During inflation, light scalar fields are subject to quantum fluctuations of order the Hubble rate. It is then natural to expect that the EW vacuum gets destabilized if the Hubble rate exceeds the size of the barrier  $h_{\rm max}  $. This can be viewed as tunnelling to the true vacuum in de Sitter space \cite{Hawking:1981fz},\cite{Shkerin:2015exa},\cite{Kobakhidze:2013tn},\cite{Joti:2017fwe}, where the fluctuations are described by the
  effective temperature $T=H/(2\pi)$ \cite{Gibbons:1977mu}. 
  The problem can also be seen from the viewpoint  of the Higgs effective potential
  in  de Sitter space: the barrier disappears at large $H$ \cite{Herranen:2014cua},\cite{Markkanen:2018bfx}.
  
   Let us consider this issue more carefully by analyzing the Higgs fluctuations \cite{Espinosa:2015qea}.  
 If the effective Higgs mass is below the Hubble rate, the evolution of its long wavelength modes is described by the Langevin equation \cite{Starobinsky:1994bd},
 \begin{equation}
 {d h \over dt}  + {1\over 3H} \, {dV (h) \over dh} = \eta (t) \;,
 \label{langevin}
 \end{equation}
where $h = \sqrt{ 2 H^\dagger H}$ is the Higgs radial mode treated as a classical field and $\eta (t)$ represents random Gaussian noise,
\begin{equation}
 \langle \eta (t) \eta(t^\prime) \rangle = {H^3 \over 4\pi^2}\, \delta (t-t^\prime) \;.
 \end{equation}
As a result,  the Higgs field experiences a random walk in a classical potential. Indeed, if the potential is neglected, one can ``square'' Eq.\,\ref{langevin} and compute its statistical average. One then finds that 
$h$ changes on the average by 
  ${\cal O}(H)$ every Hubble time $H^{-1}$. The average Higgs value then grows as the square root of time.
  It is convenient to formulate the problem in terms of the probability density of finding value $h$ after $N=Ht$ e--folds of inflation, $P(h,N)$,
which  satisfies  the Fokker-Planck equation 
\begin{equation}
 {\partial P \over \partial N} = {\partial^2  \over \partial h^2} \left(    {H^2 \over 8\pi^2}  P  \right) + {\partial  \over \partial h} \left(    {V^\prime \over 3 H^2}  P  \right) \;.
 \end{equation}
This equation can be solved numerically, while the salient features of the solution can be understood analytically.
For the Hubble rate similar to   $h_{\rm max}$, the potential contribution is subleading and the scalar is effectively free. In this case,
$P(h,N)$ satisfies the heat equation. Its solution is well known,
\begin{equation}
P(h,N)= {1\over \sqrt{ 2\pi \langle h^2 \rangle}}\, \exp \left( -{h^2 \over 2 \langle h^2 \rangle}     \right) ~,~~~ \sqrt{ \langle h^2 \rangle } = {H \over 2 \pi} \sqrt{N} \;,
 \end{equation}
 where the initial value $h=0$ has been assumed. 
 Numerical analysis shows that this Gaussian shape is maintained even for $h$ significantly above $h_{\rm max}$. Then, the probability of finding $h$ beyond the barrier after $N$ e--folds of inflation
 is\footnote{Here, the probability distribution is correctly normalized for $h$ ranging from $-\infty$ to $\infty$, although $h$ is initially taken as positive semidefinite or, in other words, $h$ and $-h$ are gauge--equivalent. Thus, for consistency,   
the bound has to be  imposed on the absolute value, $|h|$.   } 
  \begin{equation}
p\bigl(|h| > h_{\rm max }\bigr) = 1-   {\rm erf} \left(  { \sqrt{2} \pi h_{\rm max} \over H \sqrt{N}}\right) \;,
 \end{equation}
 where the error function is defined by
  \begin{equation}
 {\rm erf} (z) = {2\over {\sqrt{\pi}}} \int_0^z e^{-t^2} dt \;.
 \end{equation}
The observable Universe is composed of  $e^{3N}$ causally independent patches formed during inflation. To make sure that the Higgs field does not fall into the true vacuum in any of them during $N$ e--folds, we require
\begin{equation}
p\bigl(|h| > h_{\rm max }\bigr) \; e^{3N} < 1 \; . 
\end{equation}
Using the large $z$ asymptotics $1-  {\rm erf} (z) \simeq e^{-z^2} /(x\sqrt{\pi})$, this implies \cite{Espinosa:2015qea}
 \begin{equation}
 {H \over h_{\rm max}} < \sqrt{2\over 3} {\pi \over N} \simeq 0.04 \;. 
 \label{bound-H}
 \end{equation}
 Since the bound requires $h_{\rm max} \gg  H$, the large $z$ expansion is indeed justified. This condition gives us the criterion for vacuum stability during inflation.\footnote{In the context of eternal inflation, 
 implications of vacuum metastability have been studied in \cite{Jain:2019wxo}.}
 
 A few clarifications are in order. First, the Gaussian approximation for $P(h,N)$ breaks down at large field values, where the classical evolution takes over quantum fluctuations. 
 Only  if $h$  is above the corresponding critical value instead of  $h_{\rm max}$, does the system evolve irreversibly  to the true vacuum \cite{Hook:2014uia},\cite{Kearney:2015vba},\cite{East:2016anr}. 
 This refinement however makes an insignificant correction of about +10\%  to the bound
  (\ref{bound-H}) as shown in \cite{Espinosa:2015qea}. It also implies that there is a range of $h$ at the end of inflation, in which the fate of the field is not entirely certain: if $h$ is not far above $h_{\rm max}$, the potential favors its evolution to the true minimum, whereas thermal effects during reheating may overcome this tendency and bring it back to the stable region \cite{Espinosa:2015qea},\cite{Espinosa:2017sgp}. 
  To evaluate   viability of this option would require a detailed understanding of the preheating and thermalization processes with heavy fields coupled to the Higgs.
  
   Another point concerns the fate of the patches evolving to the true vacuum. These do not pose any  danger during inflation because the distance between the   bubbles  of true vacuum increases.
   However,  the situation changes after inflation: depending on the initial conditions for the bubble formation, many 
  such regions  expand in flat space with the speed of light eventually swallowing the entire Universe  \cite{Espinosa:2015qea}. This justifies our condition that these bubbles should not be allowed to form in the first place.
  
 Finally, the EW vacuum could in principle be destabilized by thermal effects after inflation. Indeed, while the thermal Higgs mass favors $h\sim 0$, the thermal fluctuations grow exponentially at high temperatures leading to the formation of the true vacuum bubbles. For realistic values of the Higgs and top  quark mass, however, the resulting upper bound on the reheating temperature is uninformative, $T_{\rm } \lesssim M_{\rm Pl}$ \cite{Espinosa:2007qp}.

 We conclude that stability of the EW vacuum during inflation requires the inflation scale to be quite low:  according to Eq.\,\ref{bound-H},  the Hubble rate should typically be below   $10^9$ GeV or so. 
 Traditional inflationary models, on the other hand, prefer higher values of $H$, often by orders of magnitude
 as long as $H\lesssim 10^{14}$ GeV \cite{Akrami:2018odb}, in which case the fluctuations are catastrophically  large.   
 This conclusion, however, is sensitive  to the presence of the  Higgs--inflaton and Higgs--gravity couplings which are expected on general grounds. Their effect is the subject of the next sections.

 \subsection{Cosmological challenges}

     \begin{figure}[h!]
\begin{center}
 \includegraphics[scale=0.23]{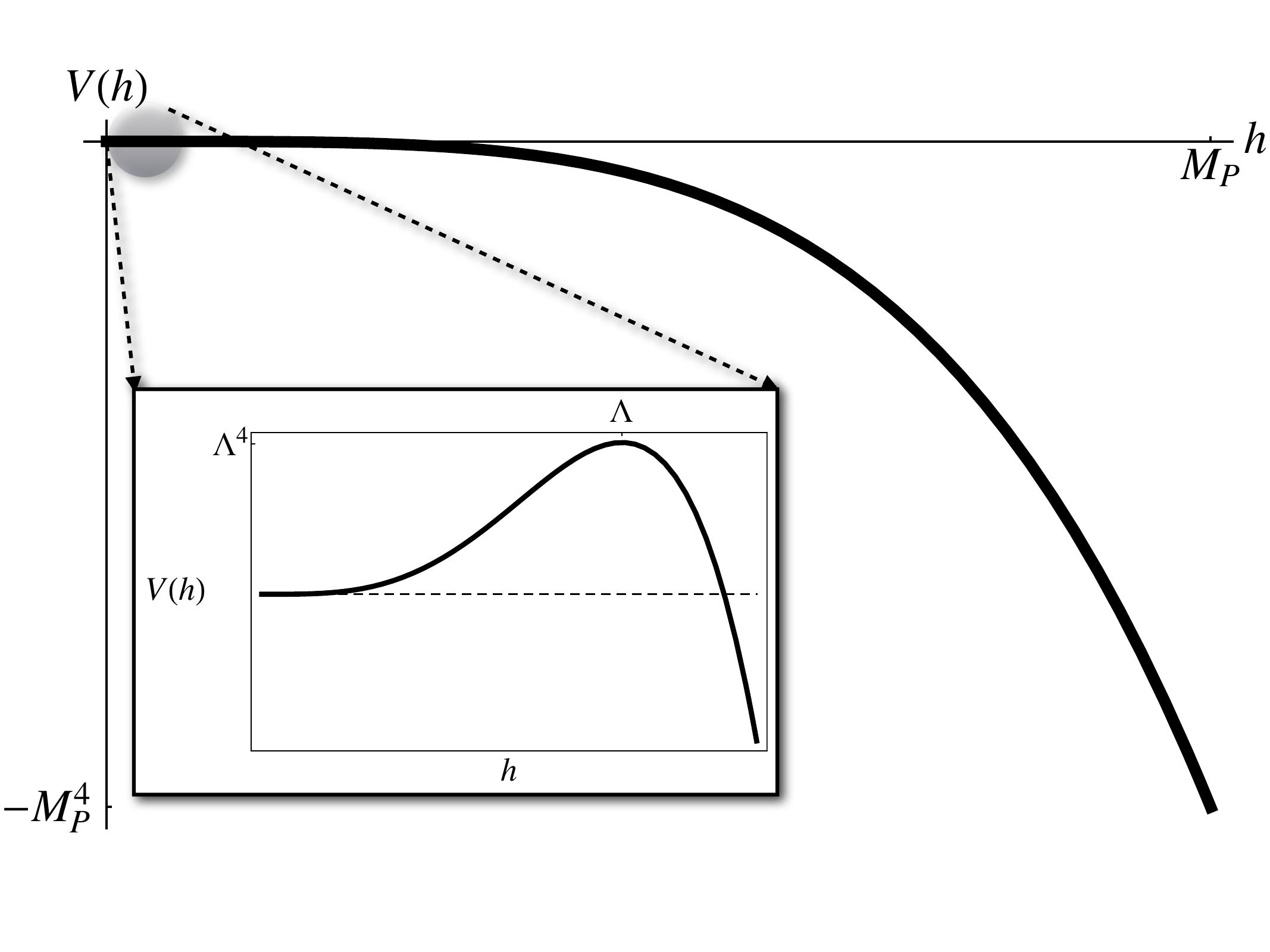} 
\caption{Schematic view of the Higgs potential. The barrier separating the two minima is located at $h \sim \Lambda \ll M_{\rm Pl}$. The figure is  from Ref.\,\cite{Lebedev:2012sy}.
\label{fig-potential}}
\end{center}
\end{figure}

 Let us formulate the cosmological challenges one faces if the EW vacuum is metastable.
 The existence of the deep minimum in the Higgs potential at large field values (Fig.\;\ref{fig-potential}) poses two  problems \cite{Lebedev:2012sy}: \\ \
 
 -- why has  the energetically disfavored vacuum at $v=246$ GeV been chosen?\\ \
 
 -- why has the system remained in this shallow vacuum despite fluctuations? \\ \

\noindent
The first problem concerns the initial conditions for the Higgs field at the beginning of inflation. The second minimum is much deeper than the electroweak one and, for most Higgs initial  values, the system would evolve to  the wrong vacuum.
In order to end up at $v=246$ GeV, the Higgs field must start its evolution at very small field values,
$h_0 < 10^{-8} M_{\rm Pl}$, while its ``natural'' range is expected to be around the Planck scale. 
Thus, one may estimate the degree of finetuning required to    be roughly $1/10^{8}$.

 The second problem concerns stability of the electroweak vacuum with respect to field fluctuations in the Early Universe \cite{Espinosa:2007qp}. The two vacua are separated by a tiny barrier which can be overcome by quantum fluctuations during inflation or preheating. 
 Unless there exists a stabilizing mechanism, such fluctuations have to be small enough. The scale of these fluctuations is tied to the scale of inflation, so vacuum stability requires 
  low scale  inflation. While this is certainly possible, it disfavors  most of the existing  inflationary models.
 
 Although the above  problems are related, they are not the same: solving the second problem with low scale inflation does not address the first problem. Conversely, setting up the right initial conditions
 does not guarantee  stability against quantum fluctuations.

 \subsection{Stabilizing the Higgs potential during inflation}
 \label{stab-infl}
 
 The problems formulated above  can be addressed within the Higgs portal framework \cite{Lebedev:2012sy}. 
 Given that the inflaton field takes on large values during inflation, the Higgs portal couplings induce an effective mass term for the Higgs field  which can drastically change its behaviour. 
 At large $\phi$, the most important inflaton--induced contribution to the Higgs potential is provided by the quartic coupling,
 \begin{equation}
\Delta V = {1\over 4} \lambda_{\phi h} \phi^2 h^2 \;.
\end{equation} 
For $\lambda_{\phi h} >0$, it has a stabilizing effect and dominates the Higgs potential when
 \begin{equation}
\phi > \sqrt{| \lambda_h | \over \lambda_{\phi h}} \, h \;.
\label{in-phi}
\end{equation} 
 If the initial values of the inflaton and the Higgs satisfiy this inequality, the Higgs field will quickly roll down to $h\sim 0$.
 We may neglect quantum gravity effects as long as  the energy density is far below ${\rm M_{Pl}}^4$ implying that 
 the Higgs field range is bounded by about ${\cal O}(10^{-1}) {\rm M_{Pl}}$.\footnote{We  assume that higher dimensional in $h$ operators may be neglected in this field range.} Above this bound, our field--theoretic description is meaningless and
 the problem we are addressing cannot even be formulated. Notably, the inflaton field is allowed to take on values above the Planck scale 
 due to flatness of its potential. To get a feeling how large $\phi$ should be to stabilize the Higgs potential, let us take $h=0.1{\rm M_{Pl}}$,
 $|\lambda_h|=10^{-2}$ and $\lambda_{\phi h}= 10^{-6}$. The required initial value of the inflaton is then 10\,M$_{\rm Pl}$ or larger. This range 
 is quite typical for chaotic inflation and much larger values are still consistent with a classical description of gravity .
 
 The above coupling should not affect the inflaton potential and spoil the inflationary predictions. 
 In particular, radiative corrections due to $\lambda_{\phi h}$ must be small enough. 
 The consequent constraint on the coupling  is strongly  model--dependent. For illustration, 
  let us consider    the quadratic inflaton potential, $V_\phi = {1\over 2} m^2_\phi \phi^2$ with 
  $m \sim 10^{-5}$M$_{\rm Pl}$.  Closing the Higgs in the loop, one finds the resulting  Coleman--Weinberg potential
   \begin{equation}
\Delta V_{\rm 1-loop}\simeq {\lambda_{\phi h}^2 \over 64 \pi^2} \phi^4 \ln {\lambda_{\phi h} \phi^2 \over m_\phi^2} \;,
\end{equation} 
 where we have  included 4 Higgs degrees of freedom. Requiring the correction not to exceed the tree potential in the last 60 e--folds of inflation,
 one finds 
 \begin{equation}
  \lambda_{\phi h} \lesssim 10^{-6} \;.
  \label{bound-lambda}
  \end{equation} 
 This bound, however, relaxes significantly in other models of inflation, e.g. those based on the non--minimal scalar coupling to gravity \cite{Lebedev:2012sy}. In these models, the inflaton quartic coupling ${1\over 4}\lambda_\phi \phi^4$
 is already present at tree level and the leading log correction $\lambda_{\phi h}^2 /(16 \pi^2) \, \ln (\mu /\mu_0)^2$ should not exceed $\lambda_\phi$. The latter can be significant, depending on the non--minimal coupling,
 and the consequent bound on $\lambda_{\phi h}$ relaxes compared to (\ref{bound-lambda}).
 
 Let us consider the evolution of the Higgs--inflaton system with a quadratic inflaton potential in more detail. If the initial Higgs value is large enough, the Higgs portal term may dominate the energy density
 of the Universe and thus affect the inflationary dynamics.
  The equations of motion together with the Friedmann equation read
 \begin{eqnarray}
\ddot{h} + 3 H \dot{h} + {\partial V \over \partial h} =0 \;, \nonumber \\
\ddot{\phi} + 3 H \dot{\phi} + {\partial V \over \partial \phi} =0 \;,
\end{eqnarray}
with the Hubble rate $H$ in {\it Planck units } (M$_{\rm Pl}=1$) given by 
\begin{equation}
3H^2= {1\over 2} \dot{h}^2 + {1\over 2} \dot{\phi}^2 + V 
\end{equation}
and 
\begin{equation}
V \simeq {1\over 4}  \lambda_{\phi h } h^2 \phi^2 +  {1\over 2} m_\phi^2 \phi^2 \;.
\end{equation}
Suppose that initially both $\dot{h}$ and $\dot{\phi}$ are small. For $\lambda_{\phi h}$ not far from the upper bound (\ref{bound-lambda}) 
and the initial Higgs value $h_0\sim 0.1$,  
 the Hubble
rate is dominated by the cross term, $H_0 \simeq \sqrt{\lambda_{\phi h}/12} \; \phi_0 h_0$. Then, taking $\phi_0$ which satisfies (\ref{in-phi}), 
one finds the hierarchy
\begin{equation}
m_\phi \ll H_0 \ll m_h^{\rm eff} \;,
\label{hierarchy}
\end{equation}
where  the effective Higgs mass is $m_h^{\rm eff} = \sqrt{\lambda_{\phi h}/2} \; \phi_0 $. Although the inflaton receives an effective mass contribution from  the $\lambda_{\phi h}$ coupling, it is small in the regime we are considering. In general, fields with masses above the Hubble rate evolve quickly while those with masses below the Hubble rate are effectively ``frozen''. Thus, the above hierarchy implies that $h$
decreases while $\phi$ undergoes a ``slow roll''. 

 The Higgs evolution at the initial stage is determined by
 \begin{equation}
\ddot{h} + \sqrt{3\over 2}\; \sqrt{\dot{h}^2 + (m_h^{\rm eff \,})^2 \; h^2 } \; \dot{h} + (m_h^{\rm eff\,  })^2\;h =0 \;,
\end{equation}
 where $m_h^{\rm eff} $  varies slowly. This equation is well known as it describes the  inflaton evolution in a quadratic potential during the inflaton oscillation epoch  (see, e.g. \cite{Mukhanov:2005sc}). 
It has a simple solution at $m_h^{\rm eff} t \gg 1$,
 \begin{equation}
h \simeq C \; { \cos (m_h^{\rm eff} t ) \over m_h^{\rm eff} t} \;,
\end{equation}
with order one $C$. 
 Since $m_h^{\rm eff} \gg H$, the asymptotic behaviour sets in after a few Hubble times. In about 10 Hubble times, $h$ reduces by an order of magnitude and at that point ${1\over 2} m^2_\phi \phi^2$
 takes over the energy density.  After that, the usual slow roll inflation takes place. The Hubble rate is almost constant and $h$ evolves according to
 \begin{equation}
\ddot{h} + 3H \dot{h} + (m_h^{\rm eff })^2 h=0 \;,
\end{equation}
with $H \simeq m_\phi \phi_0 / \sqrt{6}$.
The solutions are linear combinations of 
\begin{equation}
h=C_{\pm} \, \exp \Big(-3/2 \; H \pm \sqrt{9/4 \; H^2 -  (m_h^{\rm eff })^2 }~\Big)t \;. 
\end{equation}
Since the Higgs field  is heavy, its amplitude of oscillations  decays exponentially,
\begin{equation} 
\vert h(t) \vert \sim  e^{- {3\over 2} H t} \vert h(0) \vert \;.
\end{equation}
 The field becomes of electroweak size after about 20 $e$--folds. The inflaton, on the other hand, evolves slowly until the end of inflation. The quantum fluctuations of the Higgs field are unimportant 
 since $ m_h^{\rm eff } \gg H$ and the barrier separating the two Higgs potential minima is located at
 \begin{equation} 
h_{\rm crit} \sim \sqrt{\lambda_{\phi h} \over |\lambda_h|} \; \phi \gg H \;.
\end{equation}
The field is stuck at the origin and, therefore, 
 stability of the Higgs potential has been achieved even for sub--Planckian initial Higgs values. A numerical analysis of the Higgs--inflaton system supports this conclusion \cite{Lebedev:2012sy}.

 To summarize, we find that the Higgs evolves quickly to small field values and stays there till the end of inflation. The details of this mechanism depend on the initial field values: for instance, if $h_0$ 
 is below $10^{-2}$, the first stage in the Higgs evolution, when $h \propto 1/t$, is absent. The main ingredients  are a positive Higgs portal coupling and a large initial value of the inflaton.  
 The coupling cannot be too small: the Higgs effective mass must be above the Hubble rate during the slow roll. Thus, for a quadratic inflaton potential, the coupling must lie in the range
 \begin{equation} 
10^{-6} >\lambda_{\phi h} > 10^{-10} \;.
\end{equation}
 
 Similar conclusions hold for other {\it large field} inflation models, although the allowed range of  $\lambda_{\phi h}$ depends strongly on the specifics of the model. Note that small field models are disfavored by
 Eq.\;\ref{in-phi}. Even though the Higgs quantum fluctuations can be suppressed, such models do not address the problem of initial conditions: the deep minimum at large Higgs field values  still persists
 during (and after) inflation and the system is overwhelmingly likely to evolve there.
  
 An analogous stabilizing effect can be provided by the non--minimal Higgs--gravity coupling ${1\over 2} \xi_h h^2 R$ \cite{Espinosa:2007qp}. Since the scalar curvature $R=12 H^2$ is large in the Early Universe, this effective mass term
 can dominate the Higgs potential and lead to the field evolution described above.\footnote{The full analysis of the initial stage would require specifying the inflaton sector in order to assess  the backreaction of the Higgs on the scalar curvature.}  
 Requiring the effective Higgs mass to exceed the Hubble rate, one finds the lower bound on the coupling, 
   $\xi_h \gtrsim 10^{-1}$ \cite{Espinosa:2007qp}. Similar results have been obtained using a more sophisticated effective potential analysis in de Sitter space
   \cite{Herranen:2014cua},\cite{Markkanen:2018bfx} as well as bubble nucleation during inflation \cite{Mantziris:2020rzh}.
 
 The discussion above focuses on the leading effects. Understanding the complete dynamics of the system requires the knowledge of the UV completion as well as inclusion of more subtle effects.
 Some of them have been considered in  \cite{Kamada:2014ufa},\cite{Rodriguez-Roman:2018swn},\cite{Fumagalli:2019ohr}.
 To mention one, a departure from the exact de Sitter phase can impact significantly the stability considerations \cite{Fumagalli:2019ohr}.

  \section{Vacuum stability after inflation}
  \label{vacuum-stability-after-inflation}
 
 The issue of vacuum stability remains relevant even after inflation, especially during the inflaton oscillation epoch. In this period, the Higgs quanta production can have an  explosive character   leading to large
 field fluctuations and eventual vacuum destabilization.
 The reason is that  the Higgs portal couplings create an effective Higgs mass,
 \begin{equation}
\bigl( m_h^{\rm eff} \bigr)^2 = {1\over 2} \lambda_{\phi h} \phi^2 + \sigma_{\phi h} \phi \;,
\end{equation} 
 where $\phi$ is an oscillating background. This leads to resonant production of the Higgs field: as the inflaton goes through the origin, the effective Higgs mass turns zero and the system becomes highly non--adiabatic.  If the $ \lambda_{\phi h}$ term dominates, the resonance is parametric \cite{Kofman:1994rk},\cite{Kofman:1997yn},\cite{Greene:1997fu}, while for  a significant $\sigma_{\phi h} $ term, it is tachyonic
 \cite{Felder:2000hj},\cite{Felder:2001kt},\cite{Dufaux:2006ee}. In the latter case, the effective Higgs mass 
   squared goes negative  at $\phi <0$  leading to exponential growth of the Higgs amplitude. Analogously, the non--minimal Higgs coupling to gravity also results in   tachyonic Higgs production  after inflation
   \cite{Herranen:2015ima}.
   
  Let us note  that the physics of vacuum destabilization is semiclassical as it deals with collective phenomena. In the postinflationary Universe, 
   the main quantity that controls it is the statistical variance, $\langle h^2 \rangle$, such that if it exceeds some  critical value, fluctuations destabilize the system.  The semiclassical approach requires
   the occupation numbers to be sufficiently {\it large}, which is indeed the case during the resonance. On the other hand,   vacuum destabilization cannot be caused by a few quanta, even if they are very 
   energetic  (such as cosmic rays) \cite{Arnold:1989cb}.

   In what follows, we mostly specialize to the case of a quadratic inflaton potential in the {\it postinflationary} epoch, while many conclusions apply more generally. In this case \cite{Mukhanov:2005sc},  
  \begin{equation}
\phi (t) \simeq \Phi (t) \, \cos (mt) \;,
\end{equation} 
where  the amplitude decreases as $\Phi (t) \simeq \Phi_0 /(mt)$ and $m$ is the inflaton mass.

 Let us start by reviewing the theory of parametric resonance.

 \subsection{Parametric resonance}

 Here we follow the discussions in \cite{Mukhanov:2005sc} and \cite{Landau:1991wop} while  providing somewhat more details   to make the exposition pedagogical.
 
 In order to focus on important features of the resonance, let us neglect the Universe expansion as well as the extra Higgs couplings, $\lambda_h , \sigma_{\phi h} \rightarrow 0$.
 Then, expanding the semiclassical Higgs field in spacial Fourier modes (see Section\,\ref{resonance-basics} for details), one finds that 
  the evolution of the momentum mode $X_k$ is given by the Mathieu equation
 \begin{equation}
\ddot X_k + \omega^2 (t) X_k =0 \;,
\end{equation} 
 where
 \begin{equation}
\omega^2 (t) = {k^2 \over a^2} +  {1\over 2}\lambda_{\phi h} \Phi^2 \cos^2 mt \;,
\end{equation} 
and $a$ is the scale factor.
 Due to the periodic time dependence of $\omega (t)$,  $X_k$ can undergo resonant growth. This phenomenon is known as parametric resonance. 
 
 The physics of the resonance is largely captured by a simpler problem of wave scattering in a parabolic potential. In a semiclassical system,
 the solution to the equations of motion is given by the WKB formula
 \begin{equation}
X_k \propto {1\over \omega } \exp \left(  \pm i \int \omega \, dt\right) \;,
\end{equation} 
 where the adiabatic condition
  \begin{equation}
\left\vert {\dot \omega \over \omega^2}\right\vert \ll 1
\end{equation} 
 is assumed. For large momenta or
large enough couplings, $\sqrt{\lambda_{\phi h}} \Phi \gg m$,  this  condition is satisfied away from the inflaton zero crossing, $mt= \pi/2 + n \pi$\;.  
The solution then describes a constant number of particles with a slowly varying $\omega(t)$.
Close to (but not at) a zero crossing,
adiabaticity is violated, which can be interpreted as particle production. Let us expand the cosine close to $mt=\pi/2$. It is convenient to introduce
\begin{eqnarray}
&& k_* = \left(      \sqrt{ \lambda_{\phi h}\over 2}  m \Phi \right)^{1/2} \;, \nonumber\\
&& \tau = \left( mt - {\pi\over 2}\right) \, {k_* \over m } \;, \nonumber\\
&& \kappa = {k/a \over k_*} \;.
\end{eqnarray}
 In terms of these variables, the equation of motion takes the form ${d^2 X_k \over d \tau^2} + (\kappa^2 + \tau^2) X_k =0 \;, $
 where higher order terms in $\tau$ have been neglected. Particle creation takes place for
 \begin{equation}
 {\tau \over (\kappa^2 + \tau^2)^{3/2}} \geq 1\;,
 \end{equation}
 which can happen only for $\kappa \leq 1$ and $\tau \leq 1$. This implies, in particular, that only particles with comoving momenta below $k_*$ can be created.
 Parametric resonance is efficient for $k_* \gg m$. Let us now consider this process in more detail neglecting expansion of the Universe.

 \subsubsection{Particle creation in a parabolic potential}
 
 Consider a semiclassical field $\chi$ with the equation of motion
\begin{equation}
{d^2 \chi \over d \tau^2} + (\kappa^2 + \tau^2)\chi =0 \;.
\label{resonance-eq}
\end{equation} 
The problem is equivalent to that of wave scattering in a $negative$ parabolic potential \cite{Landau:1991wop}.
The WKB solution to the above  equation reads
\begin{equation}
\chi_{\pm} = {1 \over (\kappa^2+ \tau^2)^{1/4}} \exp \left( \pm i \int^\tau \sqrt{\kappa^2+\tau^2}\, d\tau  \right) \; ,
\end{equation} 
which can be analytically continued to complex $\tau$. 
At sufficiently large $\vert \tau \vert$, we may expand these functions in $ \kappa / \tau $,
\begin{equation}
\chi_{\pm} \simeq  \tau^{-{1\over 2} \pm {1\over 2} i \kappa^2} \exp \left(\pm {i \tau ^2 \over 2}\right) \;.
\label{chi-appr}
\end{equation} 
 
 Note that our approximate description of the resonance makes sense 
when 
  $\tau$ is large enough for the  WKB method to be applicable, $\tau > 1$, but not too large such that higher order 
  terms in $\omega (\tau)$ are subdominant, $\tau \ll k_*/m  $. In terms of $\Delta t= t-\pi/(2m)$, this requires 
   \begin{equation}
m^{-1} \gg \Delta t \gg k_*^{-1} \;,
\end{equation} 
which is a large range if the resonance is efficient.
 For the purposes of this section, we may use asymptotic expressions for $\tau \gg 1$.

   \begin{figure}[h!]
\begin{center}
 \includegraphics[scale=0.30]{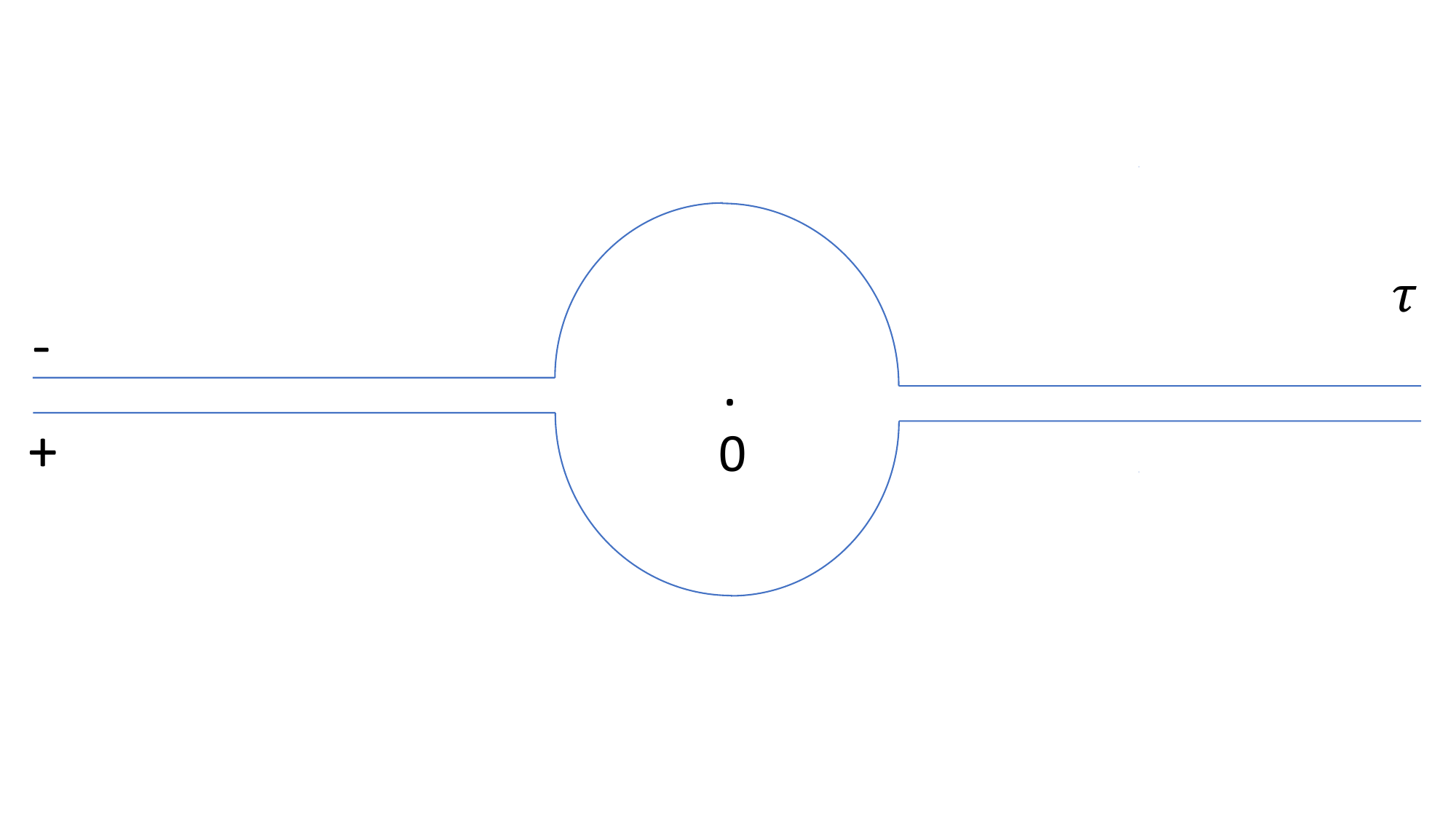} 
 \vspace{-0.7cm}
\caption{Contours for analytic continuation of the WKB solutions. The circle radius is much greater than 1.
\label{contours}}
\end{center}
\end{figure}
 Suppose we start with $\chi_+$ at $\tau \rightarrow -\infty$. At late times, $\tau \rightarrow +\infty$, the solution is a linear combination
 of $\chi_+$ and $\chi_-$ such that
 \begin{equation}
 A_+ \chi_+ \rightarrow B_+ \chi_+ + C_+ \chi_- \;, 
 \end{equation}
 where $A_\pm, B_\pm, C_\pm$ are some constants. Analogously,
 \begin{equation}
 A_- \chi_- \rightarrow B_- \chi_- + C_- \chi_+ \; .
 \end{equation}
 Since the WKB approximation works at sufficiently large $\vert \tau \vert$ in the complex plane, one can use a trick from quantum mechanics 
 \cite{Landau:1991wop}. Starting at $\tau \ll -1$, we can analytically continue our solution along an appropriate contour. For complex $\tau$,
 the exponential in  (\ref{chi-appr}) develops a real part. 
 Since the WKB approximation  corresponds to an expansion in $\hbar$ which retains the two leading terms, the exponentially suppressed contribution compared to the leading one
 must be dropped in the process of analytic continuation. This procedure is sensitive to the contour choice, in particular, how the real axis is approached. 
 Let us use the polar coordinates
  \begin{equation}
 \tau = \rho e^{i \phi} \;,
 \end{equation}
 where $\phi$ varies from $0$ to $\pi$ from right to left in the upper half plane.
 In order to relate $A_+$ and $B_+$, we can use the ``+'' contour in Fig.\;\ref{contours}. In this case, the phase varies from 0 to $-\pi$ from right to left,
  \begin{equation}
 \tau \Big\vert_{\rm right }\rightarrow \tau\Big\vert_{\rm left } \; e^{-i\pi}.
 \end{equation}
 The $\chi_+$ solution is exponentially larger than $\chi_-$ as $\phi \rightarrow 0_-$, thus the $C_+$ term must be dropped.
 The $B_+$ and $A_+$ terms match if
 \begin{equation}
 B_+ = -i A_+ e^{-\pi \kappa^2/2} \;.
 \end{equation}
Similarly, the ``-'' contour is used to determine $B_-$, in which case $ \tau \Big\vert_{\rm right }\rightarrow \tau\Big\vert_{\rm left } \; e^{i\pi}$ 
and 
\begin{equation}
 B_- = i A_- e^{-\pi \kappa^2/2} \;.
 \end{equation}
 The $C_\pm$ terms are lost in this process, but they can be found via the Wronskian
 \begin{equation}
W \equiv \dot \chi \chi^* - \chi \dot \chi^* \;.
 \end{equation}
It is easy to see that $\dot W=0$ for real $\tau$ and therefore $W$ is conserved. For the linear combination $B_+ \chi_+ + C_+ \chi_-$, the leading 
  term in $W$ at large $\vert\tau\vert$ is $2i\tau \bigl(\vert B_+\vert^2- \vert C_+\vert^2 \bigr) \vert \chi_{\pm}\vert^2$. This implies
  \begin{equation}
  \vert A_+ \vert^2 = \vert C_+ \vert^2 - \vert B_+\vert^2 \;.
 \end{equation}
 In quantum mechanics, this can be interpreted as ``flux'' conservation since the $A_+$ and $C_+ $ terms correspond to incident waves while the $B_+$ term corresponds to the reflected wave 
\cite{Landau:1991wop}.\footnote{In our convention, the direction of propagation is determined by the phase decrease along the real axis.}
 Similarly,
 \begin{equation}
  \vert A_- \vert^2 = \vert C_- \vert^2 - \vert B_-\vert^2 \;.
 \end{equation}
 Thus,
  \begin{equation}
  C_{\pm}=\sqrt {1+ e^{-\pi \kappa^2}}\,  e^{i\alpha_{\pm}} \, A_{\pm} \;,
 \end{equation}
where $\alpha_{\pm}$ are some phases.
 For a general initial configuration, by linearity we have 
  \begin{equation}
   A_+ \chi_+      +  A_- \chi_- \rightarrow (B_+ + C_-) \chi_+ + (C_+ +B_-)  \chi_- \;.
   \label{general-initial-configutration}
 \end{equation}
The relations among the coefficients remain the same as above. Indeed, Eq.\,\ref{resonance-eq} has a unique solution when $\chi$ and its derivative are fixed at some $\tau_0 \ll -1$.
This is equivalent to fixing $A_+$ and $A_-$. 
When one of them vanishes, we can use the above considerations to find
 $B_\pm , C_\pm$. Linearity of the differential equation and the initial conditions allows us to add the solutions, which makes sure 
 that the right hand side of (\ref{general-initial-configutration}) satisfies Eq.\,\ref{resonance-eq} with the right initial conditions.

Having determined how the amplitude changes upon passing the non--adiabatic region, we can interpret the result as particle creation. Our solution corresponds (approximately) to a collection 
of harmonic oscillators with frequency $\omega= \sqrt{\kappa^2 + \tau^2}$. Since $\omega$ changes slowly far from the origin, the corresponding occupation numbers  $n$ are constant in the asymptotic regions.
They are found through the harmonic oscillator relation
\begin{equation}
{\cal E} =  (n+1/2)\, \omega \;,
\end{equation}
where ${\cal E} $ is the energy stored in $\chi$, 
\begin{equation}
{\cal E} =  {1\over 2 }\vert \dot\chi \vert^2 + {1\over 2} \omega^2 \vert \chi\vert^2   
\end{equation}
averaged over a sufficiently large number of oscillations. (Here the dot denotes differentiation with respect to $\tau$.)
We thus have
\begin{equation}
 n+{1\over 2} = { {\cal E} \over \omega} \simeq \omega \vert \chi\vert^2 
\end{equation}
at large $\vert \tau\vert$. We can now compute the change in the occupation number upon passing through the non--adiabatic region,
\begin{equation}
{{n}_{\rm out} +1/2\over {n}_{\rm in} +1/2} \simeq   {  \vert    B_+ + C_-     \vert^2 +  \vert    B_- + C_+     \vert^2     \over  \vert A_+  \vert^2 + \vert A_- \vert^2} \;.
\end{equation}
 Constancy of the Wronskian requires Re $B_+ C_-^* = $Re $B_- C_+^*$, which means
\begin{equation}
{{n}_{\rm out} +1/2\over {n}_{\rm in} +1/2}  \simeq    1+ 2 e^{-\pi \kappa^2 } + {4 \vert A_+  \vert \vert A_-  \vert  \over \vert A_+  \vert^2 + \vert A_-  \vert^2 } \, e^{-\pi \kappa^2\over 2 }
\sqrt{1+e^{-\pi \kappa^2 } } \, \cos\delta \;,
\end{equation}
where $\delta$ is some phase. The term
  $1/2$ in $n+1/2$   represents ``vacuum fluctuations''.
It shows that even if one starts initially with no particles,  $A_-=0$, non--zero $\vert\chi\vert$ results in particle creation according to the above formula. Every time the system passes through $\tau=0$,
particles get created and ``flux conservation'' requires
\begin{equation}
\bigl\vert A_+ ^{(0)} \bigr\vert^2 = \bigl\vert A_+ ^{(1)} \bigr\vert^2 - \bigl\vert A_- ^{(1)} \bigr\vert^2 = \bigl\vert A_+ ^{(2)} \bigr\vert^2 - \bigl\vert A_- ^{(2)} \bigr\vert^2=...,
\end{equation}
where the superscript denotes the number of zero crossings.
Both $A_+$ and $A_-$ grow with time reaching $\vert A_+ \vert \simeq \vert A_- \vert \gg \vert A_+ ^{(0)}\vert$. The particle number grows exponentially,
\begin{equation}
 n = {1\over 2} \, e^{2\pi \mu N} ~~,~~ \mu \simeq {1\over 2 \pi} \ln \Bigl(  1+ 2 e^{-\pi \kappa^2 } + 2 \cos\delta \,   e^{-\pi \kappa^2\over 2 } \;
\sqrt{1+e^{-\pi \kappa^2 } } \Bigr) \,,
\end{equation}
where $N\gg1$ is the number of zero crossings. Here we take $n=1/2$ as the initial value. The {\it Floquet exponent} $\mu$ has the maximal value of about 0.28, at $\kappa=0$ and $\delta=0$. 
Due to the presence of the phase $\delta$, the particle number may also decrease, $\mu <0$. However, if $\delta$ takes on random values, the particle number grows on the average. 
For instance,  at  $\kappa=0$, $\mu >0$ if $\vert \delta \vert < 3\pi /4$, so the particle number increase is significantly more likely. Averaging over $\delta$ gives
\begin{equation}
 \bar  \mu \simeq {1\over 2 \pi} \ln \Bigl(  1+ 2 e^{-\pi \kappa^2 }  \Bigr) \,.
\end{equation}
The particle number grows exponentially, which can have interesting implications for vacuum stability.

\subsection{Parametric resonance and vacuum stability}

Let us apply the above quantum mechanical approach to the Higgs field in QFT and study its implications for vacuum stability.

\subsubsection{Basics}
\label{resonance-basics}

The Heisenberg representation for the Higgs field is given by \cite{Parker:1969au},\cite{Parker:2012at},\cite{Dolgov:1989us},\cite{Traschen:1990sw}
\begin{equation}
\hat h = {1\over (2 \pi)^{3/2}} \int d^3 k \left(  \hat a_k h_k(t) e^{-i {\bf k\cdot x}}  + \hat a_k^\dagger   h_k(t)^* e^{i {\bf k\cdot x}}  \right) \;,
\label{mode-expansion}
\end{equation}
where $\hat a_k$ and $\hat a_k^\dagger$ are the (time--independent) annihilation and creation operators, respectively, with the commutation relation $\left[ \hat a_k, \hat a_{k'}^\dagger \right]= \delta^{(3)}({\bf k}- {\bf k'})$ and other commutators vanishing. The comoving frame momentum ${\bf k}$ is related to the physical momentum ${\bf p}$ by ${\bf p}= {\bf k}/a$.
The rescaled momentum modes $X_k \equiv a^{3/2} h_k$ satisfy 
\begin{equation}
\ddot X_k + \omega_k^2 X_k=0 \;,
\end{equation}
 \begin{equation}
\omega_k^2 (t) = {k^2 \over a^2} +  {1\over 2}\lambda_{\phi h} \Phi^2(t) \cos^2 mt +3 \lambda_h a^{-3} \langle X^2 \rangle  + \left( {3\over 2}\right)^2 w H^2\;,
\label{Xk-EOM}
\end{equation} 
where $w$ is the equation of state parameter of the Universe, $w=p/\rho= - \left(1+ { 2\dot H \over 3 H^2}\right)$. 
We have used the Hartree approximation, that is, we have split $h(x)$ into the average field and the fluctuations, and averaged over the latter: 
$h^4 \rightarrow 6 h^2 \langle h^2 \rangle$. 
The last term in (\ref{Xk-EOM}) is small during the resonance, $\ll m^2$, and can be omitted. Note also that $w=0$
as long as the energy density is dominated by the non--relativistic inflaton.

The WKB solution for the Higgs momentum  modes is given by
\begin{equation}
X_k = {\alpha_k   \over \sqrt{2 \omega_k}} \, e^{-i\int \omega_k dt} +  {\beta_k  \over \sqrt{2 \omega_k}} \, e^{i\int \omega_k dt} \;,
\end{equation}
where $\alpha_k, \beta_k$ are  coefficients normalized as   $\vert \alpha_k\vert^2 -\vert \beta_k \vert^2=1$,
which are constant in the adiabatic regime and change only close to the inflaton zero crossing. They can be identified with 
the coefficients of the Bogolyubov transformation \cite{Bogolyubov:1958km}
\begin{equation}
\hat A_k (t)  = \alpha_k \hat a_k + \beta_k^* \hat a_{-k}^\dagger \;,
\end{equation}
which describes particle creation by a time dependent background. In terms of the time dependent annihilation and creation operators $\hat A_k (t), \hat A_k^\dagger (t)$, 
\begin{equation}
\hat h = {1\over (2 \pi )^{3/2}} \int d^3 k \left(  \hat A_k (t)g_k(t) e^{-i {\bf k\cdot x}}  + \hat A_k^\dagger   (t) g_k(t)^* e^{i {\bf k\cdot x}}  \right) \;,
\end{equation}
with $g_k(t)= 1/\sqrt{2 \omega_k a^3} \exp (-i\int \omega_k dt ) $. These operators satisfy the standard commutation relations $\left[ \hat A_k, \hat A_{k'}^\dagger \right]= \delta^{(3)}({\bf k}- {\bf k'})$, etc.
 and allow one to define the particle number operator \cite{Parker:1969au},\cite{Parker:2012at}
 \begin{equation}
 \hat n_k (t) =  \hat A_k^\dagger  (t) \hat A_k  (t) \;,
\end{equation}
  whose average over the time--independent vacuum $\vert 0 \rangle$   gives
  the particles density
   \begin{equation}
  n (t) =  \int {d^3 k \over (2\pi a)^3} \, \langle 0 \vert \hat A_k^\dagger  (t) \hat A_k  (t) \vert 0 \rangle = \int  {d^3 k \over (2\pi a)^3} \, \vert \beta_k \vert^2 \;.
  \label{nt}
\end{equation}
 Here the vacuum satisfies $a_k \vert 0 \rangle$ for all $k$ since there are no particles initially. Due to spacial translational invariance, the total 3--momentum is conserved and particles are created in pairs 
 with momenta ${\bf k}$ and ${\bf -k}$.

  The 
occupation numbers can also be found via the harmonic oscillator analogy: one can simply divide the energy of the ${\bf k }$--mode  in the comoving frame by the 
energy of a single quantum, 
 $\omega_k$,
and subtract  the vacuum contribution,
\begin{equation}
n_k = {\omega_k \over 2} \, \left(  {   \vert \dot X_k \vert^2   \over \omega_k^2} + \vert  X_k \vert^2   \right) -{1\over 2}  \simeq \vert  \beta_k\vert^2 \;,
\end{equation}
where the terms proportional to $\dot \omega_k$ have been dropped in the adiabatic approximation.
Initially there are no Higgs particles in our system,  so $\alpha_k=1$ and  $\beta_k=0$.

The vacuum expectation value of $ h^2$, that is, the variance,   is given by
\begin{equation}
\langle h^2 \rangle = {1\over (2\pi a)^3 } \int d^3 k \; \vert X_k\vert^2 \simeq  {1\over (2\pi a)^3 } \int  d^3 k \;  {n_k \over \omega_k} \,.
\end{equation}
Clearly, it grows when particle production is efficient indicating large fluctuations of the Higgs field. These fluctuations may lead to vacuum destabilization \cite{Ema:2016kpf}. 

The variance and the energy density are finite in this adiabatic approach: the vacuum contribution corresponding to $n_k =1/2$ is subtracted.  The resonance excites particles
with momenta up to $k_*$ and the corresponding occupation numbers are large, so the quanta with momenta beyond $k_*$ do not play any role and can be ``subtracted'', while 
the vacuum contribution to $n_k$ for $k< k_*$ is negligible. 

\subsubsection{Mathieu equation }
\label{Mathieu-equation}

For our purposes, the Higgs field can be treated semiclassically. Due to the resonance, the occupation numbers $n_k$ become   large
quickly and the Higgs dynamics can be extracted from classical equations of motion.

Introducing 
\begin{equation}
z\equiv {1\over 2} mt \;,
\end{equation}
we can rewrite the equation of motion in the form 
\begin{equation}
 {d^2 \over dz^2} X_k + \left[    A(k,z) +2 q(z) \cos 4 z   \right] X_k =0\;.
\end{equation}
Here
\begin{eqnarray}
&& q(z)= {\lambda_{\phi h} \Phi^2(z) \over 2 m^2} \;, \nonumber\\
&& A(k,z) = \left(  2 {k\over a m}\right)^2 + 2 q(z) + {12 \lambda_h a^{-3} \langle X^2 \rangle \over m^2} \;.
\label{q-A}
\end{eqnarray}
Note that $\omega^2$ is positive semidefinite as long as $\langle X^2 \rangle $ is negligible.
In the limit of slow Universe expansion and $\langle X^2 \rangle \rightarrow 0$, coefficients $q$ and $A$ are constant. We thus 
obtain the Mathieu equation which describes an oscillator with a periodically changing frequency. It is intuitively clear that such a  system
can exhibit resonant behavior. In particular, $q\gg 1$ corresponds to the {\it  broad} resonance regime as opposed to the narrow resonance with $q \ll 1$. In the 
former case, particles within a large range of momenta get created. 

 Since the Mathieu equation is homogeneous, the initial value of $X_k$ must be non--zero for particle production to occur. Such initial conditions are provided by the quantum fluctuations. 
 In practice, one can simulate these classically with a Rayleigh probability distribution $P(X_k) \propto \exp(-2 \omega_k |X_k|^2)$ such that the average value of $|X_k|^2$ is exactly 
 $1/(2 \omega_k)$ as in the vacuum \cite{Polarski:1995jg},\cite{Felder:2000hq}. To regularize the vacuum contribution to various physical quantities, one may set $X_k$ to zero for $k>k_*$
 as these modes do not get amplified and play no role in our discussion.

\begin{figure}[h!]
\begin{center}
 \includegraphics[scale=0.6]{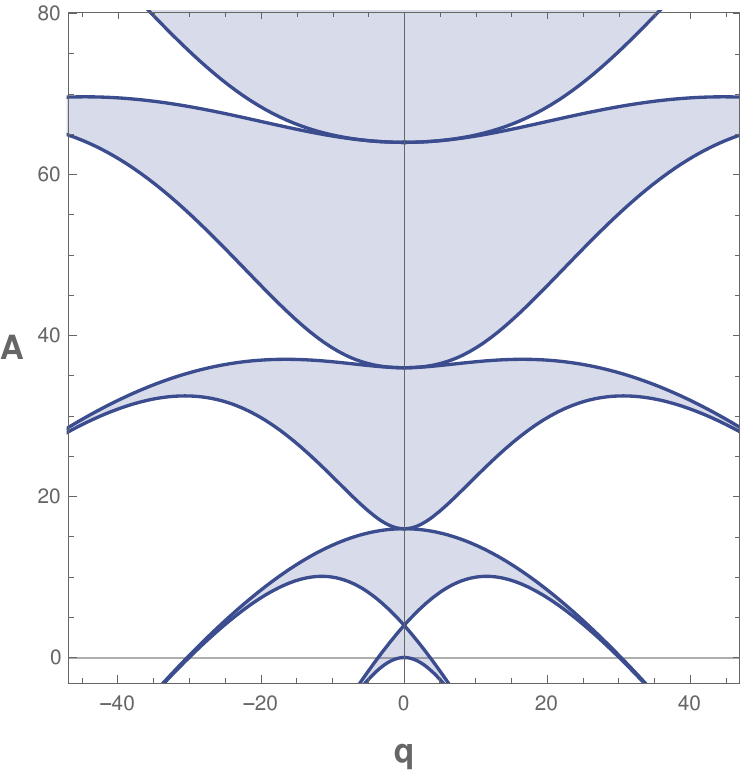} 
\caption{Stability chart of the Mathieu equation. White/dark regions correspond to unstable/stable solutions.
\label{mathieu}}
\end{center}
\end{figure}

The solutions to the Mathieu equation demonstrate drastically different behavior depending on $q$ and $A$. For some of  their values, the solutions grow exponentially, while for others 
the solutions   oscillate in time,
\begin{equation}
X_k (z)  = a_+ (k,z)  e^{\mu_k z} +  a_-(k,z) e^{-\mu_k z} \;,
\label{Floquet}
\end{equation}
where $a_\pm $ are periodic in $z$ and $\mu_k$ is the Floquet exponent, which  can be purely imaginary or have a real part depending on $q,A$. 
This behavior is conveniently represented by the stability chart shown in Fig.\,\ref{mathieu}: the dark regions correspond to ``stable'' oscillatory solutions, while in white 
regions the solutions grow exponentially. In the latter, $\mu_k$ develops a real part, Re $\mu_k \not= 0$.

In practise, the resonance is relatively short and intense so 
the Universe expansion can be accounted for adiabatically. For $\lambda_{\phi h} \Phi^2 \gg m^2$ and moderate momenta,   both $q$ and $A$ are large initially with $A \simeq 2q$. Since the inflaton amplitude decreases,  
$\Phi \propto 1/t$, the system follows a trajectory in the $(q,A)$ plane that ends at the origin. Along the way, it goes through periods with exponential growth of $X_k$ and those where $X_k$ oscillates.
When it reaches the last stability band at 
$q \sim 1$, the growth stops. The stability chart also shows that particles with large momenta corresponding $A \gg q$ are never excited since the instability bands become very narrow. In fact, the resonance is efficient if $A-2q \lesssim \sqrt{q} $ such that only the modes satisfying
\begin{equation}
{k\over a}  < k_* \sim m q^{1/4} ~~,~~ q\gg 1
\end{equation}
get excited. This agrees with our discussion of particle creation in a parabolic potential.
By the same token, 
 backreaction effects due to $\langle X^2 \rangle$ can suppress the resonance. 

The Universe expansion has the effect that the resonance becomes stochastic. Within one inflaton oscillation, the field jumps over many instability bands and its phase becomes effectively random.
The particle number can sometimes decrease while   increasing on the average. The resonant behaviour can be viewed as a collective effect due to large Bose enhancement 
of the reaction rates.
Even though 
the momenta of the created particles redshift, they remain in the resonant bands long enough
for the Bose enhancement to take effect. This is in contrast to the narrow resonance case.

 \subsubsection{Vacuum destabilization}
 
 The broad resonance is active until 
 \begin{equation}
q \simeq 1 \;,
\end{equation}
 at which point the growth of $X_k$ stops and the occupation numbers $n_k$ remain constant. The resulting field variance can be approximated  by
  \begin{equation}
\langle h^2 \rangle   \simeq  {1\over (2\pi a)^3 } \int  d^3 k \;  {n_k \over \omega_k}  \sim  {n(t) \over \sqrt{\lambda_{\phi h} / 2} \, \vert \Phi\vert}\,,
\end{equation}
 where we use  the  typical inflaton--dominated frequency  $\omega \sim \sqrt{\lambda_{\phi h} / 2} \, \vert \Phi\vert$.
 As discussed earlier, the occupation numbers grow exponentially during the resonance.
 Using the steepest descent method, (\ref{nt}) can be integrated and 
  one finds  $n(t) \propto e^{2 \bar \mu m t } \, {k_*^3 \over a^3 \sqrt{mt}}$. 
 Therefore, the Higgs fluctuations grow exponentially fast during the resonance and can lead to vacuum destabilization. The most important terms in the Higgs potential are
  \begin{equation}
V_h \simeq {1\over 4} \lambda_{\phi h }h^2 \phi^2 + {1\over 4} \lambda_h (h) h^4 \;,
\end{equation}
 where $\lambda_h (h)$ turns negative above some critical value which we can take in the ballpark of $10^{10}$ GeV. Therefore, the barrier separating the two Higgs vacua is located at
  \begin{equation}
 h_c \simeq \sqrt{\lambda_{\phi h } \over \vert \lambda_h \vert }\, \vert \phi \vert \;.
\end{equation}
It is natural to expect  that for Higgs fluctuations greater than $h_c$ the system becomes unstable. 
The situation is more subtle however: the position of the barrier is modulated by $\cos mt$ and a more careful analysis is needed. 

\begin{figure}[h!]
\begin{center}
 \includegraphics[scale=0.75]{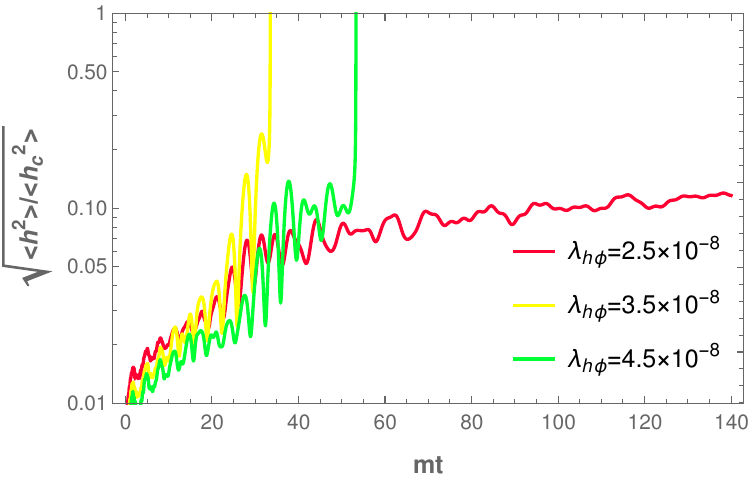} 
\caption{Time dependence of the Higgs variance during and after parametric resonance simulated with LATTICEEASY. For  $\lambda_{\phi h } > 3 \times 10^{-8}$, vacuum destabilization occurs. 
\label{var-sqrt}}
\end{center}
\end{figure}

The stability condition can be derived as follows. As $\langle h^2 \rangle $ builds up, 
  the Higgs mass term gets  dominated  by $3 \lambda_h  \langle h^2 \rangle $ 
close to each inflaton zero crossing. This is the case during a small time interval 
\begin{equation}
 \Delta t \simeq \sqrt{ 6 | \lambda_h | \langle h^2 \rangle \over \lambda_{h \phi} \Phi^2 m^2}
\end{equation}
 and the corresponding $m^2_{\rm eff} \simeq 3 \lambda_h  \langle h^2 \rangle$. The Higgs amplitude grows as $\exp (| m_{\rm eff} | \Delta t )$ during this period. As long as 
 \begin{equation}
| m_{\rm eff}| \Delta t <1,
 \label{destab}
 \end{equation}
 the growth is slow, however if this product exceeds one, the Higgs amplitude explodes quickly. 
  We are interested in vacuum stability  during the resonance, thus we impose this condition as long as $q \gtrsim 1$. Then $\lambda_{\phi h } \Phi^2  \gtrsim  2 m^2$ and we may rewrite 
 (\ref{destab}) as
 \begin{equation}
 {3 | \lambda_h |  \langle h^2 \rangle \over m^2 } 
 <1 \;.
 \end{equation}
 which is not far from  the requirement $ \sqrt{\langle h^2 \rangle}<h_c$. In fact, this condition can be obtained  directly from $V_h $ in Hartree approximation by requiring 
 $ {1\over 4} \lambda_{\phi h }h^2 \phi^2 > {3\over 2 } \lambda_h h^2  \langle h^2 \rangle$ 
 at the end of the resonance.
The above inequality  leads to the upper bound \cite{Ema:2016kpf},\cite{Enqvist:2016mqj}
 \begin{equation}
  \lambda_{\phi h } \lesssim 3 \times 10^{-8} \;
 \end{equation}
for a typical $\lambda_h = - 10^{-2}$ above the critical scale. 
Since $\langle h^2 \rangle$ is an exponential function of time and, thus, $\lambda_{\phi h } $, 
the $\lambda_h$--dependence is  only logarithmic making this result quite robust.

This estimate is confirmed by lattice simulations with LATTICEEASY \cite{Felder:2000hq}. Fig.\,\ref{var-sqrt} shows the time evolution of the Higgs variance in units of $h_c$
for a few Higgs portal  couplings. The input parameters are $m=6 \times 10^{-5}$ in Planck units ($M_{\rm Pl}=2.4 \times 10^{18}$ GeV) and $\Phi (t=0)=M_{\rm Pl}$. 
 We see   that $\lambda_{\phi h} = 3.5 \times 10^{-8}$ and $4.5 \times 10^{-8}$ lead to vacuum destabilization, that is, $\langle h^2 \rangle$ blows up before the resonance ends at $mt \sim 40 - 50$.  The corresponding critical 
 $ \sqrt{\langle h^2 \rangle}$ is not far from $h_c$, as expected. We also observe that the strength of the resonance depends non--linearly on $\lambda_{\phi h}$: for 
  $\lambda_{\phi h} = 3.5 \times 10^{-8}$, the variance grows faster than it does for   $\lambda_{\phi h} = 4.5 \times 10^{-8}$.
  
  It is important to note that the lattice simulations do not resort to the Hartree approximation, which can make a significant impact on the outcome. In the presence of quartic interactions such as
  $h^4$, the equations of motion for the different momentum modes do not decouple which makes the problem impossible to treat analytically (beyond simple approximations). Due to the coupling between 
  the different modes, the resonance usually proceeds more violently compared to the Hartree limit. A detailed discussion of lattice simulations of parametric resonance can be found in \cite{Figueroa:2016wxr}.

  \subsection{Tachyonic resonance }

  The Higgs--inflaton interaction contains, in general, a trilinear term $\phi h^2$.
   During preheating, this induces a Higgs mass term whose sign alternates in time leading to efficient particle creation \cite{Enqvist:2016mqj}. 
   This is in contrast to the parametric resonance case, where the sign is fixed.  
  The particle creation  process can be analyzed semiclassically, as before.
  With the trilinear term, the equation of motion for the Higgs momentum modes becomes
  \begin{equation}
 {d^2 \over dz^2} X_k + \left[    A(k,z) + 2p(z) \cos 2z + 2 q(z) \cos 4 z   \right] X_k =0\;,
\end{equation}
where 
 \begin{equation}
  p(z) = {2 \sigma_{\phi h } \Phi(z) \over m^2}
\end{equation}
  and the other parameters are defined in (\ref{q-A}). In the limit of slow Universe expansion, this is known as  the Whittaker--Hill equation. 
  
  \begin{figure}[h!]
\begin{center}
 \includegraphics[scale=0.45]{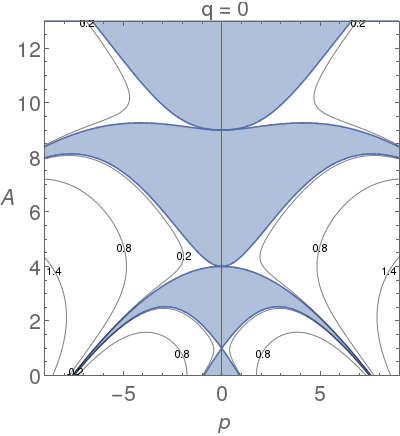} \;\;
  \includegraphics[scale=0.45]{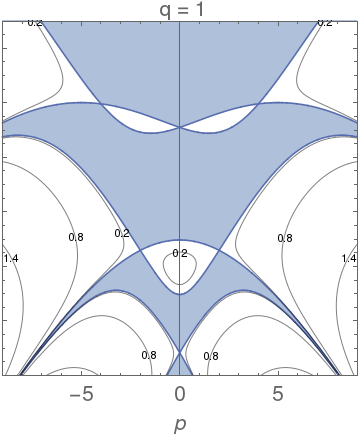} 
\caption{Examples of 2d stability charts of the Whittaker--Hill equation for $q=0$ and $q=1$. The dark/white regions correspond to stable/unstable solutions. The contours show the magnitude of the real part of the Floquet 
exponent (see Eq.\,\ref{Floquet}). The figure is from Ref.\,\cite{Enqvist:2016mqj} \copyright\, CC--BY.
 \label{WH}}
\end{center}
\end{figure}

The behaviour of the solutions of the Whittaker--Hill equation depends on $p,q$ and $A$. Since the Higgs mass  is a periodic function of time (albeit with two periods), the solution can be written in terms of the Floquet exponent as in Eq.\,\ref{Floquet}. 
 The  corresponding stability charts for fixed  $q$  are shown in Fig.\,\ref{WH}. They exhibit a rather complicated pattern compared to that of the Mathieu equation.
  
For most excited Higgs modes, one may take $A \simeq 2 q$ as before. Then, one finds that  the end of the resonance corresponds to $|p|<1, q<0.5$ with the boundary of the last stability region
described by
\begin{equation}
   q \simeq 0.5 \, (1- |p|) \;.
\end{equation}
Note that $p$ decreases slower with time than $q$ does, therefore at late times the tachyonic resonance dominates. 

For a sufficiently large initial $p$, the system stays long enough in the unstable bands for
the vacuum to be destabilized. Hence, stability imposes an upper bound $\sigma_{\phi h}$.
An estimate of this bound can be obtained as follows. 
As we saw in the previous subsection, 
one expects that as long as the $\sigma_{\phi h}$--term is greater than $|\lambda_h | h^4/4$, the system remains stable although the variance grows rapidly. In Hartree approximation, this requires
\begin{equation}
  \langle h^2 \rangle \lesssim  {|\sigma_{\phi h} | \Phi \over 3 | \lambda_h | } \;.
\end{equation}
If the fluctuations do not reach the critical value by the end of the resonance, no destabilization occurs.
Since it ends at $|p| \sim 1$, we can replace the above  condition with 
\begin{equation}
 {6 | \lambda_h |  \langle h^2 \rangle \over m^2 } 
  \lesssim 1 \;.
 \end{equation}
The Higgs variance at the end of the resonance can be estimated  as
\begin{equation}
  \langle h^2 \rangle \simeq { k_*^2 \Delta k_* \over a^3 } \, {n_{k_*} \over \omega_{k_*}}  \;,
\end{equation}
where $k_*$ is the characteristic momentum and $\Delta k_* \sim k_* / 2$ is the width of the resonant band. Towards the end of the resonance $k_* \sim m$, while $\omega_{k_*}$ can be approximated by the inflaton--induced term $\omega_{k_*} \sim \sqrt{ | \sigma_{\phi h} \Phi|}$.  
Further, 
the occupation number $n_{k_*} $ is an exponential function of time $\sim \exp (\mu_* mt)$ and the duration of the resonance 
is dictated by $|p| \gtrsim 1$. Putting all the ingredients together,
one finds \cite{Enqvist:2016mqj}
\begin{equation}
 |\sigma_{\phi h} | \lesssim 10^9 \; {\rm GeV} \;,
 \end{equation}
with the $\lambda_h$--dependence being only logarithmic. 

\begin{figure}[h!]
\begin{center}
 \includegraphics[scale=0.49]{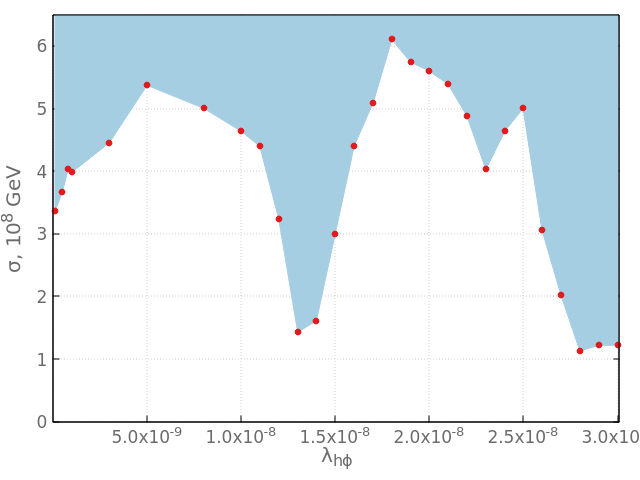} 
\caption{ Upper bound on $\sigma_{\phi h }$  from vacuum destabilization as a function of the Higgs--inflaton quartic coupling. The red dots correspond to the simulation results obtained with LATTICEEASY.
The figure is from Ref.\,\cite{Enqvist:2016mqj} \copyright\, CC--BY.
\label{sigma-bound}}
\end{center}
\end{figure}

This estimate gives the right ballpark of the bound on $\sigma_{\phi h }$, as lattice simulations show \cite{Enqvist:2016mqj}.  
However, the true bound exhibits a significant  non--linear dependence on $\lambda_{\phi h }$ due to a complicated stability band structure of the Whittaker--Hill equation. This is shown in
Fig.\,\ref{sigma-bound}.  The initial conditions for the simulation are described in the previous subsection. Note that the range of $\lambda_{\phi h }$
in the figure is such that this coupling does not lead to destabilization by itself, so   the trilinear term is the main driver  of the $\langle h^2 \rangle$ growth.

\subsection{Effect of the non--minimal Higgs--gravity coupling}

The non--minimal coupling to gravity also has an important effect on vacuum stability \cite{Herranen:2015ima}.  Suppose we add to the Lagrangian the term
  \begin{equation}
\Delta {\cal L} = {1\over 2 } \sqrt{-\hat g}\, \xi h^2 \hat R \;,
 \end{equation}
where $\hat R$ is the scalar curvature in the Jordan frame. This coupling is generated radiatively even if absent at tree level. To go over to the Einstein frame, one performs a conformal metric transformation
\begin{equation}
g^{\mu \nu} = \Omega^{-1} \hat g^{\mu\nu} ~~,~~ \Omega = 1- \xi {h^2 \over M_{\rm Pl}^2} \;,
\end{equation}
where $\hat g^{\mu\nu}$ is the Jordan frame metric. This transformation modifies the   kinetic terms and the potential
such that in the Einstein frame one finds
\begin{equation}
{\cal L}_{\rm sc} = \sqrt{-g} \left(  {1\over 2 \Omega} \, \partial_\mu \phi \partial^\mu \phi +  {1\over 2}\, {6 (\xi h)^2 + \Omega \over \Omega^2 } \,\partial_\mu h \partial^\mu  h  - {V(\phi,h) \over \Omega^2}\right) \;.
\end{equation}
For the Higgs field values far below the Planck scale,
\begin{equation}
|\xi | {h^2 \over  M_{\rm Pl}^2} ~,~ \xi^2 {h^2 \over  M_{\rm Pl}^2} \ll 1 \;,
\end{equation}
we may expand the Lagrangian in powers of $h/M_{\rm Pl}$. One then finds that the canonically normalized Higgs field $h_c$ is given by 
\begin{equation}
h_c \simeq h \left[   1+\left( \xi + {1\over 6}\right)  \, \xi \, {h^2 \over  M_{\rm Pl}^2} \right] \;,
\end{equation}
and the potential becomes  \cite{Ema:2017loe}
\begin{equation}
V\simeq {1\over 2} m^2 \phi^2 + \left( {1\over 4 }  \lambda_{\phi h }    +\xi {m^2 \over M_{\rm Pl}^2 }   \right) \phi^2 h_c^2 + {1\over 2}\sigma_{\phi h } \phi h_c^2 + {1\over 4} h_c^2 \;,
\end{equation}
where we are keeping terms up to fourth order in $h_c$. 
We see that the Higgs portal coupling $\phi^2 h_c^2 $ is generated via the non--minimal coupling to gravity when $ \lambda_{\phi h }$ is set to zero.
Even though $ \xi m^2 / M_{\rm Pl}^2$  is Planck--suppressed, $m$ is large enough to make it    comparable to $ \lambda_{\phi h }  $ considered in the previous subsections.

The equation of motion for $h_c$ reads
\begin{equation}
\ddot h_c +3 H \dot h_c - \partial_i \partial^i h_c + \left[  \left( {1\over 2} \lambda_{\phi h } +2 \xi {m^2 \over M_{\rm Pl}^2 }\right) \phi^2 - \xi {\dot \phi^2 \over M_{\rm Pl}^2} + \sigma_{\phi h } \phi  \right] h_c + \lambda_h h_c^3=0 \;.
\end{equation}
The impact of the non--minimal coupling to gravity amounts to the presence of the two Planck--suppressed terms proportional to $\xi \phi^2$ and $\xi \dot \phi^2$.
Their  effect on the Higgs dynamics can be very significant since,  for $\phi \propto \cos mt$, their size is determined by $\xi m^2/  M_{\rm Pl}^2 \sim \xi \times 10^{-10}$.
In Hartree approximation,  the parameters of the Mathieu equation in Eq.\,\ref{q-A} get shifted according to 
\begin{eqnarray}
&&  q(z)= \left( \lambda_{\phi h}  +6 \xi  \,{ m^2 \over M_{\rm Pl}^2 } \right) \, {\Phi^2(z) \over 2 m^2} \;, \nonumber\\
&&   A(k,z) = \left(  2 {k\over a m}\right)^2 + 2 q(z) - 4 \xi \,{m^2 \over M_{\rm Pl}^2 } \, {\Phi^2(z) \over  m^2} + {12 \lambda_h a^{-3} \langle X^2 \rangle \over m^2} \;.
\end{eqnarray}
For $k \simeq 0$ and $\langle h_c^2 \rangle \simeq 0$, the relation $A = 2q$ no longer holds. Due to the extra  contribution to $A$, the effective mass squared can turn negative depending on the sign of $\xi$ and therefore the resonance can be  tachyonic. For a negative $\xi$, the effect  is instead stabilizing: increasing $A$ at fixed $q$ brings one into a stable region. 
 The $p$--parameter of the Whittaker--Hill equation is not affected by the non--minimal coupling to gravity, so we may set $p=0$ for the present discussion. 
 
 \begin{figure}[h!]
\begin{center}
 \includegraphics[scale=0.45]{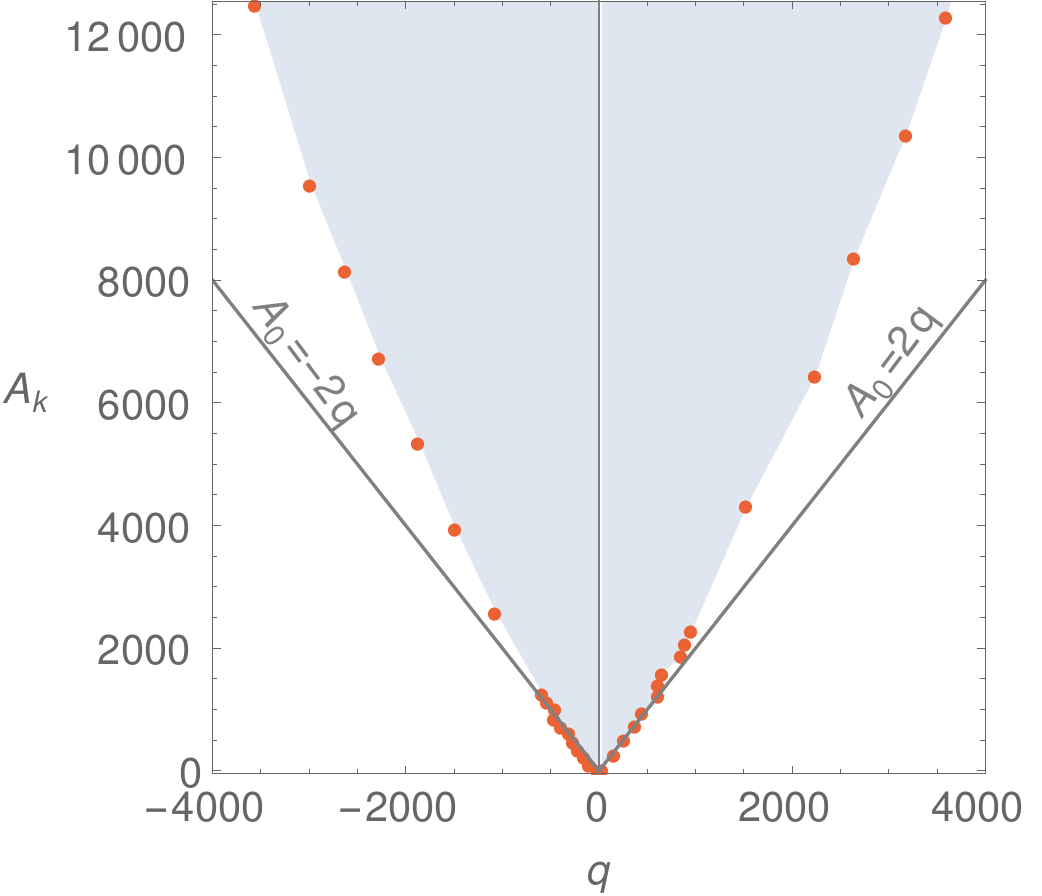} 
\caption{ Vacuum stability region in the $(q,A)$ plane. 
The initial conditions corresponding to 
points in the shaded region ensure a stable vacuum throughout preheating  with $\lambda_h (h_c) = -10^{-2}$. The red dots are obtained with LATTICEEASY. The figure is from Ref.\,\cite{Ema:2017loe}
\copyright\;IOP Publishing Ltd and SISSA Medialab Srl.  Reproduced by permission of IOP Publishing.  All rights reserved.
\label{xi-stab}}
\end{center}
\end{figure}

Vacuum stability is controlled mainly by the behaviour of $X_k$ for moderate momenta, which can be neglected in the expression for $A_k$. 
Fig.\,\ref{xi-stab} shows the region in the $(q,A)$ plane in which the EW vacuum remains stable throughout preheating. That is, the boundary of this region 
determines the maximal $|q|$ allowed for a given $A$ at the initial time.
Since it is above the $A=2q$ line, the required $\xi$ is negative. For a fixed $\lambda_{\phi h }$,
 it cannot be too large in magnitude such that $A$ remains positive. As a result, some cancellation between  the $\lambda_{\phi h }$-- and $\xi$--contributions is required, unless both of them are small.
 Allowing for such cancellations, 
 one  finds that $|\xi |$ values up to $10^4$ are in principle consistent with vacuum stability \cite{Ema:2017loe}.  In the absence the Higgs portal coupling, the stability constraint is $\xi \lesssim 10$ 
 \cite{Ema:2016kpf},\cite{Figueroa:2017slm}.

\subsection{Discussion}

The obtained bounds on the Higgs--inflaton couplings are quite robust since they depend on $\lambda_h$ only logarithmically. 
The SM instability scale $\Lambda_{\rm SM}$, which is close to $h_{\rm max}$ of Section\,\ref{SM-stability},  appears in these calculations implicitly: we are assuming that the Higgs quartic coupling at the scale $h_c \simeq \sqrt{\lambda_{\phi h}/ |\lambda_h|} \, \Phi$
(or $h_c \simeq \sqrt{\sigma_{\phi h} \Phi / |\lambda_h|} $ in the trilinear case) during the resonance 
 is negative and the destabilization occurs when
\begin{eqnarray}
 \sqrt{\langle h^2 \rangle} &\gtrsim & h_c \;, \nonumber\\
  h_c &\gg & \Lambda_{\rm SM} \;.
 \end{eqnarray}
 Clearly, this is only possible if  $\Lambda_{\rm SM}$ is below the inflationary scale. As $| \lambda_h |$ decreases and approaches zero, $h_c $ becomes so large that no fluctuation can reach it and vacuum stability gets restored.

 In our analysis, we have taken the benchmark value $\lambda_h = - 10^{-2}$. Then, according to Fig.\,\ref{var-sqrt}, $h_c \sim 10^{14}$ GeV and our considerations apply as long as
 \begin{equation}
 \Lambda_{\rm SM} \ll 10^{14} \; {\rm GeV} \;.
 \end{equation}
  In the present setting, one finds that $ \sqrt{\langle h^2 \rangle} < 10^{15}$ GeV for any $\lambda_{\phi h}$. Therefore, for $ \Lambda_{\rm SM} $ of this order or above, the system remains stable.

There are a number of effects that we have neglected. In particular, the Higgs decay into the top quarks dilutes the Higgs population and weakens the resonance. This decay is kinematically allowed since the 
Higgs mass is dominated by the large inflaton--induced term, while the Higgs VEV is small making the fermions effectively massless. Although this makes an impact on the dynamics of the resonance, the
stability bounds remain essentially unaffected \cite{Enqvist:2016mqj}. Further, we have included a single degree of freedom for the Higgs field {\it \`a la} unitary gauge. At $\langle h \rangle =0$, there are 4 Higgs degrees of freedom. Again, this complication does not  affect the above results in any significant way.

There is a degree of model dependence in our analysis.
To illustrate the effect of the Higgs portal couplings on vacuum stability, we have chosen the quadratic inflaton potential {\it during preheating}. This is not directly connected to the shape of the inflaton potential during inflation and various inflationary models can give the same small--field limit. Therefore, our results apply more generally. On the other hand, the specifics  of destabilization are sensitive to the shape of the potential during preheating. In some cases,  the potential is dominated by the quartic term during most of the preheating period. Then, the resonance is described by the Lam\'e equation
for which the instability band structure and backreaction effects are rather subtle \cite{Greene:1997fu},\cite{Abolhasani:2009nb}. This applies, in particular, to preheating in Higgs inflation \cite{GarciaBellido:2008ab}.  Other related work can be found in \cite{Kohri:2016wof},\cite{Kohri:2016qqv},\cite{Postma:2017hbk}.

While an exhaustive survey  of all the possibilities  is still lacking,
the vacuum  stability analysis for inflation driven by a non--minimal coupling to gravity has been carried out in \cite{Rusak:2018kel}. 
 The conclusion is qualitatively similar to what we find in the quadratic case: there is a coupling range, where stability is achieved both during and after inflation. The main 
 relevant parameter is $\lambda_{\phi h}/\xi_\phi $ such that for $\lambda_{\phi h}/\xi_\phi $ between $10^{-10}$ and $10^{-8}$ no destabilization occurs, although there is 
 also some dependence on $\xi_\phi$.

 \section{Stabilizing the Higgs potential via scalar mixing}
 \label{section-mixing}
 
 The  couplings that stabilize the Higgs potential during inflation can have an opposite effect during preheating. Although there is a range of couplings where stability is  achieved
 in both epochs, this tendency motivates one to look for further solutions.

 The simplest possibility    is to modify the Higgs potential in a time--independent manner. As argued in Section\;\ref{generalities}, one expects a Higgs--inflaton mixing on general grounds. This, among other things, modifies the Higgs self--coupling which, within the Standard Model,  is inferred from the Higgs mass of 125 GeV.
 Even though such a correction is small, it can completely change the asymptotic behaviour of the Higgs potential \cite{Lebedev:2012zw},\cite{EliasMiro:2012ay}. 
 
 Let us first consider the simple case of a $Z_2$ symmetric scalar potential where an SM singlet develops a large VEV. 
 
 \subsection{$Z_2$ symmetric potential with a large singlet VEV}
  
  In this Section, we illustrate the sensitivity of the Higgs potential stability to a tiny Higgs--singlet mixing.
   Consider the scalar potential (\ref{Z2-pot}), where $\phi$ is   a general SM singlet and not necessarily  the inflaton. Suppose $w \gg v$, then to leading order in $v^2/w^2$ we have 
  \begin{eqnarray}
  && m_1^2 \simeq 2 \left(   \lambda_h - {\lambda_{\phi h}^2 \over 4 \lambda_\phi}    \right) v^2 \, \label{m-coupling}\\
  && m_2^2 \simeq 2 \lambda_\phi w^2 \;, \nonumber
  \end{eqnarray}
  and the mixing angle is given by
  \begin{equation}
  \theta \simeq {\lambda_{\phi h} v\over 2 \lambda_\phi w} \;.
  \end{equation}
 The main feature that we observe is that, while  
  the mixing angle is suppressed, the lighter mass eigenvalue receives a significant correction. Since $m_1 = 125 $ GeV, 
   \begin{equation}
   \lambda_h - {\lambda_{\phi h}^2 \over 4 \lambda_\phi}  \simeq 0.13 \;.
  \end{equation}
  The resulting $\lambda_h$ is $larger$ than that in the Standard Model. This is a tree--level effect which affects 
  the boundary condition for the subsequent RG evolution of the Higgs self--coupling. 
  One finds that about a 10\% correction is sufficient to make $\lambda_h$ positive all the way to the Planck scale \cite{Lebedev:2012zw}. 
  It is remarkable that this can be achieved with tiny couplings $\lambda_{\phi h}, \lambda_\phi$ as long as $\lambda_{\phi h}^2/ \lambda_\phi$
  is substantial. As a result,   the deep minimum in the Higgs field direction is eliminated and 
  complete scalar potential stability can  be achieved for a wide range of couplings.
  
  It is instructive to look at this system from an effective field theory viewpoint  \cite{EliasMiro:2012ay}. If $\phi$ is very heavy, it can be integrated out leaving an effective Higgs potential. The form of the latter is constrained by gauge invariance, so it takes on the usual form albeit with a modified self--coupling. Specifically, let us first trade the mass parameters $m_h, m_\phi$ for the VEVs using the stationary point condition:
  \begin{eqnarray}
  &&  m_h^2 = -{1\over 2} \lambda_{\phi h } w^2 -\lambda_h v^2  \, , \nonumber\\
  &&    m_\phi^2 = -{1\over 2} \lambda_{\phi h } v^2 -\lambda_\phi w^2   \;.
  \end{eqnarray}
  Now, restore all four Higgs components and 
   use the equation of motion for $\phi$ neglecting the momentum,
   \begin{equation}
   \phi^2 = - {1\over \lambda_\phi} \left( \lambda_{\phi h }  H^\dagger H +m_\phi^2 \right) \;.
  \end{equation}
   Plugging this back into the scalar potential, one finds
   \begin{equation}
  V= \left(  \lambda_h - {\lambda_{\phi h}^2 \over 4 \lambda_\phi}  \right) \, \left( H^\dagger H - {v^2 \over 2} \right)^2  ~ + ~ {\rm const} \;.
  \end{equation}
  We thus recover the usual Higgs potential with a redefined coupling at low energies. The relation between the Higgs mass and the Higgs self--coupling remains unaffected. 
  The impact of the singlet amounts to a threshold correction: it shifts the Higgs self--coupling in the infra--red (IR) theory with respect to that in the ultra--violet (UV) theory, where both $h$ and $\phi$ are active degrees of freedom. The singlet may be very heavy as long as the threshold occurs below the SM instability scale, $m_2 \ll h_{\rm crit}$.
  
  On the other hand, if the singlet is not too heavy and remains an active degree of freedom at the TeV scale, the relation (\ref{m-coupling}) implies that the mass--coupling relation for the Higgs gets modified
  by the presence of the singlet \cite{Lebedev:2012zw}. The correction could in principle be as large as 100\% which 
   may be probed by the Higgs self--coupling measurement at the HL-LHC.

  \subsection{General potential and Higgs--inflaton mixing}
  \label{general-mixing}
 
 In general, the Higgs--inflaton potential also contains terms odd under $\phi \rightarrow -\phi$ as in Eq.\;\ref{potential}. 
 This naturally leads to the Higgs--inflaton mixing even in the absence of the inflaton VEV and thus affects the potential stability \cite{Ema:2017ckf}.

\begin{figure}[h!]
\begin{center}
 \includegraphics[scale=0.40]{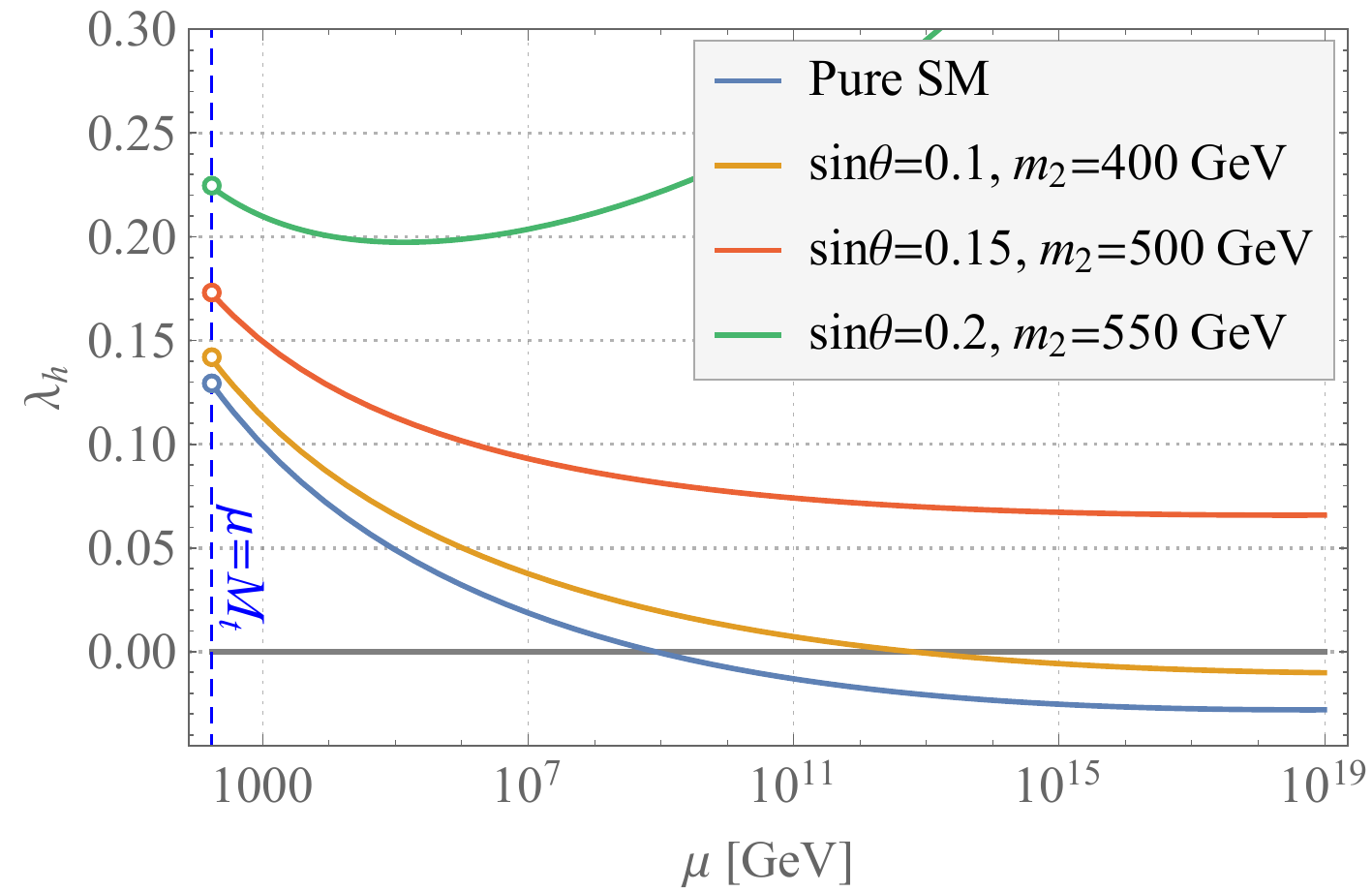} 
\caption{ Higgs quartic coupling scale dependence for different Higgs--inflaton mixing angles. The figure is from Ref.\,\cite{Ema:2017ckf}.
\label{coupling-stab}}
\end{center}
\end{figure}

 The scalar potential  is generally minimized at a non--zero inflaton VEV $\langle \phi \rangle= w $. To simplify the analysis,
 it is convenient to eliminate it by redefining
 the inflaton field \cite{Espinosa:2011ax}, 
  \begin{equation}
  \phi^\prime = \phi - w \;,
  \end{equation}
 and the dimensionful parameters,\footnote{Note the differences between our convention and those of Refs.\,\cite{Espinosa:2011ax},\cite{Ema:2017ckf}.}
  \begin{eqnarray}
  &&  b_1^\prime = b_1 + \lambda_\phi w^3 + b_3 w^2 + m_\phi^2 w \, , \nonumber\\
     && b_3^\prime = b_3 + 3\lambda_\phi w  \, , \nonumber\\
       &&  \sigma_{\phi h}^\prime = \sigma_{\phi h} + \lambda_{\phi h} w \, , \nonumber\\
         && m_\phi^{\prime \, 2}=m_\phi^2 +3 \lambda_\phi w^2 +2 b_3 w  \, , \nonumber\\
           && m_h^{\prime \, 2}=m_h^2 + {1\over 2} \lambda_{\phi h} w^2 + \sigma_{\phi h} w  \, .
  \end{eqnarray}
  The new field has a vanishing  VEV,  $\langle \phi^\prime \rangle= 0 $.
 One then proceeds to minimizing the potential along the lines of Section\;\ref{sec-z2}.
  The mass eigenstates are given by
   \beq
\left(\,\begin{matrix} h_1 \\ h_2\end{matrix}\,\right) = 
\left(\,
\begin{matrix} 
\cos\theta & -\sin\theta \\
\sin\theta & \cos\theta
\end{matrix}
\,\right)
\left(\,\begin{matrix} h-v \\ \phi^\prime\end{matrix}\,\right)\,.
\eeq
One finds the following relations among the mass eigenvalues $m_1, m_2$ and $\theta$, 
\begin{eqnarray}  
\label{eq:constr}
&& 2\lih v^2 = m_1^2\cos^2\theta + m_2^2\sin^2\theta\,, \\
&&{1\over 2} \lambda_{\phi h} v^2 + m_\phi^{\prime\, 2} = m_1^2\sin^2\theta + m_2^2\cos^2\theta \,, \\
&&\sigma_{\phi h}^\prime v = \dfrac{\sin2\theta}{2}\round{m_2^2 - m_1^2}
\,.  
\end{eqnarray}
 The first relation implies that for $m_2 > m_1$, the Higgs self--coupling is larger than that in the SM. This leads to a stable Higgs potential for a significant range of $m_2$ and $\theta$. 
 Fig.\;\ref{coupling-stab} illustrates the impact of these parameters on the RG evolution of $\lambda_h$. The stabilizing action is twofold: first, the boundary condition at the top quark mass scale  $M_t$ changes;
 second, the Higgs portal coupling contributes positively to the $\lambda_h$--running (see Section\,\ref{RG}).
 Unlike in the previous Section, here we consider substantial  mixing angles: the  upper bound from experiment is about 0.3 unless the inflaton is too light or too heavy (see Section\;\ref{Z2-pheno}).
  We see that 
  a rather small  mixing of order $10^{-1}$ is sufficient to stabilize the Higgs potential for a sub--TeV inflaton. 
 Eq.\,\ref{eq:constr} shows that even a tiny mixing can lead to the desired result for a large enough $m_2$. This is consistent with electroweak precision measurements which impose the bound on $\theta$
 (\ref{theta-log}) growing stronger only logarithmically with $m_2$. 
 
 Our conclusion is that the Higgs--inflaton mixing, which is expected on general grounds, can fully stabilize the scalar potential. It is, however, not obvious that such a mixing is allowed by cosmology, i.e.
  inflation itself. In the next subsection, we analyze this issue in detail.
 
 \subsubsection{Higgs--inflaton mixing vs cosmology}  

 The main ingredient in the following discussion is the energy scale separation between the Early Universe phenomena  and the mixing effects which occur after the electroweak phase transition. Indeed,
 in traditional models, inflation and reheating take place at large field values, while the field mixing  discussed above is relevant when the fields are close to their vacuum values. It is therefore clear that
 these phenomena are not $directly$ related. 
 
 To illustrate this point let us take the general Higgs portal Lagrangian including the non--minimal scalar--gravity couplings (\ref{L-J}) \cite{Ema:2017ckf}. At large  $h$ and $\phi$ field values, we may neglect all the dimensionful parameters leaving us with 5 variables $\lambda_i$ and $\xi_i$. This is the Higgs portal inflation set--up discussed in Section\,\ref{HiggsPortalInflation}.
 Depending on the relations among these parameters, $h$ and/or $\phi$ can drive inflation. For example,
 inflation takes place along the $\phi$ direction for 
 \begin{equation}
 \xi_\phi \gg \xi_h 
 \end{equation}
 as long as $2\lambda_\phi \xi_h - \lambda_{\phi h } \xi_\phi <0$. The scalar potential is then stabilized at $\tau\equiv h/\phi =0$.
 Taking $\xi_\phi \gg 1$, the inflationary potential in Planck units (${\rm M_{Pl}}=1$) is 
 \begin{equation}
V_E \simeq {\lambda_\phi \over 4 \xi_\phi^2} \left( 1-e^{-\sqrt{2\over 3} \chi} \right)^2\;, 
\end{equation} 
 with $\chi \simeq \sqrt{3\over 2} \ln \xi_\phi \phi^2 $.  The couplings are subject to a number of constraints. First, the CMB normalization requires $\lambda_\phi/(4 \xi_\phi^2) \simeq 10^{-10}$.
 Second, the scale of inflation $(\lambda_\phi/4 \xi_\phi^2)^{1/4} $   should not exceed the unitarity cutoff  $1/\xi_\phi$. This implies the following bounds on the running couplings:
  \begin{equation}
\xi_\phi (H) \lesssim 300 ~~, ~~\lambda_\phi (H) \lesssim 4 \times 10^{-5} \;,
\end{equation} 
 where we (loosely) identify the running scale with $H$.
 To avoid large radiative corrections to the inflaton potential, we also require $\lambda_{\phi h}^2 /16\pi^2 <\lambda_\phi$ at the inflationary scale. This results in $\lambda_{\phi h} (H) \lesssim  10^{-2}$. 
On the other hand,  the Coleman--Weinberg potential induced by $\sigma_{\phi h}$ is suppressed by $(\sigma_{\phi h}/\phi)^2$ and can be neglected.
 
 As discussed in Section\,\ref{HiggsPortalInflation}, this model predicts the spectral index and the tensor to scalar ratio that fit 
  the PLANCK data very well.   The postinflationary evolution is also ``healthy'':  a significant $\lambda_{\phi h}$ allows for efficient Higgs production and subsequent thermalization of the Higgs--inflaton system
  \cite{Ema:2017ckf}.
  Eventually,  due to the $\phi h^2 $ coupling, the inflaton decays into the Higgs pairs or other SM states, if the main channel is kinematically closed. The Higgs--inflaton mixing appears only after the electroweak phase transition making no significant impact on the Early Universe cosmology. We therefore conclude that the mixing can have a stabilizing effect on the Higgs potential while being consistent with 
  standard cosmology.

\section{Higgs and singlet condensates}
\label{condensates}

In the Early Universe, light scalars develop a background value comparable to the Hubble rate \cite{Starobinsky:1994bd}. This may apply to the Higgs \cite{Enqvist:2013kaa},\cite{Enqvist:2015sua}
as well as scalar dark matter \cite{Enqvist:2014zqa},\cite{Nurmi:2015ema}.  Let us consider under what circumstances this is possible.

Given the uncertainty in the top quark mass, one cannot exclude the possibility that the SM Higgs potential is stable at large field values.  If, in addition, the Higgs--inflaton 
interaction and the non--minimal Higgs coupling to gravity are negligible, the Higgs field is light and experiences large fluctuations during inflation.
As discussed in Section\;\ref{SM-stability}, the  evolution of its long wavelength  (``background'')  modes    is governed by the Langevin equation \cite{Starobinsky:1994bd}
\begin{equation}
 {d h \over dt}  + {1\over 3H} \, {dV (h) \over dh} = \zeta (t) \;,
 \end{equation}
where $h = \sqrt{ 2 H^\dagger H}$   and $\zeta (t)$ represents random noise,
$\langle \zeta (t) \zeta(t^\prime) \rangle = {H^3 \over 4\pi^2}\, \delta (t-t^\prime) \;.$
 However, unlike in Section\;\ref{SM-stability}, the value of the  potential plays an important role.  
The above picture applies if $h$ is ``light'' enough in this regime, $V^{\prime \prime} (h) < 9 H^2/4$, such that super--Hubble fluctuations are generated. The field experiences a random walk and its various values can be found 
according to the  (late--time) probability distribution \cite{Starobinsky:1994bd}
 \begin{equation}
 \rho (h) = C \; \exp \left(-  {2 \pi^2 \lambda_h h^4 \over 3 H^4} \right)
 \end{equation}
with the variance
\begin{equation}
\langle h^2 \rangle \simeq 0.13 \, {H^2 \over \sqrt{\lambda_h}} \;.
\label{h2}
 \end{equation}
 The corresponding  induced Higgs mass squared is given by $3 \lambda_h \langle h^2 \rangle $, which is far below $9 H^2/4$ as required by self--consistency of our approximation. Also,
 the condensate (\ref{h2}) contributes very little to the energy balance of the Universe since its fraction is suppressed by $H^2 / M_{\rm Pl}^2$.

If the Higgs initial value is large so that    $V^{\prime \prime} (h) >  9 H^2/4$, it evolves classically to smaller values as long as its energy density does not affect inflation.\footnote{This is not strictly speaking necessary before the last 60 e--folds of inflation. At the early stage, the energy density can be dominated by the Higgs field, see the discussion in Section~\ref{stab-infl}.} Then, its effective mass becomes small
and the above considerations apply. 

During inflation, the field is ``frozen'' in the sense that $\langle h^2 \rangle$ remains constant.
After inflation, the Hubble rate decreases and the Higgs evolution gets soon dominated by the classical term. The field  oscillates    around $h=0$ in a quartic potential inducing 
  an oscillating mass term for the SU(2)  gauge bosons.
  This results  in resonant production of the latter   and the condensate  decays quickly. 
  Lattice simulations show that  this occurs after ${\cal O}(10)$ oscillations \cite{Enqvist:2015sua} (see \cite{Figueroa:2015rqa} for the Abelian case). Since the Higgs energy density is very small, in standard cosmology the Higgs condensate evaporates  without leaving any significant trace.\footnote{In some cases, the Higgs VEV modulation may lead to kinematic blocking of some of the reheating channels 
thereby  inducing  observable density perturbations \cite{Karam:2020skk},\cite{Litsa:2020rsm}. The initial studies of this phenomenon can be found in \cite{Lu:2019tjj}.}

Similar considerations apply to other light scalars, except these can be stable and constitute  secluded dark matter \cite{Markkanen:2018gcw},\cite{Cosme:2018nly},\cite{Alonso-Alvarez:2018tus}. 
Although such scenarios are subject to strong isocurvature perturbation constraints, they can be viable for certain parameter choices.
Suppose we have a light scalar  $s$ with mass $m_s$ and self--interaction $\lambda_s s^4/4  $. 
At weak coupling, the field goes through the following main stages in its evolution: (1) the condensate $\langle s^2 \rangle \simeq 0.13 \, {H^2 \over \sqrt{\lambda_s}}  $
forms during inflation, (2) after inflation, the field starts oscillating around $s=0$ with a decreasing amplitude $\propto a^{-1}$,
(3) the field enters a non--relativistic regime, where the mass term dominates the potential. 
Since $s$ is stable and non--relativistic, it plays the role of cold dark matter.
If the coupling is not too small, the scalar can also thermalize due to self--interactions and its final abundance would then be  dictated by the usual freeze--out. 
The scalar condensate can account for all of the dark matter for a range of the couplings and masses. In particular, if no thermalization occurs, simple scaling arguments show that the correct relic density of DM is reproduced for \cite{Markkanen:2018gcw}
\begin{equation}
\lambda_s^{5/8} ~ {{\rm GeV} \over m_s} \sim 10^7 \times \left(   {H_{\rm end} \over M_{\rm Pl}}  \right)^{3/2} \;,
 \end{equation}
where $H_{\rm end}$ is the Hubble rate at the end of inflation. An analogous relation holds for thermal $s$ as well, although the coupling dependence of the left hand side  becomes $\lambda_s^{3/8} $.
An important feature of this scenario is the presence of isocurvature perturbations since the inflaton fluctuations are not correlated with those of $s$
\cite{Kainulainen:2016vzv}.
The PLANCK observations \cite{Akrami:2018odb} put strong constraints on such perturbations, yet tiny, $\lesssim {\cal O} (10^{-19})$, as well as quite large, ${\cal O}(1)$, couplings remain allowed \cite{Markkanen:2018gcw}. 
Similarly, one finds that a massive scalar with no self--interaction can also constitute all of the dark matter \cite{Tenkanen:2019aij}.

 \section{Postinflationary Higgs and dark matter production}
 
 Particle production after inflation constitutes an essential ingredient in our understanding of the Early Universe physics. 
 It is necessary in order to explain how the Universe became radiation--dominated and how dark matter was created.
 In general, the SM fields and dark matter can be produced at different stages of the evolution via different mechanisms.
 If the relevant couplings are strong, DM thermalizes erasing the memory of these mechanisms. This possibility 
 will be discussed in Sections\;\ref{HiggsPortalDM} and  \ref{FIMP}. On the other hand, at weak couplings, 
 the total number of DM quanta created shortly after inflation  
 remains approximately constant throughout the history of the Universe and the system retains some memory of the DM production mechanism. 
 In what follows, we focus on this possibility of  {\it non--thermal} dark matter.
 
 The only renormalizable inflaton  couplings to the Standard Model are
 \begin{equation}
 V_{\phi h} = {1\over 4} \lambda_{\phi h} \phi^2 h^2 + {1\over 2} \sigma_{\phi h} \phi h^2 \;.
 \end{equation}
 Therefore, these interactions are expected to play the leading role in reheating the Universe. 
 The minimal option to account for dark matter is to introduce a real scalar $s$  of mass $m_s$ with the interactions\footnote{One may also entertain the possibility of the inflaton and dark matter being the same field \cite{Lerner:2009xg}, however this requires tuning the inflaton mass to half the Higgs mass \cite{Lebedev:2021zdh}.}
 \begin{equation}
 V_{\phi s} = {1\over 4} \lambda_{\phi s} \phi^2 s^2 + {1\over 2} \sigma_{\phi s} \phi s^2 \;.
 \end{equation}
 Since we are interested in non--thermal DM, we may neglect possible Higgs--DM interaction as well as its self--interaction.
 The above couplings are assumed sufficiently small so that they do not induce   large loop corrections to the inflaton potential
 and DM self--coupling. On the other hand, if they are to represent the main DM production mechanism, these
 couplings  cannot be too small: the scalar condensate $\langle s^2 \rangle $ and other sources of dark matter should be subleading.
 
  The above interactions are sufficient to $fully$ describe both reheating and dark matter production.
  Depending on the model, 
   different combinations of the couplings can play the leading role.  To avoid stable or long--lived inflaton relics, it is natural to assume that $\sigma_{\phi h}$ is significant and responsible
   for producing the SM particles. Then, dark matter can be generated  primarily either via $\lambda_{\phi s}$ or $\sigma_{\phi s}$ couplings, both of which 
   yield viable options to be studied below. The results are sensitive to the inflaton potential during the {\it inflaton oscillation epoch}, which can be very different from the potential at large field values.

 Particle production can take place in different  regimes: it can be purely perturbative or semiclassical via
 parametric or tachyonic resonance. Resonant particle production  was discussed in Section\;\ref{vacuum-stability-after-inflation},
 so let us now consider the perturbative mechanisms.

 \subsection{Particle production by an oscillating background}
 
 A time--varying classical field can lead to particle production. This applies, in particular, to an oscillating inflaton background \cite{Dolgov:1989us},\cite{Shtanov:1994ce}. Let us consider perturbative production of the Higgs quanta following
 \cite{Ichikawa:2008ne} and extend that analysis to the massive Higgs case. We take 
the  Higgs--inflaton coupling to be sufficiently small so that it is outside 
  the broad resonance regime.
 To focus on the essentials of the mechanism, we also neglect the Universe expansion at the initial stage.

 \subsubsection{Quartic  Higgs--inflaton interaction}
 
 Let us start with the quartic  Higgs--inflaton interaction
 \begin{equation}
\Delta V = {1\over 4} \lambda_{\phi h} \phi^2 h^2 \;,
 \end{equation}
 with a single Higgs d.o.f.  and write 
 \begin{equation}
\phi^2(t) = \sum_{n=-\infty}^\infty \zeta_n e^{-in \omega t} \;,
 \end{equation}
 where the coefficients $\zeta_n$ are time--independent.
 We are interested in the amplitude of creating two Higgs quanta with momenta $p,q$ from the background, i.e.  the transition $|0 \rangle \rightarrow |p,q\rangle $.
  Using the Peskin--Schroeder conventions  \cite{Peskin:1995ev} for the creation and annihilation operators as well as the state normalization, we have
   \begin{equation}
-i \int_{-\infty}^\infty dt \langle f | V(t) | i \rangle = - i \,{ \lambda_{\phi h} \over 2 } \, (2\pi)^4 \delta ({\bf{p}} + {\bf{q}})  \sum_{n=1}^\infty \zeta_n \delta(E_p +E_q -n \omega).
 \end{equation}
This corresponds to the invariant amplitude for the $n$-th mode decay ${\cal M}_n = -\lambda_{\phi h} \zeta_n /2$. Taking into account the symmetry of the final state and 
integrating over the phase space $\Pi$ produces the usual factor $  |{\bf p}|/(8\pi E_{\rm cm}) \times  \theta(E_{\rm cm} -2m_h)$, where $E_{\rm cm} =n \omega$.
The reaction rate per unit volume is 
   \begin{equation}
 \Gamma = \sum_{n=1}^\infty \Gamma_n =   \sum_{n=1}^\infty  {1\over 2} \int |{\cal M}_n|^2 d \Pi_n =
 {\lambda_{\phi h}^2 \over 64 \pi  } \sum_{n=1}^\infty  |\zeta_n|^2 \sqrt{1- \left(   {2m_h \over n \omega} \right)^2} \; \theta(n\omega -2 m_h) \;.
 \label{Gamma-ss}
  \end{equation}
 In our approximation, the coefficients $\zeta_n$ are constant, so the decay is neglected. 
 Nevertheless, one can estimate the inflaton decay rate by resorting to energy conservation. A unit volume loses $\Gamma \Delta t  \langle E \rangle$ of energy in infinitesimal time period $\Delta t$,
 where $\langle E \rangle$ is the average energy of the decay products,
   \begin{equation}
  \langle E \rangle = {\sum_n  n\omega  \Gamma_n\over \sum_n \Gamma_n } \;.
    \end{equation}
 By virtue of energy conservation, this energy loss can also be expressed in terms of the inflaton decay rate $\Gamma_\phi =-{1\over \rho_\phi} \, {d\rho_\phi \over dt}$ and the inflaton energy density $\rho_\phi$,
  \begin{equation}
  \rho_\phi \Gamma_\phi \Delta t = \langle E \rangle  \Gamma \Delta t   \;.
    \end{equation}
 Hence,
  \begin{equation}
 \Gamma_\phi  =
 {\lambda_{\phi h}^2 \omega \over 64 \pi \rho_\phi  } \; \sum_{n=1}^\infty  n  |\zeta_n|^2 \sqrt{1- \left(   {2m_h \over n \omega} \right)^2} \; \theta(n\omega -2 m_h) \;.
  \end{equation}
 In the massless limit, one recovers the result of \cite{Ichikawa:2008ne}.
 Let us analyze what this implies for specific inflaton potentials.  \\ \ \\
 {\bf \underline{Quadratic inflaton potential.}} 
 Suppose $V(\phi) = {1\over 2} m_\phi^2 \phi^2$, so that $\phi(t) = \phi_0 \cos m_\phi t$, and neglect the Higgs mass, $m_h \rightarrow 0$.
 Since $\rho_\phi= {1\over 2} m_\phi^2 \phi_0^2$
 and only one term in the sum contributes  ($n=1, \omega=2m_\phi$), we get\footnote{We find a factor of 2 discrepancy with the corresponding result in  \cite{Ichikawa:2008ne}.}
  \begin{equation}
 \Gamma_\phi = {\lambda_{\phi h}^2 \phi_0^2 \over 256 \pi m_\phi  } \;.
     \end{equation}
     This formula is trivially extended to the massive Higgs case, where the decay is only allowed for $m_\phi > m_h$.
 \\ \ \\
 {\bf \underline{Quartic inflaton potential.}}  Take $V(\phi) = {1\over 4} \lambda_\phi \phi^4$. The equation of motion for the inflaton  with the initial condition $\phi(0)=\phi_0$  is solved by the Jacobi cosine,
  \begin{equation}
\phi(t)= \phi_0  \,{\rm cn} \left(\sqrt{\lambda_\phi} \phi_0 \, t, {1\over \sqrt{2}}\right)=
{\sqrt{\pi} \Gamma \left( {3\over 4}\right) \over   \Gamma \left( {5\over 4}\right)}\, \phi_0 \, \sum_{n=1}^\infty
 \left( e^{i(2n-1)\omega t} +       e^{-i(2n-1)\omega t}     \right) \, { e^{-(\pi/2) (2n-1)}   \over 1+ e^{-\pi (2n-1)}      }\;,
      \end{equation}
where
\begin{equation}
\omega = {1\over 2} \sqrt{\pi \over 6 } { \Gamma \left( {3\over 4}\right) \over  \Gamma \left( {5\over 4}\right)}\, m_\phi^{\rm eff}
\label{omega}
     \end{equation}
and 
\begin{equation}
  m_\phi^{\rm eff} = \sqrt{3 \lambda_\phi} \phi_0 \;.
     \end{equation}
The above sum is dominated by the first term with $n=1$, while the second term is suppressed by $e^{-\pi}$. Keeping just the leading term, one finds that $\phi^2(t)$ has frequency $2\omega$
and in the limit $m_h \rightarrow 0\;$\footnote{We find a factor of 2 discrepancy with the corresponding result in  \cite{Ichikawa:2008ne}.},
\begin{equation}
 \Gamma_\phi =  C \;{\lambda_{\phi h}^2  \over 64 \pi   }  \, {\phi_0 \over \sqrt{\lambda_\phi} }\,,
     \end{equation}
with $C \simeq 0.4$. This result is similar to that for the quadratic potential  if one assigns the inflaton an effective mass $m_\phi^{\rm eff}$.

In case of a heavy Higgs, $m_h > \omega$, the  decay  is exponentially suppressed. The production is dominated by the $n$--th mode, where $n$ is the closest integer to $m_h/\omega$ from above.
Taking $n \sim m_h/\omega$, the decay width scales approximately as
 \begin{equation}
 \Gamma_\phi \sim     \Gamma_\phi \Bigl\vert_{m_h=0} \; \left(  m_h\over \omega   \right)^2 e^{-2\pi \left({m_h \over \omega}-1\right)} \;.
     \end{equation}

 \subsubsection{Trilinear Higgs--inflaton interaction}
 
Let us extend the above results to  the trilinear  Higgs--inflaton interaction
 \begin{equation}
\Delta V = {1\over 2} \sigma_{\phi h} \phi h^2 \;,
 \end{equation}
 with a single Higgs d.o.f.  and expand
 \begin{equation}
\phi(t) = \sum_{n=-\infty}^\infty \xi_n e^{-in \omega t} \;.
 \end{equation}
  Repeating the decay rate calculation, one finds
 \begin{equation}
 \Gamma_\phi  =
 {\sigma_{\phi h}^2 \omega \over 16 \pi \rho_\phi  } \; \sum_{n=1}^\infty  n  |\xi_n|^2 \sqrt{1- \left(   {2m_h \over n \omega} \right)^2} \; \theta(n\omega -2 m_h) \;.
  \end{equation}
Consider now the quadratic and quartic inflaton potentials, as before.
\\ \ \\
 {\bf \underline{Quadratic inflaton potential.}}  For $V(\phi) = {1\over 2} m_\phi^2 \phi^2$ and neglecting the Higgs mass, one finds
 \begin{equation}
 \Gamma_\phi = {\sigma_{\phi h}^2   \over 32 \pi m_\phi  } \;.
     \end{equation}
This is identical to the to the decay width for the quantum process $\phi \rightarrow h h $. Again, this process is kinematically forbidden for $m_\phi <2 m_h$.
 \\ \ \\
 {\bf \underline{Quartic inflaton potential.}}  For  $V(\phi) = {1\over 4} \lambda_\phi \phi^4$ in the massless Higgs limit, we have 
 \begin{equation}
 \Gamma_\phi =  \tilde C\, {\sigma_{\phi h}^2   \over 32 \pi   } \, {1\over \sqrt{\lambda_\phi} \phi_0}\;,
     \end{equation}
 where $\tilde C \simeq 1.6$. Here we have included only the dominant contribution from the $n=1$ mode. We observe that the decay rate is similar to that of 
 the fully quantum reaction $\phi \rightarrow hh$, where $\phi$ is assigned an effective mass $ m_\phi^{\rm eff} = \sqrt{3 \lambda_\phi} \phi_0 $.
 
 The decay is allowed even for $\omega < 2m_h$, although it is exponentially suppressed. The rate scales as
 \begin{equation}
 \Gamma_\phi \sim     \Gamma_\phi \Bigl\vert_{m_h=0} \;   {2m_h\over \omega}    \, e^{-\pi \left({2m_h \over \omega}-1\right)} \;,
     \end{equation}
where we have included the dominant contribution of the mode with $n \sim 2m_h/\omega$. \\ \ \\
 The effect of the Universe expansion in the above considerations can be taken into account adiabatically by including time dependence into the oscillation amplitude $\phi_0 \rightarrow \phi_0(t)$. 
 The equations of motion for the inflaton in an expanding Universe can be solved exactly and one finds that the amplitude after a few oscillations evolves according to 
  \cite{Felder:2000hj}
 \begin{eqnarray}
&& \phi^2~ {\rm potential :}~~\phi_0 \rightarrow \phi_0 / a(t)^{3/2} ~~,~~ a(t) = a_0 \,t^{2/3} \;, \nonumber\\
&& \phi^4~ {\rm potential :}~~\phi_0 \rightarrow \phi_0 / a(t)^{}  ~~~~~\, , ~~ a(t)=a_0 \,t^{1/2} \;.
     \end{eqnarray}
 Note that, in the latter case, this also implies a time dependent frequency $\omega$ according to (\ref{omega}).
 
 The expansion also has the effect of redshifting the momenta of the decay products. Consequently, the narrow resonance resulting from the Bose enhancement of the decay amplitude
 becomes inefficient and often can be neglected (see e.g. \cite{Mukhanov:2005sc}).
 
 Finally, if the Higgs VEV can be neglected, the rates should be multiplied by 4 to account for all of the Higgs degrees of freedom.
 
 \subsection{Effect of  the non--minimal inflaton--gravity coupling}
 
 The inflaton potential during the oscillation epoch can be more complicated than $\phi^2$ or $\phi^4$. This is the case, in particular,
 in the presence of  a non--minimal inflaton--gravity coupling. Nevertheless, the particle  production  rates can be extrapolated to this 
 case as well.

 Consider inflation driven by a non--minimal inflaton--gravity coupling, as in 
 Section \ref{HiggsPortalInflation}. Let us focus on a single field case,
 \begin{equation}
 \Omega = 1 + \xi_\phi \phi^2 \;,
 \end{equation} 
 where we use the Planck units ${\rm M_{Pl}}=1$, and treat the Higgs as a heavy spectator ($\xi_h =0$). 
 After inflation, $\phi$ takes on values below the Planck scale and  we are interested in its full field range. The inflaton  kinetic function, in general, is given by $(\Omega +6 \xi_\phi^2 \phi^2 )/\Omega^2$,
 so the canonically normalized $\chi$ satisfies
 \begin{equation}
{d \chi \over d \phi} = \sqrt{  1 + \xi_\phi (1+6 \xi_\phi) \phi^2  \over  (1 +\xi_\phi \phi^2)^2 } \;.
\label{dchi/dphi}
 \end{equation}
 The solution reads  \cite{GarciaBellido:2008ab}
 \begin{equation}
  \, \chi (\phi) = \sqrt{ 1+6 \xi_\phi \over \xi_\phi} \, \sinh^{-1} \left(    \sqrt{(1+6 \xi_\phi) \xi_\phi} \, \phi  \right)
 -\sqrt{6 } \,  \sinh^{-1} \left(       {\sqrt{6} \xi_\phi    \phi \over \sqrt{1+ \xi_\phi \phi^2}}            \right)\;.
  \end{equation}
This expression applies for all $-\infty < \phi < \infty$ and  
can be simplified in various limits using $ \sinh^{-1} x = \ln \left(x + \sqrt{x^2 +1}\right) $.
 In the inflationary regime, i.e. for  $\sqrt{\xi_\phi} \phi \gg 1$, we recover  $\chi = {\sqrt{ 1+ 6 \xi_\phi \over \xi_\phi}} \ln \sqrt{\xi_\phi} \phi $.

Let us consider in more detail the large non--minimal coupling limit.
 For $6\xi_\phi \gg 1$, we have 
 \begin{equation}
|  \chi | = \sqrt{3\over 2} \ln (1+ \xi_\phi \phi^2) ~~,~~ V_E(\chi)= {\lambda_\phi \over 4 \xi_\phi^2 }\, \left(     1- e^{ \sqrt{2\over 3} |\chi |} \right)^2
    \end{equation}
 in the entire field range of $\phi$ as long as $1/(6 \xi_\phi)$ can be neglected. Since the exact expression assures  that the sign of $\chi$ coincides with the sign of $\phi$,
 in the limit $ \xi_\phi \phi^2 \ll 1$, we have $\chi \simeq \pm \sqrt{3\over 2} \xi_\phi \phi^2$. At very small $\phi$, however,  terms of order $1/(6 \xi_\phi)$ become important and
 the above approximation breaks down. In this case, one finds instead 
 $\chi \simeq \phi$ which can also be seen directly from (\ref{dchi/dphi}).
 Thus, after inflation we have
 \begin{equation}
\chi \simeq \left \{
  \begin{tabular}{ccc}
 $\phi$  &  for & $\phi^2 \ll {1\over 6 \xi_\phi^2 } ~,$ \\
  $\pm \sqrt{3\over 2} \xi_\phi \phi^2$ & for   & ${1\over 6 \xi_\phi^2 } \ll \phi^2 \ll  {1\over  \xi_\phi } ~.$
  \end{tabular}
\right. 
\end{equation}
  The corresponding scalar potential in these regimes is
\begin{equation}
V_E(\chi) \simeq \left \{
  \begin{tabular}{ccc}
 ${\lambda_\phi \over 4}   \chi^4$  &  for & $|\chi |\ll {1\over 2 \xi_\phi } ~,$ \\
  $ {\lambda_\phi \over 6 \xi_\phi^2 } \chi^2 $ & for   & ${1\over 2 \xi_\phi } \ll |\chi | \ll  1  ~.$
  \end{tabular}
\right. 
\end{equation}   
The potential  shape appears quite peculiar: at larger field values, it is quadratic, while at smaller values, it is quartic.   This entails different particle production regimes at 
early and late times. Our previous results for the production rates, therefore,  apply to the case at hand as long as the oscillation amplitude falls into an appropriate field range.

 It is also possible that dark matter itself has a substantial non--minimal coupling to gravity. 
 Since the curvature oscillates in the inflaton oscillation epoch, a tachyonic instability ensues resulting in efficient dark matter production
 even in the absence of the direct inflaton--DM coupling \cite{Fairbairn:2018bsw}.

 \subsection{Resonant vs non--resonant dark matter production and reheating }

Dark matter and SM radiation can be produced by entirely different mechanisms. Consider the set--up where the dominant Higgs--inflaton coupling is trilinear, while the leading 
DM--inflaton coupling is quartic,
\begin{equation}
V_{\phi h} = {1\over 2} \sigma_{\phi h} \, \phi h^2 ~~,~~ V_{\phi s} = {1\over 4} \lambda_{\phi s} \phi^2 s^2 \;,
\end{equation}
where $s$ is a scalar DM candidate. Suppose that the Higgs--DM and DM self--interaction are negligible such that it never thermalizes. In this case, the DM abundance
is determined by a short period of  DM production right after inflation, while the Universe reheats via perturbative inflaton decay into the Higgses.
Depending on the couplings, the dynamics of the system can be quite complicated due a non--trivial resonance structure and rescattering effects.
Thus, lattice simulations are often necessary to make reliable predictions \cite{Khlebnikov:1996mc}. Related studies have appeared in \cite{Figueroa:2016wxr}, while much of  the following material
 relies on the analysis of  \cite{us}. Further improvements can be facilitated by the  advanced computational tools of \cite{Figueroa:2020rrl},\cite{Figueroa:2021yhd}.

 \subsubsection{$\phi^2$ inflaton potential}
 
 Let us assume the quadratic inflaton potential {\it during preheating}, $V_\phi = {1\over 2} m_\phi^2 \phi^2$, where $m_\phi$ is a free parameter (not directly related to inflationary
 predictions), and consider light dark matter, $m_\phi \gg m_s$.
 For small enough $\sigma_{\phi h} $ and significant $\lambda_{\phi s}$, the leading particle production mechanism after inflation is the parametric resonance whose strength is controlled
 by (see Section\,\ref{Mathieu-equation})
 \begin{equation}
q= {\lambda_{\phi s} \Phi^2 \over 2 m_\phi^2} \gg 1 \;,
\end{equation}
where $\Phi$ is the time--dependent amplitude of inflaton oscillations,  $\phi(t) \simeq \Phi(t) \cos\, m_\phi t$ and $\Phi (t) = \Phi_0 /(m_\phi t) = \Phi_0 /a^{3/2}$. \footnote{ In the relation $m_\phi t = (a/a_0)^{3/2}$, we take $a_0=1$ for simplicity.
By means of $dt = da/ (aH)$, this implies $\Phi_0 \sim 1$ in Planck units, which we
assume  in this section. A generalisation is straightforward \cite{us}.   }
 The resonance is active until either $q$ reduces to about 1 or the produced DM induces a large inflaton mass (``backreaction'') \cite{Kofman:1997yn},
 \begin{eqnarray}
  q &\simeq& 1~~~~~~~~~~~~({\rm weak~coupling}) \;,\nonumber\\
  \sqrt{\lambda_{\phi s} \langle s^2 \rangle  \over 2 m_\phi^2} &\sim& 1 ~~~~~~~~~~~~({\rm strong~coupling}) \;.
 \end{eqnarray}
 In the first case, the energy density of the Universe is dominated by the non--relativistic inflaton and the total number of the DM quanta remains constant
 after the resonance has ended. In the second case,
 the system can be highly relativistic with the inflaton and DM contributing roughly 50\%  each to the energy balance. 
 Due to rescattering,
 the total number of DM particles can still grow  significantly
 after the end of the resonance, although the growth rate is much slower than that during the resonance. To evaluate the resulting DM output, one has to resort to lattice simulations.
 Since the inflaton is heavy, the system becomes non--relativistic and inflaton--dominated some time after the end of rescattering. Subsequently, the perturbative inflaton
 decay into the Higgses reheats the Universe.

 The   dark matter abundance   is normally parametrized by $Y = n/s_{\rm SM}$, where $n$ is the DM number density and $s_{\rm SM}$ is the SM entropy density,
 $s_{\rm SM}= 2\pi^2 g_{*s} T^3/45$, with   $g_{*s}$  being the number of degrees of freedom contributing to the entropy. The observed DM density corresponds to
 $Y_\infty = 4.4. \times 10^{-10} \; {\rm GeV}/m_s$.
 The abundance $Y$ remains constant after reheating and thus can be evaluated at the reheating  stage defined by 
 \begin{equation}
\Gamma (\phi \rightarrow h_i h_i) \simeq H_R ~~~,~~~ \Gamma (\phi \rightarrow h_i h_i) = {\sigma_{\phi h}^2 \over 8\pi m_\phi} \;,
\end{equation}
 where $H_R$ is the Hubble rate at reheating and 
 4 Higgs d.o.f. have been taken into account. 
  At this point, the inflaton energy density converts into radiation,  
   \begin{equation}
H_R = \sqrt{\rho \over 3} = \sqrt{\pi^2 g_* \over 90} T_R^2 \;
\end{equation}
 in Planck units, $M_{\rm Pl}=1$, and 
  $g_*$ being the number of d.o.f. contributing to the energy density. 
 
 In order to cover both the weak and strong coupling regimes, let us express $Y$
   directly in terms of the simulation output. The result depends on the duration of the relativistic phase after the resonance. Denoting by $a_e, a_*$ and $a_R$ the scale factors
   at the end of the simulation, at the beginning of the non--relativistic epoch and at reheating, respectively, the system evolves according to 
 \begin{equation}
a_e  \stackrel{\rm { rel}}  \longrightarrow         a_*     \stackrel{\rm nrel} \longrightarrow a_R \;,
\end{equation}
where in the first period  the equation of state is $w\simeq 1/3$, while in the second $w\simeq0$.\footnote{In reality, the equation of state and its evolution are more complicated \cite{Antusch:2020iyq}, yet
this does not affect the results in a significant way.}
During the relativistic phase, the DM contribution to the total energy density can be substantial, up to about 50\%, whereas in the second phase  it is negligible since DM remains relativistic.
Denoting 
\begin{equation}
\rho_e (s) =\delta\; \rho_e (\phi) \;,
\end{equation}
where $\rho_e (\phi)$ and $\rho_e (s)$ are the energy densities of the inflaton and dark matter at the end of the simulation, 
the corresponding Hubble rate   is given by $H_e = \sqrt{1+\delta}\,\sqrt{ \rho_e(\phi) /3} $. Since only the inflaton energy gets converted into SM radiation, 
the Hubble rate scaling 
  $H \propto a^{ - 3 (w+1)/2} $ implies 
\begin{equation}
H_R = {H_e  \over \sqrt{1+\delta} }\;   {a_e^2 \over a_*^2}  \; {a_*^{3/2} \over a_R^{3/2}} \;.
\end{equation}
Solving for $a_R$, one finds that the coupling $\sigma_{\phi h}$ required by the correct DM abundance is given by
\begin{equation}
\sigma_{\phi h} \simeq  1.6 \times 10^{-8} \; \sqrt{m_\phi}  \; {H_e^2 \over (1+\delta) \;n_e} \, {a_e\over a_*} ~~\left(   {{\rm GeV}  \over m_s} \right) 
\label{sigma-DM}
\end{equation}
 in Planck units, assuming $g_* \simeq 107$. Here $H_e$ and $n_e$ are the simulation output, while $a_e/a_*$ can be estimated by considering the average energy of the inflaton quantum:
 for example, one may require $p_e/(a_*/a_e) \sim m_\phi$, where $p_e$ is the typical particle energy or momentum at the end of the simulation. At weak coupling, $a_e/a_*$ is close to 1,
 while at stronger coupling this ratio is about 1/few or below. Since it is computationally challenging to  observe the system over a long period, $a_e/a_*$ can only be estimated and thus 
  introduces   uncertainty.
 
 When the coupling is sufficiently strong, the system reaches the state of quasi--equilibrium by the end of the simulation. In this case,  the above formula simplifies and takes on a  $universal$  form.
 Since the energy density is distributed almost equally between the inflaton and dark matter, $\delta \simeq 1$ and $3 H_e^2 /\left[(1+\delta)n_e\right]$ gives an average energy of the DM quantum
 at the end of the simulation. In quasi--equilibrium, it is similar to  the average energy of the inflaton quantum. The latter multiplied by $a_e/a_*$ gives $m_\phi $ by definition, so
 \begin{equation}
\sigma_{\phi h} \simeq  5 \times 10^{-9} \; m_\phi^{3/2} ~~\left(   {{\rm GeV}  \over m_s} \right) \;.
\label{sigma-DM-simplified}
\end{equation}
 The result is independent of the coupling and initial conditions, and also applies to other inflaton potentials, e.g. $\phi^4$. This is natural since memory of the initial condition is erased in equilibrium. 
 Note that the crucial assumption is that the inflaton and DM reach a quasi--equilibrium state in the relativistic regime, so the coupling must be strong enough and 
 the system must contain sufficient energy to make the inflaton and DM relativistic.

 At weaker coupling, an order of magnitude of the required $\sigma_{\phi h}$ can be estimated using the theory of broad parametric resonance. 
 Since DM makes  only a small contribution to the energy density,  the Hubble rate is determined by the inflaton potential, 
 $H = {1\over \sqrt{6}} m_\phi \Phi $.
 The number density of $s$ produced by the resonance can be approximated by \cite{Kofman:1997yn}
 \begin{equation}
n \sim {k_*^3 \over 64 \pi^2 a^3 \sqrt{\pi\mu m_\phi t }} \;e^{2\mu m_\phi t} \; ,
\end{equation}
where $k_* = \left(\sqrt{\lambda_{\phi s}/2} \;m_\phi \Phi_0 \right)^{1/2}$ and $\mu$ is the effective Floquet exponent. The scale factor and the corresponding $m_\phi t$ at the end of the resonance are found
by requiring $q \simeq 1$. Noting that $H^2 /n$ remains constant after that, one finds
\begin{equation}
\sigma_{\phi h} \sim 5 \times 10^{-7} \; {\sqrt{m_\phi \over \lambda_{\phi s}}} \, \Phi_0 \; e^{-2\mu \sqrt{\lambda_{\phi s}\over 2} \, {\Phi_0 \over m_\phi}}   ~~\left(   {{\rm GeV}  \over m_s} \right) \;.
\label{sigma-DM-simple}
\end{equation}
 in  Planck units. This result is   sensitive to the exact value of $\mu$ and thus only yields a ballpark estimate. For  the special case of $\phi^2$--inflation, $\Phi_0 \simeq 1$ and 
 $m_\phi \simeq 5\times 10^{-6}$, while the typical value of $\mu$ is between 0.15 and 0.175. This gives, for instance, $\sigma_{\phi h}\sim 10^{-13} - 10^{-12}$ at $\lambda_{\phi s} =10^{-7}$, $m_s=1$ GeV.
The resulting reheating temperature is relatively low,
 $T_R \sim 10^2 \sigma_{\phi h} \ll m_\phi$. 

\begin{figure}[h] 
\centering{   
\includegraphics[scale=0.46]{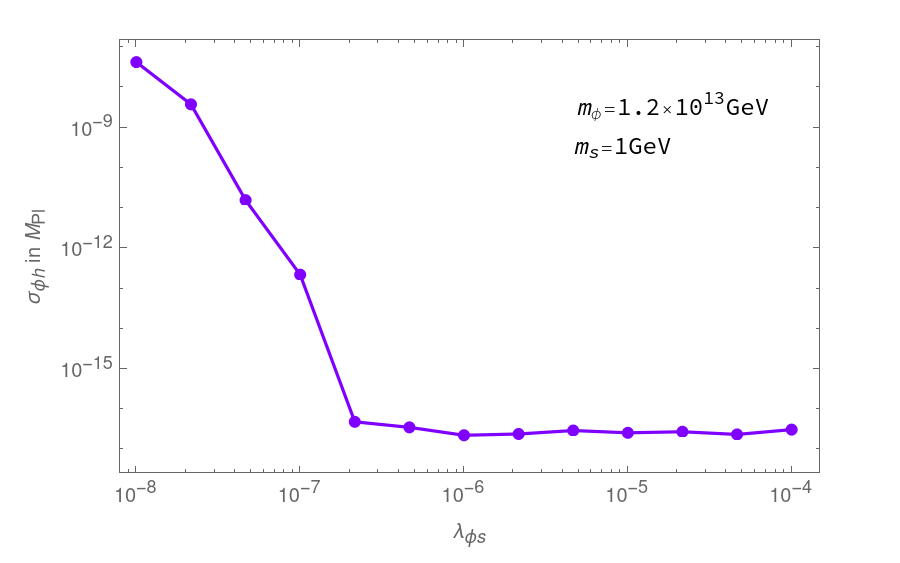} }
\caption{ \label{sigma-lambda}
$\sigma_{\phi h} $ vs $\lambda_{\phi s}$ producing the correct DM relic abundance for a $\phi^2$ potential, based on LATTICEEASY simulations. 
$\sigma_{\phi h} $ is given in units of  $M_{\rm Pl} =2.4 \times 10^{18}$ GeV and the initial inflaton value is chosen to be $\Phi_0 \simeq M_{\rm Pl}$.
The area above the curve is excluded by overabundance of dark matter.
}
\end{figure}

 Fig.\,\ref{sigma-lambda} shows the couplings producing the right amount of dark matter with mass 1 GeV based on Eq.\,\ref{sigma-DM} and lattice simulations. 
 The results for other DM masses are obtained by a simple rescaling as in Eq.\,\ref{sigma-DM}. 
 The weak coupling regime is reasonably well
 described by (\ref{sigma-DM-simple}), while for $\lambda_{\phi s}$ above $10^{-7}$ the behaviour becomes qualitatively different. The inflaton zero mode gets destroyed through rescattering and the energy
 balance is shared almost equally between the inflaton and dark matter. Thus, increasing  $\lambda_{\phi s}$ does not increase the DM output leading to a plateaux in the figure.
 This result is consistent with Eq.\;\ref{sigma-DM-simplified}.
 
 The trend shown in the figure is clear: a smaller $\sigma_{\phi h}$ leads to delayed  inflaton decay and thus dilutes the DM density. Therefore, the area above the curve leads to over--abundance 
 of dark matter and is  ruled out.

The above considerations neglect certain subleading effects which could, in general, affect the DM abundance. In particular,
 at weak coupling, the inflaton zero mode continues oscillating after the resonance and hence leads to further DM production. However, as discussed in the next subsection, this effect is modest  and can be neglected for our purposes. Also, a substantial $\sigma_{\phi h}$ results in  Higgs production via tachyonic resonance at early times. For the values of $\sigma_{\phi h}$ shown in the figure, this does not affect the
 DM abundance in any significant way.

 The range of allowed couplings is limited by the following factors. Large values of $\lambda_{\phi s } \gtrsim  {\cal O}(10^{-3})$ generally lead to 
 significant radiative corrections to the inflaton potential, while yet stronger couplings can result in dark matter thermalization via induced DM self--coupling  
   of order $\lambda_{\phi s }/(4\pi)^2$.  On the other hand,
 very small couplings $\lambda_{\phi s } < 10^{-8}$ do not lead to (broad) resonant DM production. 
 While this possibility is viable, it requires a perturbative instead of semiclassical  treatment, which we discuss next.
 \\ \ \\
  {\bf \underline{Non--resonant dark matter production.}}
  At weaker couplings, $q\lesssim 1$, there is no resonant enhancement of dark matter production. Nevertheless, 
 $s$  is still produced by a time--varying inflaton background and  can account for all of  the observed DM. 
 According to Eq.\,\ref{Gamma-ss}, the process is allowed kinematically when the induced DM mass
 $\sqrt{\lambda_{\phi s}/2 } \, \Phi$
  is below $m_\phi$ and 
 the production rate per unit volume is
  \begin{equation}
\Gamma (\phi \phi \rightarrow ss) \simeq  {\lambda_{\phi s}^2 \over 1024 \pi}  \Phi^4\;,
\end{equation}
 where we have used $\zeta_1 =\Phi^2/4$, $\omega = 2m_\phi$, and neglected the narrow resonance effects which are insignificant
 in an expanding Universe. The corresponding $s$ number density is found via the Boltzmann equation,
  \begin{equation}
\dot n + 3Hn = 2\, \Gamma (\phi\phi \rightarrow ss) \;.
\end{equation}
 At weak coupling, the energy density is dominated by the non--relativistic inflaton and 
  $\Phi =\Phi_0 \,a^{-3/2}$.  The Boltzmann equation is then easily integrated given an appropriate boundary condition. The solution
  has the form $a^3 \, n(t) = \alpha - \beta \, a^{-3/2}$, with constant $\alpha,\beta$. 
  
  Let us consider two coupling regimes. If the coupling is substantial, $s$ is produced initially via the resonance and after that,
  when $q\lesssim 1$, 
  it is produced perturbatively. On the other hand, if the coupling is very weak, $q(0)<1$, no resonance gets activated and perturbative DM 
  production takes place from the start. 
  In the first case, 
  the boundary condition is $n=0$ at $q=1$ and one finds
   \begin{equation}
n(t) = {\sqrt{2} \lambda_{\phi s}^{3/2} \Phi_0^3 \over 512 \pi a^3}
\end{equation}
 at late times. In our convention  $a= (mt)^{2/3}$ with $a=1$ at the start of the inflaton oscillation epoch.\footnote{We take $a_0=1$ for simplicity, implying $\Phi_0 \sim 1$ in Planck units.
A generalisation is straightforward \cite{us}.}
  One can verify that the above  density is much smaller than that created by the resonance, so it can indeed be neglected in our analysis.

 In the second case, $n=0$ at $a=1$, so
    \begin{equation}
n(t) = { \lambda_{\phi s}^{2} \Phi_0^4 \over 512 \pi m_\phi a^3}
\end{equation}
 for $a\gg 1$. As before, the reheating scale factor is found via $H_R \simeq \Gamma$ and the observed abundance of dark matter is reproduced for
 \begin{equation}
\sigma_{\phi h} \simeq 4.4 \times 10^{-6} \;   {m_\phi^{7/2} \over \lambda_{\phi s }^2 \Phi_0^2}   ~~\left(   {{\rm GeV}  \over m_s} \right) 
\end{equation}
 in  Planck units.
 
 Depending on $m_s$, the required $\sigma_{\phi h} $ may be large, $\sigma_{\phi h} \Phi_0/m_\phi^2 \gg 1$, such that the Higgs production via tachyonic resonance becomes
 important. In particular, this can destroy the classical inflaton background and invalidate our calculation. However, dark matter production peaks right after the inflaton starts oscillating,
 while the Higgs backreaction sets in significantly later. Therefore, the total number of produced DM quanta does not change significantly, while the system turns relativistic for some
 time introducing a factor analogous to $a_e/a_*$  in (\ref{sigma-DM}).

 We note that as long as $\lambda_{\phi s} \gtrsim 10^{-10}$, the effective mass of $s$ during $\phi^2$--inflation is greater than the Hubble rate. As a result, it behaves as a heavy field 
 frozen at the origin in field space and no significant  condensate $\langle s^2 \rangle$ gets generated. In this regime, $\lambda_{\phi s}$ is indeed the main source of
 dark matter production. For other inflationary potentials, the corresponding bound on $\lambda_{\phi s} $ is model dependent.

  \subsubsection{$\phi^4$ inflaton potential}
 
 Due to approximate conformal invariance, particle production in the potential $V_\phi = {\lambda_\phi \over 4} \phi^4$ exhibits qualitatively different features \cite{Greene:1997fu}.
 Although a massless inflaton is not a phenomenologically viable option, its small mass can be neglected at the initial stages. In this case, $\phi$ satisfies the equation of motion
 \begin{equation}
 \ddot\phi + 3H \dot \phi + \lambda_\phi\, \phi^3 =0 \;.
 \end{equation}
 The solution oscillates with a decreasing amplitude. After a few oscillations, it can be written in a closed form in terms of the conformal time $\eta$ defined by 
 $d\eta =  dt/a(t)$. 
 Introducing   a rescaled field $\varphi = a \phi$, one finds that $\varphi$ satisfies a simple ``Minkowski space''  equation
 $\varphi^{\prime\prime}_{\eta \eta} + \lambda_\phi\, \varphi^3 =0 $, where a negligible term $a^{\prime\prime}_{\eta \eta}/a$ has been omitted. 
 This is an equation for an elliptic function, which implies 
  \begin{equation}
\phi(t) = {\Phi_0 \over a(t)} \; {\rm cn} \left( x, {1\over \sqrt{2}} \right) \;
 \end{equation}
 where $x=(48 \lambda_\phi)^{1/4} \sqrt{t} $. The scale factor satisfies $a(0)=1$ and $a(t) \propto \sqrt{t}$ shortly thereafter.
  The oscillation period in $x$ is $T = \Gamma^2(1/4)/\sqrt{\pi} \simeq 7.4$ and the Jacobi cosine is dominated by the first term in its expansion, $\propto \cos {2\pi\over T} x$ .
  
  The dark matter field $s$ can be quantized treating $\phi$ as a classical background, as in Section\;\ref{resonance-basics}. In particular, the expansion (\ref{mode-expansion})
  applies to $s$ as well.  Denoting by $s_k$ the corresponding momentum $k$--modes, it is convenient to define a ``comoving'' mode $X_k (t) \equiv  a(t) s_k(t)$
 in analogy to   $\varphi = a \phi$. The equation of motion for $X_k$ in the $\phi$--background can be approximated by 
   \begin{equation}
X_k^{\prime\prime} + \left[     \kappa^2  + {\lambda_{\phi s} \over 2 \lambda_\phi} \;  {\rm cn}^2 \left( x, {1\over \sqrt{2}} \right)  \right] \,X_k=0 \;,
 \label{Lame}
 \end{equation}
  where the prime denotes differentiation with respect to $x$ and
     \begin{equation}
    \kappa^2  = {k^2 \over \lambda_\phi \Phi_0^2} \;.
     \end{equation}
  This belongs to the class of Lam\'e equations. Since the effective mass term is periodic, the Floquet analysis applies and the $X_k$ evolution is determined by a stability chart
  in terms of $\kappa^2$ and $\tilde q = {\lambda_{\phi s} /( 2 \lambda_\phi})$, an example of which is shown in
   \cite{Greene:1997fu}.
   It is important that the growth of $|X_k|$ is controlled by the {\it ratio} of the couplings $\tilde q$, unlike in the case of the Mathieu equation. For 
  ${\lambda_{\phi s} /( 2 \lambda_\phi}) >1$,  the Floquet exponent is significant, ${\cal O}(10^{-1})$, leading to fast amplitude growth. For ${\lambda_{\phi s} /( 2 \lambda_\phi}) <1$,
  the Floquet exponent scales as $10^{-1} \, {\lambda_{\phi s} /( 2 \lambda_\phi})$ and becomes negligible at small $\lambda_{\phi s}$.
  
  The conformal nature of the system leads to qualitatively different particle production features compared to those in the $\phi^2$ case. 
  The ratio ${\lambda_{\phi s} /( 2 \lambda_\phi}) $, which plays the role of 
  the $q$--parameter of the Mathieu equation, does not decrease in time and therefore the resonance always ends for a different reason, namely, the backreaction of produced particles \cite{Greene:1997fu}.
  The latter transfer the energy from the coherently oscillating field thereby reducing its amplitude and, furthermore, induce a mass term.
  The $\phi^4$ resonance can also be narrow, yet efficient. Indeed, at  ${\lambda_{\phi s} /( 2 \lambda_\phi}) \ll 1$, the Lam\'e equation is very well approximated by the Mathieu equation with $q\ll 1$.
  Thus, even in an expanding space, the particle number grows exponentially, although the effect is only seen on a relatively long timescale. In contrast, the narrow resonance 
  for a $\phi^2$ potential is inefficient due to the redshifting of the produced particle momenta or reduction of $q$.

  Another important difference compared to the $\phi^2$  case is that inflaton oscillations lead to production of the $\phi$--quanta. Quantizing the fluctuations above the $\phi$--background,
  one finds that the corresponding $k$--modes satisfy the   Lam\'e equation (\ref{Lame}) with a special value ${\lambda_{\phi s} /( 2 \lambda_\phi}) =3$.
  Although the Floquet exponent is quite small in this case, the $\phi$--quanta production is more efficient than production of DM quanta for ${\lambda_{\phi s} /( 2 \lambda_\phi}) \lesssim 1/3$.
  This process stops due to backreaction roughly when $\langle \phi^2 \rangle \simeq 0.05 \,\Phi_0^2$, making the DM output suppressed.

 In realistic constructions, the inflaton mass is non--zero and the resonance structure becomes more complicated  \cite{Greene:1997fu}. For small enough $m_\phi$, the $\phi^4$--potential approximation 
  is applicable until the oscillation amplitude reaches some  limiting value, typically around $m_\phi / \sqrt{\lambda_\phi}$.

\begin{figure}[h] 
\centering{   
\includegraphics[scale=0.46]{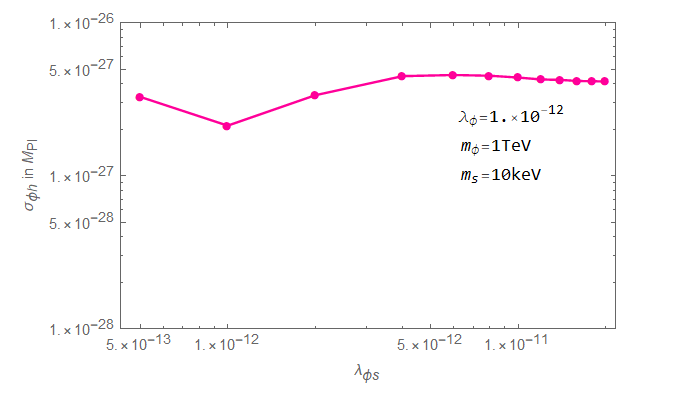} }
\caption{ \label{sigma-lambda-phi4}
$\sigma_{\phi h} $ vs $\lambda_{\phi s}$ producing the correct DM relic abundance for a $\phi^4$ potential, based on LATTICEEASY simulations. 
$\sigma_{\phi h} $ is given in units of  $M_{\rm Pl} =2.4 \times 10^{18}$ GeV and the initial inflaton value is chosen to be $\Phi_0 \simeq 1.7 \,M_{\rm Pl}$.
The area above the curve is excluded by overabundance of dark matter.
}
\end{figure}

  In practice, the dynamics of the DM production are complicated by various effects such as backreaction and rescattering, hence lattice simulations are often necessary for a reliable analysis.
  The simulations can be trusted outside the narrow resonance regime so that the occupation numbers are sufficiently large.
  This is the case roughly for 
   $\lambda_{\phi s} \gtrsim 0.5\,\lambda_\phi$. 
Then, about  half of the inflaton energy gets transferred to dark matter and the total number of the DM quanta remains constant at the end of the simulation. The zero inflaton mode, i.e. the classical background,
gets largely destroyed by that time, while the characteristic energy of the DM and inflaton quanta becomes of order $\sqrt{\lambda_\phi} \phi \times  (\lambda_{\phi s} / 2\lambda_\phi)^{1/4}$.

An example of the coupling set producing the correct DM relic abundance for $m_\phi =1$\;TeV is shown in Fig.\;\ref{sigma-lambda-phi4}. 
To calculate the required $\sigma_{\phi h}$ in terms of the simulation output, one can again use Eq.\;\ref{sigma-DM}, which for substantial couplings simplifies to Eq.\,\ref{sigma-DM-simplified}.
As before, the system goes through the relativistic and non--relativistic stages. During the first period, the energy density scales as $a^{-4}$. Even though DM acquires large induced mass $\propto \sqrt{\lambda_{\phi s}} \phi$, far above its momentum,
it redshifts just like the momentum does. As a result, the total energy density scales as if the system were  relativistic. When the characteristic momentum redshifts to small enough values, the bare inflaton mass becomes relevant and the system
turns non--relativistic.   The relativistic period lasts much longer than it does  in the $\phi^2$ case and  one finds $a_e \ll a_* \ll a_R$.
  Dark matter does not get diluted in the long first stage, which generally leads to a dark Universe. Therefore, a subsequent, rather long non--relativistic period is required implying a very small $\sigma_{\phi h}$, of order eV for the parameters of Fig.\;\ref{sigma-lambda-phi4}.
  The corresponding reheating temperature is low, $50$ MeV, but still consistent with the cosmological data \cite{Hannestad:2004px}.
   
  Fig.\;\ref{sigma-lambda-phi4} shows that $\sigma_{\phi h}$ is almost independent of $\lambda_{\phi s}$, at least for  ${\lambda_{\phi s} /( 2 \lambda_\phi}) > 1$. This is expected from 
  Eq.\,\ref{sigma-DM-simplified}, since the system enters the state of quasi--equilibrium. One may verify that these results are also independent of $\lambda_\phi$.
  For lower couplings, the simulations are less reliable due to a limited running time and 
  the apparent dip around $\lambda_{\phi s} \sim 10^{-12}$ may in fact be an artefact  of the numerical approximations.

The   trend  shown in Fig.\;\ref{sigma-lambda-phi4}  persists at larger values of  the inflaton--DM coupling until  $\lambda_{\phi s} \sim 4\pi \sqrt{\lambda_\phi} $, where it starts inducing a significant radiative  correction to the inflaton self--coupling. Also, at large $\lambda_{\phi s} $,
the effect of a non--zero $m_\phi$ on the resonance becomes tangible \cite{Greene:1997fu}.
For weak inflaton--DM couplings, the system enters a narrow resonance regime and the classical
 simulation becomes unreliable. At $\lambda_{\phi s} / 2\lambda_\phi \ll 1$, a perturbative analysis  is expected to apply, at least on a moderate time scale,
 i.e.   when the Bose enhancement is not very strong.

 The values of $\sigma_{\phi h}$ lying above the curve in Fig.\;\ref{sigma-lambda-phi4} are excluded since they lead to a ``dark Universe'' with overabundant dark matter.
 In this case, the non--relativistic inflaton--dominated period is too short and does not provide for enough dilution of the  DM density. It is remarkable  that even tiny values of $\sigma_{\phi h}$
 above ${\cal O}$(eV) are ruled as long as $m_\phi=1$\;TeV and ${\lambda_{\phi s} /( 2 \lambda_\phi}) > 1$.

Finally, dark matter can be considered heavy during inflation as long as $\sqrt{\lambda_\phi} \phi \lesssim \sqrt{\lambda_{\phi s}} $, so that no appreciable  DM condensate forms.
In this case, the effect described in   Section\;\ref{condensates} is insignificant. If, however, $s$ is light, the condensate contribution to the DM abundance can be neglected if the
self--coupling $\lambda_s$ is not too small (but still insufficient for thermalization).

 \subsection{Matter production via standard  inflaton decay}
 
 Perhaps the simplest possibility to reheat the Universe and produce dark matter is to employ the trilinear couplings
 \begin{equation}
V_{\phi h} = {1\over 2} \sigma_{\phi h} \, \phi h^2 ~~,~~ V_{\phi s} =   {1\over 2} \sigma_{\phi s} \, \phi s^2 \;,
\end{equation}
which induce the processes $\phi \rightarrow ss$, $\phi \rightarrow h_i h_i $.
 At small enough couplings, the tachyonic resonance effects can be neglected and the  decay is perturbative (see \cite{Moroi:2020bkq} for a recent discussion).
For a quadratic inflaton potential, $V_\phi = {1\over 2} m_\phi^2 \phi^2$, the tachyonic resonance is weak if $\sigma_{\phi h} \Phi_0 /m_\phi^2 \ll 1$,
$\sigma_{\phi s} \Phi_0 /m_\phi^2 \ll 1$.
This also implies that the inflaton decay into $hh$ and $ss$ is kinematically allowed.  In more complicated systems, e.g. with complex scalars, the DM abundance may be  determined by an interplay of different effects \cite{Heurtier:2017nwl}, one of which is the direct inflaton decay.

 The DM abundance is controlled by $\phi \rightarrow ss$, while the dominant inflaton decay mode is $\phi \rightarrow h_i h_i $,
 \begin{equation}
\Gamma (\phi \rightarrow ss) = {\sigma_{\phi s }^2 \over 32 \pi m_\phi }  ~~,~~ \Gamma_{\rm tot} \simeq {\sigma_{\phi h }^2 \over 8 \pi m_\phi } \;,
\end{equation}
with $\sigma_{\phi s } \ll \sigma_{\phi h }$.
$Y$ is given by the number of the DM quanta accumulated up to reheating, where the inflaton energy converts into that of SM radiation. The DM number density  $n$ in this period
is found from the Boltzmann equation
\begin{equation}
 \dot n +3Hn  = 2 \Gamma(\phi \rightarrow ss) \; n_\phi \;,
 \end{equation}
 where $n_\phi$ is the inflaton number density, $n_\phi = {1\over 2} m_\phi \phi^2$.
The effect of the decay
on the inflaton amplitude can be neglected until the reheating stage,
so the DM number grows linearly with time,
\begin{equation}
 n (t) =  {1\over a^3 }\;\int dt \; 2 \Gamma(\phi \rightarrow ss)  \; n_\phi  \;a^3 =2 \Gamma(\phi \rightarrow ss) \, n_\phi  t \;,
 \end{equation}
with the boundary condition $n(0)=0$.
Trading the time variable for the scale factor, $m_\phi t = a^{3/2}$ for $a\gg1$,
\footnote{As before, we take $a_0=1$ for simplicity, implying $\Phi_0 \sim 1$ in Planck units.
A generalisation is straightforward \cite{us}.}
 and requiring $H_R \simeq \Gamma_{\rm tot}$, one finds that at the reheating point
$Y \simeq 0.3 \,{ \Gamma(\phi \rightarrow ss) \Phi_0 \over  \Gamma^{1/2}_{\rm tot } m_\phi } $. The correct DM relic density is then reproduced for
\begin{equation}
 \sigma_{\phi h} \simeq 4 \times 10^7 \; {\sigma_{\phi s}^2 \Phi_0 \over m_\phi^{3/2}} \; {m_s \over {\rm GeV}}
 \end{equation}
in Planck units.

To get a feeling for the numerical values, consider the input parameters typical for quadratic inflation,
 $\Phi_0 \sim 1$, $m_\phi \sim 5 \times 10^{-6}$   and  take $m_s =1$ GeV. Then $\sigma_{\phi h} \simeq  4 \times 10^{15 } \sigma_{\phi s}^2 $,
implying that the Higgs decay mode dominates for  
 $\sigma_{\phi s} > 3\times 10^{-16}\sim$ 1 TeV.
The reheating temperature is roughly $T_R \sim 10^2 \sigma_{\phi h}$, far above the required few MeV.
To avoid   strong tachyonic resonance, one must ensure $\sigma_{\phi h} \Phi_0 /m_\phi^2 \ll 1$, which for the above parameters implies $\sigma_{\phi h } \ll 3\times 10^{-11}$.
Therefore, there is a viable range  of $\sigma_{\phi h }$ between $10^{-15}$ and $10^{-11}$ for our parameter choice.

It is important to note  that we require the absence of tachyonic resonance in order for simple perturbative estimates to be applicable. Nevertheless, large $\sigma$--terms can lead to viable models, yet
a dedicated lattice study would normally be needed in this case.

 Finally, in all of the above considerations, we have assumed that the portal couplings are the main actors in producing (dark) matter. In addition, there are ever--present gravity contributions which may or may not be 
 significant. Indeed,
  the Universe expansion itself creates a setting for particle production \cite{Parker:1969au},\cite{Grib:1976pw}. If the expansion is fast enough, it can result in strong non--adiabaticity generating field  modes for which $\dot \omega /\omega^2 >1 $. In the scalar case, the two relevant quantities are the particle mass and the non--minimal coupling to gravity. Both of them lead to particle production and the effect is particularly strong for non--conformally coupled  ($\xi_s \not= -1/6$) or  heavy states, 
 which can, for example,  constitute superheavy dark matter (WIMPzillas) \cite{Kuzmin:1998kk}--\cite{Chung:1998bt}. However, only a tiny fraction of the inflaton energy gets converted into dark matter \cite{Kuzmin:1998kk}, unlike
 for DM production via strong parametric or tachyonic resonance, and the contribution to the observed DM abundance can be completely neglected for a light  conformally coupled scalar.
  Recent analyses show that the gravitational production  mechanism is also efficient for vector dark matter  \cite{Graham:2015rva},\cite{Ema:2019yrd} as well as very light scalars \cite{Alonso-Alvarez:2018tus}.
It should be noted that, 
 in this class of  models, dark matter density perturbations are not directly correlated with those of the inflaton, hence they  are often subject to significant isocurvature perturbation constraints.
After inflation, 
 dark matter can also be efficiently produced 
  via its   non--minimal  coupling to gravity  $\xi_s$ \cite{Fairbairn:2018bsw}, when the curvature oscillations induce tachyonic resonance. This mechanism can account for all of the DM in a wide range of masses,
  depending on $\xi_s$,
  and should in general be superimposed with the Higgs portal DM production.

   To summarize this section, the simplest renormalizable  Higgs portal and inflaton--DM  couplings are sufficient to describe both reheating and non--thermal dark matter production.    This mechanism can be very efficient such that
   even tiny couplings can produce the right amount of dark matter and SM radiation. On one hand, it raises a theoretical question as to how such small couplings can be justified.
   This issue  can only be addressed in a more fundamental theory, presumably incorporating gravity. On the other hand, it also shows that such couplings should not  be ignored
   and treated carefully  since even their
   tiny values can make a difference. The prospects of directly observing dark matter in this class of models are clearly rather dim.

 \section{Higgs portal dark matter}
 \label{HiggsPortalDM}
 
 In this section, we focus on the Higgs coupling to dark matter, which was taken negligible in our previous considerations. Such a
 coupling can be very significant and responsible for production of thermal dark matter as well as its observable signatures.

 The Higgs field plays a special role in probing the dark  sector. Already at the renormalizable level,
 it can couple  to dark scalar states  which carry no Standard Model quantum numbers \cite{Silveira:1985rk}.
 The reason is that $H^\dagger H $ is a unique SM dim-2 operator which is both gauge and Lorentz invariant.   
  The lowest dimension  interactions between the Standard Model fields  and the SM singlets  $\phi, V_\mu, \chi$ of spin 0, 1, 1/2, respectively, 
 possessing $Z_2$ parity 
 are given by 
  \begin{eqnarray}
 &&-\Delta {\cal L}_{\rm scal} = {1\over 2} \, \lambda_{\phi h} H^\dagger H \, \phi^2 \; , \\
 &&-\Delta {\cal L}_{\rm vect} = {1\over 2}\, \lambda_{ Vh } H^\dagger H \, V_\mu V^\mu \; , \label{h-v} \\
 &&-\Delta {\cal L}_{\rm ferm} = {1\over 2 \Lambda}\,  \lambda_{\chi h}  H^\dagger H  \, \left(  \bar\chi \chi  +  c\; \bar\chi i\gamma_5 \chi \right) \label{h-chi}\; ,
 \end{eqnarray}
 where $\phi$ is real, $\chi$ can be either Dirac or Majorana;  $ \lambda_{\phi h}, \lambda_{ Vh } , \lambda_{\chi h}, c$ are real dimensionless couplings, $\Lambda$ is a scale, and  
 we have allowed for CP violation in the fermion interactions.
 The assumed parity symmetry $f_\alpha \rightarrow - f_\alpha$  ($f_\alpha = \phi, V_\mu, \chi$) makes these fields stable and, thus, dark matter candidates. 
 In addition to the above Higgs portal terms, the dark fields generally possess self--interaction as well as mass terms.
 
Among these couplings, only the scalar one is renormalizable. The vector and fermion interactions arise 
 from integrating out heavy states. In the fermion case, this is manifested in the interaction being dimension 5. The vector coupling
 violates gauge invariance, so the corresponding coupling must be proportional to the gauge field mass \cite{Lebedev:2011iq}. This makes the corresponding Higgs portal interaction 
  higher dimensional, in fact,  dimension 6. Therefore, the couplings (\ref{h-v}),(\ref{h-chi}) should be treated within effective field theory.
  
  Dark matter phenomenology can be studied in terms of the effective couplings. In this case, one must make sure that, in the corresponding parameter range,
  the effective field theory approximation  is legitimate. In particular, the energy of the relevant processes must be below the unitarity cuf--off \cite{Lebedev:2011iq},\cite{Arcadi:2020jqf}.  This implies, for example,
  that the vector cannot be too light, otherwise the LHC scattering processes as well as $h \rightarrow VV$ would ``blow up''.

  The effective  approach to dark matter of different spins   was pursued originally in  \cite{Kanemura:2010sh},\cite{Djouadi:2011aa}.  In particular,
  Ref.\;\cite{Djouadi:2011aa} has found that light thermal dark matter ($m_{\rm DM}< 60$ GeV) is in conflict with the LHC Higgs measurements due to the efficient invisible decay mode.
  The most recent analysis from {\textsf{GAMBIT}} \cite{Athron:2018hpc} concludes that all the DM candidates remain viable, although 
  the available parameter space has shrunk considerably owing  to the  improved direct DM detection limits. 
   In the vector case,
  the viable window corresponds to 
   the dark matter mass   above a few TeV with   ${\cal O}(1)$ Higgs portal coupling, apart from the narrow resonance strip with $m_{\rm DM} \simeq m_{h_0}/2$, where $m_{h_0}$ is the SM Higgs mass.
  In the fermion case, the allowed DM mass spans a wide range from the resonance, $m_{\rm DM} \simeq m_{h_0}/2$, to multi--TeV, as long as sufficient CP violation is present \cite{LopezHonorez:2012kv},\cite{deSimone:2014pda}. 
  The scalar case   will be discussed below in detail.

  In what follows, instead of the effective couplings,
    we will  consider the simplest UV complete models leading to the above   interactions.
  This has some advantages. For example, it   uncovers the  non--trivial origin of the parity symmetry in the vector case. Also,
  it allows for DM annihilation channels which are absent in the effective theory, yet essential for model viability (see Section\;\ref{secluded}). 
  One finds that phenomenology of the UV complete models is healthier, while 
  the results in the effective field theory limit are obtained 
    by decoupling the heavy states.

   A review 
   of various 
  aspects of Higgs portal dark matter including collider phenomenology can be found in \cite{Arcadi:2019lka}.
An effective field theory approach to dark matter of different spins is discussed in \cite{Duch:2014xda},\cite{Criado:2020jkp}.

 \subsection{Scalar dark matter}

 The minimal model of Higgs portal dark matter is a scalar extension of the Standard Model with $Z_2$ parity \cite{Silveira:1985rk}.
 We will consider both a real and a complex scalar options, with  the latter 
 being particular interesting phenomenologically.
 The simplest models and their close relatives
 have been studied extensively in the literature \cite{McDonald:1993ex}-\cite{Claude:2021sye} and remain viable to date,
 although their allowed parameter shrivels due to constant pressure from direct DM detection experiments.

 \subsubsection{Minimal real scalar dark matter}
 \label{sec-min-dm}

 The simplest option to account for dark matter is to add a single new degree of freedom, that is, a real scalar $\phi$ \cite{Silveira:1985rk}. To ensure  stability of DM, one imposes a $Z_2$ symmetry,
 \begin{equation}
 \phi \rightarrow -\phi \;.
 \end{equation}
 This is a discrete symmetry which may be a remnant of gauge symmetry at high energy. For instance, when a field with even charge under a gauged U(1) develops a VEV, a $Z_2$ subgroup remains unbroken.
 Examples of this type appear in realistic string constructions \cite{Lebedev:2007hv}. For our purposes, the origin of the $Z_2$ is unimportant and, in fact, 
 it may not even be exact:
 a tiny violation of this parity 
 would be compatible with the data.
The  corresponding scalar potential is given by Eq.\;\ref{Z2-pot}, which we reproduce here for convenience,
 \begin{eqnarray}
&& V(\phi, h) = {1\over 4} \lambda_h h^4 + {1\over 4} \lambda_{\phi h }h^2 \phi^2 + {1\over 4} \lambda_\phi \phi^4  +
{1\over 2} m_h^2 h^2 + {1\over 2} m_\phi^2 \phi^2   \;, \nonumber
\end{eqnarray}
 except now $\phi$ does not develop a VEV in order to preserve $Z_2$. The potential minimization is trivial and the dark matter mass $m$ is given by\footnote{In principle, the dark matter mass is temperature dependent, e.g. via $v=v(T)$. However, typically its effect on the relic abundance  is insignificant: the freeze--out temperature is roughly $m/20$. For heavy DM,
 it may be above the EW transition temperature, but in this case $m_\phi^2 \gg  \lambda_{\phi h }v^2/4$.  }
 \begin{equation}
 m^2 = m_\phi^2  + {1\over 2} \lambda_{\phi h }v^2  \;,
 \end{equation}
 where $v \simeq 246$ GeV.  Shifting $h \rightarrow v+h$ in the above potential, we obtain the Higgs--DM interaction terms
  \begin{equation}
 V_{\phi h} =  {1\over 2} \lambda_{\phi h }v \, h \phi^2 +{1\over 4} \lambda_{\phi h }h^2 \phi^2 \;,
 \end{equation}
 where we now denote the physical Higgs boson by $h$.

  As long as DM self--interaction can be neglected, the model is described by 2 parameters:  
  \begin{equation}
  m~~, ~~ \lambda_{\phi h } ~~.
  \end{equation}
  This means the model is very predictive and {\it a priori} it would be difficult to fit all the data with just these two parameters.
  The most important constraints on the model are imposed by requiring the correct DM relic density,  absence of the signal in direct detection experiments and the LHC Higgs measurements.
  In addition, for light enough DM, indirect DM detection places a further significant constraint.
   Let us consider these aspects of DM phenomenology separately.
   \\ \ \\
   {\bf {\underline {DM relic density.}}} If the Universe temperature is high enough and the Higgs portal coupling is sufficiently strong, dark matter reaches thermal equilibrium. In this case, 
   the relic DM density can be calculated in terms of $m$ and $ \lambda_{\phi h }$. Note, however, that this approach relies on thermalization of dark matter which may have never been achieved
   for low reheating temperatures.
   In fact, the standard cosmology only requires temperatures to be above a few MeV for successful nucleosynthesis. Furthermore, the DM density calculation can be 
   substantially different in non--standard cosmologies \cite{Hardy:2018bph}.
   
   Assuming the standard cosmological history and 
   a thermal momentum distribution for DM,
   its number density $n(t)$ evolves according to the Boltzmann equation \cite{Kolb:1990vq}
   \begin{equation}
   {d n \over dt} + 3Hn = - \langle \sigma v_r \rangle \, \left[  n^2 -n^2_{\rm eq}\right] \, .
   \end{equation}
   Here, $H$ is the Hubble rate determined by the effective number of degrees of freedom $g_*$ at temperature $T$,
   \begin{equation}
  H= \sqrt{\pi^2 g_* \over 90} \, {T^2 \over M_{\rm Pl}}  \;,
   \end{equation}
 $\langle \sigma v_r \rangle $ is the thermal average of the annihilation cross section times the relative velocity $v_r$, and 
   $n_{\rm eq}$ is the equilibrium number density at  $T$.
   The thermal average in the {\it non--relativistic } limit  is defined by
   \begin{equation}
   \langle \sigma v_r \rangle = 
   {\int {d^3 {\bf p_1} }  {d^3{\bf p_2}  } \,f(p_1) f(p_2)\; \sigma v_r  \over
   \int {d^3 {\bf p_1} }  {d^3{\bf p_2} } \, f(p_1) f(p_2)   } ~~,~~ f(p) = e^{- (E-\mu)/T}\;,
   \end{equation}
   where $\mu $ is the time--dependent effective chemical  potential and 
   \begin{equation}
  n=  \int {d^3 {\bf p} \over (2\pi)^3} \, f(p) \;.
   \end{equation}
  For a given $T(t)$ which is dictated by entropy conservation,  the evolution of $n(t)$ is determined by the time dependence of $\mu$.
  The equilibrium value corresponds to $\mu=0$ and 
   \begin{equation}
  n_{\rm eq}=  \left(    { mT   \over 2\pi}   \right)^{3/2} \; e^{- {m/T}} \;.
   \end{equation}

   The above Boltzmann equation is valid for dark matter consisting of one species. There is no phase space symmetry factor  associated with identical particles in the initial state since it gets cancelled by a factor of 2 
   stemming from the DM number change in a single reaction. For DM consisting of particles and anti--particles, the  Boltzmann equation for the total number density contains an extra factor of 1/2
   in front of the cross section \cite{Gondolo:1990dk}.

    The solution to the above Boltzmann equation is well known: $n(t)$ tracks closely $n_{\rm eq}(t)$ until the reaction rate becomes slower than the Hubble rate. At this point, the effective chemical
    potential becomes significant and the DM ensemble falls out of  (chemical) equilibrium with the SM thermal bath. For WIMPs, this occurs at  temperature $T\sim m/20$.
    The total number of DM particles remains approximately constant and is conveniently expressed in terms of the DM abundance $Y$,
    \begin{equation}
    Y = {n \over s_{\rm SM} } ~~,~~ s_{\rm SM} ={2\pi^2 \over 45} \, g_{*s} \, T^3 \;,
    \label{Y}
    \end{equation} 
    where $s_{\rm SM}$ is the SM entropy density and 
    $g_{*s}$ is the effective number of degrees of freedom contributing to the entropy. As long as the DM contribution can be neglected, the SM thermal bath entropy is conserved, $s_{\rm SM} a^3=\,$const,
    which determines the temperature evolution. 
     
    The cosmological and astrophysical data  \cite{Ade:2015xua}  require that in the present epoch
     \begin{equation}
    Y_\infty =  4.4 \times 10^{-10} \; \left( {{\rm GeV}\over m} \right) \;.
    \end{equation} 
    The number density at freeze--out is determined by the 
     the annihilation cross section, so for a given $m$ the relic abundance constraint fixes  the Higgs portal coupling $\lambda_{\phi h }$.

\begin{figure}[t!] 
\centering{  \vspace{-3cm}
\includegraphics[scale=0.52]{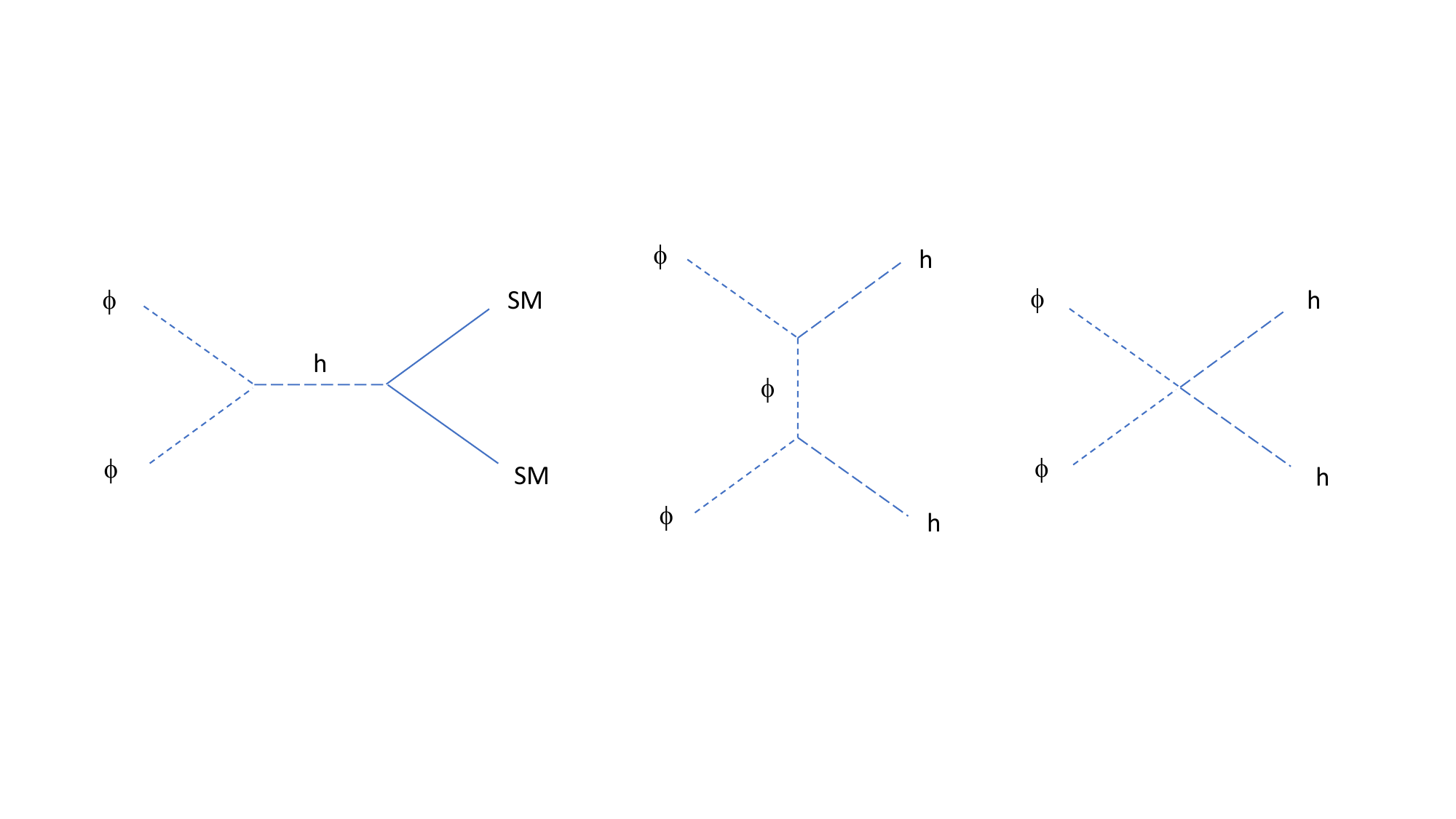}
 \vspace{-3cm} }
\caption{ \label{scalar-ann}
Leading diagrams for scalar DM annihilation.
}
\end{figure}

    There are a number of annihilation channels $\phi \phi \rightarrow$ SM as shown in Fig.\;\ref{scalar-ann}. 
    For light dark matter, the $s$--channel Higgs exchange dominates with the relevant DM coupling being ${1\over 2} \lambda_{\phi h } v\, h \phi^2 $. For example, the annihilation cross section into fermions of mass $m_f$ with $N_c^f$ colors is given by
      \begin{equation}
   \sigma v_r = N_c^f \; {\lambda_{\phi h }^2 m_f^2 \over 4\pi}\; {1\over (4 m^2 -m_{h_0}^2)^2} \; \left(    1- {m_f^2 \over m^2}        \right)^{3/2}                 \;.
    \end{equation} 
  This expression exhibits resonant enhancement of DM annihilation close to $2m \simeq m_{h_0}$, where $m_{h_0}$ is the SM Higgs mass.
   In this case, the pole is regularized by including the Higgs decay width in the propagator: $(4 m^2 -m_{h_0}^2)^2 \rightarrow (4 m^2 -m_{h_0}^2)^2+ m_{h_0}^2 \Gamma_h^2  $,
  which can otherwise be neglected. Even a small $\lambda_{\phi h } \ll 1$ is sufficient to obtain the right relic abundance.
  
  Extension to all final states in the $s$--channel can be done by subsuming the corresponding contributions into the modified Higgs width. Indeed, both the decay width and 
  $\sigma v_r$ include the same final state contributions and phase space integral, while differing in the center-of-mass energy. One finds \cite{Burgess:2000yq}
      \begin{equation}
    \sigma v_r = {\lambda_{\phi h }^2 v^2 \over m\, (4m^2 - m_{h_0}^2)^2   } \, \Gamma (m_{h_0}^* = 2m) \;,
    \end{equation}  
      where $\Gamma (m_{h_0}^* = 2m) $ is the decay width of the SM Higgs boson with mass $2m$.
      
      Practical calculations are often performed with the software package  {\texttt{micrOMEGAs}} \cite{Belanger:2008sj}. It takes into account all the annihilation channels and performs numerical integration of the Boltzmann equation.
      A parameter space scan shows that, away from the resonance region, EW/TeV masses   are consistent with the relic DM abundance constraint for $\lambda_{\phi h }$ of order 
      ${\cal O}(10^{-1})$ to $   {\cal O}( 1) $
      (see e.g., \cite{Djouadi:2011aa}). Close to the resonance, the coupling can be much smaller, $10^{-4}-10^{-3}$, however the relic density computation becomes more involved
      due to kinetic decoupling of DM from the SM plasma \cite{Binder:2017rgn}.\footnote{Different stages of the WIMP freeze--out have been analyzed in detail in \cite{Bringmann:2006mu}.}
      \\ \ \\
      {\bf \underline{Direct DM detection.}} 
      Dark matter can potentially be detected via its interactions with nuclei. Such scattering is mediated by the Higgs boson, whose interactions are spin--independent. The
       signal rate is determined by a coherent sum of the nucleon--DM scattering  amplitudes and grows with the size of the nucleus. The result is conveniently parametrized 
       in terms of the spin--independent nucleon--DM 
      scattering cross section.

      Scattering of dark matter off nucleons proceeds via Higgs exchange in the $t$--channel. The corresponding  momentum transfer is suppressed by the DM velocity
      $v_{\rm DM} \sim 10^{-3}$, so it does not exceed ${\cal O}(100$\,MeV)
      and can be neglected. In this case,
      the Higgs boson can be ``integrated out'' resulting in the following contact interaction with light quarks $q$ and gluons,\footnote{This result assumes the standard Yukawa couplings, which
      is not necessarily the case for light quarks. With Higgs--dependent Yukawa couplings \cite{Giudice:2008uua}, the amplitude can be significantly enhanced \cite{Bishara:2015cha}. }
         \begin{equation}
   \Delta {\cal L}^{\rm eff} = { \lambda_{\phi h } \over 2 m_{h_0}^2} \; \phi^2 \; \left( \sum_q m_q \bar q q - {\alpha_s \over 4\pi} G_{\mu\nu}^a G^{\mu\nu\, a}
   \right) \;,
    \end{equation}  
      where $G^a_{\mu\nu} $ is the gluon field strength and $\alpha_s$ is the SU(3) fine structure constant.
      Here, the heavy quarks have been integrated out using the conformal anomaly according to Shifman, Vainshtein and Zakharov \cite{Shifman:1978zn}: 
      a heavy quark contribution $m_Q \bar Q Q$ gets replaced with $- \alpha_s /12\pi \; G_{\mu\nu}^a G^{\mu\nu\, a}$.
      To get the dark matter interaction with nucleons, one takes the matrix element of the quark and gluon operators over  the nucleon state,
      \begin{eqnarray}
      && \langle N | m_q \bar q q     |  N \rangle = m_N f_q   \nonumber \\
      &&  \langle N |    G_{\mu\nu}^a G^{\mu\nu\, a}     |  N \rangle = - {8 \pi \over 9 \alpha_s} m_N f_{TG}  \;,
      \end{eqnarray}
      where $f_q , f_{TG}$ are some parameters determined from low energy experiments \cite{Ellis:2008hf},\cite{Crivellin:2013ipa},\cite{Hoferichter:2015dsa} 
      and lattice simulations \cite{Junnarkar:2013ac}. This results in the effective DM--nucleon interaction
   \begin{equation}
   \Delta {\cal L}^{\rm eff}_{\phi N} = { \lambda_{\phi h } m_N f_N   \over 2 m_{h_0}^2}              \; \phi^2 \;  \bar N N  \;,
       \end{equation}  
    with 
     \begin{equation}
 f_N = \sum_{u,d,s} f_q + {2\over 9 } f_{TG} \simeq 0.3 \;.
       \end{equation}  
    Numerically, the sum over the light quarks contributes about 0.1 to $f_N$, while the gluon term contributes about 0.2. Thus, the nucleon--Higgs coupling is dominated by the gluons or, more
    precisely, the heavy quarks.  
    
    The resulting DM--nucleon scattering cross section is \cite{Kanemura:2010sh}
    \begin{equation}
    \sigma^{\rm SI}_{\phi -N}= { \lambda_{\phi h }^2  f_N^2   \over 4 \pi m_{h_0}^4}\, {m_N^4 \over (m+m_N)^2}  \;.
    \end{equation}
   The current XENON1T upper bound  on $  \sigma^{\rm SI}$  is of order $10^{-46}$\;cm$^2$ for electroweak DM masses and reaches the minimum  $4\times 10^{-47}$\;cm$^2$ at $m \simeq 30$\;GeV 
    \cite{Aprile:2018dbl}. This imposes a strong constraint on the model. In particular, apart from the very narrow resonance region,  it pushes the thermal DM to a multi--TeV range. 
    More generally, it rules out the Higgs portal couplings above $10^{-2}-10^{-1}$ for $m $ between 100 GeV and 1 TeV \cite{Arcadi:2019lka}.
    \\ \ \\
    {\bf \underline{Invisible Higgs decay.}}
    A strong constraint on light dark matter is imposed by the LHC bound on invisible Higgs decay \cite{Djouadi:2011aa}.
    The SM Higgs decay width is dominated by the small $b$--quark Yukawa coupling  which makes it very narrow, $\Gamma_h \sim 4$ MeV. Thus, the decay branching ratios are sensitive to non--standard  modes
    including  $h \rightarrow \phi\phi$, when  
   kinematically allowed.
  In our case,    the invisible decay width is given by
       \begin{equation}
\Gamma (h \rightarrow \phi\phi) =  {\lambda_{\phi h}^2 v^2 \over 32\pi m_{h_0}} \; \sqrt{1- {4m^2 \over m_{h_0}^2}} \;.
       \end{equation}  
       Close to the kinematic threshold $m \simeq m_{h_0}/2$,  collider 
       DM production via  Higgs decay should include a momentum--dependent width as described in  \cite{Heisig:2019vcj}.
    
  A recent detailed analysis of the relevant LHC constraints can be found in \cite{Arcadi:2019lka}.
    The direct bound on the invisible Higgs decay branching ratio is ${\rm BR} (h \rightarrow {\rm inv}) < 0.24$ at 95\% CL \cite{Khachatryan:2016whc}.   The indirect bound is stronger: the value adopted in 
    \cite{Arcadi:2019lka} is 0.2, while the bound based on the overall Higgs signal strength $\mu$ used in  \cite{Huitu:2018gbc} is close to 0.1. The reason is that the number of produced 
    Higgs bosons 
     which subsequently  decay into known 2--body final states agrees very well with the Standard Model prediction, within 10\%. A significant ${\rm BR} (h \rightarrow {\rm inv}) $ would spoil
     this agreement by reducing  $\mu$, which leads  to a stronger bound.
    
    In case of thermal dark matter, the required Higgs portal coupling is  above $10^{-1}$ which leads a large invisible branching ratio, typically above 80\%. Thus, for either choice of the 
     ${\rm BR} (h \rightarrow {\rm inv}) $ constraint, light dark matter is ruled out 
   \cite{Djouadi:2011aa}. For non--thermal DM, this possibility remains and the invisible decay provides one with a sensitive channel to probe  parameter space with $m < m_{h_0}/2$.
 \\ \ \\
 {\bf \underline{Indirect DM detection.}}
 Dark matter can be detected indirectly via products of its annihilation in objects with large DM density, for instance the Galactic Center or dwarf galaxies. Non--observation of a significant 
 signal places further bounds on the model. Excellent reviews of indirect DM detection can be found in \cite{Slatyer:2017sev},\cite{Gaskins:2016cha}.
 
 The main constraint comes from gamma and X--rays produced in the annihilation either directly or through cascade decays of the final state. The relevant observable is the photon 
 flux at the detector within the energy range $dE$ and solid angle $d\Omega$,  
 \begin{equation}
{d N_\gamma \over dE \,dt\, d\Omega} = {A\over 4\pi} \; \left(   {dN_\gamma \over dE } \right)_0 \; \kappa \,{\langle \sigma v_r \rangle \over  m^2} \; \int_0^\infty \rho^2_{\rm DM} dr \;.
\end{equation}
Here $N_\gamma$ is the photon number, $A$ is the detector area, $(dN_\gamma / dE)_0$ is the photon number within the energy range  $dE$ emitted {\it per annihilation}, and $\kappa=1/2$ 
for self--conjugate DM (as in our case) or $1/4$ for DM consisting of particles and anti--particles. The last factor is the DM density squared integral along the line of sight defined by the angular variables
 $(\varphi , \theta)$, often called the $J$--factor,
 \begin{equation}
J (\varphi , \theta ) \equiv \int_0^\infty \rho^2_{\rm DM} (\varphi , \theta, r ) \,dr \;,
\end{equation}
where $r$ is the distance from the detector. It is a function of the DM density profile, e.g. Navarro-Frenk-White (NFW)  \cite{Navarro:1995iw} or Einasto \cite{Einasto:1965czb}.

Although the above formulae are  quite    elementary, it is worth clarifying them in order to understand their limitations. The annihilation rate per unit volume is given by the average cross section and the DM number density, $\Gamma = 
\langle \sigma v_r \rangle \, n_{\rm DM}^2$. 
Since the photon emission is isotropic, the fraction $A/(4\pi r^2)$ of the photons reaches the detector.  
Each elementary volume $dV=r^2 dr d\Omega$ thus contributes $\Gamma \,dV\, A/(4\pi r^2)\, (dN_\gamma / dE)_0$ detector photons per unit time and $dE$. Integrating over $r$ along the line of sight
and recalling that in the non--relativistic limit $\rho_{\rm DM} = m\,n_{\rm DM}$, one obtains the above results. The $\kappa$--factor takes into account the phase space symmetry of the initial state as well as the difference between 
the total DM density and the density of its annihilating components. For identical particles, there is just one component and $\kappa =1/2$. For DM consisting of particles and anti--particles, there is no symmetry factor,
but the annihilation rate is  $\Gamma = 
\langle \sigma v_r \rangle \, n_{+} n_-$, where $n_+ = n_-=n_{\rm DM}/2$ are the particle and antiparticle densities, respectively. Hence, $\kappa =1/4$ in this case. 
 
Although the $J$--factor is larger in the Galactic Center, the dwarf spheroidal galaxies impose a stronger constraint on DM annihilation \cite{Ackermann:2015zua}. The latter contain fewer baryons leading to 
a smaller background and better control over a possible signal. The resulting bounds on the model parameters have been discussed in \cite{Cline:2013gha},\cite{Beniwal:2015sdl},\cite{Athron:2017kgt},\cite{Athron:2018hpc},
yet the XENON1T direct detection constraints and the LHC bound supersede those from indirect detection. In practice, the indirect detection constraints are conveniently
incorporated using 
{\texttt{micrOMEGAs}} \cite{Belanger:2008sj}.
\\ \ \\
 {\bf \underline{Combined constraints.}}
The latest comprehensive analysis of all of the constraints has been performed within the  {\textsf{GAMBIT}} framework \cite{Athron:2018ipf}. It assumes thermal production of DM and takes into account the LHC, direct and indirect DM detection bounds. In addition, it requires perturbativity of the couplings at least up to the scale max$(m,m_t)$, which  bounds
$\lambda_{\phi h}$ from above by  $\sqrt{4\pi}$ in this energy range. At $m \gg m_t$, the Higgs portal coupling necessary for the right DM relic abundance becomes large, so the requirement of perturbativity
of $\lambda_{\phi h}$ and $\lambda_h$ sets the limit $m<4.5$ TeV \cite{Athron:2018ipf}. Among the other constraints, the most prominent role is played by the direct DM detection bound from XENON1T,
which rules out much of the parameter space. It gets weaker at $m> 1$ TeV \cite{Arcadi:2019lka} such that the allowed parameter space is skewed towards large DM masses. The constraint is also loose around $m\simeq m_{h_0}/2$,
where a very small $\lambda_{\phi h}$ is sufficient to obtain the correct relic DM abundance.
\begin{figure}[h!] 
\centering{   
\includegraphics[scale=0.72]{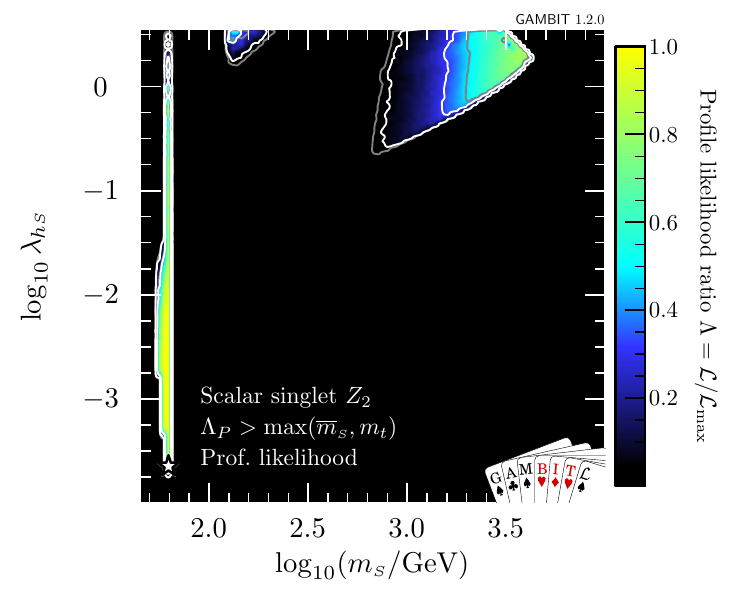}
 }
\caption{ \label{scalarZ2}
Dark matter mass vs the Higgs portal coupling in the scalar DM model subject to all of the constraints. The color coding corresponds to the profile likelihood, with the light regions being more likely. 
The white contours  indicate $1\sigma$ and $2\sigma$ confidence regions.
In terms of the convention in the text, $m_S \equiv m$, $\lambda_{hS} \equiv \lambda_{\phi h}$. The figure is from Ref.\;\cite{Athron:2018ipf}  \copyright\, CC--BY.
}
\end{figure}

Fig.\;\ref{scalarZ2} shows  two notable (light--colored) regions where all of the constraints are satisfied \cite{Athron:2018ipf}. The first one is the resonant annihilation region at $m \simeq m_{h_0}/2$, while the other
corresponds to multi--TeV DM with order one coupling. The constraints are also satisfied for  $m$ close to $ m_{h_0}$, although the allowed parameter range is small.
The profile likelihood has been computed by varying the model parameters as well as the nuisance parameters, e.g. the DM local density and mean velocity,
nuclear matrix elements, etc.\footnote{Ref.\,\cite{Athron:2018ipf} also treats the relic DM density constraint as an upper bound: $Y( t \rightarrow\infty)  \leq Y_{\rm obs}$.  The lower edges of the allowed regions 
correspond to the observed value (see also \cite{Athron:2017kgt}). } The scanning procedure is described in detail in \cite{Athron:2018ipf}.

The conclusion is that, apart from rather contrived possibilities $m\simeq m_{h_0}/2$ and  $m\simeq m_{h_0}$, 
the scalar DM mass  is pushed into the multi--TeV range by the XENON1T bounds. 
Yet, it cannot be too heavy due to perturbativity constraints such that the allowed parameter range will largely be probed by upcoming direct detection experiments.
On the other hand,  the LHC prospects for observing scalar DM are not very promising: its production via an off--shell Higgs is impeded by a small cross section ($< 1$\;fb) \cite{Djouadi:2011aa}. 
The on--shell Higgs decay can be a useful probe, e.g. via monojet events \cite{Djouadi:2012zc}, but its kinematic reach is limited.

 \subsubsection{Pseudo--Goldstone dark matter} 
 
 As we have seen, the minimal Higgs portal model favors   a multi--TeV dark matter mass. 
 This is no longer the case  if one adds one more degree of freedom and makes the scalar  $\phi $ complex.
 In this Section, we  consider an interesting possibility of pseudo--Goldstone dark matter, which naturally satisfies the direct DM detection
 constraints even for light masses \cite{Gross:2017dan}.
 
Consider a scalar potential invariant under a global U(1): $\phi \rightarrow e^{i \xi} \phi$. Upon spontaneous symmetry breaking, this entails a massless Goldstone boson. To make 
the model phenomenologically viable, let us add  soft (dim--2) explicit  symmetry breaking resulting  in a massive pseudo--Goldstone boson. 
The potential then consists of two pieces \cite{Gross:2017dan},
 \begin{eqnarray}
 && V_0 = m_h^2 |H |^2 + m_\phi^2 |\phi |^2 + \lambda_h |H|^4 + \lambda_{\phi h} |H|^2 |\phi  |^2 + \lambda_\phi |\phi |^4 \;, \nonumber \\
 && V_{\rm soft} = - {\tilde m_\phi^{2} \over 4} \, \phi^2 + {\rm h.c.}
 \end{eqnarray}
 The absence of higher dimensional U(1) breaking terms can be justified by treating the couplings as spurions, as we discuss below.
 All the parameters except for $\tilde m_\phi^2$ are real, while $\tilde m_\phi^2$ can be made real and positive by a U(1) rotation. In this case, the potential depends on the phase of $\phi$ 
 as $-\cos (2 {\rm Arg}\,\phi)$ and is minimized at ${\rm Arg} \, \phi =0$.

 An important feature of the above potential is that after spontaneous symmetry breaking it retains the CP--like symmetry,
 \begin{equation}
 \phi \rightarrow \phi^* \;.
 \end{equation}
 This makes the imaginary part of $\phi$ stable and, thus, a dark matter candidate. Expressing
 \begin{equation}
\phi = {1\over \sqrt{2}} \left(  \varphi + i\chi +  w      \right) \;,
 \end{equation}
 where the VEV $w$ is real, one finds that $\varphi$ mixes with the SM Higgs producing two CP--even mass eigenstates $h_1, h_2$ as detailed in Section\;\ref{sec-z2}.
 The formulae for the mixing angle and the mass eigenvalues  (\ref{tantwotheta}),(\ref{evalues}) remain the same. The pseudoscalar component $\chi $ does not mix
 with the scalars  and  plays the role of dark matter with mass
  \begin{equation}
m_\chi = \tilde m_\phi \;.
 \end{equation}

\begin{figure}[h] 
\centering{   
\includegraphics[scale=0.25]{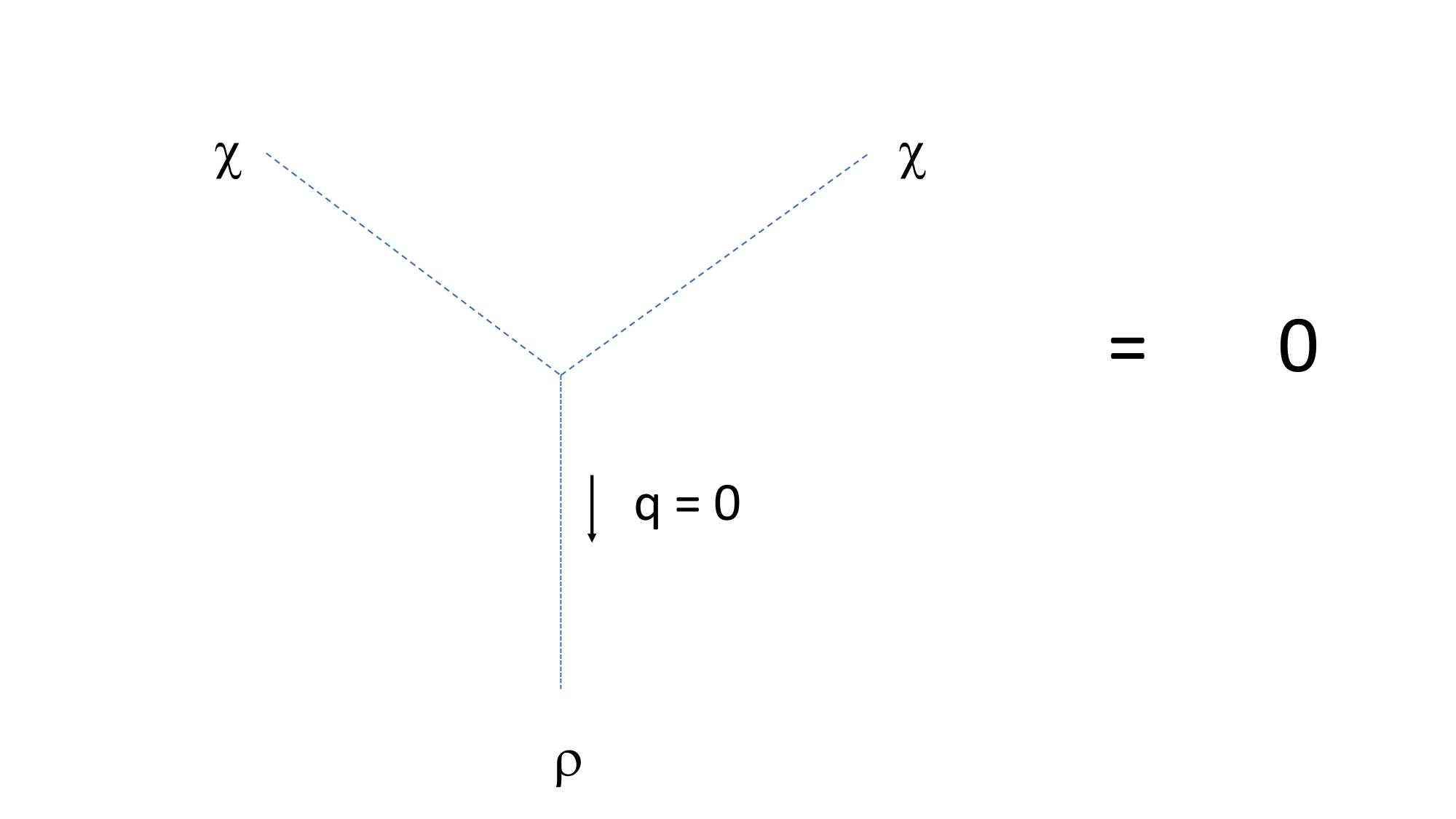}
 }
\caption{ \label{goldstone-vertex}
The pseudo--Goldstone boson vertex vanishes for $\chi$ on--shell and zero momentum transfer.
}
\end{figure}

 The interactions relevant to DM--nucleon scattering calculations are
 \begin{equation}
 \Delta {\cal L}_{\chi - f} =
 {1\over 2 w} \, \chi^2 \; \left( h_1 \; m_1^2 \, \sin\theta - h_2 \; m_2^2 \, \cos\theta\right) - \sum_f {m_f \over v} \; \bar f f \, \left( h_1 \cos\theta + h_2 \sin\theta\right) \;.
 \label{L-chi-f}
 \end{equation}
 The scattering proceeds via the $t$--channel $h_1$ and $h_2$ exchange. One immediately concludes that the two contributions cancel each other in the limit of zero momentum transfer,
  \begin{equation}
 {\cal A} \propto \sin\theta \, \cos\theta \; \left(   {m_2^2 \over t -m_2^2} -    {m_1^2 \over t -m_1^2} \right) \approx 0 \;,
  \end{equation}
where the neglected terms are of order $t/m_{1,2}^2 \sim {\bf {q}}^2/m_{1,2}^2 $, where ${\bf {q}}$ is the momentum transfer.
 This remarkable cancellation has a simple explanation. The pseudo--Goldstone $\chi$ inherits the Goldstone boson property that the trilinear vertex vanishes in the limit of zero momentum transfer,
 as long as the {\it massive} $\chi$ is on--shell. This is most easily seen in polar coordinates,
 \begin{equation}
 \phi = {1\over \sqrt{2}} \, \rho \, e^{i\chi} \;,
 \end{equation}
 such that the $\chi \, \chi \,  \rho \, (q=0) $ vertex vanishes when   $(\Box + m_\chi^2)\,\chi =0$, as shown in Fig.\;\ref{goldstone-vertex}. This property, however, is not respected by
 higher dimensional U(1)--breaking operators, e.g. $\phi^4$. As long as the latter are suppressed, the above amplitude is vanishingly small.

The main contribution to the direct detection amplitude is generated at 1--loop level. Indeed, radiative corrections induce dim--4  U(1)--breaking operators
  \begin{equation}
\Delta V = \lambda_{\phi h}^\prime | H |^2 \phi^2 + \lambda_\phi^\prime \phi^4 + \lambda_\phi^{\prime \prime} |\phi |^2 \phi^2 + {\rm h.c.}
 \end{equation}
 The couplings are all real  so that the symmetry $\phi \rightarrow \phi^* $ is unbroken and vanish as $m_\chi \rightarrow 0$. Their explicit expressions can be found in \cite{Gross:2017dan}, while for our purposes it suffices to know that
 they are bounded by ${\cal O} (\lambda_{\phi h} \lambda_\phi /16 \pi^2)$ or ${\cal O} ( \lambda_\phi^2 /16 \pi^2)$. The resulting DM--nucleon scattering cross section is of order \cite{Gross:2017dan}
  \begin{equation}
\sigma_{\chi - N}^{\rm SI} \sim {\sin^2 \theta \over 64\pi^5 } \, {m_N^4 f_N^2 \over m_1^4 v^2} \, {m_2^8 \over m_\chi^2 w^6} 
 \end{equation}
 for EW scale masses, while for light $\chi$ it is suppressed further. It does not exceed $10^{-49} $\;cm$^2$, whereas the current XENON1T upper bound is $4\times 10^{-47}$\;cm$^2$
 at the most sensitive point. 
 Similar results are obtained by a careful analysis of $\chi - N$ scattering at 1 loop \cite{Azevedo:2018exj},\cite{Ishiwata:2018sdi},\cite{Glaus:2020ihj}.
  As a result, absence of direct DM detection does not impose any significant bound on the model parameters. In fact, a more stringent constraint is set by
 perturbative unitarity: the process $h_2 h_2 \rightarrow h_2 h_2$ requires $\lambda_\phi \lesssim 4\pi/3$ \cite{Chen:2014ask}, which can be translated into a bound on $w$ for a fixed $m_2$.

 The DM annihilation amplitude does not exhibit  the same suppression since the momentum transfer in this case is significant, $2 m_\chi$. Therefore, it is easy to reconcile the direct DM detection bounds
 with the required DM annihilation cross section. While $m_\chi < m_1/2$ is essentially excluded by the Higgs invisible width measurements, the indirect  detection 
 constraint is insignificant \cite{Huitu:2018gbc} such that  most of the remaining mass range is available.

\begin{figure}[h] 
\centering{   
\includegraphics[scale=0.78]{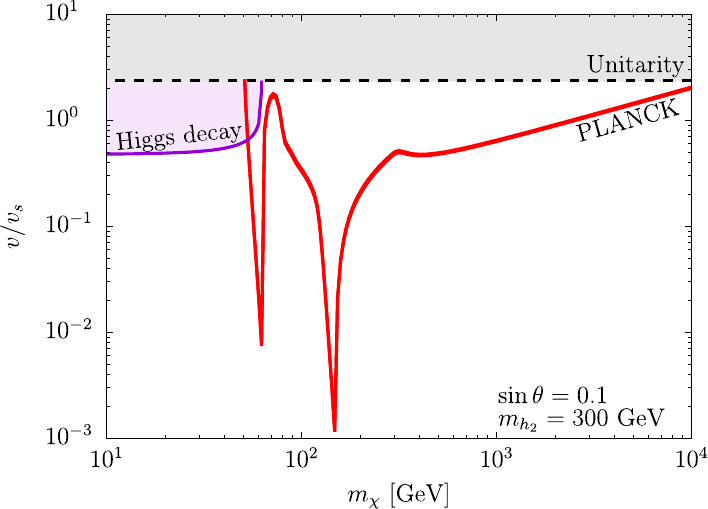}
 }
\caption{ \label{fig-goldstone-plot}
Constraints on pseudo--Goldstone dark matter \cite{Gross:2017dan}. The red band corresponds to the correct thermal relic abundance as measured by PLANCK. The purple shaded region is excluded by invisible Higgs decay,
while in the grey region perturbative unitarity is violated. In terms of the convention in the text, $v_s \equiv w$. 
{\it Figure credit:} reprinted figure with permission from C.~Gross, O.~Lebedev and T.~Toma,
Phys. Rev. Lett. \textbf{119}, no.19, 191801 (2017);
doi:10.1103/PhysRevLett.119.191801. Copyright (2017) by the American Physical Society.
}
\end{figure}

 An example of the allowed parameter space is shown in Fig.\;\ref{fig-goldstone-plot}. The vertical axis $v/w$ represents the strength of the DM coupling to $h_1, h_2$  (see Eq.\,\ref{L-chi-f}).
 The red band is consistent with the thermal DM relic abundance \cite{Ade:2015xua}. 
It features  the characteristic dips associated with resonant annihilation through $h_1$ and $h_2$.  Most points within the band satisfy all the other constraints such that the 
allowed DM mass ranges from about 60 GeV to 10 TeV.
 
 Some of the parameter space can be probed at HL--LHC \cite{Huitu:2018gbc}. The most promising channel  appears to be VBF production of $h_1$ and $h_2$ with their subsequent  (on--shell) decay  
 into a pair of $\chi$'s. For the $h_2$ mode, the invisible decay branching ratio can be large and its kinematic reach extends further than that for $h_1$ does, making it a particularly interesting probe 
 of pseudo--Goldstone dark matter. 
 While the thermal DM relic appears elusive to the LHC searches (apart from the resonance regions), the non--thermal option can be probed efficiently for  $\sin\theta \gtrsim 0.1$ and 
  $m_2 \,, \, m_\chi < m_2/2 $
 in the electroweak range. The indirect DM detection constraint plays a significant role in this case \cite{Huitu:2018gbc}, while future missions can become a sensitive probe of the model \cite{Arina:2019tib}.

 Let us now discuss how this set--up can arise from a more fundamental theory. We have neglected the U(1) breaking terms of dimension higher than 2. This can be justified if the underlying, possibly gauge,  U(1) symmetry is broken at a high scale  by a VEV of another scalar $\Phi$ with even U(1) charge $q_\Phi$, while the charge of $\phi$,   $q_\phi$, is odd. The residual parity symmetry $\phi \rightarrow -\phi$ requires that the low energy effective theory contain
 only even powers of $\phi$. The higher dimensional operators are suppressed by a higher power of the   effective theory scale $\Lambda$. Indeed, the allowed operators have the form 
 \begin{equation}
 {\phi^{2k} \, \Phi^l \over \Lambda^{2k+l-4}}\;,
 \end{equation}
 with integer $k,l$. Define
 \begin{equation}
 \epsilon \equiv  {\langle \Phi \rangle \over \Lambda} ~~~,~~~ n \equiv -2 \, {q_\phi \over q_\Phi} \;,
 \end{equation}
 where $n $ can be chosen integer and $\epsilon \ll 1$. Then
 \begin{equation}
 \tilde m_\phi^2 \sim \langle \Phi \rangle^2 \, \epsilon^{n-2} ~~,~~ \lambda_{\phi h}^\prime \sim \lambda_\phi^{\prime\prime} \sim \epsilon^n ~~,~~ 
  \lambda_\phi^{\prime} \sim \epsilon^{2n} \;.
 \end{equation}
 Clearly, for small enough $\epsilon$, the leading term is $\tilde m_\phi^2$, while the others are suppressed. The higher dimensional terms can violate CP leading to DM decay. However,
 a very small $\epsilon$ makes the decay  time much longer than the age of the Universe such that dark matter can be treated as stable for all practical purposes  \cite{Gross:2017dan}.
 
 Various aspects of the pseudo--Goldstone dark matter  and its generalizations  have been studied in \cite{Alanne:2018zjm}--\cite{Abe:2020ldj}. 
 In particular, the  gravitational wave output during a phase transition  in such setups was considered in \cite{Kannike:2019wsn},\cite{Kannike:2019mzk},\cite{Alanne:2020jwx}.
In \cite{Abe:2020iph},\cite{Okada:2020zxo} it was noted that 
 the U(1) can be identified with the B--L symmetry. 
 Possible venues to generalize the set--up to non--Abelian  symmetries were explored in \cite{Karamitros:2019ewv}.

 Finally, let us make a general note that scalar dark matter in the Higgs portal framework can have properties significantly deviating from those of the standard thermal DM. Some examples are presented in 
 \cite{Ponton:2019hux},\cite{Azatov:2021ifm}.

 \subsection{Vector dark matter}
 
 The concept of gauge symmetry plays a central role in the Standard Model. It is thus reasonable to assume that it also features in the hidden sector. As we show below, 
 stability of dark matter can be a natural consequence of gauge symmetry. 
 
 Vector dark matter stability stems from symmetries of the Lie groups, including their outer automorphisms. When combined with the minimal field content, they result in  ``custodial''  symmetries of the Lagrangian
which stabilize dark matter \cite{Hambye:2008bq},\cite{Lebedev:2011iq},\cite{Gross:2015cwa}. This phenomenon is analogous to  the appearance of the baryon number symmetry in the SM: 
SU(3)\,$\times$\,SU(2)\,$\times$\,U(1) gauge symmetry and the SM field content 
 automatically imply a  conserved baryon number (apart from the anomaly) which makes the proton stable. In our case, some of the gauge fields are stable and play the role of dark matter.

 Suppose the hidden sector is endowed with gauge symmetry. To avoid appearance of light or massless states, let us introduce the minimal number of hidden ``Higgs'' multiplets $\phi_i$ to break this gauge symmetry completely.\footnote{Massless states can also be avoided via confinement in the hidden sector, see e.g. \cite{Hambye:2009fg}.} Such multiplets couple naturally to the Standard Model  through the Higgs portal. At the renormalizable level, 
 the coupling has to be quartic since
 the trilinear terms are forbidden by gauge symmetry. For the U(N) dark sector symmetry and $\phi_i$ in fundamental representation, the Higgs portal couplings have the form 
  \begin{equation}
V_{\phi_i h}= \lambda_{\phi_i h} \, H^\dagger H \phi^\dagger_i \phi_i \;,
\label{V-portal}
    \end{equation}
 where $i$ labels the hidden Higgs multiplets. If $\phi_i$ with different $i$ have the same quantum numbers under other possible symmetries, the mixed terms involving $\phi_i^\dagger \phi_j$
are also allowed. 
 One general consequence of these couplings is that the Higgs mixes with the hidden scalars  because both $H$ and $\phi_i$ develop non--zero VEVs. These states then mediate interactions between
 dark and observable matter.
 
In what follows, we focus on unitary gauge groups in analogy with the Standard Model, although other possibilities can be equally viable.
Let us start with   the simplest case of an Abelian symmetry group.

\subsubsection{U(1) gauge symmetry}

Consider a U(1) gauge theory with a   charged scalar $\phi$ \cite{Lebedev:2011iq},
 \begin{equation}
 {\cal L_{\rm dark}}= -{1\over 4} F_{\mu\nu} F^{ \mu\nu} + (D_\mu \phi)^\dagger D^\mu \phi -V \;,
 \end{equation}
 where $F_{\mu\nu}$ is the field strength tensor of the gauge field $A_\mu$ and $\phi$ is assigned  charge +1/2 (for an easier comparison with the non--Abelian case).
 Suppose $\phi$ develops a VEV,
  \begin{equation}
  \langle \phi \rangle= {1\over \sqrt{2}}~ \tilde v \;, 
  \end{equation}
  which can always be taken real. 
   The imaginary part of $\phi$ gets absorbed  by the gauge field that now becomes massive,
    \begin{equation}
   m_A={ \tilde g \tilde v \over 2} \; , 
   \end{equation}
   where $\tilde g$ is the gauge coupling. The real part of $\phi$ is a physical scalar, which in the unitary gauge can be written as
     \begin{equation}
    \phi={1\over \sqrt{2}} \; \bigl(\rho +\tilde v \bigr) \;,
   \end{equation}
   such that $\rho$ is canonically normalized. This theory has the following gauge--scalar interactions:
 \begin{eqnarray}
&& \Delta {\cal L}_{\rm s-g}= {\tilde g^2\over 4} \tilde v \rho \; A_\mu A^{ \mu} +
 {\tilde g^2\over 8} \rho^2 \; A_\mu A^{\mu} \;. 
 \end{eqnarray}
 We observe that  the system possesses the $Z_2$ symmetry 
\begin{equation}
A_\mu \rightarrow - A_\mu ~,
\end{equation} 
which corresponds to the usual charge conjugation. In the Lie group language, it acts as an outer automorphism by conjugating the U(1) group elements. 
In terms of the original fields, the symmetry transforms  $\phi \rightarrow \phi^*$ and $A_\mu \rightarrow - A_\mu$, which leaves 
 both the Lagrangian and the vacuum invariant. 
 
Since the vertices always contain pairs of $A_\mu$, the massive gauge field is stable and can constitute dark matter. It interacts with the SM fields via the scalar mediators: due to the Higgs portal coupling 
(\ref{V-portal}),
$\rho$ mixes with the Higgs $h$ producing mass eigenstates $h_1$,$h_2$ as in Section~\ref{sec-z2},
 \begin{eqnarray}
&& \rho= - h_1 \; \sin\theta + h_2 \; \cos\theta \;, \nonumber\\
&& h = h_1 \; \cos\theta + h_2 \; \sin\theta \;.
 \end{eqnarray}
Interactions mediated by $h_1$ and $h_2$ allow for the usual DM freeze--out and also produce a direct detection signal, which we discuss in Section~\ref{vector-pheno}. 
If the mixing angle is sufficiently small and $h_2$ can be  integrated out, we recover the effective Higgs--vector interaction (\ref{h-v}).

The above considerations assume that the U(1) does not mix with the SM hypercharge, that is,  no $Z_2$--violating $F_{\mu\nu}F^{\mu\nu}_Y$ term is present in the Lagrangian. This is justified if the U(1) is secluded, i.e. there are no fields charged under both the SM gauge group and the dark sector symmetry.
In this case, if the U(1) and U(1)$_Y$ are orthogonal initially so that there is no tree level kinetic mixing, such a mixing will not be generated radiatively. Settings of this type are common in string theory, e.g.
in the $E_8 \times E_8$ heterotic string \cite{Gross:1984dd}, where the SM gauge group originates from one $E_8$, while the dark symmetry comes from the other $E_8$. Similar considerations  apply to field theoretic 
constructions based on  group products $G_1 \times G_2$ which contain no fields  transforming under both groups, but possess a  Higgs portal--like coupling $|s_1|^2 |s_2|^2$,
where $s_1$ and $s_2$ are scalars transforming under $G_1$ and $G_2$, respectively. Appropriate symmetry breaking within each sector  brings us to the framework considered here.

The model at hand can be viewed as an effective field theory limit of a more fundamental framework.
In general, higher dimensional operators may affect stability of dark matter. This depends on whether or not the UV completion preserves the charge conjugation symmetry ${\cal C}$.
The effective theory is built out of gauge invariant operators 
\begin{equation}
{O}\left (F_{\mu\nu} , D_\mu, \phi, \phi^* \right)  \;.
\label{non-ren-O}
\end{equation}
If ${\cal C}$ is violated, operators like 
\begin{equation}
c \; \left[  \phi^* D_\mu \phi \right] \,  \left[  \phi^* D^\mu \phi \right]  + {\rm h.c.}
\label{c-viol}
\end{equation}
with complex $c$ induce linear couplings of the gauge field to $\rho$. 
In this case, $A_\mu$ decays via $\rho - h$ mixing, although its lifetime can be very long depending on the suppression scale of the dim-6 operator. 
This leads to an interesting option of long--lived dark matter.

Also, if  the matter sector at low energies  is enlarged, the stabilizing $Z_2$ can be broken. For instance, introduction of the second charged scalar generally entails both the presence of a linear in $A_\mu$ coupling and a scalar--pseudo-scalar mixing. In this case, stability of dark matter is lost.
 
 Our conclusion here is that a very simple renormalizable model  with a secluded U(1) and a single dark Higgs leads to stable
  dark matter. The stability is guaranteed by gauge symmetry and  the minimal field content.

\subsubsection{SU(2) gauge symmetry}

The above considerations are extended to the SU(2) case \cite{Hambye:2008bq} in a straightforward manner. It is sufficient to introduce a single dark Higgs doublet $\phi$ in order to break the 
 symmetry  completely.
 The relevant Lagrangian is 
 \begin{equation}
 {\cal L_{\rm dark}}= -{1\over 4} F_{\mu\nu}^a F^{a \; \mu\nu} + (D_\mu \phi)^\dagger D^\mu \phi -V\;,
 \end{equation}
where $a=1,2,3$ is the adjoint SU(2) index. At the minimum of the potential, $
\phi$ develops a VEV $\tilde v$ and the field can be written (in the unitary gauge) as
\begin{equation}
\phi = {1\over \sqrt{2}}\; \left( 
\begin{matrix}
 0 \\
 \rho +\tilde v
\end{matrix}
\right)~,
\end{equation}
where $\rho$ is real. 
As in the U(1) case, the vector field mass in terms of the gauge coupling $\tilde g$ is given by $m_A= \tilde g \tilde v/2$. The SU(2) gauge interactions can be split into two terms,
\begin{eqnarray}
&& \Delta {\cal L}_{\rm s-g}= {\tilde g^2\over 4} \tilde v \rho \; A_\mu^a A^{ a\; \mu} +
 {\tilde g^2\over 8} \rho^2 \; A_\mu^a A^{ a\; \mu} \;, \nonumber \\
&& \Delta {\cal L}_{\rm g-g} = - \tilde g \epsilon^{abc} (\partial_\mu A_\nu^a) A^{\mu \;b}
A^{\nu \;c} -{\tilde g^2\over 4} \left( (A_\mu^a A^{\mu \;a})^2 -
 A_\mu^a A_\nu^a \; A^{\mu \;b} A^{\nu \;b} \right) ~. 
\end{eqnarray} 
Although the triple gauge vertex breaks the $Z_2$ of the previous section, the Lagrangian preserves a custodial SO(3),
\begin{equation}
A_\mu \rightarrow {\cal O} A_\mu \;,
 \end{equation}
 with ${\cal O} \in$ SO(3).
 This symmetry contains analogs of the $Z_2$ parity which reflect two of the gauge fields. Since only the gauge fields transform under SO(3) and  are degenerate in mass, they
 are stable.  In many respects, such dark matter behaves similarly to the Abelian vector DM, except it experiences  self--interaction.  An advantage over the U(1) case is that the kinetic mixing 
 with the SM gauge bosons is forbidden by gauge invariance.
  
  An interesting feature of non--Abelian dark sectors is that some DM components  remain stable in the presence of arbitrary non--renormalizable operators.
  As long as gauge invariance is preserved
  by integrating out  heavy states, 
  the effective theory is built out of operators  
  \begin{equation}
{O}\left (F_{\mu\nu}^a , D_\mu, \phi, \phi^\dagger \right)  \;.
\label{non-ren-O-su2}
\end{equation}
SU(2) invariance requires the number of $\phi$'s to be equal to the number of $\phi^\dagger$'s. Thus, the system has a global U(1) symmetry 
\begin{equation}
\phi \rightarrow  e^{i \zeta } \phi \;.
\end{equation}
A combination of this U(1) and gauge transformation ${\cal U}= {\rm diag} (e^{i \zeta } , e^{-i \zeta } )$ remains unbroken by the vacuum. Therefore, after spontaneous symmetry breaking,
the system  possesses  a global U(1)$^\prime$  $\sim$ SO(2) and, generally,  the custodial symmetry gets reduced by higher-dimensional operators:
\begin{equation}
{\rm SO(3) \rightarrow SO(2)} \;.
\end{equation}
Under this SO(2),
  \begin{eqnarray}
&&A_1 \rightarrow A_1 \; \cos 2\zeta + A_2 \; \sin 2\zeta \;, \nonumber \\
&&A_2 \rightarrow -A_1 \; \sin 2\zeta + A_2 \; \cos 2\zeta \;,
\end{eqnarray}
  while all other fields are invariant. As a result, $A_{1,2}$ remain stable and constitute dark matter.  
  On the other hand, $A_3$ decays, for instance, via an analog of operator  (\ref{c-viol}) which breaks SO(3). Decaying SU(2) dark matter can produce an interesting gamma ray signature \cite{Arina:2009uq} .

\subsubsection{SU(3) and larger groups }

The SU(3) case \cite{Gross:2015cwa} is more complicated as it involves multiple dark Higgses. One may instead use a larger group representation, e.g. a {\bf 6}--plet, to break the symmetry, however, the corresponding scalar potential 
is too constrained and the breaking is not complete. The minimal option to break SU(N) to nothing employs N-1 dark scalar multiplets in the fundamental representation \cite{Gross:2015cwa}.
 
Consider an SU(3) gauge theory with 2 scalar triplets $\phi_1$, $\phi_2$,
\begin{equation}
 {\cal L}_{\rm dark} = - \frac12 \textrm{tr} \{G_{\mu \nu} G^{\mu \nu}\} + |D_\mu \phi_1|^2 + |D_\mu \phi_2|^2 -V_{} \,.
  \end{equation}
The potential  $V = V_{\rm portal} + V_{\rm dark}$ includes  the Higgs portal terms,
\begin{equation}
V_{\rm portal}= \lambda_{H11} \, |H|^2 |\phi_1|^2 + \lambda_{H22} \, |H|^2 | \phi_2|^2 - ( \lambda_{H12} \, |H|^2 \phi_1^\dagger \phi_2 + {\rm h.c.})\;,
  \end{equation}
as well as the purely dark scalar potential,
\bal
V_{\rm dark} &=
m_{11}^2 |\phi_1|^2
+ m_{22}^2 |\phi_2|^2
- ( m_{12}^2 \phi_1^\dagger \phi_2 +   {\rm h.c.})
\nn \\ 
& 
+ \frac{\lambda_1}{2} |\phi_1|^4
+ \frac{\lambda_2}{2} |\phi_2|^4
+ \lambda_3 |\phi_1|^2 |\phi_2|^2
+ \lambda_4 | \phi_1^\dagger\phi_2 |^2
\nn \\ 
& 
+ \left[
\frac{ \lambda_5}{2} ( \phi_1^\dagger\phi_2 )^2
+ \lambda_6 |\phi_1|^2
( \phi_1^\dagger\phi_2)
+ \lambda_7 |\phi_2|^2
( \phi_1^\dagger\phi_2 )
+ {\rm h.c.} \right] \,.
\eal
SU(3) gauge freedom allows one to remove 5 degrees of freedom of $\phi_1$ and 3 degrees of freedom of $\phi_2$, so that they can be put in the form
\begin{equation} 
\label{unitarygauge}
\phi_1={1\over \sqrt{2}} \,
\left( \begin{array}{c}
0\\0\\v_1+\varphi_1
\end{array} \right) \,,
\quad 
\phi_2= {1 \over \sqrt{2}}\,
\left( \begin{array}{c}
0\\v_2+\varphi_2\\(v_3+\varphi_3) + i (v_4+\varphi_4)
\end{array} \right) ~,
\end{equation}
where  $v_i$ are real VEVs and $\varphi_{1-4} $ are real scalars. 
This gauge choice is analogous to the unitary gauge in the Higgs sector of the Standard Model, $H^T= (0,v+h)/\sqrt{2}$.

The Lagrangian possesses an accidental global U(1),
\begin{equation}
\phi_i \rightarrow e^{i \xi} \phi_i \;.
\end{equation}  
Even though it gets broken by the VEVs, a combination of this U(1) and a gauge transformation
\begin{equation}
{\cal U} = {\rm diag} \left(  e^{2i \xi},e^{-i \xi},e^{-i \xi}\right)
\end{equation}  
remains  preserved. This global U(1)$^\prime$ symmetry guarantees that some of the gauge bosons are stable. Indeed, the gauge fields transform as
\begin{equation}
A^\mu \rightarrow {\cal U}  \; A^\mu \; {\cal U}^\dagger \;, 
\end{equation} 
so,  in the Gell--Mann basis for the SU(3) generators,   $A^\mu_{3,6,7,8}$ remain invariant while $A^\mu_{1,2,4,5}$ transform non--trivially,
\begin{eqnarray}
&&A_1 \rightarrow A_1 \; \cos 3\xi + A_2 \; \sin 3\xi \;, \nonumber \\
&&A_2 \rightarrow -A_1 \; \sin 3\xi + A_2 \; \cos 3\xi \;,
\end{eqnarray}
where the Lorentz index has been suppressed. 
The same relation holds for the $(A_4, A_5)$ pair up to the replacement $1\rightarrow 4$, $2\rightarrow 5$. The scalars are all neutral under U(1)$^\prime$, so  the lighter pair out of $(A_1,A_2)$ and $(A_4,A_5)$
has to be stable. Note that U(1)$^\prime$ ensures $m_{A_1}= m_{A_2}$ and $m_{A_4}= m_{A_5}$.

The potential is quite complicated. To simplify its analysis, let us assume that CP is conserved in the dark sector: the couplings and the VEVs of $\phi_{1,2}$ are real ($v_4=0$).
In this case, the system possesses an extra $Z_2$ analogous to charge conjugation, i.e. it conjugates the group elements and the scalar fields. Under the $Z_2$,
 \begin{eqnarray}
 A_{1,3,4,6,8} &\rightarrow& - A_{1,3,4,6,8}\;, \nonumber\\
 \varphi_4 &\rightarrow& - \varphi_4 \;,
\end{eqnarray}
while the other fields are invariant.
This symmetry ensures that additional states are stable. Their identity depends on the   mass spectrum which defines which decays are allowed kinematically.

A complication in the SU(3) case is that, in our gauge, some of the scalars mix kinetically with the gauge bosons, i.e. terms of the type $\kappa_{ai} A^\mu_a \partial_\mu  \phi_i$
appear in the Lagrangian. These are eliminated by a ``gauge--like'' transformation $A^\mu_a \rightarrow A^\mu_a + \omega_{ai} \partial^\mu \phi_i$, where $ \omega_{ai}$ are constants depending on the VEVs. 
 Such a transformation does not affect the gauge field masses and kinetic terms, so we will keep  the same notation for the transformed fields, $A^\mu_a$. 
 
 Let us make a further simplification and set $v_3=0$. Then, the gauge boson masses read 
  \begin{equation}
m^2_{A^1}=m^2_{A^2}=\frac{\tilde g^2 }{4}v_2^2\;,
\quad
m^2_{A^4}=m^2_{A^5}=\frac{\tilde g^2 }{4}v_1^2\;,
\quad
m^2_{A^6}=m^2_{A^7}=\frac{\tilde g^2 }{4}(v_1^2+v_2^2) \;.
\label{mA1}
\end{equation}
The components corresponding to the diagonal generators mix. The mass matrix is diagonalized by
\bal
\left( \begin{array}{c}
A'^3_\mu\\
A'^8_\mu
\end{array} \right)
=
 \left( \begin{array}{c}
\cos \alpha \, A^3_\mu + \sin \alpha \, A^8_\mu\\
\cos \alpha \, A^8_\mu - \sin \alpha \, A^3_\mu
\end{array} \right)\,,
\quad
\textrm{where}
\quad
\tan2 \alpha = \frac{\sqrt{3} v_2^2}{2 v_1^2-v_2^2} \;,
\eal
and the eigenvalues  are
 \begin{equation}
m^2_{A'^3}=\frac{\tilde g^2 v_2^2}{4} \Big(1-\frac{\tan \alpha}{\sqrt{3}}\Big) \,,
\qquad
m^2_{A'^8}=\frac{\tilde g^2 v_1^2}{3} \frac{1}{1-\frac{\tan \alpha}{\sqrt{3}}} \,,
\label{A3prime-mass}
\end{equation}
For $v_1 > v_2$, the lightest gauge bosons are
\begin{equation}
A_1, A_2, A_3^\prime \;.
\end{equation}
In the simplest case, these  constitute dark matter. As long as $A_3^\prime $ is lighter  than the pseudoscalar $\varphi_4$,  their stability is ensured by 
the  U(1)$^\prime \times Z_2$ symmetry. Note that although $A_{1,2}$ are somewhat heavier than 
$A_3^\prime$, their decay is forbidden by  U(1)$^\prime$.

 Dark matter can also be composed of stable components with different spins: if $\varphi_4$ is light, we have  multicomponent DM consisting of $A_1,A_2, \varphi_4$.
 While the scalars all generally mix with the Higgs, $\varphi_4$ does not as long as CP is conserved in the dark sector. 
 In all cases, DM interacts with the SM via exchange of the 
 Higgs--like scalars which are mixtures of $h$ and $\varphi_{1-3}$. 
 The list of the interaction vertices for the canonically normalized fields can be found in  \cite{Gross:2015cwa}. The resulting phenomenology is rather involved \cite{Arcadi:2016kmk}.

In the most general case, the scalar potential contains complex couplings and CP is thus broken. Then, only U(1)$^\prime$ remains and dark matter consists of
\begin{equation}
A_1, A_2 \;.
\end{equation}
This scenario is quite interesting phenomenologically since it allows for pair annihilation of $A_1$ and $A_2$  into the lighter $A_3^\prime$. The latter then decays into the SM states,
facilitating the ``secluded dark matter'' scenario which we discuss later. In this case, the direct DM detection rate is effectively decoupled from the DM annihilation rate.
It is interesting that this scenario is realized regardless of the relation between $v_1$ and $v_2$: one of the diagonal generator fields is always the lightest, so the above  argument applies to the other regimes up to the field relabelling
\cite{Arcadi:2016kmk}.

Higher dimensional operators do not affect stability of dark matter since U(1)$^\prime$ remains unbroken. The argument parallels that for the SU(2) case  (cf.\,Eq.\,\ref{non-ren-O-su2}).

 Generalization to the  SU(N) case is quite straightforward \cite{Gross:2015cwa}: to break SU(N) completely, one introduces N-1 fundamentals. 
 They can be parametrized  in a way similar to (\ref{unitarygauge}) such that N-1 components of the first N--plet are rotated away and so on. One row of all of the multiplets can be set to zero,
 which always entails an unbroken global U(1).
 The lightest gauge fields 
are associated with an SU(2) block that couples to the smallest VEV. In the simplest case, these states constitute  dark matter whose 
  phenomenology is similar to that of SU(3) dark matter.

Finally, it should be noted that breaking SU(N) to nothing is not strictly speaking necessary: the   unbroken gauge group  can confine  at low energies and  thus lead to acceptable phenomenology. 
The logarithmic coupling running creates a scale hierarchy in the dark sector such that 
  one encounters  ``dark radiation'' \cite{Ko:2016fcd} and/or heavy non--WIMP dark matter \cite{Buttazzo:2019iwr},\cite{Buttazzo:2019mvl}.

\subsubsection{Phenomenology of vector dark matter}
\label{vector-pheno}

{\it  \underline{Abelian dark matter}.}  Let us start with the U(1) case.
The relevant terms in the Lagrangian are
\bal
{\cal L} &\supset {1\over 2} m_A^2 A_\mu A^\mu + {\tilde g \, m_A \over 2} 
\left(-h_1 \sin\theta + h_2 \cos\theta \right) A_\mu A^\mu \nonumber \\
&+
{\tilde g^2 \over 8 } \left( h_1^2 \sin^2 \theta -2 h_1 h_2 \sin\theta \;\cos\theta +
h_2^2 \cos^2 \theta \right) A_\mu A^\mu \;.
\label{DMinteractions}
\eal
The scalars $h_1, h_2$ with masses $m_1,m_2$ couple to the SM fields just like the Higgs does, up to the suppression factors $\cos\theta$ and $\sin\theta$, respectively.

It is instructive to highlight the differences between  the scalar and vector DM.
For many applications one may use the 
 the non--relativistic limit $|{\bf p}_A| \ll m_A$, in which case vector DM behaves as 3 distinct massive scalars with no inter--species coupling. Compared to single species scalar DM with  the same density,
 the vector DM annihilation cross section is therefore smaller   by a factor of 3 while the scattering rate off nuclei remains the same. On the other hand, the collider phenomenology can be very different: 
 production of the longitudinal components of the vector at high energies  is enhanced by the derivative (Goldstone) couplings. Thus, the decay $h_2 \rightarrow AA$ with $m_2 \gg m_A$ is much more efficient for the vectors.

The DM scattering on nucleons is mediated by  the $t$--channel exchange of  $h_{1,2}$.
 The corresponding  spin--independent cross is \cite{Gross:2015cwa}
\begin{equation}
\sigma^{\rm SI}_{A-N}= { \tilde g^2 \, \sin^2 2\theta \over 16 \pi} \; {m_N^4 f_N^2 \over v^2} \;
{ (m_{2}^2 - m_{1}^2 )^2  \over m_{1}^4 m_{2}^4} \;,
\end{equation}
where $m_N$ is the nucleon mass and $f_N \simeq 0.3$ parametrizes the Higgs--nucleon coupling.
Although the $h_1$ and $h_2$ contributions come with opposite signs, normally there is no significant cancellation between them and the amplitude  is dominated by the $h_1$ exchange.
For electroweak/TeV masses and $\theta \sim 10^{-1}$, the stringent direct DM detection bounds constrain 
the dark gauge coupling  to be ${\cal O}(10^{-1})$ or below. These relax for much heavier DM, in which case the experimental limits are loose.

\begin{figure}[h] 
\centering{
\includegraphics[scale=0.296]{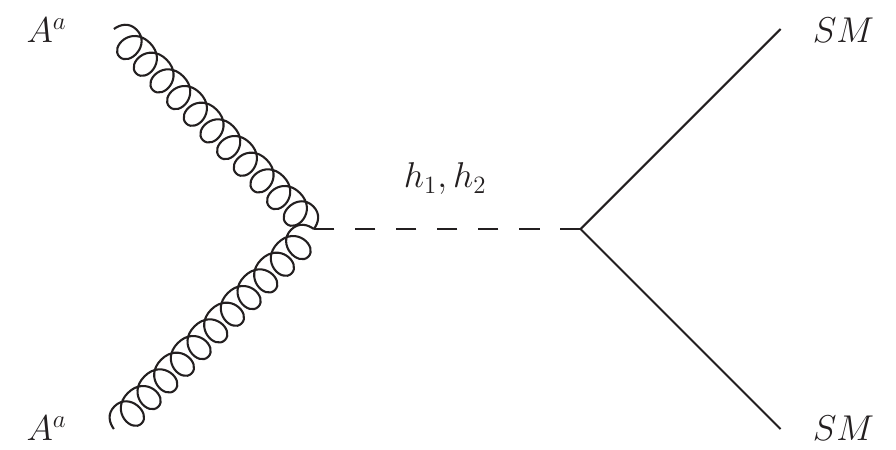}
\includegraphics[scale=0.296]{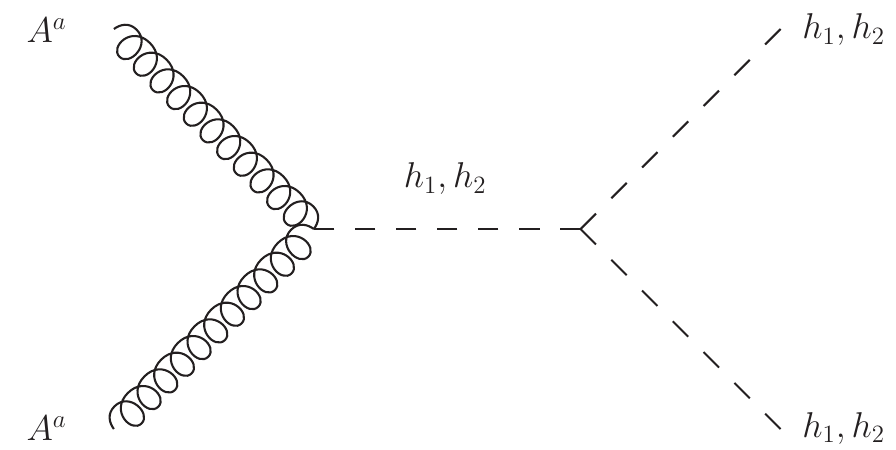}
\includegraphics[scale=0.296]{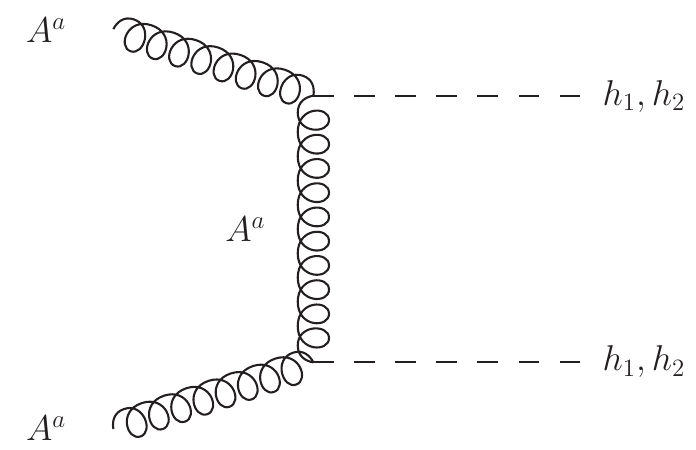}
\includegraphics[scale=0.296]{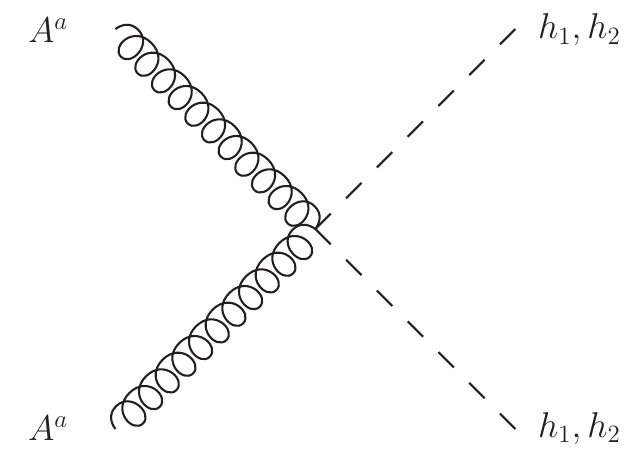}
 }
\caption{ \label{vector-ann}
Leading diagrams for vector DM annihilation \cite{Gross:2015cwa}.
}
\end{figure}

 The DM relic abundance depends strongly on its production mechanism. Let us assume that DM reaches thermal equilibrium with the SM thermal bath, while keeping in mind  that non--thermal 
 production can also be viable. In this case, the abundance is dictated by the annihilation cross section, which is characteristic of WIMP dark matter.  
 The  annihilation process   is mediated by $h_1,h_2$ or 
 produces   $h_1$ and $h_2$ in the final state. The leading diagrams are shown in Fig.\;\ref{vector-ann}. The main $s$--channel  becomes less efficient as $\sin\theta $ decreases.
 For instance, the annihilation cross section into fermions having $N_c^f$ colors is given by
 \begin{equation}
  \sigma v_r \Bigl\vert_{AA \rightarrow \bar f f}=
 \frac{\tilde{g}^2 N_c^f }{48 \pi v^2} \; \sin^2 {2 \theta} \;   \frac{{m_f^2 m_A^2 \left(m_{1}^2-m_{2}^2\right)}^2}{{\left(m_{1}^2-4 m_A^2\right)}^2 {\left(m_{2}^2-4 m_A^2\right)}^2} \; {\left(1-\frac{m_f^2}{m_A^2}\right)}^{3/2} \;,
 \end{equation} 
 while the results for the other final states can be found in \cite{Arcadi:2016kmk}.
  Thus, in order to reconcile 
 the required   cross section with the direct detection constraints, one has to resort to resonant enhancement. Indeed, for $2m_A \sim m_{2}$, the $s$--channel annihilation becomes very efficient even at small couplings. For a heavy $h_2$, the resonance is broad and no significant fine--tuning is necessary to satisfy the bounds \cite{Gross:2015cwa}. On the other hand, the $h_1$ resonance is narrow and the resonant enhancement is at work for $2m_A$ very close to $m_{1}$.  Another interesting regime is $m_A > m_{2}$, in which case pair annihilation into $h_2$'s can dominate, while the direct detection amplitude gets suppressed by small $\theta$. This ``secluded DM'' option will be discussed in the next section. 
 
 Indirect DM detection bounds do not impose any significant additional constraints unless DM is light, $m_A \lesssim m_W$ \cite{Gross:2015cwa}. 
 For $m_A < 62$ GeV, the invisible Higgs decay sets a strong bound on the Higgs--DM coupling. At  $m_A \ll m_1, m_2$, it requires roughly   $\tilde g \, \theta \lesssim 10^{-4} \;m_A/ {\rm GeV}$, which   rules out thermal  DM \cite{Djouadi:2011aa}.

 Altogether, one finds that for electroweak/TeV masses, (thermal) vector dark matter is viable for  $2m_A \sim m_{2}$ and $\tilde g \,\theta \sim {\cal O}(10^{-2})$. Smaller mixing angles are allowed if the secluded option is realized.

Both thermal and non--thermal DM options can to some extent be probed at the LHC. The DM production is mediated by $h_1$ and $h_2$, and manifests itself as missing energy, $E_T$.
 The most promising channels are the monojet and $h_{1,2}$--VBF production with missing energy,
 \begin{eqnarray}
&& pp \rightarrow h_{1,2} \; j  ~\rightarrow  AA\; j \nonumber ~,\\
&& pp \rightarrow h_{1,2}\; jj  \rightarrow AA \; jj \;.
 \end{eqnarray}
 These are efficient if the scalars decay into DM on--shell. The $h_2$ decay appears more interesting as
it has a larger kinematic reach. The corresponding decay width is 
 \begin{equation}
  \Gamma (h_2 \rightarrow AA) = {\tilde g^2  \cos^2\theta \;m_{2}^3 \over 128 \pi m^2_A} \left( 1-{4 m_A^2 \over m_{2}^2} + {12 m_A^4 \over m_{2}^4}  \right) 
  \sqrt{1- {4 m_A^2 \over m_{2}^2}} \;.
 \end{equation} 
The consequent  decay branching ratio is calculated using 
$\Gamma(h_2 \rightarrow {\rm SM}) = \sin^2 \theta \; \Gamma_{\rm SM} (m_h = m_{2})$ as well as the $h_2 \rightarrow h_1 h_1$ width given by 
Eq.\,\ref{gamma211}. To evaluate the signal strength, one also needs 
the  $h_2$ production cross section, which is easily obtained by rescaling the SM result with $m_h = m_2$ by $\sin^2 \theta$.
The cut--based monojet study \cite{Kim:2015hda} concludes that only the light $h_2$ range with $ m_2 \lesssim 300$ GeV can be probed for a realistic $\sin\theta$. Similar results are obtained in the VBF channel \cite{Chen:2015dea}, while the corresponding analysis  in the decoupling limit $m_2 \rightarrow \infty$ can be found in  \cite{Endo:2014cca}.
 The prospects, however, are likely to  improve when modern statistical techniques, e.g. machine learning, are employed (see the discussion below Eq.\,\ref{multiHiggs}).

\begin{figure}[h] 
\centering{
\includegraphics[scale=0.34]{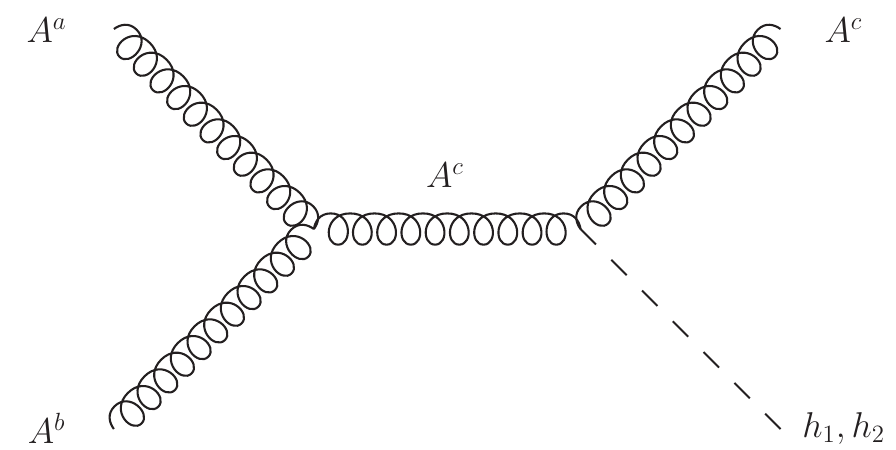} \hspace{0.8cm}
\includegraphics[scale=0.34]{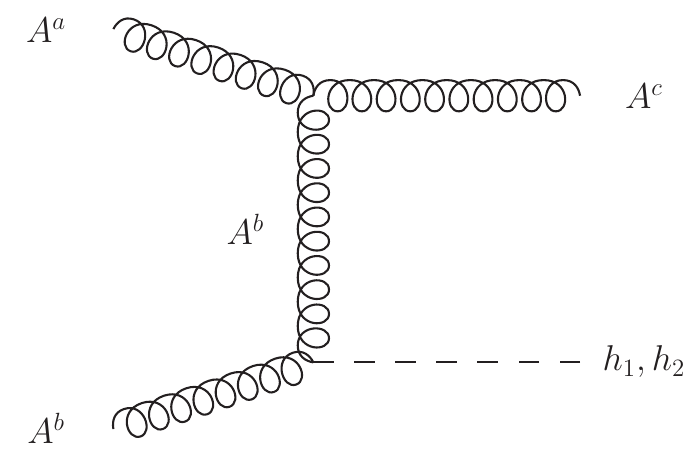}
 }
\caption{ \label{semi-ann}
Semi--annihilation of vector DM \cite{Gross:2015cwa}.
}
\end{figure}

Various aspects of Abelian vector DM, including phenomenology of models with extended scalar sectors,  have been studied in \cite{Farzan:2012hh}-\cite{Azevedo:2018oxv}. For example, in addition to the usual 
constraints, one may require the Higgs portal coupling to stabilize the Higgs potential \cite{Duch:2015jta}. Another variation is to relax the assumption of thermalized DM and focus on its freeze--in production
  \cite{Duch:2017khv}. Loop corrections to the direct detection amplitude have been computed in \cite{Glaus:2019itb}.
\\ \ \\
\noindent
{\it \underline{Non--Abelian dark matter}.} Many properties of Abelian DM are retained up to the summation over the group indices, $A_\mu A^\mu \rightarrow A_\mu^a A^{\mu,a}$, $a=1,..,N^2-1$. 
DM contains multiple components  which can only annihilate with its own species 
resulting in a
  smaller by a factor of $1/({N^2-1})$  annihilation cross section into the SM final states. Yet, DM self--interaction  brings in  new ``semi--annihilation'' channels \cite{Hambye:2008bq},\cite{DEramo:2010keq}, where the
DM number reduces by one, Fig.\,\ref{semi-ann}. However, their effect is not very significant in the allowed parameter space.

DM self--interaction also allows for $3\rightarrow 2$ processes which can be efficient  if the coupling is large enough \cite{Choi:2017zww},\cite{Choi:2019zeb}. 
A scalar analog of this phenomenon will be discussed in Section\;\ref{phi4}.
For yet larger coupling values,
the dark sector undergoes confinement and the massive vectors can be described by  interpolating operators of the type $\phi^\dagger T^a i \overleftrightarrow{D}_\mu \phi$, whose 
phenomenology is discussed in 
\cite{Hambye:2009fg}.

At weak coupling, the main difference from the Abelian case is that there are gauge fields with different masses which generally leads to a broader spectrum of reactions. 
Some of them allow for very efficient DM annihilation even at tiny $\theta $ \cite{Arcadi:2016qoz}, which we discuss in the next section.
Regarding  collider physics, larger non--Abelian sectors generally entail more complicated cascade decays, including those with multi--Higgs final states and missing $E_T$ \cite{Flores:2019hcf}.
For example, the processes 
  \begin{eqnarray}
&& pp \rightarrow h_{n} \;    \rightarrow  \tilde A  \tilde A \rightarrow AA \; h_1 h_1 \nonumber ~,\\
&& pp \rightarrow h_{n}\;  \rightarrow \tilde A  \tilde A      \rightarrow AA \; h_1 h_1 h_1      
\label{multiHiggs}
 \end{eqnarray}
are only possible for dark groups larger than SU(2). Here $h_n $ and $\tilde A$ denote collectively the scalars that mix with the Higgs and unstable gauge fields, respectively.
Again, these channels are efficient if $h_n$ decays into gauge bosons on--shell. At the HL--LHC, the two Higgs mode can be studied with the help of the Boosted Decision Trees technique. 
For $h_1$'s decaying to $\bar b b + \gamma\gamma$ and $\bar b b + WW$, the model can be probed in the regime $m_{h_n} \gg m_A $ with $m_{h_n} \lesssim 600$ GeV and substantial
$\sin\theta \sim 0.3$ \cite{Flores:2019hcf}. For a fully hadronic $h_1$ decay, $h_1 h_1 \rightarrow \bar b b \bar b b$, and heavy $h_n$, one may use jet substructure analysis to probe $m_{h_n}$ up to 1 TeV or so
\cite{Blanke:2019hpe}. The three Higgs final state, on the other hand, is likely to be of little relevance at the LHC  \cite{Flores:2019hcf}.

Various  aspects as well as extensions of the SU(2)  DM  model have been studied in 
\cite{Arina:2009uq},\cite{Hambye:2013dgv}-\cite{Nomura:2020zlm}. For example, adding non--renormalizable operators generally makes some of DM  unstable, 
whose decay can produce gamma ray lines \cite{Arina:2009uq}. 
One may also impose additional constraints on the model, e.g. classical scale invariance \cite{Hambye:2013dgv},\cite{Carone:2013wla},\cite{Karam:2015jta},\cite{Khoze:2016zfi}.
Finally,   variations of the SU(3) vector DM has  been studied in \cite{Karam:2016rsz},\cite{Poulin:2018kap}.

 A recent update of the parameter space analysis of vector dark matter, including the XENON1T constraint as well as the prospects for observing invisible Higgs decay, 
  has appeared in \cite{Arcadi:2021mag}.

\subsubsection{``Secluded'' dark matter: evading  direct detection}
\label{secluded}

The main idea of secluded DM is to decouple direct DM detection from DM annihilation  
using metastable mediators
\cite{Pospelov:2007mp},\cite{Pospelov:2008jd}. This allows one to obtain the correct DM relic density with no pressure from the ever-improving direct detection bounds.

 In our case, the direct detection amplitude goes to zero as $\theta \rightarrow 0$, while the annihilation can still be efficient if there are 
states lighter than DM. In particular, the channel \cite{Arcadi:2016qoz}
\begin{equation}
A_1 A_1 \, , \, A_2 A_2 \rightarrow A_3^\prime A_3^\prime
\end{equation}
is always open kinematically for SU(3) and larger groups. Here the group indices refer to the SU(2) block which couples to the smallest VEV (cf.~Eqs.\,\ref{mA1},\,\ref{A3prime-mass}).
The diagonal component mixes with the other Cartan generators resulting in a lighter mass eigenstate $A_3^\prime$.
If CP is broken in the dark sector, $A_3^\prime$ is unstable and decays into  (on-- or off--shell)  scalars which mix with the Higgs. The relevant vertices are  listed in the Appendix of  \cite{Gross:2015cwa}.
The above annihilation process is controlled by the gauge coupling and the mass difference between $A_{1,2}$ and $A_3^\prime$. It is insensitive to $\theta$, so the correct relic abundance can be obtained 
without violating the direct detection bounds as long as $\theta \ll 1$.

\begin{figure}[h]
\begin{center}
\includegraphics[width=6.8 cm]{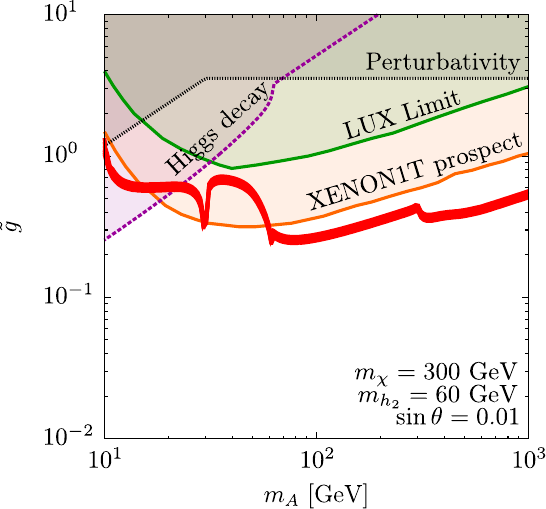}
\includegraphics[width=6.8 cm]{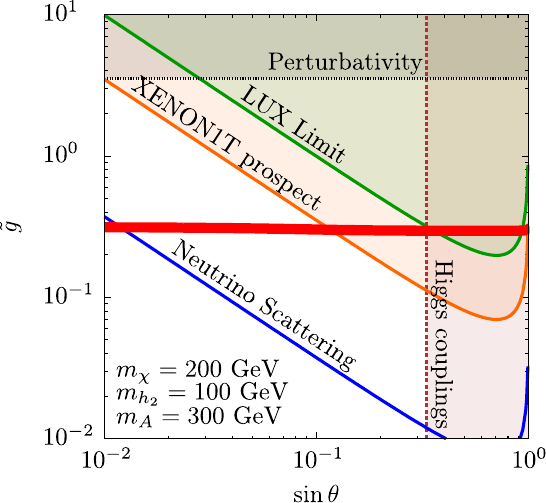}
\end{center}
\caption{ SU(3) vector dark matter constraints in the ``secluded'' regime. 
 The red band indicates the correct DM relic  density. The other curves mark the following constraints:
grey -- perturbativity, purple -- invisible Higgs decay (${\rm BR}_{\rm inv} <0.11$), dark red -- Higgs couplings ($\sin\theta<0.33$), green and orange --   direct DM detection constraints. The blue line represents the direct detection event rate corresponding to the background   neutrino scattering off nuclei. The figure is from Ref.\,\cite{Arcadi:2016qoz}.
 }
\label{fig:dm2}
\end{figure}

In addition to the above channel, there may exist further analogous processes which survive at $\theta \ll 1$. If $h_2$ is lighter than DM,  the channel
\begin{equation}
A_1 A_1 \, , \, A_2 A_2 \rightarrow  h_2 h_2
\end{equation}
is available even for U(1) and SU(2) dark  gauge groups. 
The corresponding cross section in the limit $\theta \ll 1$ and $m_{1} \ll m_A,m_{2}$ is given by \cite{Arcadi:2016qoz}
\bal
 \sigma{v_r}=
\frac{\tilde{g}^4}{576\pi m_A^2} \;  
\frac{11m_{2}^8-80m_{2}^6m_A^2+240m_{2}^4m_A^4-320m_{2}^2m_A^6+176m_A^8}
{\left(4m_A^2-m_{2}^2\right)^2\left(2m_A^2-m_{2}^2\right)^2}  \; \sqrt{1-\frac{m_{2}^2}{m_A^2}} \;,
\eal
while the total DM annihilation cross section is rather complicated. This process is sufficient to reconcile the relic density and direct detection constraints in much of the parameter space with small $\theta$
\cite{Arcadi:2016qoz}. 

A representative  summary of the constraints for the SU(3) case is shown in Fig.\,\ref{fig:dm2}. In addition to $AA \rightarrow
h_2 h_2$, another channel $AA \rightarrow \chi \chi$, where $\chi$ is the mostly pseudoscalar state,   contributes to DM annihilation  when kinematically allowed. 
On the other hand, the channel $AA \rightarrow A_3^\prime A_3^\prime$ is always open.
The red band produces the correct  (thermal) DM relic density and exhibits characteristic dips  at $ m_A \simeq m_2/2 , m_2 , m_\chi$ corresponding to the resonance and kinematic thresholds.
Further constraints include perturbativity of scalar and gauge couplings, bounds on the Higgs couplings and direct detection constraints \cite{Akerib:2015rjg},\cite{Aprile:2015uzo}.\footnote{Here, the XENON1T ``prospect'' constraint assumes 
$\sigma_{A-N}^{\rm SI} \lesssim 10^{-46}$\;cm$^2$ at the most sensitive point  $m_A \simeq 40$ GeV, which is not far from  the current bound.  } 
We conclude that all of the above constraints can be satisfied for
$\theta \ll 1$ in a broad range of DM masses.  The plot does not include, however, indirect detection constraints which are relevant for light DM and exclude some of the parameter space (see e.g \cite{Poulin:2018kap}).
Finally, such secluded DM is notoriously difficult to observe at colliders: the small $\theta$ makes its production cross section highly suppressed.

Our conclusion  is that vector dark matter provides a viable WIMP dark matter candidate. Within the Higgs portal framework, its stability is a natural consequence of gauge symmetry 
and  
the  minimal matter content.

\subsection{Fermion dark matter}

Dark matter can be composed of fermion fields. 
A simple and viable option is provided by an SM singlet fermion $\chi$ \cite{Kim:2006af},\cite{Kim:2008pp}. Although a pair of such DM particles cannot couple directly to the Higgs field 
  at the renormalizable level, it can couple to a scalar singlet  which mixes with the Higgs due to  the usual Higgs portal interaction.
  Redefining the  scalar field and performing a chiral rotation on the fermion, the relevant Lagrangian can be brought to the form 
  \begin{equation}
- \Delta {\cal L} = m_\chi \bar \chi \chi + \lambda_\chi\, \phi \, \bar \chi \chi +  \lambda_\chi^{\rm cp} \,\phi \, \bar \chi i \gamma_5 \chi \;,
\label{fermionDM}
\end{equation}
where $\phi$ has a zero VEV and $m_\chi$, $\lambda_\chi$, $\lambda_\chi^{\rm cp}$ are real.
The renormalizable interactions, in general, contain a CP violating coupling \cite{LopezHonorez:2012kv}, which in our convention we identify  with $ \lambda_\chi^{\rm cp}$,
while the scalar potential can contain odd powers of $\phi$.
The explicit form of the above transformations can be found in \cite{LopezHonorez:2012kv},\cite{Beniwal:2015sdl}. For our purposes, it is not necessary and we may parametrize 
the observables in terms of $m_\chi ,   \lambda_\chi ,  \lambda_\chi^{\rm cp}$  as well as the scalar sector parameters:
the Higgs--singlet mixing angle $\theta$, and the scalar masses $m_1, m_2$ 
  as in Section\; \ref{general-mixing}. Decoupling $h_2$ at small enough $\theta$, we recover the effective Higgs--fermion interaction (\ref{h-chi}).

In the fermion case, stability of dark matter  stems from a conserved fermion number corresponding to the symmetry
\begin{equation}
\chi \rightarrow e^{i \kappa} \chi \;,
\end{equation}
which can be either global or gauge.
 If there is no conserved number or parity,  a coupling to the leptons  $H^c \bar l \chi$ would be allowed making  $\chi$ 
  unstable although possibly long--lived. It is thus natural to take $\chi$ to be a Dirac fermion. In what follows, we will consider both the Dirac and Majorana fermion options,
  with the latter possessing a 
   Majorana parity $\chi \rightarrow -\chi$.

\subsubsection{Dirac fermion dark matter}

Consider the Dirac fermion case.  
DM annihilation   into SM fermions 
proceeds via $s$--channel exchange of the two scalars.
The amplitudes proportional to $\lambda_\chi$ and $\lambda_\chi^{\rm cp\; }$ do not interfere and 
the corresponding cross section 
  is given by 
\begin{equation}
 \sigma v_r   = N_c^f \, {m_\chi^2 m_f^2  \sin^2 2\theta \over  8 \pi v^2} \, \left[      v_r^2   \,\lambda_\chi^2 /4+   \lambda_\chi^{\rm cp\; 2}  \right] \,
{ (m_2^2-m_1^2)^2  \over (4m_\chi^2 -m_1^2)^2  (4m_\chi^2 -m_2^2)^2} \, \left(  1-{m_f^2 \over m_\chi^2}    \right)^{3/2} \;,
\end{equation}
 where subleading terms in DM velocity $v_r$ have been dropped. This expression agrees with the corresponding result in \cite{Kim:2008pp}.
The contribution of the $\lambda_\chi$--coupling is velocity suppressed, $v_r^2 \sim T_{\rm fo}/m_\chi \sim 1/20$. This is easily understood from parity or CP considerations: a
fermion--antifermion pair has parity $(-1)^{l+1}$, where $l$ is the orbital angular momentum.  The $s$--wave state, therefore, cannot annihilate into a parity--even scalar  $\phi$ if parity is conserved. 
The coupling $\lambda_\chi^{\rm cp}$ lifts this condition. 
The same considerations apply to other SM final states.

While the CP--even contribution to DM annihilation is suppressed, for direct DM detection the situation is reversed: the CP--violating coupling can be neglected.
The bilinear $ \bar \chi i \gamma_5 \chi$ vanishes in the static limit, so the contribution 
 of the $\lambda_\chi^{\rm cp}$ coupling to the DM--nucleon scattering cross section is suppressed by the current dark matter velocity squared $\sim 10^{-6}$. Thus, the result is completely dominated by the
 $\lambda_\chi$--coupling,
 \begin{equation}
\sigma^{\rm SI}_{\chi-N} = {\lambda_\chi^2  \sin^2 2\theta  \,m_N^2 f_N^2 \over 4\pi v^2 } \, {m_\chi^2 m_N^2 \over (m_\chi + m_N)^2} \, { (m_2^2 -m_1^2)^2 \over m_1^4 m_2^4} \;.
\end{equation}
This  agrees with the  cross section quoted in  \cite{Baek:2011aa}. The presence of the CP--violating term has a beneficial effect: it increases the annihilation cross section while having no impact on the
direct DM detection  \cite{LopezHonorez:2012kv}. As a result, for $\lambda_\chi^{\rm cp} > \lambda_\chi$, much of the parameter space is consistent with the  thermal DM constraints.
A similar interplay is observed for the axial DM coupling $\bar \chi \gamma_\mu \gamma_5 \chi $ to a $Z^\prime$ \cite{Lebedev:2014bba}.

For light DM, the invisible Higgs decay imposes an important constraint on the model. One finds  
\begin{equation}
 \Gamma (h_1 \rightarrow \chi \chi)= { \sin^2 \theta \, m_1 \over 8 \pi} \;
\left[     \lambda_\chi^{\rm cp\; 2}  +  \lambda_\chi^2 \, \left(   1-{4 m_\chi^2 \over m_1^2}   \right)  \right] \, \sqrt{  1-{4 m_\chi^2 \over m_1^2}   }\, .
\end{equation}
An analogous expression holds for $h_2 \rightarrow \chi \chi$ up to the replacement $m_1 \rightarrow m_2$, $\sin\theta \rightarrow \cos\theta$. 
We observe that 
close to the resonance regions, $m_\chi \simeq m_{1,2}/2$, the contribution of $\lambda_\chi$ to the invisible scalar decays is velocity suppressed and the CP--violating coupling dominates.
Such regions are of particular interest since  the relic density constraint is easily compatible with the direct detection bound near  the resonances.

Phenomenology of the CP--conserving limit of the model has been studied in \cite{Kim:2008pp},\cite{Baek:2011aa}. Most results carry over to the Majorana case, as discussed below.
 All the annihilation channels are included via the {\texttt{micrOMEGAs}} package \cite{Belanger:2008sj}.
The thermal DM relic abundance is compatible with the direct detection bounds around the resonances $m_\chi \simeq m_{1,2}/2$ and when $\chi \chi \rightarrow h_2 h_2$ is efficient.
In the latter case, $\theta \ll 1$ realizes the  ``secluded'' DM option.  The presence of the CP--violating coupling $\lambda_\chi^{\rm cp} > \lambda_\chi$
widens significantly the allowed parameter space making the constraints easily 
compatible \cite{LopezHonorez:2012kv},\cite{Ghorbani:2014qpa}. The conclusion is that the EW/TeV DM masses  and EW cross sections are generally allowed such that $\chi$ behaves as a WIMP.

The LHC Higgs measurements strongly constrain and essentially rule out thermal DM with $2 m_\chi < 125$ GeV \cite{Djouadi:2011aa}, unless the DM annihilation proceeds via
$\chi \chi \rightarrow h_2 h_2$  \cite{LopezHonorez:2012kv} with $\theta \ll 1$. 
In this case, one requires a light $h_2$ which normally entails an efficient $h_1 \rightarrow h_2 h_2$  decay. This mode can however be suppressed by tuning  $\lambda_{\phi h}$ for a given  $\sin \theta$ 
\cite{Falkowski:2015iwa}.

 Similar considerations apply to fermion dark matter of spin 3/2 \cite{Chang:2017dvm}.

\subsubsection{Majorana fermion dark matter}

If $\chi$ is a Majorana fermion with mass $m_\chi$, one must replace in Eq.\;\ref{fermionDM},
\begin{equation}
m_\chi \rightarrow  {1\over 2 } m_\chi \;.
\end{equation}
 Keeping the  interaction terms the same as in the Dirac case, one finds the following modifications.

 The annihilation rate increases by a factor of 4 since the amplitude receives two identical contributions,\footnote{It appears that $\sigma v_r$ quoted in \cite{Djouadi:2011aa} applies to the Dirac instead
 of  Majorana DM.}
\begin{equation}
  \sigma v_r  \Bigl\vert_{\rm Maj} = 4 \;  \sigma v_r  \Bigl\vert_{\rm Dir}  \;.
\end{equation}
One should keep in mind, however, that in the reaction rate calculation,
 the integral over the initial state phase space gets halved due to 2 identical particles and each reaction reduces the fermion number by 2.

 The velocity suppression of the CP--even amplitude still holds:  while the fermion and anti-fermion intrinsic parities are opposite,
 the Majorana intrinsic parity is $\pm i$. Thus, the $s$-wave is still P--odd  and its annihilation into $\phi$ via $\lambda_\chi$
 is forbidden.

\begin{figure}[h] 
\centering{   
\includegraphics[scale=0.78]{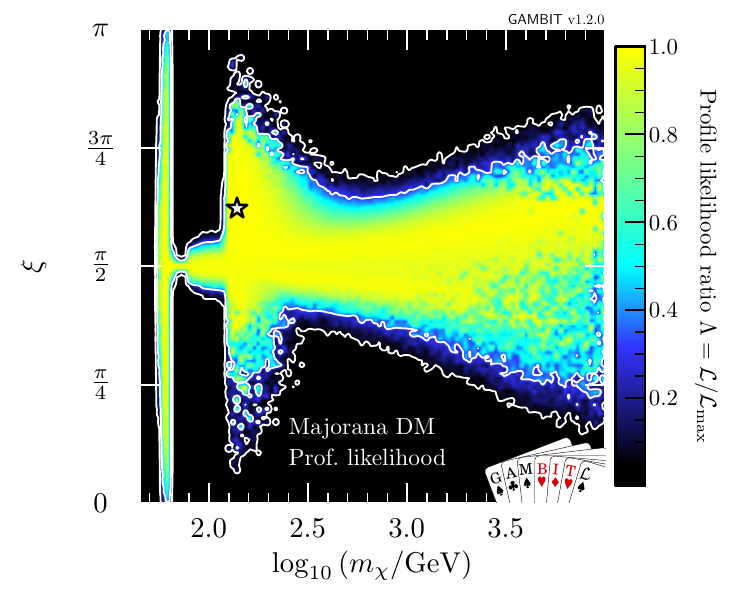}
 }
\caption{ \label{fig-majorana}
Profile likelihood for Majorana dark matter in terms of the CP--violating phase and DM mass. The CP--phase is defined by $\xi \equiv \arctan \left( \lambda_\chi^{\rm cp}/ \lambda_\chi^{}\right) $
and the effective field theory limit $m_2 \rightarrow \infty$ is assumed. The constraints and scanning procedure are as those for Fig.\;\ref{scalarZ2}. 
The figure is from Ref.\;\cite{Athron:2018hpc} \copyright\, CC--BY.
}
\end{figure}

The direct detection cross section also increases by a factor of 4 since the amplitude doubles,
\begin{equation}
\sigma^{\rm SI}_{\chi-N}  \Bigl\vert_{\rm Maj} = 4 \; \sigma^{\rm SI}_{\chi-N} \Bigl\vert_{\rm Dir} \;.
\end{equation}
Concerning the invisible width, the amplitude increase is partially compensated by the phase space symmetry factor such that
\begin{equation}
 \Gamma (h_1 \rightarrow \chi \chi)  \Bigl\vert_{\rm Maj} = 2 \;  \Gamma (h_1 \rightarrow \chi \chi) \Bigl\vert_{\rm Dir} \;.
\end{equation}
These results agree with those quoted in  \cite{Kanemura:2010sh} in the limit $m_2 \gg m_1$,  $\lambda_\chi^{\rm cp\; } \rightarrow 0$.

Altogether, the effect of replacing a Dirac fermion with a Majorana one is non--trivial and does not simply amount to a rescaling 
of the couplings.  In general, if DM consists of particles and distinct from them antiparticles, the annihilation cross section required
 to obtain the same total number density as in the case of identical particles gets doubled  (see Section\,\ref{sec-min-dm}) \cite{Gondolo:1990dk}. Thus, for Majorana particles
 it is easier to reconcile the relic density constraint with the direct detection bound.

Phenomenology of Majorana DM has been studied extensively in \cite{Kanemura:2010sh},\cite{Athron:2018hpc},\cite{LopezHonorez:2012kv},\cite{Esch:2013rta},\cite{Fairbairn:2013uta}.
Most results of these analyses apply to the Dirac case as well. 
The conclusion is that   the  minimal fermionic Higgs portal dark matter is a viable WIMP candidate, especially if CP violation is allowed in the dark sector. This is 
illustrated in Fig.\;\ref{fig-majorana}, which shows the profile likelihood in terms of the CP--violating phase $\xi \equiv \arctan \left( \lambda_\chi^{\rm cp}/ \lambda_\chi^{}\right) $ and $m_\chi$ in
the effective field theory limit $m_2 \rightarrow \infty$ 
\cite{Athron:2018hpc}.
It assumes thermal dark matter  subject to  all the relevant constraints as in Fig.\;\ref{scalarZ2}. 
The color coding shows how likely a particular parameter choice is.
We observe that $m_\chi$ is allowed to be anywhere above the resonance up to about 10 TeV.
Ref.\;\cite{Athron:2018hpc} concludes that  
 future direct and indirect   DM detection experiments will probe large portions of the favored parameter space.

To summarize, all of the Higgs portal DM candidates presented here remain viable, especially if one goes beyond the effective field theory framework. The UV complete models have certain
favorable features, some of which are not captured by the effective approach. These include, for instance, new efficient DM annihilation channels into metastable states as well   
as natural cancellations among different contributions to the direct DM detection amplitude. 
In most cases, we find that the framework will further be explored by, first and foremost,
the ever--improving direct DM detection experiments.  Certain regions of parameter space, especially for light dark matter,  can also be probed at the LHC and via indirect DM detection.

\section{ Feebly interacting dark scalars}
\label{FIMP}

While WIMPs may  offer a natural explanation for the observed dark matter relic abundance, the lack of their direct detection signal motivates one to explore further options.
Dark matter may be very weakly interacting and
 its density could be determined by mechanisms other than the standard freeze--out.
In this section, we discuss the 
cosmology of  scalars with  a feeble Higgs portal coupling
 \begin{equation}
V_{\phi h} = {1\over 2}\,  \lambda_{\phi h} H^\dagger H \, \phi^2 \;,
\end{equation}
$  \lambda_{\phi h} \ll 1$.
Here, $\phi$ can play the role of (non-WIMP) dark matter if $\langle \phi \rangle =0$. For a non--zero VEV, it is an unstable scalar which 
may couple to dark matter. In the latter case, there are both quartic and trilinear Higgs--singlet interactions.
Although, in general, one expects an independent  $H^\dagger H \, \phi$ term in $V_{\phi h}$, its effect can be 
mirrored by that of $\lambda_{\phi h} \langle \phi \rangle \,  H^\dagger H \, \phi$, so we will not consider it separately.

The dark scalars may be produced by the SM thermal bath via the Higgs portal coupling. If the coupling is very small, the dark sector never thermalizes and 
the relic abundance is determined by ``freeze--in'' rather than freeze--out. This   applies as well  to unstable dark scalars which could subsequently decay into DM or SM states.
It is also possible that $ \lambda_{\phi h}$ is completely negligible, in which case the singlet must be produced shortly after inflation, for instance,
through its small coupling to the inflaton.
 In this limit, the dark matter thermodynamics is determined by the self--coupling $\phi^4$ \cite{Carlson:1992fn} and its initial density.

Unlike in the WIMP case, much of the relevant dynamics can take place at relativistic energies, which requires going beyond the standard non--relativistic 
approach.

\subsection{Relativistic treatment}

Most dark matter studies employ the non--relativistic approximation to the reaction rates. 
This approach is often justified. 
For example, the WIMP freeze--out occurs at temperatures far below its mass such that  the
corresponding relic abundance can be computed in the non--relativistic regime.
However, for feebly interacting scalars, this is not the case and interesting physics can occur at
high temperatures where the non--relativistic approximation breaks down. 
It is therefore necessary to develop a fully relativistic approach to dark matter dynamics.

The main ingredient in these studies is the relativistic   $a \rightarrow b$ reaction rate per unit volume \cite{Kolb:1990vq},
\begin{equation}
\Gamma_{a\rightarrow b} = \int \left( \prod_{i\in a} {d^3 {\bf p}_i \over (2 \pi)^3 2E_{i}} f(p_i)\right)~
\left( \prod_{j\in b} {d^3 {\bf p}_j \over (2 \pi)^3 2E_{j}} (1+f(p_j))\right)
\vert {\cal M}_{a\rightarrow b} \vert^2 ~ (2\pi)^4 \delta^4(p_a-p_b) . 
\label{Gamma}
\end{equation}
Here $p_i$ and $p_j$ are the initial and final state momenta, respectively.
${\cal M}_{a\rightarrow b}$ is the  QFT  $a \rightarrow b$  transition amplitude.
For convenience, we absorb in $\vert {\cal M}_{a\rightarrow b} \vert^2$ both
 the {\it  initial and final} state phase space symmetry factors.
 $f(p)$ is the momentum distribution function, which for scalars  in  kinetic equilibrium takes the  Bose--Einstein form  
 \cite{Bernstein:1988bw},\cite{Bernstein:1985th},\cite{Kolb:1990vq}
\begin{equation}
f(p)= {1 \over \exp^{E-\mu\over T} -1 } \;,
\end{equation}
where $\mu$ is the {\it effective} chemical potential. This distribution maximizes entropy for a given energy and particle number.
The final state factor $1+ f(p_j)$ accounts for the Bose enhancement of the reaction rate due to degenerate states.

In this section, we are primarily interested in scalar systems with {\it no conserved quantum numbers} and no CP--violation. 
The latter implies, in particular,  $\vert {\cal M}_{a\rightarrow b} \vert=\vert {\cal M}_{b\rightarrow a} \vert$.  
In full thermal equilibrium, the absence of conserved numbers means that the 
  effective chemical potential vanishes and 
\begin{equation}
 \prod_{i\in a} f(p_i) \prod_{j\in b} \bigl(1+f(p_j)\bigr) =  \prod_{j\in b} f(p_j) \prod_{i\in a} \bigl(1+f(p_i)\bigr) 
\end{equation}
 by virtue of  energy conservation. Therefore,
 \begin{equation}
\Gamma_{a\rightarrow b} = \Gamma_{b\rightarrow a} \;,
\end{equation}
which signifies detailed balance. In this situation, the particle number is conserved. More generally, at non--zero effective chemical potential,
\begin{equation}
\Gamma_{a\rightarrow b} =  \exp \left( \sum_{i\in a} \mu_i /T - \sum_{j\in b}  \mu_j /T \right) \; \Gamma_{b\rightarrow a} \;.
\label{Gamma-b-a}
\end{equation}
This applies, in particular, to scalars in kinetic equilibrium and leads to particle production or absorption. 
For example,  the dark matter freeze--out regime belongs to this category.

The evolution of the number density $ n=  \int {d^3 {\bf p} \over (2\pi)^3} \, f(p)  $  of a given species is described by the 
 Boltzmann equation. In case the number change is due to the reactions
  $a\rightarrow b$ and  $b\rightarrow a$, it reads
   \begin{equation}
  {d n \over dt} + 3Hn =            \Delta N \;  (   \Gamma_{a\rightarrow b} - \Gamma_{b\rightarrow a} ) \;,
\end{equation}
where $\Delta N $ is the particle number change in the reaction  $a\rightarrow b$ and $H$ is the Hubble rate.
This equation is to be supplemented with the entropy conservation condition for a closed system,
 \begin{equation}
 s \,a^3 = {\rm const} \;,
\end{equation}
where $s$ is the entropy density and $a$ is the scale factor.
The two equations are necessary in order  to determine the two unknowns,  $T(t)$ and $\mu (t)$. 
In thermal equilibrium, the right hand side of the Boltzmann equation vanishes and the total particle number is conserved,
  \begin{equation}
 {d\over dt } \, n a^3 = 0 \;,
\end{equation}
while the effective chemical potential remains zero.  In the ultra--relativistic regime, $n(T) = \zeta (3) /\pi^2 \; T^3$ and $T \propto 1/a$.

Maintaining thermal equilibrium in an expanding Universe requires that the reactions be faster than the expansion,
which can be formulated as
\begin{equation}
 \Gamma_{a\rightarrow b} \gg 3nH \;.
\end{equation}
In this case, we can treat our system ``adiabatically'': 
  at every point in time, the particle distribution is the Bose--Einstein one with time--dependent temperature and zero chemical potential.
  Solving the Boltzmann equation explicitly, one finds that, 
  at later stages of the system evolution, 
this condition gets violated: the number changing reactions become too slow, $ \Gamma_{a\rightarrow b} \sim 3nH$. Although  elastic scattering processes keep the particles in kinetic equilibrium
with a well defined temperature, a non--zero effective chemical potential develops and the system departs from full thermal equilibrium. 
  This signifies ``freeze--out''. It can take place both in the relativistic and non--relativistic regimes, depending on the mass and the coupling. 
  While before freeze--out the particle density can well
  be approximated by the corresponding equilibrium value, after freeze--out the total particle number is approximately conserved and $n \propto 1/a^3$. The result is conveniently parametrized in
  terms of $Y=n/s_{\rm SM}$ as in Eq.\,\ref{Y}, which remains constant.

 The freeze--out phenomenon is to be contrasted with  another mechanism dubbed ``freeze--in''. The hidden sector may have never reached thermal equilibrium due to feeble couplings. Instead, the SM thermal bath would constantly produce the dark quanta via the Higgs portal coupling. When the plasma cools down and the process becomes inefficient, the total number of dark quanta ``freezes--in''
 and can again be parametrized in terms of $Y$. In what follows, we will consider both of these possibilities.

The relevant processes such as  particle production and freeze--out  can occur at high temperatures, where the non--relativistic approximation breaks down. At the next step, we
derive the relativistic expression for the reaction rate in terms of the corresponding cross section. 

\subsubsection{Generalizing the Gelmini--Gondolo formula}
\label{Gelmini-Gondolo}

The well--known Gelmini--Gondolo formula \cite{Gondolo:1990dk} expresses the reaction rate in terms of the cross section using the Maxwell--Boltzmann approximation. In what follows, we generalize this result to the Bose--Einstein
distribution, which is essential for relativistic systems. We follow closely the analysis of Ref.\,\cite{Arcadi:2019oxh}. Its generalization to asymmetric reactions can be found in Ref.\,\cite{DeRomeri:2020wng}.

Consider a $2\rightarrow n$ reaction, where the final state may be composed of a different species.  
Denoting the initial state momenta by $p_1$ and $p_2$, we may express the reaction rate 
  in terms of the cross--section $\sigma (p_1,p_2)$ as
\begin{equation}
\Gamma_{2 \rightarrow n} = (2\pi)^{-6} \int d^3 {\bf p_1} d^3 {\bf p_2} ~f(p_1) f(p_2) ~\sigma (p_1,p_2) v_{\rm M \o l} \;,
\end{equation}
with the M\o ller velocity and the distribution function given by 
\begin{equation}
v_{\rm M\o l}= {F(p_1,p_2) \over E_1 E_2} \equiv { \sqrt{(p_1 \cdot p_2)^2 - m_1^2 m_2^2} \over E_1 E_2 } \;,
\end{equation}
\begin{equation}
f(p)= {1 \over \exp^{u\cdot p  -\mu \over T} -1 } ~ ~, ~~u=(1,0,0,0)^T\;.
\end{equation}
Here we use a covariant expression for the Bose--Einstein distribution in a frame with a 4--velocity $u$. 
In
 the gas rest frame, $u=(1,0,0,0)^T $.  For the cases of interest,
the initial particles are identical, $m_1=m_2=m$. 
 The cross section is defined by
\begin{equation}
\sigma (p_1,p_2)= {1\over 4 F(p_1,p_2)} \int \vert {\cal M}_{2\rightarrow n} \vert^2 \,(2\pi)^4 \delta^4 \left(p_1+p_2 -\sum_i k_i\right)
\prod_i {d^3 {\bf k}_i \over (2 \pi)^3 2E_{k_i}}  \; \bigl(      1+ f(k_i)     \bigr) \;,
\label{sigma-def}
\end{equation}
where ${\cal M}_{2\rightarrow n} $ is the QFT scattering amplitude, except in our convention  we absorb all the phase space symmetry factors directly into  $ \vert {\cal M}_{2\rightarrow n} \vert^2$.
This cross section also includes the final state Bose enhancement factors $1+f(k_i)$.

Apart from the distribution functions, the reaction rate and $\sigma (p_1,p_2)$ are composed of Lorentz--invariant factors. 
In many cases, the cross section is easiest calculated  in the center-of-mass (CM) frame of the two colliding particles. For example, one can employ the CalHEP tool  \cite{Belyaev:2012qa}  for computing $\sigma^{\rm CM}(p_1,p_2)$,
while absorbing the factors  $1+f(k_i)$ into the momentum--dependent vertices. 

Let us now convert the expression for the rate $\Gamma_{2 \rightarrow n}$
 into the CM frame. 
For a given pair $p_1,p_2$,  it is the frame where   $p_1+p_2$ has only zero spacial components. 
Introduce
\begin{equation}
p={p_1+p_2\over 2} ~,~k={p_1-p_2\over 2} ~,
\end{equation}
such that
\begin{equation}
{d^3 {\bf p_1}\over 2E_1} {d^3 {\bf p_2}\over2 E_2} = 
2^4 d^4p ~d^4k ~\delta\Big((p+k)^2-m^2\Big) ~\delta\Big((p-k)^2-m^2\Big)
\;.
\end{equation}
Any time--like vector $p$ can be written as
\begin{equation}
p=\Lambda(p)~\left(
\begin{matrix}
&E& \\
&0&\\
&0&\\
&0&
\end{matrix}
\right),
\end{equation}
where $\Lambda(p)$ is a Lorentz transformation. 
The explicit parametrization in terms of rapidity $\eta$ and angular coordinates $\theta,\phi$ is 
\begin{eqnarray}
&& p^0= E \cosh \eta, \nonumber\\
&& p^1=E \sinh \eta \sin \theta \sin \phi ,\nonumber\\
&& p^2=E \sinh \eta \sin \theta \cos \phi ,\nonumber\\
&& p^3=E \sinh \eta \cos \theta  .\nonumber\\ 
\end{eqnarray}
In other words,   in the convention  $p=(p^0, p^3, p^2, p^1)^T$,  the Lorentz transformation is given by 
\small
\begin{eqnarray}
\Lambda(p)&=&\left(
\begin{matrix}
&1 & 0& 0& 0& \\
&0 & 1& 0& 0&  \\
&0 & 0& \cos\phi & -\sin\phi&  \\
&0& 0& \sin\phi & \cos\phi &
\end{matrix}
\right)
\left(
\begin{matrix}
&1 & 0& 0& 0& \\
&0 & \cos\theta & -\sin\theta & 0&  \\
&0 & \sin\theta & \cos\theta & 0&  \\
&0& 0& 0 & 1 &
\end{matrix}
\right)
\left(
\begin{matrix}
&\cosh \eta & \sinh \eta& 0& 0& \\
&\sinh \eta & \cosh \eta & 0 & 0&  \\
&0 & 0 & 1 & 0&  \\
&0& 0& 0 & 1 &
\end{matrix}
\right),\nonumber \\
 \Lambda(p)^{-1}&=&
\left(
\begin{matrix}
&\cosh \eta & -\sinh \eta& 0& 0& \\
&-\sinh \eta & \cosh \eta & 0 & 0&  \\
&0 & 0 & 1 & 0&  \\
&0& 0& 0 & 1 &
\end{matrix}
\right)
\left(
\begin{matrix}
&1 & 0& 0& 0& \\
&0 & \cos\theta & \sin\theta & 0&  \\
&0 & -\sin\theta & \cos\theta & 0&  \\
&0& 0& 0 & 1 &
\end{matrix}
\right)
\left(
\begin{matrix}
&1 & 0& 0& 0& \\
&0 & 1& 0& 0&  \\
&0 & 0& \cos\phi & \sin\phi&  \\
&0& 0& -\sin\phi & \cos\phi &
\end{matrix}
\right). \nonumber
\end{eqnarray}
\normalsize
The $p$-integration measure   becomes
\begin{equation}
d^4p= \sinh^2 \eta \; E^3dE ~d\eta ~d\Omega_p \;,
\end{equation}
where $\Omega_p$ is the solid angle in $p$-space.
Now, apply the same  Lorentz transformation $\Lambda(p)$ to  vector $k$,
\begin{eqnarray}
  k= \Lambda(p)~k' &\xrightarrow{\text{ drop~the~prime}}& \Lambda(p)~k ,\nonumber\\
 d^4k=d^4k' &\xrightarrow{\text{ drop~the~prime}}&  d^4k \equiv dk_0 \; \vert {\bf k}\vert^2 \, d \vert {\bf k}\vert  d\Omega_k \;,
\end{eqnarray}
where $\Omega_k$ is the corresponding solid angle. 
We drop the prime   for convenience, remembering that $k$  is now in 
{the CM frame}.

Because of the $\delta$-functions, the $k$-integral reduces to that over the solid angle $\Omega_k$. Therefore, for any $G(p_1,p_2)$, 
\begin{equation}
\int {d^3 {\bf p_1}\over 2E_1} {d^3 {\bf p_2}\over2 E_2} ~G(p_1,p_2)=
2 \int_m^\infty  dE    ~\sqrt{E^2-{m^2 }}  ~E^2 \int_0^\infty  d\eta  ~\sinh^2 \eta  ~ \int d\Omega_p ~d\Omega_k ~G(p_1,p_2) .
\end{equation}
In the integrand, one must set $k_0=0, \vert {\bf k}\vert=\sqrt{E^2-m^2}$ in $k$-dependent quantities. Note that $E$ is the particle energy in the CM frame.

The cross section is  computed in the CM frame, so the final state momenta $k_i$ must be transformed to that frame as $k_i = \Lambda (p) \, k_i^\prime  $. Dropping the prime for convenience, we 
have for the final state Bose enhancement factors,
\begin{equation}
1+f(k_i)= 1+ {    1  \over e^{ (k_i^0 \cosh \eta + k_i^3 \sinh \eta- \mu)/T}  -1   }
\label{1f}
\end{equation}
since $(\Lambda^{-1}u) \cdot k_i = k_i^0 \cosh \eta + k_i^3 \sinh \eta$. It is thus  clear that  $\sigma^{\rm CM}(p_1,p_2)$ depends on $E$ and $\eta$ only. Therefore, the angular integration over $\Omega_p$
and $\Omega_p$ can be performed explicitly. Using 
\begin{eqnarray}
&&u\cdot p_1= (\Lambda^{-1}u) \cdot (p+k)= E\cosh\eta + \sqrt{E^2 -m^2} \sinh\eta  ~\cos\theta_k \;,\nonumber\\
&&u\cdot p_2= (\Lambda^{-1}u) \cdot (p-k)= E\cosh\eta - \sqrt{E^2 -m^2} \sinh\eta  ~\cos\theta_k \;,
\end{eqnarray}
for $p,k$ in the CM frame and 
   $k^3= \vert {\bf k}\vert \cos \theta_k$, we find that the only non--trivial integral is that over $\theta_k$ and 
\begin{eqnarray}
 \Gamma_{2\rightarrow n}  &=&  
 {4 T \over \pi^4} \int_m^\infty dE ~E^3 \sqrt{E^2-m^2} \int_0^\infty d\eta \,{    \sinh \eta \over e^{ 2(E  \cosh\eta -\mu)/T }-1}~
\ln {  \sinh    {E\cosh\eta + \sqrt{E^2 -m^2} \sinh\eta -\mu \over 2T}    \over
\sinh   {E\cosh\eta - \sqrt{E^2 -m^2} \sinh\eta -\mu \over 2T}  } \nonumber \\
& \times & \sigma^{\rm CM}(E,\eta) \;.  
\label{Gamma2n}
\end{eqnarray}
We emphasize that, in our convention, 
  $\sigma^{\rm CM}(E,\eta)$ includes the Bose enhancement  factors for the final state as well as the phase space symmetry factors for both {\it initial and final} states. $E$ is half the CM energy.
  The rate for the inverse reaction $\Gamma_{n\rightarrow 2} $ is obtained via  relation (\ref{Gamma-b-a}).
  
  Eq.\,\ref{Gamma2n} generalizes the well known Gelmini--Gondolo result valid in the non--relativistic limit.\footnote{This calculation can trivially be adapted to the Fermi--Dirac statistics.}
   At low $T$, one may replace the Bose--Einstein distribution with the Maxwell--Boltzmann one 
  and neglect the final state quantum statistical factors. Setting $\mu=0$, we recover
  \begin{eqnarray} 
&&\Gamma_{2\rightarrow n} \simeq  {2 T\over \pi^4}  \int_m^\infty dE ~\sigma^{\rm CM}(E)~ E^2(E^2-m^2) ~ K_1(2E/T) ,
\label{MBresult}
\end{eqnarray}
where $K_1(x)$ is the modified Bessel function.

At high temperature, the bare mass $m$ receives a significant thermal correction,
\begin{equation}
m^2 \rightarrow m^2_{\rm eff} = m^2 + c \,T^2 \;,
\end{equation}
 where $c$ is a coupling--dependent constant. In this regime, $m$ in the expression for the rate (\ref{Gamma2n}) should be replaced with $m_{\rm eff}$. This is required by the correct high temperature behaviour 
 $ \Gamma_{2\rightarrow n} \propto T^4$ and also regularizes the infrared divergence as $m \rightarrow 0$.
 For example, at $T \gg m$,
 \begin{equation}
 \Gamma_{2\rightarrow 2} \propto T^4 \; \ln {T \over m_{\rm eff}} \rightarrow {\rm const}\times  T^4 \;,
\end{equation}
so the bare mass is indeed irrelevant only if the thermal mass contribution is included.

We note that the difference between the Maxwell--Boltzmann and Bose--Einstein results becomes particularly significant at $T \gg m$. Depending on the number of external legs and the coupling, the latter can lead to {\it orders of magnitude} enhancement of the rate. The effect is more pronounced for  larger $n$ and smaller coupling: the Bose--Einstein distribution peaks at low energies and  the thermal mass
controls how low the energy can be. For example, the Bose--Einstein enhancement of $\Gamma_{2\rightarrow 4}$ in the $\phi^4$--model to be studied below reaches two orders of magnitude already at $T/m =10$ and weak coupling \cite{Arcadi:2019oxh}.

The effects of quantum statistics 
in different contexts have been studied, for instance,  in \cite{Dolgov:1992wf},\cite{Drewes:2015eoa},\cite{Adshead:2016xxj},\cite{Olechowski:2018xxg}.

\subsection{Decoupled scalar dark matter}
\label{phi4}

In order to understand the dynamics of scalar dark matter, it is instructive to consider an extreme case of ``decoupled'' DM,
\begin{equation}
\lambda_{\phi h} \rightarrow 0\;,
\end{equation}
with the simplest  $Z_2$--symmetric  potential,
\begin{equation}
V(\phi) = {1\over 2} m^2 \phi^2 + {1\over 4} \lambda_\phi \phi^4 \;,
\end{equation}
where $m^2, \lambda_\phi >0$.
In this case, dark matter would be produced in the early Universe  through its small coupling to the inflaton, which however is unimportant for its subsequent evolution. The dynamics
are determined by the DM mass and its self--coupling. For certain parameter choices, it can lead to the correct relic density  and satisfy all the other observational constraints.
Some of the main features of the model have been discussed in \cite{Carlson:1992fn},\cite{Bernal:2015xba}, while its general relativistic treatment has been presented in \cite{Arcadi:2019oxh}.

\begin{figure}[h] 
\centering{  
\includegraphics[scale=0.39]{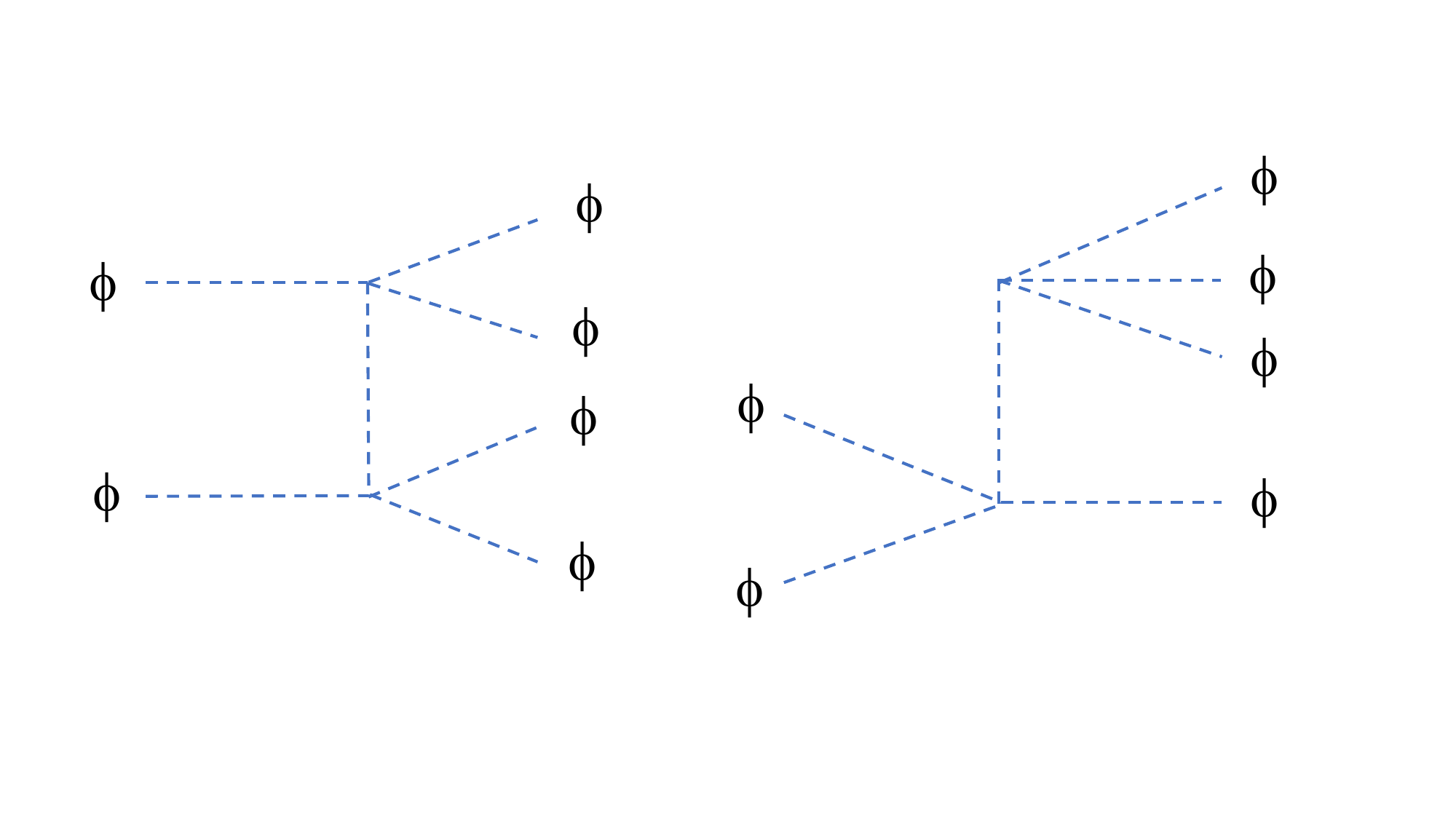}
 \vspace{-0.5cm} 
}
\caption{ \label{diag-phi4}
Leading particle number changing process in the $\phi^4$--theory at weak coupling.
}
\end{figure}

We are interested in thermodynamics of dark matter.
That is, DM 
   is assumed to have reached some sort of thermal equilibrium,
   \begin{equation}
   f(p) ={1\over e^{E-\mu \over T} -1 }\;,
   \end{equation}
    be it kinetic or  full thermal equilibrium. In the former case, the particle number is conserved and  elastic scattering 
   only redistributes the energy. In case of thermal equilibrium, particle number changing processes are also efficient and, since there is no conserved quantum number,  chemical potential vanishes.
  The elastic  $2\rightarrow 2$ scattering     occurs at leading order in the coupling, while inelastic processes are first allowed at second order (Fig.\,\ref{diag-phi4}).
 The efficiency of a given process  is determined by comparing the corresponding reaction rate to the expansion rate,
 \begin{eqnarray}
 && \Gamma_{2\rightarrow 2} \gtrsim 3nH ~~ \Rightarrow ~~{\rm kinetic~equilibrium} ~,\nonumber\\
 &&  \Gamma_{2\rightarrow 4} \gtrsim 3nH ~~ \Rightarrow ~~{\rm thermal~equilibrium} ~,
 \label{therm-cond}
 \end{eqnarray}
  where further  order one coefficients have been neglected.
  For small couplings, the system only reaches kinetic equilibrium.
   In this case, the total DM particle number is constant and determined by the initial conditions, while the temperature is controlled by the average initial energy.
   In general, the dark sector temperature can be very different from that of the observable sector.
   The chemical potential is read off from the corresponding number density, e.g. in the ultrarelativistic regime $T \gg m$, 
   \begin{equation}
n= {T^3 \over \pi^3 } \;{\rm Li}_3(e^{\mu/T}) \;,
\label{n-mu}
\end{equation}
where ${\rm Li}_3(x)$ is the third {degree}  polylogarithm, ${\rm Li}_3(x)= \sum_{n=1}^\infty 
x^n/n^3$. The system is dilute for $\mu <0$.

   In what follows, we treat $T$ and $\mu$ as independent model parameters. The first step is to determine the thermalization conditions. For the $2\rightarrow 2$ process,
    the cross section defined by (\ref{sigma-def}) can be computed explicitly. In the CM frame,
   \begin{equation}
 \sigma^{\rm CM}_{2\rightarrow 2}(E,\eta)=   {1\over 4F(p_1,p_2)} \int d\Omega_{\bf k_1} ~ { \vert {\bf k_1} \vert \over (2\pi)^2 8 E }~
\vert  {\cal M} \vert^2 ~\Big(1+ f(k_1)\Big)~ \Big(1+f(k_2)\Big) \;,
\end{equation}
where $E$ is the particle energy and 
the phase space symmetry factors for the initial and final states have been  included directly in the amplitude, $\vert  {\cal M} \vert^2 = (6 \lambda_\phi)^2 /(2 ! 2!)$.\footnote{In the limit $f(k_i) \rightarrow 0$,
this differs from the standard QFT cross  section \cite{Peskin:1995ev} by a factor of $1/2!$.} Using (\ref{1f}) and computing the angular integral, we obtain
 \begin{equation}
  \sigma^{\rm CM}_{2\rightarrow 2}(E,\eta)= {1\over 2!2!} \times {9\lambda^2_\phi T \over 16 \pi E^2 \sqrt{E^2-m^2}  \sinh\eta}
~ {1 \over 1-e^{ -{2E \over T} \cosh \eta}} ~
\ln {  \sinh    {E\cosh\eta + \sqrt{E^2 -m^2} \sinh\eta \over 2T}    \over
\sinh   {E\cosh\eta - \sqrt{E^2 -m^2} \sinh\eta \over 2T}  }~,
\end{equation}
where we have set $\mu=0$. A non--zero $\mu$ is trivially included by replacing  $E\cosh\eta \rightarrow E \cosh\eta -\mu$.
The corresponding reaction rate is then computed by plugging this expression into Eq.\,\ref{Gamma2n}.
The thermal mass contribution is accounted for by replacing
 \begin{equation}
m^2 \rightarrow m^2 + {\lambda_\phi\over 4 } T^2 \;
\label{thermal-mass}
\end{equation}
in all of these expressions. Without it, the rate becomes divergent as $m\rightarrow 0$.
 
The  $2\rightarrow 4$ cross section cannot be presented in a compact form. Instead, one may use numerical evaluation with CalcHEP  \cite{Belyaev:2012qa} by absorbing the Bose factors 
$1+f(k_i)$ into a redefined vertex. The CalcHEP output is then integrated numerically to yield the rate. 

\subsubsection{Thermalization constraints}
\label{subsec-therm}

\begin{figure}[h] 
\centering{  
\includegraphics[scale=0.89]{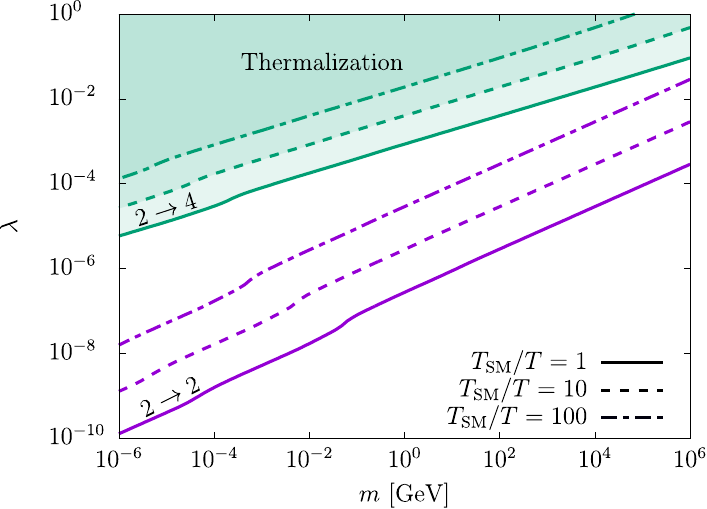}
}
\caption{ \label{thermalization-plot}
Thermalization constraints in the $\phi^4$--model at $\mu \sim 0$. In the areas above the purple (green) lines, the coupling is large enough for kinetic  (full thermal) equilibrium. 
In terms of the convention in the text, $\lambda \equiv 6 \lambda_\phi$.  
 The figure is from Ref.\;\cite{Arcadi:2019oxh} \copyright\, CC--BY.
}
\end{figure}

The thermalization constraints on the coupling can be determined as follows.
Let us assume that the average energy of dark matter in the Early Universe was large enough compared to its mass and it had sufficient time to thermalize.
 The consistency of this assumption is then verified    by analyzing the conditions (\ref{therm-cond}). If, for a given $\lambda_\phi$, there exists a temperature at which one or both of inequalities 
 (\ref{therm-cond}) are satisfied, thermalization is said to occur. To determine this temperature, one maximizes the ratio
  \begin{equation}
{\Gamma_{2\rightarrow n} \over 3nH} \rightarrow {\rm max}
\label{Tmax}
\end{equation}
with respect to $T$  for a fixed $\lambda_\phi$, which has a non--trivial effect through the thermal mass contribution. 
If the maximal value is greater than one, the thermalization assumption is consistent.

In order to analyze (\ref{Tmax}), we need a relation between the dark temperature $T$ and the observable sector temperature $T_{\rm SM}$.
Let us focus on the regime in which the dark contribution to the energy density of the Universe is small. This is the case when $T < T_{\rm SM}$ and the number of degrees
of freedom  in the dark sector is much smaller than that in the SM sector ($g_*$). 
Then, the Universe expansion is controlled by the observable sector and its entropy conservation $s_{\rm SM} a^3 = {\rm const} $. 
When the change in the number of SM d.o.f. can be neglected,
we have 
\begin{equation}
  a= {{\rm const} \over T_{\rm SM}} ~~,~~H= \sqrt{\pi^2 g_* \over 90}~ {T_{\rm SM}^2\over M_{\rm Pl}} \;.
 \end{equation}
The dark sector entropy is conserved  separately, $s a^3 ={\rm const}$. In the relativistic regime, $s \propto T^3$ and,  neglecting   
   the SM d.o.f. change, $T \propto T_{\rm SM}$. It is thus convenient to parametrize  the system in terms of 
\begin{equation}
\xi \equiv {  T_{\rm SM} \over T} \;,
\end{equation}
which is (almost) constant at $T \gg m$.

The resulting thermalization constraints at $\mu \sim 0$ for different $\xi$ are shown in Fig.\,\ref{thermalization-plot}.  
Their qualitative behavior can be understood as follows. One finds numerically that
the ratio ${\Gamma_{2\rightarrow 4} /( 3nH) }$ is maximized at $T \sim 2m/\sqrt{\lambda_\phi}$. The main 
dependence on the coupling and temperature in the relativistic regime is captured by the relations 
 $n \propto T^3$ and $\Gamma_{2\rightarrow 4} \propto \lambda_\phi^4 T^4$.  
 Setting ${\Gamma_{2\rightarrow 4} /( 3nH) }$ to one, 
 this results in $\lambda_\phi \propto
 m^{2/9} \xi^{4/9}$, which describes  approximately  the trend observed in Fig.\,\ref{thermalization-plot}.
 Analogous arguments apply to kinetic equilibrium considerations.
 
We observe that  thermal equilibrium requires tangible couplings, e.g. $10^{-4}-10^{-3}$ for $m \sim 1$ GeV, while kinetic equilibrium is reached for much smaller couplings.
There is a large range of $\lambda_\phi$ in which only kinetic equilibrium is possible.

Finally, we have set $\mu \sim 0$ so that the number density is taken to be close to the equilibrium value. 
For dilute systems, $\mu <0$, the density and the reaction rates are lower leading to stronger constraints on the coupling.

\subsubsection{Freeze--out and relic abundance}

 The number density evolution is governed by the Boltzmann equation,
   \begin{equation}
  {d n \over dt} + 3Hn =            2 \;  (   \Gamma_{2\rightarrow 4} - \Gamma_{4\rightarrow 2} ) \;.
  \label{Boltzmann-2}
\end{equation}
 In thermal equilibrium, the right hand side vanishes although the individual terms are large.
   The relic abundance  of   thermalized dark matter is determined by  freeze--out. At some point, $\Gamma_{2\rightarrow 4}$ and $\Gamma_{4\rightarrow 2}$ become slower than the Hubble
  expansion, and the particle number remains approximately constant. This can happen both in the relativistic and non--relativistic regimes, with the latter allowing for a semi--analytical treatment.
  \\ \ \\
  {\bf \underline{Non--relativistic freeze--out}.}
  In this regime, the Bose--Einstein distribution function can be replaced with the Maxwell--Boltzmann one, $f(p)= e^{- (E-\mu)/T}$,
  and the final state quantum statistical factors can be set to one.
   The chemical potential dependence then  factorizes as
  \begin{equation}
\Gamma_{2\rightarrow 4}= 
e^{-2 \mu/T} \Gamma_{4\rightarrow 2}= e^{2 \mu/T} \Gamma_{4\rightarrow 2}(\mu=0)  \;.
\end{equation}
  The $4\rightarrow 2$ reaction proceeds with the initial particles effectively at rest, so it is convenient to define 
 \begin{equation}
\sigma_{4\rightarrow 2}v^3 \equiv  
{1\over 2 E_{k_1}    2 E_{k_2} 2 E_{k_3} 2 E_{k_4}     } \int {  d^3{\bf p_1} \over  (2\pi)^3 2 E_1 } {  d^3{\bf p_2} \over  (2\pi)^3 2 E_2 }
\vert {\cal M}_{4\rightarrow 2} \vert^2 ~ (2\pi)^4 \delta^4\left( \sum_i p_i- \sum_j k_j\right) \;,
\label{sigmav3}
\end{equation}
since it  is momentum independent in the non-relativistic limit $k_i \simeq (m,\vec{0})^T$. Here, the phase space symmetry factor $1/(2!4!)$ is absorbed in $\vert {\cal M}_{4\rightarrow 2} \vert^2$,
as usual. Then, to leading order in the non--relativistic expansion,   the reaction rate  reads
  \begin{equation}
\Gamma_{4\rightarrow 2}(\mu=0)= 
  (2\pi)^{-12}   \int   \prod_i \left( d^3{\bf k_i}~ e^{-E_{k_i}/T} \right) \sigma_{4\rightarrow 2}v^3=
  \langle    \sigma_{4\rightarrow 2}v^3  \rangle  n_{\rm eq}^4 \;,
   \end{equation}
   where $\langle ... \rangle$ denotes a thermal average at $\mu=0$ and $ n_{\rm eq} =(2\pi)^{-3} \int d^3 {\bf p} f(p)\vert_{\mu=0}$. 
   Trading the chemical potential for $n$ via $e^{\mu/T} = n/n_{\rm eq}$, we get the Boltzmann equation in the form 
    \begin{equation}
{dn \over dt} +3Hn = 2 \langle  \sigma_{4\rightarrow 2}  v^3 \rangle    
\bigl( n^2 n_{\rm eq}^2 -n^4\bigr) \;,
\label{boltzmann1}
\end{equation}
where $\langle  \sigma_{4\rightarrow 2}  v^3 \rangle $ is temperature independent. An explicit calculation shows that (see Appendix C of  \cite{Arcadi:2019oxh})
\begin{equation}
{\langle  \sigma_{4\rightarrow 2}  v^3 \rangle =  {1\over 2!4!} \;  { \sqrt{3} \, 81 \,\lambda_\phi^4 \over \;16\, \pi m^8 } \;. } 
\label{NRsigma}
\end{equation}

In order to determine $n$ and $T$ as functions of time (or $T_{\rm SM}$), we need a second equation, which follows from entropy conservation,
\begin{equation}
{s \over s_{\rm SM}} = {\rm const} \;.
\label{eq-entropy}
\end{equation}
The SM sector entropy density is given by 
\begin{equation}
 s_{\rm SM}={2 \pi^2  g_{*s} \over 45}  \; {T_{\rm SM}^3}   \;, 
 \label{SSM}
 \end{equation}
where  $g_{*s}$ is  the number of degrees of freedom contributing to the entropy.\footnote{The evolution of $g_*$ with $T_{\rm SM}$ can be found in Ref.\,\cite{Kolb:1990vq}.} At $T \ll m$, the dark matter entropy density  and the number density can be approximated by
\begin{equation}
  s_{\rm }\simeq  {m-\mu \over T} ~n ~~~,~~~ n=\left(  {mT \over 2\pi}  \right)^{3/2} e^{- {m-\mu\over T}} \;.
  \label{s-n}
   \end{equation}
Using these equations, one eliminates $m-\mu$   in favor of $n$ and $T$. Then, the Boltzmann equation supplemented by the entropy conservation condition can be solved numerically
(see \cite{Arcadi:2019oxh} for further details).

\begin{figure}[h!]
\begin{center}
\includegraphics[scale=0.6]{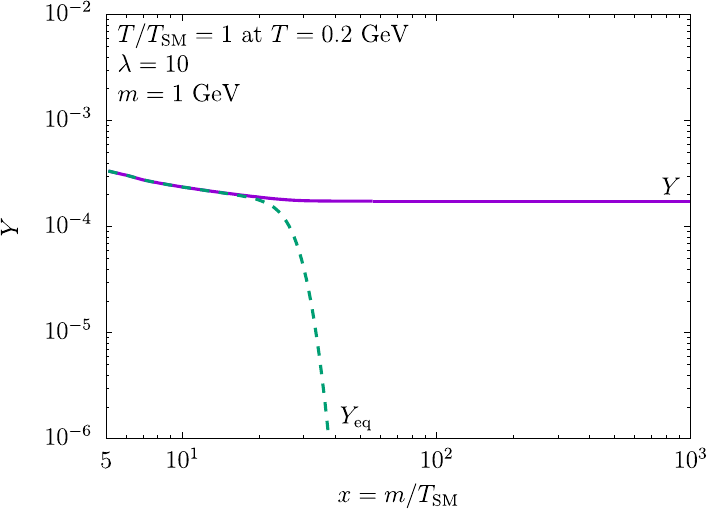}
\includegraphics[scale=0.6]{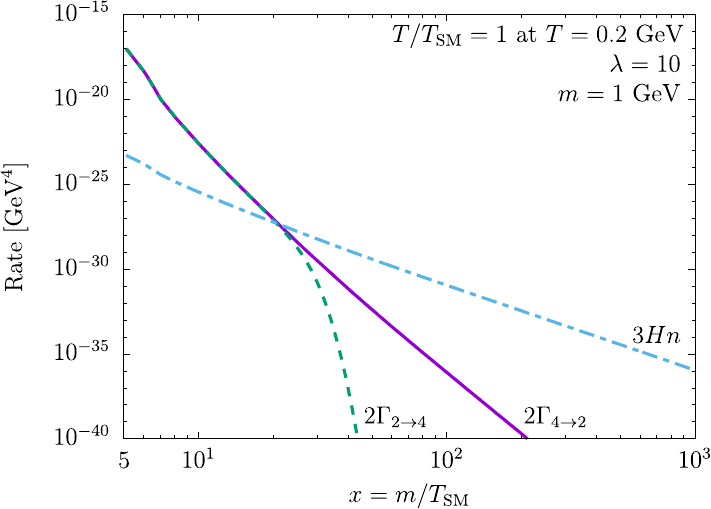}
\includegraphics[scale=0.6]{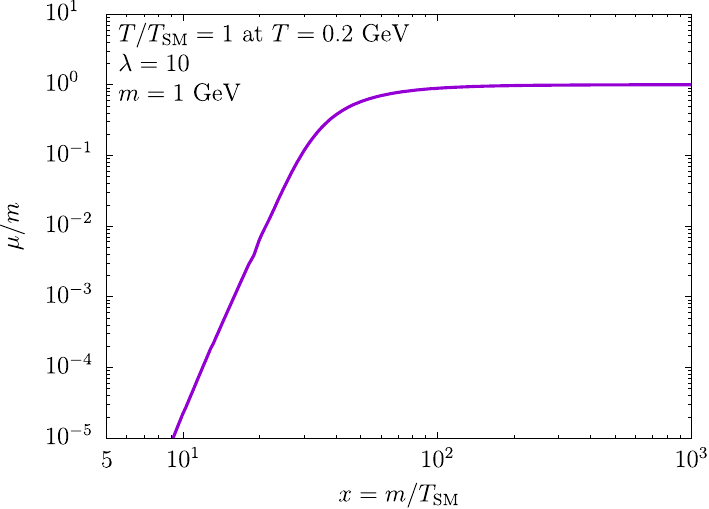}
\includegraphics[scale=0.6]{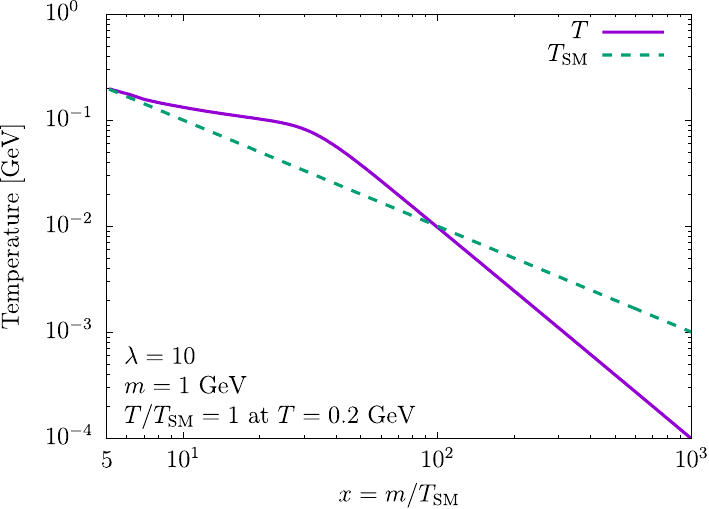}
\end{center}
\caption{  Evolution of thermodynamic quantities with $m/T_{\rm SM}$ for non--relativistic freeze--out.  In terms of the convention in the text, $\lambda \equiv 6 \lambda_\phi$.  
 The figure is from Ref.\;\cite{Arcadi:2019oxh} \copyright\, CC--BY.
\label{fig:NR}}
\end{figure}

The result is conveniently presented in terms of $Y \equiv {n\over s_{\rm SM}}$, which is proportional to the total number of the DM quanta. Using $m/T_{\rm SM}$ as the time variable, one
finds the solution shown in Fig.\;\ref{fig:NR}   for a representative choice of model parameters. At the initial point, $\mu $ is set to zero signifying thermal equilibrium.  
We observe that $Y$ tracks closely its equilibrium value $Y_{\rm eq}$ until the freeze--out point, where the particle number changing processes become slower than the expansion
and a substantial $\mu$ develops. After that, $Y$ remains almost constant.
It should be noted that 
  before freeze--out, the solution  $n$ does not exactly coincide with the equilibrium value $n_{\rm eq}$, but is very close to it. This is because the Maxwell--Boltzmann $n_{\rm eq}$ does not solve the Boltzmann equation: the left hand side does not vanish indicating the total particle number change.  For the true solution, both sides of the Boltzmann equation do not vanish. The particle number 
   reduction
is provided by the $4\rightarrow 2$ process being slightly more efficient than its inverse. It also leads to ``heating'' of the dark sector prior to freeze--out as required by entropy conservation \cite{Carlson:1992fn},
\begin{equation}
T \propto {1 \over \bigl| \ln T_{\rm SM}\bigr|} \;.
\end{equation}
This behaviour is seen   in the right lower panel of Fig.\;\ref{fig:NR}.

 After freeze--out, both the entropy and particle number are conserved. Eq.\,\ref{s-n} then implies $m-\mu \propto T$, which  combined with (\ref{eq-entropy}) yields
 \begin{equation}
 T \propto T_{\rm SM}^2 \;,
 \end{equation}
 the trend observed in Fig.\;\ref{fig:NR}. 
 \\ \ \\
  {\bf \underline{Relativistic freeze--out}.} At $T\gsim m$, the Maxwell--Boltzmann limit is not applicable and one should use the Bose--Einstein distribution. This leads to a number of complications. In particular,
  the chemical potential dependence does not factorize, although the relation $\Gamma_{2\rightarrow 4}= 
e^{-2 \mu/T} \Gamma_{4\rightarrow 2}$ is still valid. Also, it is not meaningful to factorize out $ \langle  \sigma_{4\rightarrow 2}  v^3 \rangle  $ since it is $\mu,T$--dependent.
Nevertheless, the Boltzmann equation in the form (\ref{Boltzmann-2}) together with the entropy conservation condition can be solved numerically. The DM entropy density is now
given by the general expression (see e.g., \cite{Kolb:1990vq}),
\begin{equation}
s= {\rho+p -\mu n \over T} \;,
\end{equation}
where $\rho$ is the energy density and $p$ is the pressure. 
 
 The resulting solutions exhibit qualitatively similar behaviour to that in the non--relativistic case except their dependence on $T_{\rm SM}$ is milder and there is {\it no ``heating'' phase} in the DM evolution
 \cite{Arcadi:2019oxh}.
 An important feature of these solutions  is that freeze--out cannot occur at very high temperatures, $ T \gg m/ \sqrt{\lambda_\phi} $. In this regime, $\Gamma_{2\rightarrow 4} \propto T^4$, 
 while $nH \propto T^5$ for a given $\xi$. Therefore, if $\Gamma_{2\rightarrow 4} \gtrsim nH$ initially (as required by thermalization), it will hold at lower temperatures thereby forbidding  
 freeze--out. The situation changes when the thermal mass becomes comparable to the bare mass, $\lambda_\phi T^2 \sim m^2$, and the scaling of $\Gamma_{2\rightarrow 4} $ 
 gets modified. 
 As in the previous case, the total DM particle number remains almost constant  after freeze--out defined by $2\Gamma_{2\rightarrow 4} \simeq 3nH$. This allows one to determine the relic DM density
 directly from the freeze--out temperature. 
   
\begin{figure}[h!]
\begin{center}
\includegraphics[scale=0.66]{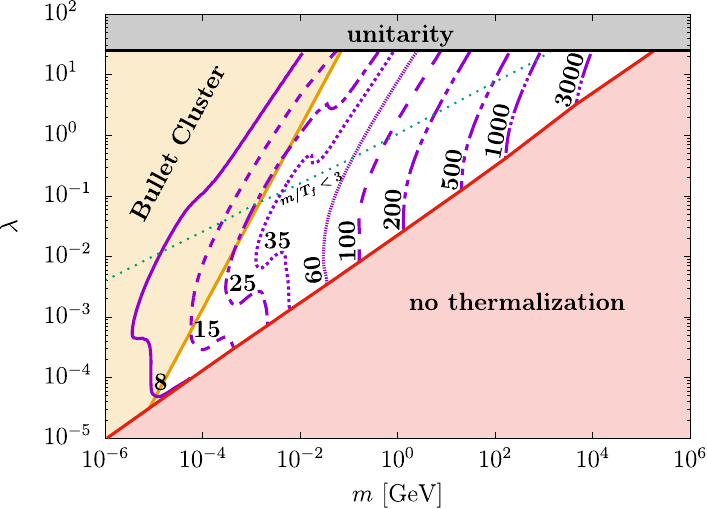}
\includegraphics[scale=0.65]{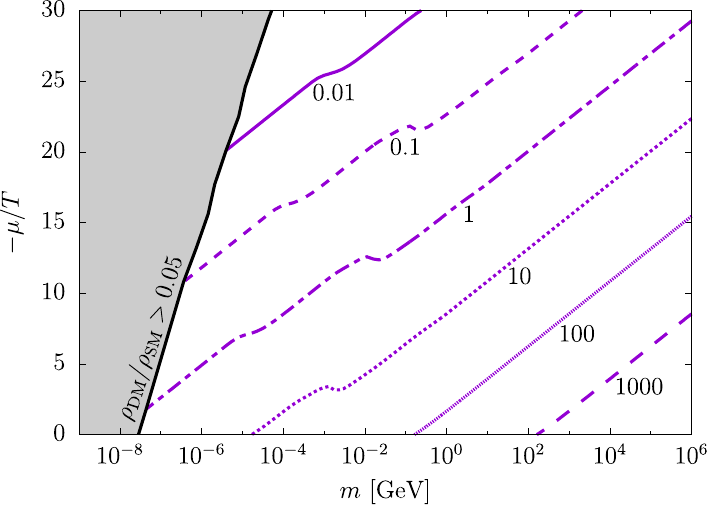}
\end{center}
\caption{   Allowed parameter space in the $\phi^4$--model. Along the curves, the correct DM relic abundance is reproduced for a fixed $\xi \equiv T_{\rm SM}/T$ at freeze--out. 
The shaded areas are excluded. {\it Left:} Self--coupling vs mass for thermal DM ($\lambda \equiv 6 \lambda_\phi$). {\it Right:}   Effective chemical potential vs mass for DM in kinetic equilibrium.
 The figures are  from Ref.\;\cite{Arcadi:2019oxh} \copyright\, CC--BY.
\label{fig:param-space}}
\end{figure}

 Fig.\,\ref{fig:param-space} delineates parameter space producing the correct relic abundance,
 $Y_\infty = 4.4 \times 10^{-10} \,{\rm GeV}/m$. In the left panel, DM is thermalized and has the right relic abundance along the curves marked by $\xi$ at freeze--out. The shaded areas are excluded by perturbative unitarity, significant self--interaction in conflict with the Bullet Cluster observations and non--thermalization.
 The green dashed line separates the regions with non--relativistic (above the line) and relativistic (below the line) freeze--out.
 
 For non--relativistic freeze--out, simple arguments show that the correct relic abundance is reproduced along the lines with $\lambda_\phi \propto m\, \xi^{-7/4}$, where $\xi $ is evaluated at freeze--out
 and logarithmic terms have been neglected. The kinks  observed in the Fig.\,\ref{fig:param-space} appear due to the abrupt change in the SM degrees of freedom. In the relativistic regime, the curves 
 tend to vertical lines since $Y$ is mostly determined by the temperature ratio and becomes rather insensitive to $\lambda_\phi$. 
Too small couplings are however inconsistent with the thermalization assumption. On the other hand, large couplings violate either the Buller Cluster bound on DM self--interaction, 
$\sigma/m \lesssim 1$ cm$^2$/g \cite{Markevitch:2003at},
with $\sigma =9 \lambda_\phi^2 /(32 \pi m^2)$, or perturbative unitarity in $\phi \phi \rightarrow \phi \phi$ scattering, $\lambda_\phi \lesssim 4\pi/3$.
 
 The right panel of Fig.\,\ref{fig:param-space} displays allowed parameter space for DM that only reaches kinetic equilibrium, which corresponds to weaker coupling and/or a dilute DM gas. 
 In this case, the total particle number is determined by the initial conditions and can be expressed in terms of the effective  chemical potential 
 via Eq.\,\ref{n-mu}. We are interested mostly in dilute dark matter, which can easily evade the cosmological  constraints,  and thus take $\mu <0$. 
 For positive $\mu$, its magnitude is bounded by $m$ to avoid a singular distribution function, whereas a negative $\mu$ can be arbitrarily large in magnitude in our model,
 unlike $\mu$ in models with non--trivial antiparticles  \cite{Haber:1981fg},\cite{Haber:1981ts}.
   In the plot, $-\mu/T$ and $\xi$ are  fixed at $T \gg m$. Since $\mu \propto T \propto T_{\rm SM}$  (away from the SM thresholds) as required by entropy and particle number conservation
 in this regime, the ratios remain almost constant.

 For large $-\mu$,  the system becomes very dilute since the number density scales as $n \propto e^{\mu/T}$. In order to treat DM thermodynamically, one must make sure that $\Gamma_{2 \rightarrow 2}
 \gtrsim 3nH$ at a given $\mu$, which requires couplings much larger than those in Fig.\,\ref{thermalization-plot} \cite{Arcadi:2019oxh}. 
 The fixed relic density curves are produced assuming that the requisite coupling is available and 
 correspond approximately to $e^{-\mu /T} \propto m/ \xi^3$. 
 We exclude low values of $|\mu | /T$ for a given $\xi$ since, in this case, relativistic DM makes a significant contribution to the overall energy density  ($>$\,5\%) and thus violates one of our assumptions. This
 constraint also helps evade the BBN bounds on the extra relativistic degrees of freedom. It is interesting that the dark sector is allowed to be  much hotter than the observable one, $\xi \ll 1$, as
 long as it is sufficiently dilute.
 
 We note that dark matter in this model  can be both cold and warm. The latter is defined as being semi--relativistic at the temperature $T_{\rm SM} \sim 300$ eV corresponding to small scale structure formation. 
 Whether or not our DM candidate is relativistic at this stage depends both on its mass and 
 $\xi$. Clearly, close to the excluded region, it tends to be warm or even hot. A recent analysis \cite{Egana-Ugrinovic:2021gnu} shows that the Lyman--$\alpha$ constraint extends the excluded area
 somewhat further depending on $\lambda_\phi$, although determining its exact boundary  requires a dedicated study.

We conclude that the simplest scalar $\phi^4$--model can account for the observed dark matter and exhibits non--trivial thermodynamic behaviour. 
Its main feature, i.e. that the number--changing ``cannibalistic'' processes are essential for the DM dynamics, is inherited by more complicated set--ups \cite{Hochberg:2014dra},\cite{Bernal:2015ova},\cite{Pappadopulo:2016pkp}.

\subsection{Higgs portal freeze--in    }

A small Higgs portal coupling  can source gradual production of dark states, which never reach thermal equilibrium. This is known as the ``freeze--in'' mechanism 
\cite{McDonald:2001vt},\cite{Kusenko:2006rh},\cite{Hall:2009bx}.
It can operate via decay, fusion as well as  $2\rightarrow 2$ processes in the SM thermal bath. The produced scalar can  itself constitute dark matter or decay to dark matter subsequently. 
In the Higgs portal framework, this mechanism is often studied within the non--relativistic approximation 
as in  \cite{Yaguna:2011qn},\cite{Chu:2011be}, while here we will follow
the relativistic formulation of \cite{DeRomeri:2020wng},\cite{Lebedev:2019ton}.

Let us consider a $Z_2$ symmetric scalar potential (\ref{Z2-pot}) with $\lambda_{\phi h} \ll 1$. 
Suppose  both $h$ and $\phi$ develop VEVs, $v$ and $w$, respectively. Expanding the potential around the VEVs, i.e. replacing
$h \rightarrow h+v$, $\phi\rightarrow \phi+w$ in (\ref{Z2-pot}) notwithstanding abuse of notation,
one finds the following interaction terms 
\begin{equation}
\Delta V_{\phi h} =
    {v\over 2} \lambda_{\phi h} h \phi^2 + {w\over 2} \lambda_{\phi h} \phi h^2 + w \lambda_\phi \phi^3 
    + {1\over 4} \lambda_{\phi h} h^2 \phi^2 + {1\over 4 } \lambda_\phi \phi^4\;.
\label{vw}
\end{equation}
This potential contains odd powers of $\phi$, so the effect of the $\sigma_{\phi h}$--term in the more general potential  (\ref{potential}) can largely be captured in this set--up.

Focussing on the small mixing regime,  $\theta \ll 1$, we may neglect $\theta^2$ terms and  treat $h$ and $\phi$ as mass eigenstates with masses squared (cf. Eq.\,\ref{evalues})
\begin{equation}
m_1^2 \simeq 2 \lambda_h v^2 ~~,~~m_2^2 \simeq 2 \lambda_\phi w^2 
\end{equation}
 and 
 \begin{equation}
\theta \simeq {\lambda_{\phi h} \over \sqrt{4 \lambda_h \lambda_\phi}} \; {m_1 m_2 \over m_2^2 -m_1^2} \;.
\end{equation}
For $m_1$ close to $m_2$, this approximation fails. When $m_1$ and $m_2$ are substantially different, $|\theta|$ is bounded by min$\{ m_1/m_2 , m_2/m_1\}$. 
As long as the masses are  sufficiently hierarchical and $\theta \ll 1$,
the trilinear couplings in (\ref{vw}) give the leading contributions to
  the $h \rightarrow \phi\phi $ and $hh \rightarrow \phi $ processes (see Section\,\ref{Z2-pheno}).

Particle production can take place in different regimes, i.e. at high and low temperatures.  
At high temperature, the potential receives significant thermal corrections. Their main effect is captured by the thermal mass contributions (see e.g., \cite{Petraki:2007gq},\cite{Laine:2016hma}),
 \begin{equation}
m_h^2 \rightarrow m_h^2 + c_h T^2 ~~,~~ m_\phi^2 \rightarrow m_\phi^2 + c_\phi T^2 \;,
\label{ther-mass}
\end{equation}
where
\begin{eqnarray}
&& c_h \simeq {3 \over 16} g^2 + {1\over 16} g^{\prime 2} + {1\over 4} y_t^2 + {1\over 2} \lambda_h \;, \nonumber\\
&& c_\phi= {1\over 4} \lambda_\phi + {1\over 6} \lambda_{\phi h}\;,
\end{eqnarray}
which assumes that the SM and $\phi$ are in thermal equilibrium, and  
includes  contributions of 4 Higgs d.o.f. In the freeze--in regime, $\phi$ is {\it not thermalized}, so the $\lambda_\phi$--contribution is absent,
yet the thermal Higgs contribution to $c_\phi$, i.e.  $\Delta c_\phi =\lambda_{\phi h}/6$, can affect the dynamics
 and the vacuum structure. At small $\lambda_{\phi h}$, this effect is, however, unimportant in practice.

The   high temperature   minimum is normally  at $v=w=0$.\footnote{Here we take $\lambda_{\phi h} >0$, although the main results are very similar for both signs. This is because
particle production is more efficient at lower temperatures, where the  effects of the phase transition in the $\phi$ direction are irrelevant.}
 The transition to non--zero VEVs takes place at the corresponding critical temperatures:
$v=0 \rightarrow v\not=0$ at $T_c^v$ and $w=0 \rightarrow w\not=0$ at $T_c^w$. An 
adequate description of the transition requires non--perturbative input and is quite complicated  \cite{Gould:2021dzl},\cite{Schicho:2021gca}.
We will therefore make a simplification:
since the dark scalar abundance 
depends rather weakly on the critical temperature,   for our purposes it suffices to approximate
(at $\theta \ll 1$):
\begin{eqnarray}
&& T_c^v = {  \vert m_h \vert \over  \sqrt{c_h}  } ~~,
~~ T_c^w = {  \vert m_\phi \vert \over  \sqrt{c_\phi}  }      \;, \nonumber\\
&& -m_h^2 = {\lambda_{\phi h }\over  4 \lambda_\phi}\; m_2^2 +{1\over 2} m_1^2 \;, \nonumber\\
&&     -m_\phi^2 = {\lambda_{\phi h }\over  4 \lambda_h}\; m_1^2 +{1\over 2} m_2^2 \;.
\label{T-crit}
\end{eqnarray}
 Above these thresholds, some of the vertices vanish which affects  the  available production mechanisms.

 \subsubsection{Dark scalar production and non--thermalization constraints}

The dark scalar can be  produced in the SM thermal bath via $hh\rightarrow \phi$, $h\rightarrow \phi\phi$, and $hh\rightarrow \phi\phi$ reactions,  at leading order.
The reaction rate calculation proceeds as in Sec.\,\ref{Gelmini-Gondolo}, except the final state Bose enhancements factors can be neglected,
\begin{equation}
1+ f(k_i) \rightarrow 1 \;.
\end{equation}
This is because the density of $\phi$ is assumed to be far below its equilibrium value, which imposes an important consistency check.
For a single Higgs d.o.f., an explicit calculation of the leading contributions yields \cite{Lebedev:2019ton},\cite{DeRomeri:2020wng}
  \begin{eqnarray}
\Gamma_{hh \rightarrow \phi} &=& {\lambda_{\phi h}^2 w^2 m_2 T \over 32 \pi^3  }\; \theta(m_2- 2m_1) \int_0^\infty  d\eta {\sinh\eta \over e^{m_2 \cosh \eta \over T}-1} \;
\ln {   \sinh { m_2 \cosh \eta +  \sqrt{m_2^2-4m_1^2}  \sinh\eta  \over 4T}  \over 
 \sinh { m_2 \cosh \eta -  \sqrt{m_2^2-4m_1^2}  \sinh\eta   \over 4T}}  \;, \nonumber \\
  \Gamma_{h \rightarrow \phi \phi} &=&  { \lambda_{\phi h}^2 v^2 m_1^2 \over 64 \pi^3   } \sqrt{1-{4m_2^2 \over m_1^2}} \; \theta(m_1 -2m_2)
 \; \int_1^\infty dx \; { \sqrt{x^2-1} \over e^{{m_1\over T} x} -1}  \;,  \\
  \Gamma_{hh\rightarrow \phi \phi} &=&    { \lambda_{\phi h}^2 T \over 64 \pi^5} 
  \int_{m_1}^\infty dE ~E \sqrt{E^2-m_2^2} \int_0^\infty d\eta {    \sinh \eta \over e^{{2E\over T} \cosh\eta }-1}~
\ln {  \sinh    {E\cosh\eta + \sqrt{E^2 -m_1^2} \sinh\eta \over 2T}    \over
\sinh   {E\cosh\eta - \sqrt{E^2 -m_1^2} \sinh\eta \over 2T}  } \;. \nonumber
\end{eqnarray}
These rates include appropriate phase space symmetry factors and are to  be multiplied by the number of Higgs d.o.f. in a given phase. 
The masses $m_1$ and $m_2$ include thermal corrections, which become particularly important in the symmetric phase at $T \gg T_c^{u,w}$, 
\begin{equation}
m_1 \simeq \sqrt{c_h} \,T ~~,~~m_2 \simeq \sqrt{c_\phi} \, T\;.
\end{equation}
As emphasized before, the thermal masses are important for the correct reaction rate scaling at high $T$ or $m \rightarrow 0$.
In the broken phase, their effect is less significant and can be found via 
(\ref{ther-mass}) and (\ref{T-crit}). Here, we neglect 
the gauge boson contribution to $\phi$--production  in the broken phase.\footnote{The gauge boson contributions have been considered in \cite{Heeba:2018wtf}, although in a different 
parametric regime. For example, for a light $\phi$, the production mode $h \rightarrow \phi \phi$ dominates unless $\lambda_\phi$ is very small such that $\theta > \lambda_{\phi h}$ 
and $tg \rightarrow t \phi$ becomes significant. For a heavy $\phi$, the fusion mode $hh\rightarrow \phi$ dominates even at small $\lambda_\phi$.}

 The number density of $\phi$ is computed from the Boltzmann equation, assuming zero initial  abundance,
  \begin{equation}
 \dot{n} +3nH = 2 \hat \Gamma_{h\rightarrow \phi \phi} + 2\hat \Gamma_{hh \rightarrow \phi\phi} + \hat \Gamma_{hh \rightarrow \phi}  
  \;,
\end{equation}
where
 \begin{eqnarray}
 && \hat\Gamma_{h\rightarrow \phi \phi}  =    \theta(T_c^v-T) \; \Gamma_{h\rightarrow \phi \phi} \;, \\
 && \hat\Gamma_{hh \rightarrow \phi\phi} =    ( 4 - 3 \theta(T_c^v-T)) \; \Gamma_{hh \rightarrow \phi\phi}   \;, \\
 && \hat \Gamma_{hh \rightarrow \phi}   =   ( 4 - 3 \theta(T_c^v-T)) \, \theta(T_c^w -T )\;\Gamma_{hh \rightarrow \phi}   \;.
 \end{eqnarray}
 The $\theta$--functions account for (the main effects of) the   phase transitions and the change in the Higgs d.o.f. 
 In practice, the dependence on $T_c^w$ is weak due to the smallness of the Higgs portal coupling, and 
 can normally be omitted. The inverse reaction rates are neglected due to the low density of $\phi$.
   
   The freeze--in mechanism requires 
   \begin{equation}
   n(T) < n_{\rm eq} (T)
   \end{equation}
at all temperatures down to $T \lesssim m_2/3$, where $\phi$ becomes non--relativistic. This condition ensures non--thermalization of $\phi$, i.e. suppression
of the inverse reactions $\phi\phi \rightarrow hh$, etc. At sufficiently low temperatures, the $\phi$ production rate drops and its abundance freezes in.
(The above condition is violated eventually, but no thermalization can then occur.)

\begin{figure}[h!]
\begin{center}
\includegraphics[scale=0.32]{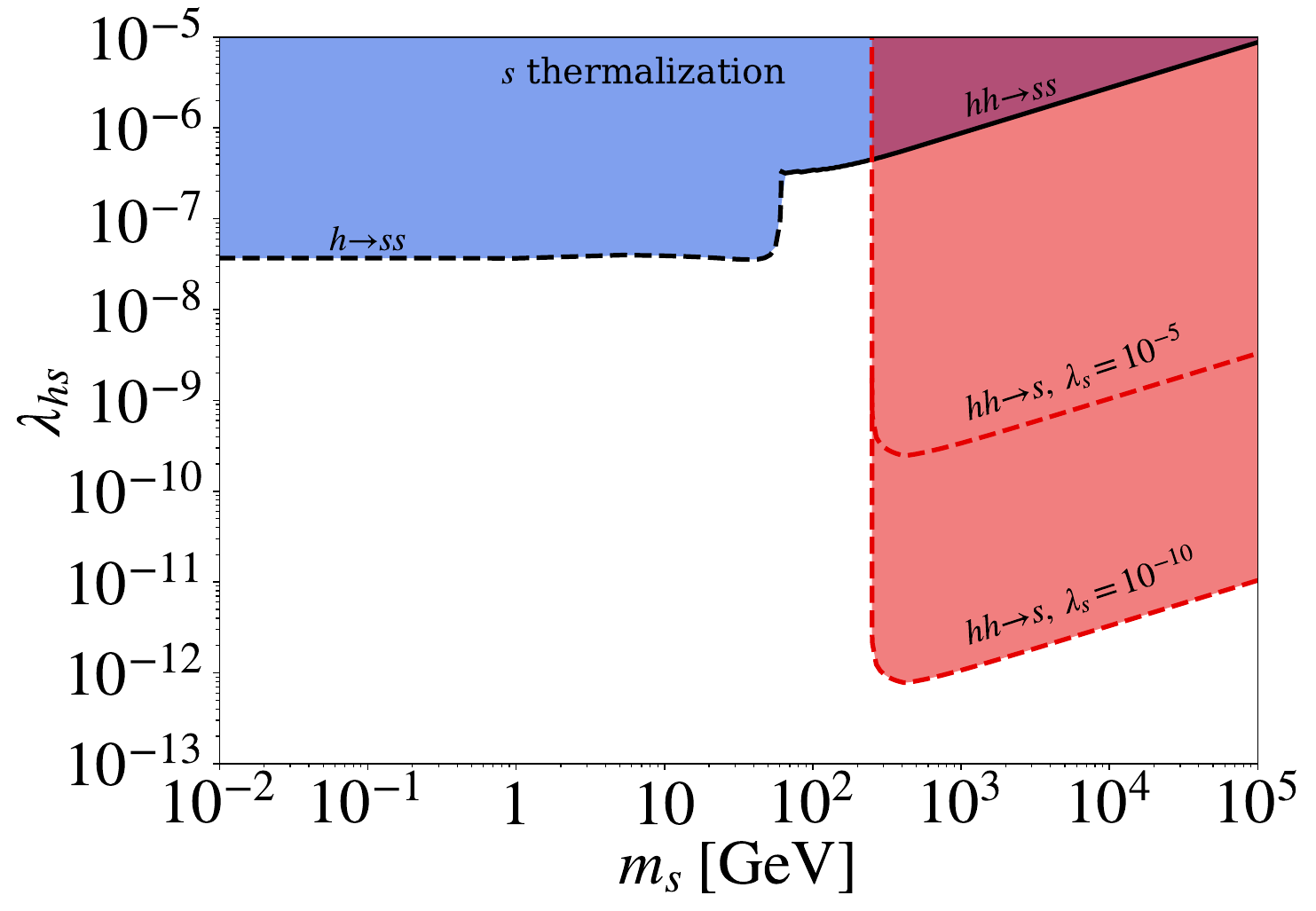}
\includegraphics[scale=0.32]{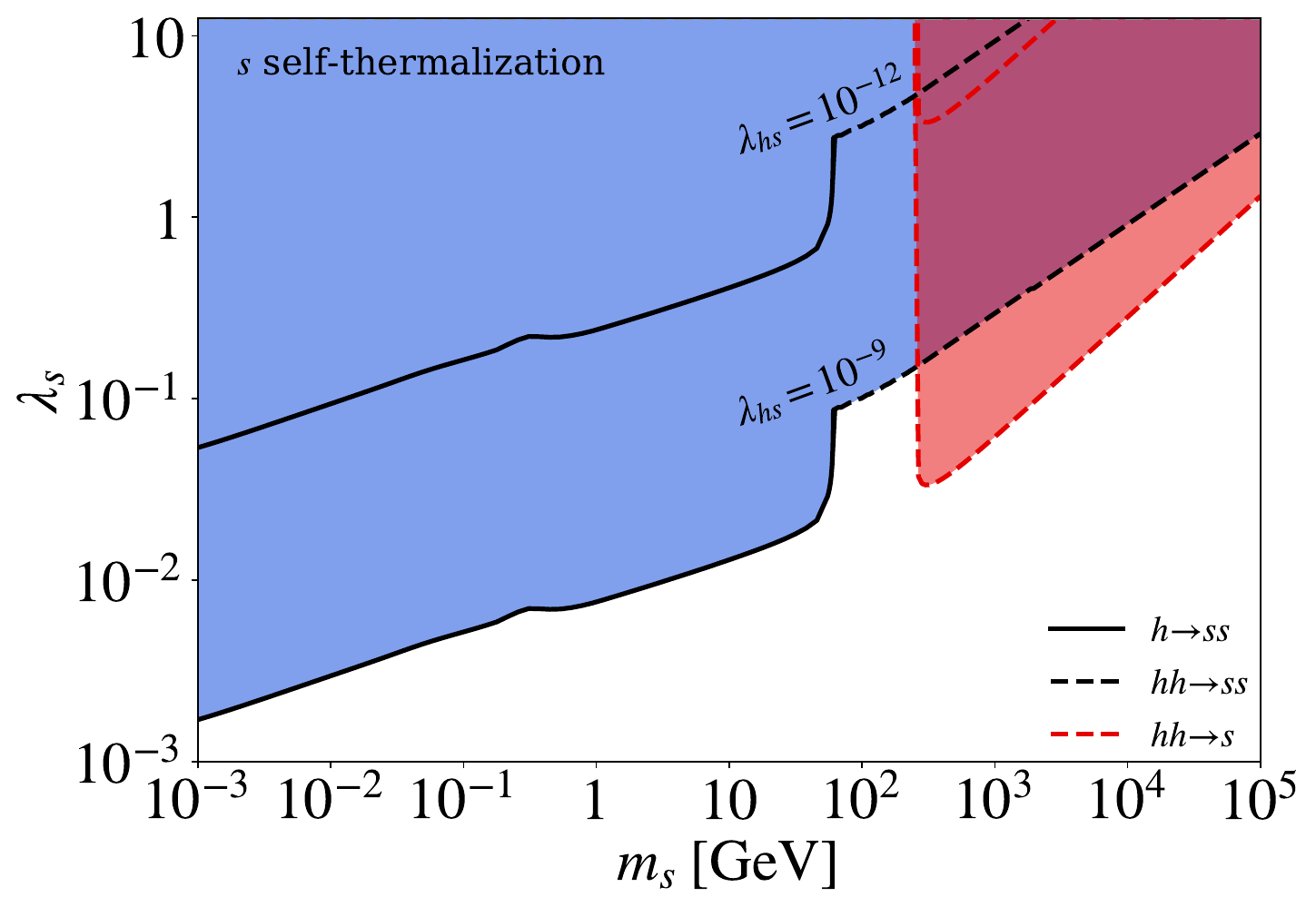}
\end{center}
\caption{   Non--thermalization constraints for Higgs portal freeze--in production. The shaded areas are excluded in the limit of small mixing.
In terms of the notation in the text, $s \equiv \phi$, $m_s \equiv m_2$, $\lambda_s \equiv \lambda_\phi$, $\lambda_{hs} \equiv \lambda_{\phi h}$.
The figures are from Ref.\,\cite{DeRomeri:2020wng} \copyright\, CC--BY.
\label{freeze-in-therm}}
\end{figure}

Solving the Boltzmann equation, one finds the constraints on the couplings shown in Fig.\,\ref{freeze-in-therm}, left panel. For light $\phi$, the decay mode $h\rightarrow \phi \phi$ sets the strongest
constraint, while for heavy $\phi$, the fusion mode $hh\rightarrow \phi$ normally dominates.  The rate of the latter depends $w$, which can be traded for $\lambda_\phi$ via $w=m_2/\sqrt{2 \lambda_\phi}$.
Thus, the exclusion contours are labelled by the corresponding $\lambda_\phi$.

The constraints can be understood semi--analytically \cite{DeRomeri:2020wng}.
 The bound on $\lambda_{\phi h} $ at $m_2 \ll m_1$ is independent of $m_2$,
 \begin{equation}
\lambda_{\phi h}(h\rightarrow \phi \phi) < 4 \times 10^{-8}\;,
\end{equation}
since, in this regime,  $\Gamma_{h \rightarrow \phi\phi}$ is independent of $m_2$  and $\phi$--production stops around $T \sim m_1/5 \sim 25 $ GeV, when the Higgs bosons become
depleted  in the SM plasma. 
For larger $m_2 \gtrsim m_1$,
the   reaction $hh \rightarrow \phi\phi$ grows more significant. At high temperature, the reaction rate scales as $T^4$ resulting in 
 $n(T) \propto T^2$ (see Section\,\ref{freeze-in-dm}). The non--thermalization constraint is then most severe at 
  the lowest temperatures consistent with this scaling, $T \sim m_2$.
This yields
 \begin{equation}
\lambda_{\phi h}(hh \rightarrow \phi\phi) < 6 \times 10^{-8}\; \sqrt{ m_2\over {\rm GeV}}\;.
\end{equation}
The fusion channel $hh \rightarrow \phi$ is more complicated. 
Consider the regime $m_2 \gg 2 m_1$.
There are two relevant factors that control the fusion  efficiency: the Higgses must have enough energy to produce $\phi$, which sets a lower bound on $T$, and the Higgs thermal mass
may not be too large for the process to be allowed kinematically, which in turn sets an upper bound on $T$. One finds numerically  that the process operates mostly in the temperature window between
$m_2$ and about $m_2/6$, and
  \begin{equation}
\lambda_{\phi h}(hh \rightarrow \phi) < 6 \times 10^{-9}\; \sqrt{ \lambda_\phi m_2\over {\rm GeV}}\;.
\end{equation}
 Here, the appearance of $\lambda_\phi$ can be understood from the reaction rate scaling as $\lambda_{\phi h}^2/\lambda_\phi$ for a fixed $m_2$.

The right panel of Fig.\,\ref{freeze-in-therm} displays non--thermalization constraints on the self--coupling of $\phi$, which we have neglected in 
 the Boltzmann equation.
If   $\lambda_\phi$ is substantial, the dark sector can ``self--thermalize'' as discussed in Section\,\ref{subsec-therm}. This would invalidate the freeze--in approach
and thus imposes an extra constraint, which can be estimated as follows.
 The reaction $\phi \phi \rightarrow  \phi\phi\phi\phi$  is important if it is faster than the Hubble expansion. In Section\,\ref{subsec-therm}, we considered this reaction for $n(T) \sim n_{\rm eq}(T)$,
 whereas now the density is much lower. In the present case, the final state Bose enhancement factors can be neglected and we may approximate
  \begin{equation}
   \Gamma_{\phi \phi \rightarrow  \phi\phi\phi\phi} \simeq n^2 \langle \sigma_{24} v_{\rm rel} \rangle \;,
   \end{equation}
where  $\sigma_{24}$ is the  $\phi \phi \rightarrow  \phi\phi\phi\phi$  QFT cross section and $v_{\rm rel}$ is the relative (M\o ller) velocity. 
In the relativistic regime,   the cross section can be approximated by 
$\sigma_{24}( s) \sim 10^{-4 } \lambda_\phi^4 \ln^2 ( s/2m_2^2) / s$, where $s$ is the Mandelstam variable.
If $2 \Gamma_{\phi \phi \rightarrow  \phi\phi\phi\phi} \gsim 3 nH$, the dark sector populates quickly and thermalizes. Noting that $\phi$ production is dominant at late times and
$ \Gamma_{\phi \phi \rightarrow  \phi\phi\phi\phi} / (nH)$ is maximized at the lowest $T$ consistent with the relativistic rate scaling, we thus find the non--thermalization criterion in the form
  \begin{equation}
{  n \langle \sigma_{24} v_{\rm rel} \rangle  \over H} \Biggr\vert_{T\sim m_2}  \lesssim 1\;.
\label{s-self-therm}
\end{equation}
Here, the cross section is evaluated at $\sqrt{s} \sim 2T$ and $v_{\rm rel}$ can be set to 2. The number density $n(T)$ is a function of the Higgs portal coupling obtained by solving numerically the 
Boltzmann equation. Clearly, for larger $\lambda_{\phi h}$, the constraint on $\lambda_\phi$ is more severe.
The resulting bounds in Fig.\,\ref{freeze-in-therm} are thus marked by $\lambda_{\phi h}$ and depend on the mode dominating $\phi$--production.

The above considerations are limited by a number of factors: the small mixing approximation breaks down for $m_1 \sim m_2$, the final state quantum statistical factors  become significant 
around the border of the excluded regions, and phase transitions can have a non--trivial effect on particle production. In particular, in  vicinity of the electroweak crossover, the SM Higgs mass
$m_{h_0}$ reduces to about 15 GeV \cite{DOnofrio:2015gop}, which enables the fusion mode $hh \rightarrow \phi$ even for a relatively  light $\phi$.  
Although it operates only in a small window of temperatures, $\phi$--production is very efficient, being boosted 
by    Bose enhancement   at $T/m_{h_0} \gg 1$ \cite{DeRomeri:2020wng}.

The obtained bounds are important for self--consistency of freeze--in production. They are relevant to models where $\phi$ itself plays the role of  dark matter  or decays into dark matter pairs, e.g.
the right--handed neutrinos \cite{Kusenko:2006rh},\cite{DeRomeri:2020wng}.

\subsubsection{Dark matter abundance}
\label{freeze-in-dm}

The dark scalar $\phi$ is stable when $w=0$ such that the $Z_2$ symmetry remains unbroken. In this case, it can constitute dark matter.
Stable $\phi$'s are pair--produced in the SM plasma 
  via $h\rightarrow \phi\phi$ and $hh\rightarrow \phi\phi$, while  the fusion channel is no longer available.

To understand  qualitative behaviour of the produced DM abundance,
it is instructive to solve the Boltzmann equation in the ultra--relativistic regime, where the rates exhibit  simple scaling.
Consider a reaction with $N$ particles in the final state, $i \rightarrow N$, whose rate scales as $T^l$. Using approximate conservation of the SM entropy, $g_{s*} a^3 T^3={\rm const}$, one 
can trade the time variable for $T$. Neglecting time dependence of $g_{s*}$, we have
\begin{equation}
 T {dn \over dT} -3n + c T^{l-2}=0 \;,
 \end{equation}
where 
 \begin{equation}
 c\equiv {N \; \Gamma_{i \rightarrow N} \over H T^{l-2}} \;.
 \end{equation}
Since $H \propto T^2$, $c$ is constant in this regime.
Assuming that the initial DM density is zero at temperature $T_0$, the solution is given by
 \begin{equation}
  n(T) = {c \over 5-l}\; T^3 \; \Bigl(T^{l-5}- T_0^{l-5}\Bigr) \;,
   \end{equation}
   while, for a special case $l=5$,  the solution reads $n(T)=cT^3 \ln {T_0\over T}$.
If  $l \leq 4$, the late time density is insensitive to $T_0$:
 \begin{equation}
  n(T) \simeq {c \over 5-l}\; T^{l-2} \;.
   \end{equation}
   On the other hand, $l\geq 5$ leads to the ``UV freeze--in'', i.e. the DM abundance
   dominated by the early time production,
\begin{equation}
  n(T) \simeq {c \over l-5}\; T^{3} \; T_0^{l-5}\;,
   \end{equation}
while, for $l=5$, $n(T)=cT^3 \ln {T_0\over T}$. This scaling is generated by non--renormalizable operators.

Now let us  specialize to the case of interest. As before, denote the scalar mass eigenvalues by $m_1$ and $m_2$, although the mixing is  absent. 
For a heavy $\phi$,   $hh\rightarrow \phi\phi$ dominates DM production. Then,  $l=4$, $n(T) \propto T^2$,  and the DM abundance  $Y =n/s_{\rm SM} \propto 
\lambda_{\phi h}^2  T^{-1}$ is determined by the
lowest temperature consistent with the relativistic scaling, $T \sim m_{2}$.  The observed relic abundance constraint $Y_\infty = 4.4 \times 10^{-10} \,{\rm GeV}/m_2$ then 
imposes a mass--independent bound on $\lambda_{\phi h}$ \cite{Lebedev:2019ton},
\begin{equation}
 \lambda_{\phi h} \simeq 2.2 \times 10^{-11} 
   \end{equation}
for $g_{s*} \simeq 107$. This value has been obtained by solving the Boltzmann equation numerically.

For a light $\phi$, the decay mode $h\rightarrow \phi\phi$ is more important. It operates at moderate temperatures $T \lesssim v$, when the Higgs VEV is non--zero. The corresponding 
$n(T)$ can be``eyeballed''  by  neglecting the Higgs bare mass so that $m_1 \propto T$. Then, $l=2$, $n(T) \propto T^0$, and $Y \propto \lambda_{\phi h}^2  T^{-3}$.
The relativistic scaling applies for $T$ not far from the Higgs mass,
 so the abundance is independent of the DM mass. We thus obtain the Higgs portal coupling producing the right DM abundance,
 \begin{equation}
 \lambda_{\phi h} \simeq 1.2 \times 10^{-11} \; \sqrt{{\rm GeV} \over m_2}\;,
   \end{equation}
where the coefficient has been determined numerically.

\begin{figure}[h!]
\begin{center}
 \includegraphics[scale=0.65]{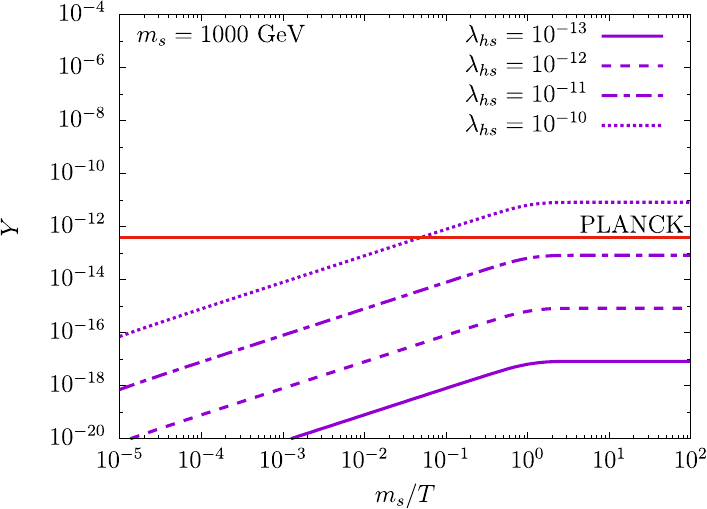}
 \includegraphics[scale=0.65]{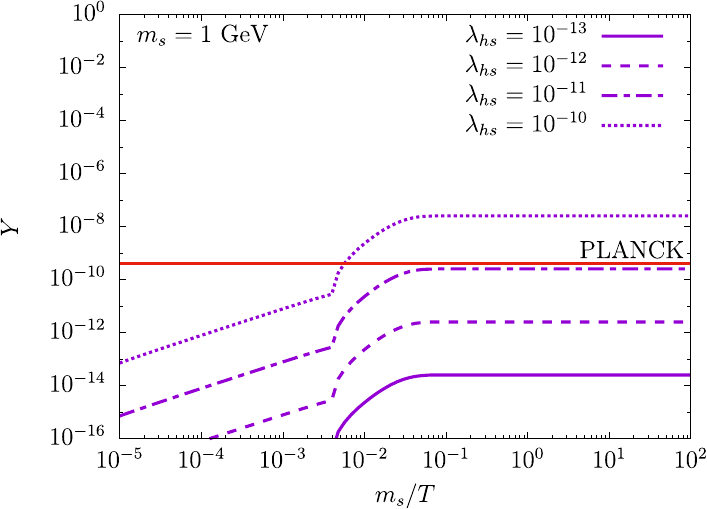}
\end{center}
\caption{   Numerical solutions to the Boltzmann equation for different Higgs portal couplings.
The observed DM relic abundance is given by the red line (PLANCK). {\it Left:} 
DM production is dominated by Higgs annihilation. 
{\it Right:} DM production is dominated by Higgs decay.
In terms of the notation in the text, $m_s \equiv m_2$, $\lambda_{hs} \equiv \lambda_{\phi h}$. The figure is from Ref.\,\cite{Lebedev:2019ton}.
\label{Y(T)}}
\end{figure}

The numerical solutions to the Boltzmann equation for different couplings and  masses are presented in Fig.\,\ref{Y(T)}.
In the left panel,  DM is heavy and the $hh \rightarrow \phi\phi$ process dominates.  We observe that the DM abundance $Y$ accumulates steadily until $T \sim m_2$, where it freezes in.
In the right panel,  $m_2 \ll m_1$ and the Higgs annihilation channel is responsible for DM production only at high temperatures. It gets overtaken by Higgs decay at electroweak $T$, after the phase transition.
The DM abundance freezes in at $T \sim 25 $ GeV. The mechanism operates in a wide range of the DM masses above the usual warm DM bound of
${\cal O}(1)$\,keV, whose exact value depends on further details such as the velocity distribution (see e.g. \cite{Kamada:2019kpe},\cite{Huo:2019bjf}).

 We see that the observed DM relic abundance can be produced by the SM plasma via a very small Higgs portal coupling. Although a theoretical rationale for the size of this
 coupling would be welcome, in any event,  the freeze--in mechanism remains a viable and interesting option.

Dark matter  freeze--in   within   the Higgs portal and closely related models has been studied, with various degrees of sophistication, in \cite{Heeba:2018wtf},\cite{Blennow:2013jba}-\cite{Chianese:2020yjo}. For example, Ref.\,\cite{Heeba:2018wtf}
considers freeze--in production of long lived Higgs portal DM. The relevant constraints on feebly interacting DM have been reviewed in \cite{Bernal:2017kxu}.
Related relativistic analyses have recently appeared in \cite{Bandyopadhyay:2020ufc},\cite{Biondini:2020ric}. Finally, the {\texttt{micrOMEGAs}} tool 
has been updated to include the freeze--in option \cite{Belanger:2018ccd}.

\subsection{Constraints on a decaying Higgs portal scalar: an example}

A light scalar mixed with the Higgs is subject to a number of cosmological constraints. 
These depend on the abundance of the scalar and hence its production mechanisms, as well as 
its decay rate.
The analysis of the general potential (\ref{potential})  is very complicated, so 
it is instructive to consider a simple ``super--renormalizable'' Higgs portal of Refs.\,\cite{Piazza:2010ye},\cite{Fradette:2018hhl}.
The $\phi$--dependent part of the potential contains only two terms,
\begin{equation}
 V_\phi = {1\over 2} m_\phi \phi^2 + {1\over 2} \sigma_{\phi h} \, \phi h^2 \;,
\end{equation}
where the unitary gauge has been assumed. This potential implies a non--zero Higgs--singlet mixing.
At  small $\sigma_{\phi h}$,  
the mass eigenvalues are
 $m_1, m_2 \simeq  m_\phi $ and the mixing angle  
  is given by 
 \begin{equation}
 \theta \simeq { \sigma_{\phi h}  v \over m_1^2 -m_2^2  } \;.
\end{equation}
Denoting the mass eigenstates {\it again} by $h,\phi$, one finds that the leading in $\theta$ interactions 
contain vertices $\phi h^2$ and $\phi h^3$. In particular, the $\phi h^2$ term comes with the coefficient
 \begin{equation}
\lambda_{\phi h h} = {\sigma_{\phi h} \over 2} - 3 \lambda_h v \theta  \simeq -\sigma_{\phi h} \;,
\end{equation}
 where the second equality is only valid  for $m_2^2  \ll m_1^2$.
 
If the initial abundance of $\phi$ is negligible, it is mainly produced via freeze--in in the SM thermal bath.
 At small $\theta$, 
 the $\phi$ production mechanisms are very different from what we have considered so far. Since only linear in $\phi$ vertices are present, the fusion mode is still available, yet
 it operates for heavy $\phi$ exclusively.\footnote{At the electroweak crossover, the Higgs boson becomes lighter which briefly opens 
 up the reaction $hh \rightarrow \phi$ for $m_\phi \gtrsim 30$ GeV. However, the dynamics involved are non--perturbative making 
the predictions less reliable. }
  If $\phi$ is light, it gets produced at electroweak temperatures via scattering processes like $tg \rightarrow t \phi$, $Zh \rightarrow \phi h $, etc.
 \cite{Berger:2016vxi},\cite{Heeba:2018wtf},\cite{Fradette:2018hhl}.
A combination of these channels yields  \cite{Fradette:2018hhl}
 \begin{equation}
 Y\sim {\rm few} \times 10^{11} \, \theta^2 
\end{equation}
for the abundance of the scalar. This applies to $m_\phi < 100$ GeV and small enough mixing angles. For $\theta > 10^{-6} $, the scalar thermalizes with the Standard Model and its
abundance is then controlled by the usual freeze--out. A conservative estimate  of Ref.\,\cite{Fradette:2018hhl} yields $Y \sim 10^{-3}$ at freeze--out.

The second ingredient in our analysis is the decay rate of $\phi$. This can be deduced from the decay  rate of the SM Higgs boson with the same mass, up to the factor of  $\sin\theta$ at the vertex. 
For example, a very light scalar can only decay into photons. The process is mediated by loops of charged particles. Since its characteristic energy scale is low, there is substantial QCD uncertainty 
stemming from the treatment of   light quarks. Their contribution is incorporated via pion and kaon loops. Depending on the prescription, the resulting decay rate can change by a factor of a few.
At $m_\phi \ll 2 m_e$, it can be approximated by 
\begin{equation}
\Gamma (\phi \rightarrow \gamma\gamma) = {   \alpha^2 m_\phi^3      \,  \theta^2  \over 256 \pi^3 v^2 } \; C^2  \;,
\end{equation}
where $C$  is a constant which is taken to be close to 2 in \cite{Fradette:2018hhl}. A naive summation over all the quarks and leptons  as well as  the $W$--bosons \cite{DeRomeri:2020wng}  in the loop gives a somewhat larger value.

\begin{figure}[h!]
\begin{center}
 \includegraphics[scale=0.5]{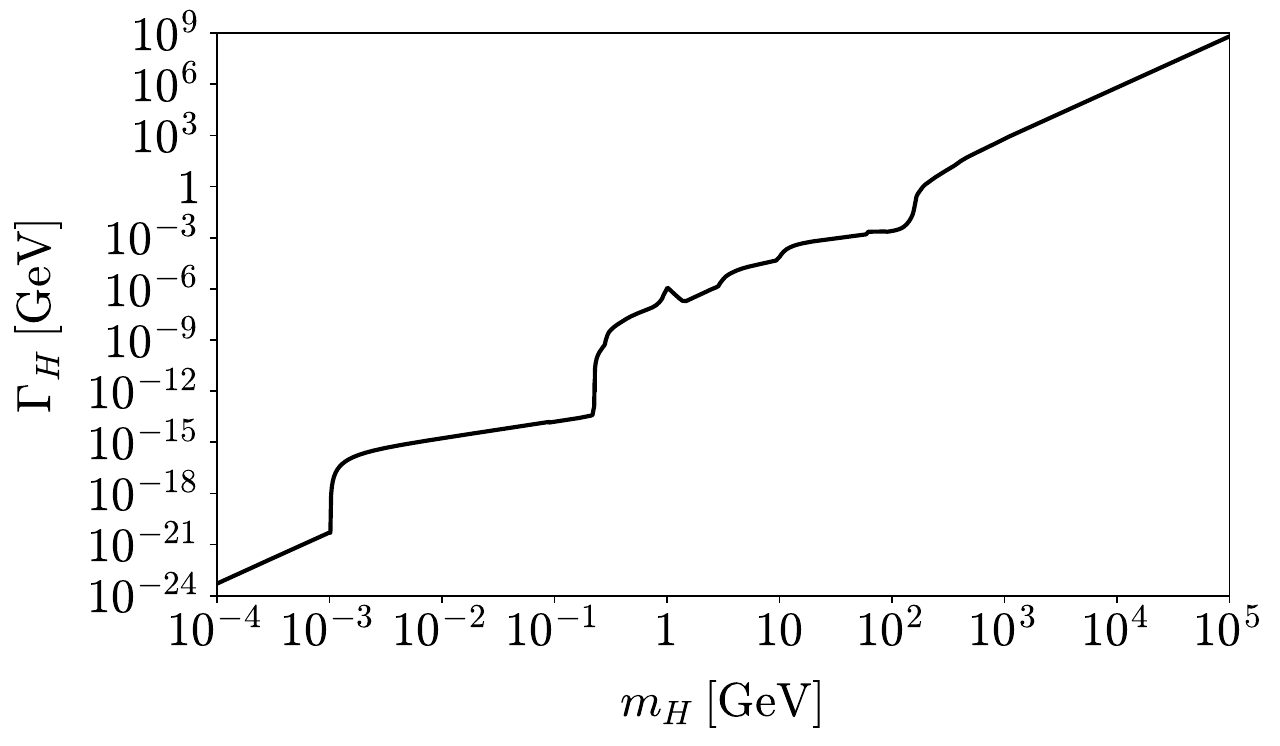}
\end{center}
\caption{    The Standard Model Higgs decay width $\Gamma_H$ versus the Higgs mass $m_H$.
\label{decay-rate}}
\end{figure}

The Higgs decay width calculations and discussion can be found in
\cite{Spira:1995rr}-\cite{Boiarska:2019jym}. Refs.\,\cite{Fradette:2017sdd},\cite{Winkler:2018qyg}, \cite{Boiarska:2019jym} consider specifically phenomenology of a scalar which mixes with the Higgs. The SM
Higgs decay width for different Higgs masses is shown
 in Fig.\,\ref{decay-rate}. As mentioned above, it is subject to tangible uncertainty in some regimes, e.g.  at  very small  and  $ {\cal O}(1) $\;GeV  
 Higgs masses, where light mesons play an important role.   
   The corresponding $\phi$--decay width is obtained by rescaling $\Gamma_H$ with $\theta^2$.

The Higgs--singlet mixing angle is constrained, first of all, by particle experiments. A light $\phi$ can be  emitted in various flavor--changing transitions which agree very well with the Standard Model.
Such processes are not suppressed by small Yukawa couplings since $\phi$ can be attached to the top quark or  $W$ in the   loop.
Depending on $\theta$, $\phi$ can manifest itself  either as missing energy or as a peak in the invariant mass distribution of its decay products, e.g. $\ell^+ \ell^-$.
For $m_\phi < 2 m_\mu$, the most sensitive mode is $K^+ \rightarrow \pi^+ + {\rm inv.}$ Combined with other transitions such as  $K \rightarrow \pi e^+ e^-$ and alike, it yields the constraint \cite{Andreas:2010ms}
\begin{equation}
\theta \lesssim 10^{-4} \;.
\end{equation}
In Ref.\,\cite{Andreas:2010ms}, this bound was obtained for a light pseudoscalar, yet analogous considerations apply to the scalar case giving a similar result
(see also \cite{Flacke:2016szy}).  
For a heavier $\phi$, $B$--decays, e.g. $B \rightarrow K \mu^+\mu^-$,   impose the strongest constraints. The corresponding bound relaxes to  about $\theta \lesssim 10^{-3}$ depending on the exact value of $m_\phi$
\cite{Schmidt-Hoberg:2013hba},\cite{Flacke:2016szy}. Couplings of a yet  heavier $\phi$ are constrained by LEP and the LHC as discussed in  Section\,\ref{Z2-pheno}.

\begin{figure}[h!]
\begin{center}
 \includegraphics[scale=0.5]{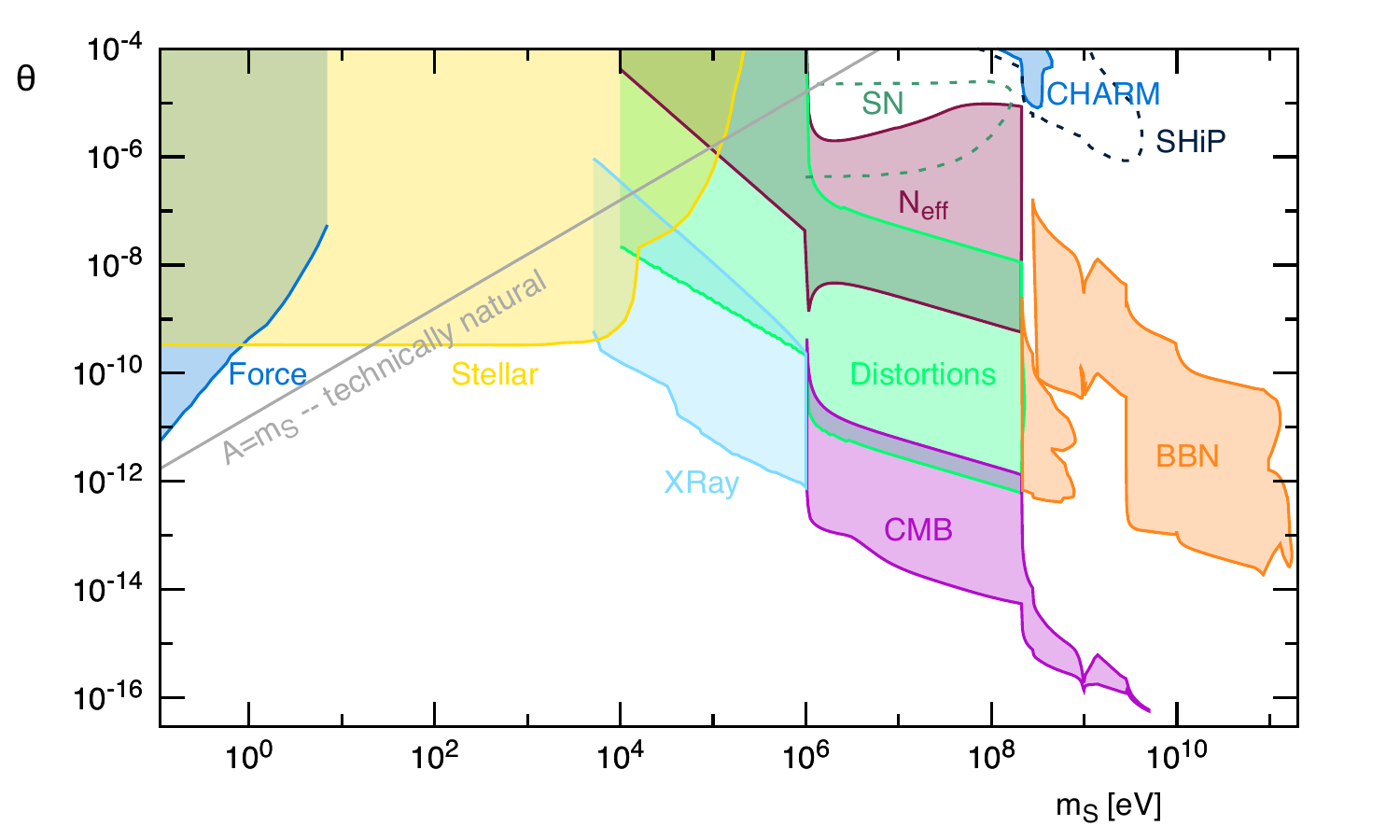}
\end{center}
\caption{   Constraints on the Higgs--singlet mixing  $\theta$   for the super--renormalizable Higgs portal. In terms of the notation in the text, $m_S \equiv m_\phi$.  
 The figure is from Ref.\,\cite{Fradette:2018hhl} \copyright\, CC--BY.
\label{singlet}}
\end{figure}

Further constraints are imposed by various cosmological considerations depending on the abundance of $\phi$. These were first considered in  \cite{Berger:2016vxi} and subsequently
refined in \cite{Fradette:2018hhl}, whose exposition we follow below. The summary is presented in Fig.\,\ref{singlet}.
 In the GeV mass range, the most important constraints stem from the Big Bang Nucleosynthesis. Particles with lifetime $\tau > 0.1 $ sec decaying into hadrons affect the abundance of light nuclei, which
 agrees well with the Standard Model. For example, pions and Kaons interact strongly with the nuclei and can dissociate ${}^4$He, in particular. For longer $\phi$ lifetimes, the effect of meson decay into electromagnetic showers becomes more important,  leading to photodissociation of light nuclei.  These considerations exclude a range of $\theta$ above $10^{-14}$. 
 A small mass region  around $300-500$ MeV with $\theta \sim 10^{-5 } -10^{-4}$ is excluded by the CHARM beam dump experiment \cite{Bergsma:1985qz}.
 Below the pion threshold, CMB measurements impose the strongest constraints.
 Late time energy injection can ionize hydrogen, distort the blackbody radiation spectrum and add an extra relativistic component at decoupling, which normally  is parametrized in terms of
 $N_{\rm eff} = 3.04 \pm 0.33$ \cite{Ade:2015xua}.
 For $m_\phi < 2m_e$, the lifetime of $\phi$ is longer than the age of the Universe, yet its slow decay creates a diffuse X--ray background \cite{Essig:2013goa} constrained by the INTEGRAL satellite \cite{Bouchet:2008rp}. For $m_\phi < 5 $\;keV, the tightest constraints come from stellar cooling considerations \cite{Hardy:2016kme}, while for sub-eV masses the fifth force measurements become
 important \cite{Piazza:2010ye},\cite{Kapner:2006si}. The dashed contours  in Fig.\,\ref{singlet} show the projected sensitivity of the SHiP experiment \cite{Alexander:2016aln} and
 an order-of-magnitude estimate of the supernova energy loss constraint \cite{Krnjaic:2015mbs}.

 In conclusion, light scalars mixing with the Higgs are subject to a range of cosmological and particle physics constraints. It is interesting that these   exclude tiny mixing angles of order $10^{-14}$
 in some mass intervals. For the scalar mass below 1 MeV, a combination of laboratory and astrophysical constraints requires the mixing angle to be below $10^{-9}-10^{-10}$ at least.

\section{Other cosmological implications of the Higgs portal}

A number of important aspects of Higgs portal physics remain beyond the scope of this review. 
The most conspicuous missing item  is the subject of  cosmological phase transitions. This is  particularly important for electroweak baryogenesis,
notoriously  precluded in the Standard Model.
One of the ingredients which would make it possible is the strong first order electroweak phase transition, i.e. 
\begin{equation}
{v(T_c) \over T_c} >1 \;,
\end{equation}
where $T_c$ is the critical temperature and $v(T_c)$ is the temperature dependent Higgs VEV.
The singlet extended Standard Model allows for this possibility \cite{Profumo:2007wc},\cite{Espinosa:2011ax}, both
with a general scalar potential of
Eq.\,\ref{potential} and its $Z_2$--symmetric version. Most studies of this system are based on perturbative calculations, 
while only very recently first non--perturbative analyses   have appeared  \cite{Gould:2021dzl},\cite{Schicho:2021gca},\cite{Niemi:2021qvp}
showing significant limitations of the perturbative approach.
This subject is also interesting in the context of gravitational wave detection \cite{Gould:2019qek}.

 Another well known aspect of the Higgs portal  concerns the fundamental  issue of a scale hierarchy between the Higgs mass and the Planck scale \cite{Dvali:2003br},\cite{Graham:2015cka}.
 The Higgs VEV much below the theory cutoff scale  $M$ can, for example,  be generated via cosmological evolution of the ``relaxion'' field $\phi$ with the 
 potential \cite{Graham:2015cka}
 \begin{equation}
 V_{\phi} = \bigl(-M^2 + g \phi\bigr) \;H^\dagger H + \Lambda^4 \, \cos (\phi /f) + \Delta V(g \phi) \;,
 \end{equation}
 where $g,f$ are dimensionful parameters and $\Lambda$ is an analog of the QCD  scale. The term $\cos (\phi /f)$ is generated by QCD instantons in
 analogy with the usual axion potential.  
 At large initial values of $\phi$, the Higgs mass term is positive leading to a vanishing VEV, while for small $\phi$, $v\not= 0$.  
 Below the QCD scale, the $\phi$ potential is modulated  by  the cosine with a Higgs--dependent amplitude  such that, in an appropriate parameter range,   a cosmological slow roll can  lead to $\phi$ being 
 stuck in a local minimum with a hierarchically small Higgs VEV.
 This approach is 
  a subject of ongoing discussion and continued research.
  
  To summarize, there are interesting avenues to address some of the outstanding particle physics problems via the Higgs portal.
  Each of these exciting research areas 
  deserves a separate review.

 \section{Conclusion}

  In this review, we have discussed cosmological aspects of the Higgs portal couplings, focussing
  on inflation, vacuum stability and dark matter.  
  Such couplings are expected on general grounds in quantum field theory 
  and  can make an important  
  impact on the evolution of the Universe, even if their numerical values are tiny. In particular, the Higgs portal couplings can solve the cosmological problems associated with   apparent vacuum metastability.
  They can drive the reheating process  after inflation and 
    be responsible  
 for dark matter production. 
 These couplings can also 
  provide the necessary annihilation channel for dark matter, making the WIMP paradigm consistent with observations.

  The Higgs portal can be explored further via collider experiments and astrophysical observations. On the collider front, the typical signatures would include Higgs--mediated
  production of  dark states that escape detection and manifest themselves as missing energy, possible invisible Higgs decay and Higgs--singlet mixing.
   Looking for such footprints  would  likely require painstaking data analysis as well as  large integrated luminosity at the LHC.
  WIMP--like Higgs portal dark matter can further be probed via direct and indirect detection experiments. On the other hand,
   if dark matter interacts very weakly with the Higgs field,  this would be challenging 
 and one could instead look for other astrophysical signatures, for example,  of its possible self--interaction.

 Altogether, the Higgs portal cosmology is an exciting  and still developing research  field  with potential to surprise.
  \\ \ \\
 {\bf Acknowledgements.} I am thankful to Jong--Hyun Yoon and Dimitrios Karamitros for producing some of the figures and helpful discussions.

\end{document}